\documentclass[11pt]{article}
\usepackage{amsmath}
\usepackage{amsfonts}
\usepackage{mathrsfs}
\usepackage{scalerel}
\usepackage{float}
\usepackage{slashed}
\usepackage{cite}
 \usepackage{tabu}
 \usepackage{soul}

\usepackage[colorlinks=true,
linkcolor=Blue, 
citecolor=Blue,
filecolor=Blue,
urlcolor=Blue,
linktoc=page, %%%
pdfstartview=FitV,
bookmarksopen=true]{hyperref}

\usepackage[dvipsnames]{xcolor}
\usepackage{eso-pic}% http://ctan.org/pkg/eso-pic
\usepackage{datetime}

\topmargin - 0.8 in

\def\a{\alpha}
\def\b{\beta}
\def\g{\gamma}
\def\d{\delta}
\def\e{\epsilon}
\def\s{\sigma}
\def\l{\lambda}
\def\p{\partial}
\def\r{\rightarrow}
\def\D{\mathcal{D}}
\def\O{\mathcal{O}}

\def\H{\mathcal{H}}
\def\X{\mathcal{X}}
\def\J{\mathcal{J}}
\def\L{\mathcal{L}}

\newcommand{\bi}{\begin{itemize}}
\newcommand{\ei}{\end{itemize}}

\newcommand{\be}{\begin{equation}}
\newcommand{\ee}{\end{equation}}

\newcommand{\bea}{\begin{eqnarray}}
\newcommand{\eea}{\end{eqnarray}}

\numberwithin{equation}{section}
\makeatletter
\renewcommand{\@seccntformat}[1]{%
  \csname the#1\endcsname.\ }
\makeatother

\title{
\vskip-1mm
Classical $J\bar T$ symmetries --  three ways -- and a precision \\[4pt] holography check \vskip5mm}

\author{ Silvia Georgescu$^{\S, \dag}$   and 
Monica Guica$^{\dag,\ddag,\sharp }$\vspace{1mm}\\
\\\vspace{1mm}
${}^\S$\emph{\small 
Department of Mathematics, King’s College London, Strand, London, WC2R 2LS, United Kingdom} \\  \vspace{1mm}
${}^\dag$\emph{\small Universit\'e Paris-Saclay, CNRS, CEA,} 
\emph{\small Institut de Physique Th\'eorique, 91191 Gif-sur-Yvette, France} \\ %\vspace{1mm}
 \vspace{1mm}
${}^\ddag$\emph{\small Institute of Physics, Ecole Polytechnique Fed\'erale de Lausanne, CH-1015 Lausanne, Switzerland} \\ 
${}^\sharp$\emph{\small Theoretical Physics Department, CERN, CH-1211 Geneva 23, Switzerland }}

\date{}

\begin{document}

\maketitle

%\AddToShipoutPictureBG*{
 % \AtPageUpperLeft{
  %  \hspace{\paperwidth}
   % \raisebox{-\baselineskip}{
    %  \makebox[0pt][r]{\today ~~ \currenttime ~~~~~}
%}}}

\abstract{
\vskip2mm

\noindent The $J\bar T$ deformation  is a fully tractable irrelevant deformation of two-dimensional  CFTs, which yields a UV-complete QFT that is local and conformal along one lightlike direction and non-local along the remaining one.  Such QFTs  are interesting, in particular,  as  toy models for the holographic dual to (near)-extremal black holes. %\textcolor{blue}{- the so-called Kerr/``CFT'' correspondence.} 
Despite its non-locality, the deformed theory has been shown to posess a (Virasoro-Kac-Moody)$^2$ %$\times$ Virasoro-Kac-Moody
 symmetry algebra, whose action  on the non-local side is non-standard. In this article we present, in a unified way, three different perspectives on the classical symmetries of $J \bar T$ - deformed CFTs: Hamiltonian, Lagrangian and holographic,  showing how the peculiar action of the symmetries can be recovered in each of them. %\textcolor{violet}{
 The perfect match  we obtain  between the Lagrangian and holographic results constitutes a precision check of the holographic dictionary proposed in \cite{Bzowski:2018pcy},  which we slightly generalize and  improve. We also comment on the interpretation of the Comp\`ere-Song-Strominger boundary conditions from the $J\bar T$ viewpoint. }

\tableofcontents

\section{Introduction}

Asymptotic symmetries are a proeminent tool in the bottom-up approach to non-AdS holography, lying at the heart of the celestial holography programme \cite{Pasterski:2016qvg} (see \cite{
Strominger:2017zoo,Raclariu:2021zjz,Pasterski:2021rjz,McLoughlin:2022ljp,Donnay:2023mrd} for reviews), as well as that of a proposed holographic description of (near)-extremal black holes known as the 
``Kerr/CFT'' correspondence \cite{Guica:2008mu,Compere:2012jk}. It is thus important to carefully understand  their action on the asymptotic space-time - namely, the boundary conditions that determine them - in order to draw the appropriate conclusions about the putative holographically dual boundary theory. 

The line of  investigation leading to this work has been motivated by a puzzle raised in the Kerr/CFT context. There, a bottom-up asymptotic symmetry group (ASG) approach  finds that the extreme\footnote{While the original correspondence was proposed for extremal black holes, it soon became clear that it can be extended \cite{Bredberg:2009pv,Matsuo:2009sj,Castro:2009jf} to near-extremal ones. It has since been understood that in black holes too close to extremality, quantum  corrections \cite{Ghosh:2019rcj, Iliesiu:2020qvm,Heydeman:2020hhw},  and, more recently, higher-derivative effects \cite{Horowitz:2023xyl} become important, so our comments refer only to near-extremal  black holes outside this regime, whose classical geometry can be trusted. Note also that the Kerr/CFT approach is focused on explaining the leading entropy at extremality, rather than departures from it. In particular, excitations in the non-AdS$_2$ directions are of central importance.    } Kerr spacetime admits asymptotic symmetries whose algebra is one copy of a centrally-extended Virasoro algebra, thus suggesting that the extreme Kerr black hole  is described by  (a chiral half of) a two-dimensional CFT \cite{Guica:2008mu} (the same applies to its higher-dimensional and charged generalisations, see e.g. \cite{Hartman:2008pb,Lu:2008jk}).  Nonetheless, scattering experiments off the Kerr black hole \cite{Bredberg:2009pv,Becker:2010jj}, as well as explicit string theoretical constructions  \cite{El-Showk:2011euy}, clearly show that the holographic dual is not a local field theory; rather, it is a  two-dimensional QFT that is local and conformal on the left and non-local on the right, which is obtained via an irrelevant deformation of a CFT$_2$ by an operator of dimension $(1,2)$. The string theory construction suggests the resulting theory is UV-complete, despite its non-locality.  The above-mentioned puzzle is that the Virasoro symmetry found by the ASG analyses we cited resides precisely on the non-local side\footnote{One way to circumvent this puzzle has been to posit that the dual theory could be a warped CFT \cite{Hofman:2011zj}, namely a QFT with $SL(2,\mathbb{R}) \times U(1)$ global symmetry that gets enhanced to a  left-moving Virasoro $\times\; U(1) $ Kac-Moody algebra. Such theories do appear to have a universal entropy \cite{Detournay:2012pc}. Nonetheless, note the non-locality must still be present, as it is inherent to warped AdS.  We further comment on the emergence of warped CFTs in holography in our discussion in appendix \ref{appendixCSS}.}. 

A priori, there are two possible resolutions to this puzzle. The more prozaic one is that the Virasoro ASG is just an artifact of the low-energy approximation, and is in fact not present at the level of the full theory. %(The results of \cite{} do not, however, support this view). 
The more interesting option is that the dual boundary theory does, in fact, possess a Virasoro symmetry, which must coexist, somehow, with the non-locality. Clearly, such a scenario would put severe constraints on both the structure of the non-locality\footnote{The non-locality of the holographic dual of extreme rotating black holes is indeed  of a restricted kind, in that holographic two- and three- point functions take precisely the form of CFT$_2$ momentum-space correlators, but with dimensions shifted to particular momentum-dependent functions \cite{Bredberg:2009pv,Becker:2010jj}. } and the action of the symmetries.

%This appears to be in direct contradiction with the results of the ASG analyses, which consistently exhibit a Virasoro symmetry algebra on the non-local side\footnote{Are the ASG analyses wrong\footnote{This could easily be the case, given arbitrariness of boundary conditions.}? Or, perhaps, the dual non-local field theory has a very special structure that allows the coexistence of Virasoro and non-locality? Unfortunately, untractable.}. 

This puzzle could finally be tackled with the advent of $J\bar T$ - deformed CFTs \cite{Guica:2017lia}, a \emph{solvable} set of non-local, UV-complete two-dimensional theories that have precisely the ``QFT structure'' of the  Kerr/``CFT'' correspondence - namely, they are local and conformal on the left and non-local on the right - and can thus be considered as toy models for it.  Thanks to their solvability, it was indeed possible to show explicitly that  $J\bar T$ - deformed CFTs %the theory (whose non-locality structure is of a very special/restricted kind {\color{ForestGreen}(more comments about what we mean here by non-locality and quasi-locality?)})
  realise the second option above, namely they allow for a set of Virasoro-Kac-Moody symmetry generators on the non-local side (in addition to the standard, expected Virasoro-Kac-Moody  on the local one). Nonetheless, if one considers not just the generators' algebra, but also their action on the fields -  which implements certain field-dependent coordinate transformations - one finds that 
  this  symmetry is realised in a non-standard fashion. 
  %\textcolor{blue}{(which reflects the non-locality of the underlying theory)}
    Namely,  the generators that act more naturally\footnote{By this we mean that the field variation they induce depends (quasi)locally on the symmetry parameter. See \eqref{rightafftransf} vs. \eqref{rightaffineflowedaction} for a concrete example of this action, and section \ref{section4:Lagrangian} for our definition of quasi-local.}  on the fields in the theory are not the ones (denoted with a tilde below) that satisfy the (Virasoro-Kac-Moody)$^2$ algebra, but rather a set of generators related to them by a ``spectral flow by the right-moving Hamiltonian'' %\footnote{{\color{ForestGreen}}} 
%{\color{ForestGreen}(leave $k,R$ arbitrary?  \textcolor{red}{yes, but i rescale $L$} is the bar visible?) \textcolor{red}{Is this correct? (factor $R_v$)}
%\textcolor{red}{Is this correct? (factor $R_v$)} {\color{ForestGreen}(I think the L is rescaled by $R_v$ as in eq 3.2 in 2110.07614, it's good because in this way we don't need to introduce $R_v$ now)}

\be \label{quasilocalgen1}
L_m = \widetilde L_m + \l H_R \widetilde J_m +  \frac{\l^2 k H_R^2}{8\pi }\delta_{m,0} \;, \;\;\;\;\;\; J_m =  \widetilde J_m + \frac{\lambda k}{4\pi}H_R \delta_{m,0}
\ee

\be \label{quasilocalgen2}
\bar L_m = \widetilde{\bar{L}}_m + \l : H_R \;\, \widetilde{\!\!\bar{J}}_m :+ \frac{\l^2 k H_R^2}{8\pi }\delta_{m,0} \;, \;\;\;\;\;\; \bar J_m =  \; \widetilde{\!\!\bar{J}}_m + \frac{\lambda k}{4\pi} H_R\delta_{m,0}
\ee
It follows that the algebra of the right-moving $\bar L_m, \bar J_m$ generators is a non-linear modification of the Virasoro-Kac-Moody one. 
%
%that, consequenty, do not satisfy a VKM algebra on the right. Thus, in contradistinction with standard CFT, the generators of the VKM algebra are not the same as those that act naturally/quasi-locally on the fields in the theory. 
%
% More precisely/in a nutshell, the Virasoro generators do not correspond to the natural Fourier basis of generators of the field-dependent  symmetries of the theory, as would be the case in a standard CFT, but differ from the latter by what can be called a spectral flow by the right-moving Hamiltonian. It is the Fourier-basis generators (whose algebra is not Virasoro, but a non-linear modification thereof) that are the natural symmetry generators of the theory.
Defining (non-local) $J\bar T$  analogues of primary operators via the Ward identities with respect to the untilded generators above, 
  the interplay of the two generator bases can be used to compute all of their correlation functions exactly, obtaining precisely the structure  previously observed from the gravity side \cite{Guica:2021fkv}.
  
It is of course an open question whether the mechanism by which $J\bar T$  - deformed CFTs reconcile Virasoro symmetry and non-locality is also the one at play in the Kerr/``CFT'' correspondence, but we think it's an essential one.   Since, in the extremal black hole - or, more generally, warped AdS$_3$ - context,  the absence of a tractable holographic dual forces one to rely almost exclusively on bulk methods to infer the properties of the dual QFT, a   proper understanding of how this type of symmetries act on the dual spacetime, and how the associated non-linear charge algebra is recovered, appears  essential  for finally elucidating this type of holographic correspondence.   

The main goal of this article is to provide a global picture for the 
 classical extended symmetries of $J\bar T$ - deformed CFTs, by unifying  three different perspectives: Hamiltonian, Lagrangian, and holographic. In particular, we show  that their holographic realisation as asymptotic symmetries of AdS$_3$ with certain mixed boundary conditions  precisely reproduces the structure found on the field theory side. This is an important precision check   of both the holographic dictionary and of the asymptotic symmetry group computation which, as we will see, requires certain special features of the asymptotic symmetry generators in order to reproduce the non-linear terms in the commutators of $\bar L_n, \bar J_n$ above.

This article generalises and unifies several previous works on the extended symmetries of $J\bar T$ - deformed CFTs;  here is a short summary of how it relates to this previous literature. %fits in the global picture. 
The original  analysis of \cite{Guica:2020uhm}, while capturing much of the physics, misses an important subtlety concerning the zero mode of the field-dependent coordinate, which is essential for the charges to have a well-defined action on the phase space and a consistent algebra. This zero mode was later found in \cite{Guica:2020eab} via a Hamiltonian analysis, in which the consistency of the charges' action on phase space was used as an input. The present article's first contribution is to streamline the derivation of \cite{Guica:2020eab}, showing how the field-dependent coordinate with the correct zero mode simply emerges from the $J\bar T$  flow of the Virasoro generators, with no further assumptions. We then translate the action of the conserved charges on the basic fields from the well-understood Hamiltonian formalism to the Lagrangian one. We find that the main effect of the zero mode is to supplement  the na\"{i}ve field-dependent symmetries put forth in \cite{Guica:2020uhm} by a large affine transformation acting simultaneously on the left- and the right-movers, whose coefficient is configuration-dependent, in that it is proportional to a certain conserved charge. This large affine ``compensating'' transformation is required by the consistency of the charges' action  on phase space (which is partly reflected in their algebra), and not merely  by the  invariance of the action under the given transformation, which was the only requirement of \cite{Guica:2020uhm}. %This trait  appears characteristic of these types of symmetries. 
It is this Lagrangian structure that is most closely reproduced by the holographic analysis. 

Next, we turn to the dual spacetime  analysis. The holographic dual of a $J\bar T$ - deformed holographic CFT is AdS$_3$ gravity with mixed boundary conditions for the metric and the Chern-Simons gauge field(s) dual to the $U(1)$ current(s). The holographic dictionary was 
 derived in \cite{Bzowski:2018pcy} for the case of a purely chiral $U(1)$ current - however, certain aspects were not fully understood. Here, we slightly generalise the dictionary to the case of a \emph{non-chiral} $U(1)$ current - for which, as shown in \cite{LeFloch:2019rut}, the field-theory analysis is conceptually simpler - and show that the previous issues  clear up.  
 Boundary conditions in hand, we perform a careful asymptotic symmetry group analysis of this spacetime.  We find that the periodicity of the gauge parameters after the 
  field-dependent  transformations requires the presence of  a compensating  large affine transformation, with an \emph{identical}  
  structure  to that we found in the Lagrangian analysis in field theory. This piece is essential for reproducing the non-linear terms in the algebra of the asymptotic charges, which we  %\textcolor{red}{True?} {\color{ForestGreen}(yes)} 
 carefully compute.  %the asymptotic symmetry algebra, \textcolor{violet}{  taking into account/essential role played by integrability}, we are able to reproduce both choices of ASG. \textcolor{red}{True?}
   The result  differs from that  of \cite{Bzowski:2018pcy}, who did not uncover the above subtleties related to periodicity, and also relied 
    on the representation theorem of \cite{Barnich:2010eb} to infer the charge algebra, rather than computing it directly, as we do here.% As we show, this representation theorem does not quite hold in the case at hand, due to the fact that the  modified Lie bracket algebra  of the symmetry parameters does not close.

We are thus able to fully reproduce the peculiar features of the symmetries\footnote{
We note that these peculiar features, namely the presence of compensating large  transformations with charge-dependent coefficients,  are extremely similar to those of $T\bar T$ - deformed CFTs, another exactly solvable irrelevant deformation  of two-dimensional CFTs \cite{Smirnov:2016lqw,Cavaglia:2016oda}. The most up-to-date analysis of the extended $T\bar T$ symmetries from a field-theoretical perspective is \cite{Guica:2022gts} and, from a holographic one, the appendix of \cite{Georgescu:2022iyx}.}, as well as the non-linear symmetry algebra, from a largely independent spacetime analysis, where the only input from the dual field theory are 
 the low-energy effective action and the boundary conditions for the bulk fields. This constitutes a highly non-trivial precision test of the proposed holographic dictionary.
We can then use this dictionary to interpret other proposed holographic dualities for AdS$_3$ with non-standard boundary conditions, such as the Comp\`ere-Song-Strominger (CSS) one \cite{Compere:2013bya}. While the allowed metrics of \cite{Compere:2013bya} coincide with a truncation of the space of metrics we consider, the gauge field sector is entirely absent in their analysis, leading to missing contributions to the conserved charges, which in turn result in entirely different conclusions about symmetry  structure of the dual theory. 
We therefore conclude that there is no subsector of the $J\bar T$ - deformed holographic CFT  whose low-energy dual bulk description is pure $3d$ Einstein gravity with CSS boundary conditions.

This article is organised as follows. In section \ref{section2:review}, in order for the article to be  self-contained, we provide a  brief review of what is currently known about the symmetries of $J\bar T$ - deformed CFTs. In section \ref{section3:Hamiltonian}, we review and streamline the derivation of \cite{Guica:2020eab} of the tilded generators  \eqref{quasilocalgen1} - \eqref{quasilocalgen2} in the Hamiltonian formalism, emphasizing the emergence of the field-dependent coordinate from the flow. In section \ref{section4:Lagrangian}, we study the symmetry transformations in the Lagrangian formalism.  In section \ref{section5:holography}, we review and slightly generalise the holographic dictionary of \cite{Bzowski:2018pcy}, and perform various checks that  our new proposal reproduces all expected results in this more general setup. The asymptotic symmetry group analysis is performed in section \ref{section6:asg}, and we end with a discussion in section \ref{section7:discussion}.   Various  adjacent issues are discussed in the  appendices.

%{\color{blue}[$J\bar T$ - deformed CFTs: QFT prototype of the Kerr/CFT correspondence. Since much of the evidence for the latter relies on the symmetries, important to properly understand the extended $J\bar T$ symmetries. The important takeaway is that the extended symmetry algebra on the non-local side  depends on the basis of generators one chooses; in a natural Fourier basis for the latter, the algebra is \emph{non-linear}. This can be seen both on the field theory and the gravitational side via a careful analysis of the symmetry generators. 

%In this article, we collect and expand the known facts about classical symmetries of $J\bar T$ - deformed CFTs: we review and streamline the Hamiltonian results of \cite{}, develop the Lagrangian interpretation of these symmetries and finally rederive their holographic implementation, showing that the algebra one obtains is in fact non-linear. ]}

\section{Brief review of $J\bar T$ - deformed CFTs}\label{section2:review}

In this section, we review what is known about $J\bar T$ - deformed CFTs and their conserved currents, following \cite{Guica:2020uhm,Guica:2020eab}. The action is defined via the flow equation  %\emph{Sign!}
\be \label{flact}
\p_\l S = \int d^2 x \, \O_{J\bar T}
\ee
where the $J\bar T$ operator is defined from the coincidence limit  %\emph{Notation?}

\be
 \lim_{\varepsilon \r 0} \e^{\a\b} J_\a(x) T_{\b V} (x+\varepsilon) = \O_{J\bar T}(x) + \ldots
\ee
where $J_\a$ is a $U(1)$ current, $T_{\a V}$ is the generator of right-moving translations\footnote{The time and space coordinates of the QFT will be denoted as $t,\s$, and the null coordinates as $U,V = \s \pm t$. Our convention for the $\e$ tensor is $\e^{\s t} =1$ or, in null coordinates, $\e^{UV} =- 2$. } and  the $\ldots$ stand for total derivative terms. This is a particular case of a general  construction by Smirnov and Zamolodchikov \cite{Smirnov:2016lqw}, who showed that  the operator obtained from the coincidence limit of an antisymmetric combination of the components of two conserved currents is well-defined at the quantum level, up to the above-mentioned total derivatives. An important property of such %Smirnov-Zamolodchikov 
operators %(which follows from their definition above)
 is that their expectation value  in energy eigenstates factorizes
\vskip-2mm
\be
\langle n | \O_{J\bar T} | n\rangle =  \e^{\a\b} \langle n | J_\a  | n\rangle  \langle n | T_{\b V}   | n\rangle
\ee 
 One may also define the deformation at the level of the Hamiltonian as 
\be
\p_\l H = -\int d\s \, \O_{J\bar T} \label{hamflow}
\ee
While these two definitions are equivalent at the classical level \cite{Kruthoff:2020hsi}, quantum-mechanically it is less clear whether they are the same; note, in particular, that generic total derivative ambiguities in the $J\bar T$ operator may be dropped from \eqref{flact}, but only spatial derivaties are irrelevant for \eqref{hamflow}. 

% in the first definition total derivative  are irrelevant, whereas only spatial ones are irrelevant in the second. 

The  two equations above may be used to derive a flow equation for the energies  of the individual energy levels of the deformed theory in finite size (i.e., a circle of circumference $R$), which can be solved exactly. %This is one of the most important observables that were studied in this sytem. 
The solution for the deformed right-moving energy reads %\emph{Factors!}

\be
E^{[\l]}_R \equiv \frac{E^{[\l]} - P}{2} = \frac{4\pi}{\l^2 k} \left( R - \l J_0 + \sqrt{(R-\l J_0)^2 - \frac{\l^2 k R}{2\pi} E_R^{[0]}} \right) \label{defspec}
\ee
where $E_R^{[0]}$ and $J_0$ are the undeformed right-moving energy and, respectively,  the
 left-moving charge, while  $k$ is the level of the $U(1)$ current\footnote{In our conventions, $k=2\pi$ for a free boson. We will sometimes use $\hat k = \frac{k}{2\pi}$ to ease the notation. }.
% 
  %\textcolor{red}{Understand vs di Francesco!} {\color{ForestGreen}(this agrees with Polchinski, after we take into account the difference between our conventions and his conventions)}}. % For a free boson, we have $k=2\pi$ in our conventions. \emph{True? Shall we set $k=2\pi$, or what?} 
 Note that the momentum $P$ - being quantized in units of $2\pi/R$ - is the same as in the undeformed theory. 

\subsection{The $J\bar T$ - deformed free boson: Lagrangian analysis \label{fbanalysislagr}}

In the classical theory, one can simply solve the flow equation for the action or for the Hamiltonian. Concentrating on a $J\bar T$ - deformed free boson for simplicity, the flow equation \eqref{flact}, where the components of the stress tensor and the shift current are computed using the Noether method, fully determines the deformed action to be %\emph{Check sign!} 

%{\color{ForestGreen} (now with this factor and the expressions for the currents from the next page in \ref{flact} we get 2.1)} \textcolor{red}{I don't think this is correct, the 2 is in the Lagrangian, not the action.}

\be
S= -  \int dU dV \frac{\p_U \phi \p_V \phi }{1-\l \p_V \phi} \label{fbact}
\ee
%where $U,V = \s \pm t$ are null coordinates. 
The same action is obtained if one uses instead the topological current \eqref{topcurr} to define the deformation.  It is easy to note this action is invariant under $U \r F(U)$, where the  $F$ is an arbitrary periodic function. This invariance may also be seen by 
%This theory/action clearly has $SL(2,\mathbb{R})_L$ invariance, which we expect to see enhanced to a full left-moving Virasoro.  To see this, one may
 computing the components of the left-moving stress tensor %{\color{ForestGreen}(here $\hat{k}=1$ everywhere, correct? we have some problems with this in section 4)}
\be \label{freebosonHL}
T_{UU} = \frac{(\p_U \phi)^2}{(1-\l \p_V \phi)^2} \;, \;\;\;\;\;\; T_{VU} = 0
\ee
and noting, using the  the equation of motion 
\be \label{jtbarscalareom}
\p_V \left(\frac{\p_U \phi}{1-\l \p_V \phi}\right) =0
\ee
that $T_{UU}$ is ``holomorphic'' (i.e., a function of $U$ only) on-shell. Consequently,  the charges $\int_0^R d\s f(U) T_{t U}$ are conserved on-shell for  arbitrary periodic functions $f(U)$, and correspond to the infinite left-moving conformal symmetries of the system. The components of the  right-moving\footnote{By ``right-moving stress tensor'' we mean the generator of translations in the right-moving null direction $V$, which is not exactly right-moving itself, in that the on-shell solution for $T_{\a V}$ is not just a function of $V$.}   stress tensor read

\be \label{freebosonTv}
T_{UV} =\frac{\lambda\partial_U\phi (\partial_V\phi)^2}{(1-\lambda\partial_V \phi)^2} \;, \;\;\;\; T_{VV} = \frac{(\partial_V\phi)^2}{1-\lambda\partial_V \phi}
\ee
The theory also posseses two conserved $U(1)$ currents: the shift current $J_\a$ associated with the symmetry of \eqref{fbact} under constant shifts of $\phi$, and  the topologically conserved current $ J_{top}^\a = \e^{\a\b} \p_\b \phi$. The components of these currents are %{\color{ForestGreen}(so we are using the convention $\epsilon^{UV}=-2$)}
\begin{align}\label{freebosoncurrents}
J_{\mbox{\tiny{$U$}}}^{sh}&=\frac{\partial_U\phi}{(1-\lambda\partial_V\phi)^2} \hspace{1cm}J^{sh}_{\mbox{\tiny{$V$}}}=\frac{\partial_V\phi}{1-\lambda\partial_V\phi}\\
J^{top}_{\mbox{\tiny{$U$}}}&=\partial_U\phi\hspace{2.4cm}J^{top}_{\mbox{\tiny{$V$}}}=-\partial_V\phi \label{topcurr}
\end{align}
We will be denoting the associated conserved charges as 
\be
Q_0 = \int_0^R d\s\, (J_{\mbox{\tiny{$U$}}}^{sh} -J_{\mbox{\tiny{$V$}}}^{sh}) \equiv J_0 + \bar J_0 \;, \;\;\;\;\;\;\; w = \int_0^R d\s \,  (J_{\mbox{\tiny{$U$}}}^{top} - J_{\mbox{\tiny{$V$}}}^{top}) \equiv J_0 - \bar J_0 \label{defchzm}
\ee 
Note that the following linear combination corresponds to a chiral conserved current 

\be
K_\a = \frac{1}{2} (J_{\mbox{\tiny{$\alpha$}}}^{sh} + J_{\mbox{\tiny{$\alpha$}}}^{top} - \l T_{\a V})\;, \;\;\;\; \mbox{or} \;\;\;\;\;\;  K_U  =   \frac{\p_U \phi}{1-\l \p_V \phi} \;, \;\;\;\; K_V  =  0   \label{defK}
\ee
We will denote its associated conserved charge as $Q_K$, which equals $J_0 + \l E_R/2$. Note that
%{\color{ForestGreen}those solution is given by $\phi(U,V)=f(U)+g(V-\lambda f(U))$, for arbitrary $f,g$.}
%In particular, 
if we multiply this current by an arbitrary function of $U$, the result is still conserved. These generate the infinitely extended left-moving affine symmetries. One may as well compute 
 the components of the ``right-moving'' current %\emph{Check sign!} {\color{ForestGreen}(checked)}
\be
\bar K_\a = \frac{1}{2} (J_{\mbox{\tiny{$\alpha$}}}^{sh} - J_{\mbox{\tiny{$\alpha$}}}^{top}  - \l T_{\a V})\;, \;\;\;\; \mbox{or} \;\;\;\;\;\;  \bar K_U  = \frac{\lambda \partial_U\phi\partial_V \phi}{1-\lambda\partial_V\phi}  \;, \;\;\;\; \bar K_V  =   \partial_V\phi \label{defKb}
\ee
In \cite{Guica:2020uhm}, it was observed that the ratios of the components of the right-moving stress tensor and of the right-moving  current are the same
\be
\frac{T_{UV}}{T_{VV}} = \frac{\bar K_U}{\bar K_V}  = \frac{\l \p_U \phi}{1-\l \p_V \phi} \equiv -\frac{\p_U v}{\p_V v} \label{firstdefnv}
\ee
so, if one introduces a ``field-dependent coordinate'' $v$ as above, then the currents $\bar f(v)\,  T_{\a V}$ and $\bar \eta(v) \bar K_\a$ are conserved for arbitrary functions $\bar f, \bar \eta$. %\textcolor{blue}{One may also show that the action is invariant under the associated set of transformations (given by \eqref{} without the large affine term).} 
They should thus correspond to  right-moving field-dependent symmetries. Clearly, the general  solution to \eqref{firstdefnv} is $v = V - \l \phi+ const.$

The construction of the field-dependent right-moving symmetries, as presented, 
 may appear somewhat ad-hoc. Also, note that this construction does not fix the constant term in $v$. This term turns out to have a very important effect on the algebra of the charges; as explained at length in \cite{Guica:2020eab}, the na\"{i}ve choice $v=V-\l \phi$ leads to an algebra that is inconsistent with the deformed spectrum \eqref{defspec} and the quantization of the shift charge \eqref{defchzm}.
%\footnote{\textcolor{blue}{A similar problem arises in $T\bar T$, where the  zero modein question  does not even appear to exist. }}
% \textcolor{red}{Would it have otherwise looked consistent (closed)?}{\color{ForestGreen}(I think so. I didn't check all the commutators, but for the ones I checked the algebra was closed. In this case also the action is actually invariant.)} 
The correct prescription is to remove the zero mode of the field $\phi$ from  $v$.

 One would thus like to have a first-principles approach to this problem, where the field-dependent symmetries, including the correct expression for the field-dependent coordinate, follow from some basic definition. This basic definition was put forth in \cite{Guica:2020eab} (see also \cite{Guica:2021pzy}) and is naturally formulated in Hamiltonian language. Before discussing it, let us  review a few basic facts about the $J\bar T$ deformation in the Hamiltonian formalism.

\subsection{Hamiltonian formulation of classical $J\bar T$ - deformed CFTs \label{hamformjtb}} 

As emphasized in \cite{Jorjadze:2020ili}  for $T\bar T$, and extended in \cite{Guica:2020uhm} to $J\bar T$, the   flow equation  \eqref{hamflow} that defines the theory in the Hamiltonian picture is particularly easy to solve  in the classical limit  for arbitrary classical seed conformal field theories, whose fields we   denote as $\phi_i$.

For $J\bar T$, one assumes the theory has a $U(1)$ symmetry, which is modeled as a shift  symmetry of one of the scalar fields, $\phi$. This amounts to assuming that the Hamiltonian density, $\H$, can depend on $\phi'\equiv \p_\s \phi$, but not $\phi$.  The corresponding shift current has components
\begin{align}
J_{\mbox{\tiny{$t$}}}^{sh} =\pi \hspace{1cm}J_{\mbox{\tiny{$\sigma$}}}^{sh} =\partial_{\phi'}\mathcal{H}
\end{align}
There is also a topologically conserved current, $J_{\mbox{\tiny{$\alpha$}}}^{top} $, whose components are given by interchanging $\pi \leftrightarrow \phi'$ in the above. It is convenient to introduce the following combinations of the time components of these two currents 
%{\color{ForestGreen}(factors of $\hat{k}$ or set them to 1?)} \textcolor{red}{I'd say set to one right now, then reinstate in $\H_R$. }
\begin{align}
\mathcal{J}_{\pm}&=\frac{\pi\pm\phi'}{2}
\end{align}
Of course, this only leads to a $U(1)$ current algebra with (rescaled) level $\hat k=1$; nonetheless, the level can be easily set to the desired value by rescaling $\J_+$ by $\hat{k}^{1/2}$, which we will view as a corresponding rescaling of $\phi, \pi$.  
The  components of the stress tensor are given in terms of the deformed Hamiltonian density by
\be
T_{tt} = \H \;, \;\;\;\; T_{t\s} = \mathcal{P}  = \sum_i \pi_i \phi'_i \;, \;\;\;\;\; T_{\s t} = \partial_{\pi_i}\mathcal{H}\partial_{\phi'_i}\mathcal{H} \;, \;\;\;\;\; T_{\s\s} = \pi^i\partial_{\pi_i}\mathcal{H}+\phi'^i\partial_{\phi'_i}\mathcal{H}-\mathcal{H} \label{stresstham}
\ee 
The undeformed theory has  Hamiltonian density $\H^{[0]}$ and momentum density $\mathcal{P} $, which encode all the independent components of the undeformed stress tensor. Together with current densities $\J_\pm $, they generate a (Witt-Kac-Moody)$^2$ symmetry algebra with level $\hat{k}=1$ %\emph{Correct?} {\color{ForestGreen}(ok)} % All the components of the stress tensor are determined from these.
in our conventions. Since we are modeling the left/right affine symmetries via a single boson $\phi$, we are effectively assuming that the undeformed Hamiltonian has a decoupled free boson sector, $\H^{[0]} = \frac{1}{2} (\pi^2+\phi'^2) + \H^{[0]}(other \, fields)$. 

%, assumed to have a $U(1)$ shift symmetry in $\phi_1 \equiv \phi $.  starts from a theory with Hamiltonian density $\H^{(0)}$, momentum density $\mathcal{P}$ and current densities $\J_\pm $, so the initial theory has Virasoro-Kac-Moody symmetry squared. All the components of the stress tensor are determined from these.
%
% 
%%where $\phi'=\partial_{\sigma}\phi$. 
%The components of the topologically conserved current are given by interchanging $\pi \leftrightarrow \phi'$
%\begin{align}
%\tilde{J}_{t}=\phi'\hspace{1cm}\tilde{J}_{\sigma}=\partial_{\pi}\mathcal{H}
%\end{align}
%%We denote by $\mathcal{J}_{\pm}$ the time components of the combinations $J^{\pm}=\frac{J+\tilde{J}}{2}$:
%For later convenience, we introduce the combinations (of their time components)
%\begin{align}
%\mathcal{J}_{\pm}&=\frac{\pi\pm\phi'}{2}
%\end{align} 
%%and define the left/right Hamiltonian densities as
%
%\be
%\H_{L,R} = \frac{\H \pm \mathcal{P}}{2} \;, \;\;\;\;\;\; \mathcal{P} = \sum_k \phi_k' \pi_k
%\ee
The local (unintegrated) classical flow equation 
\be
\p_\l \H = - \O_{J\bar T} = J_t T_{\s V} - J_\s T_{t V}
\ee 
with the currents given as above, can be solved exactly for any initial seed, in spite of its non-linearity \cite{Kruthoff:2020hsi}. The  (right) Hamiltonian density of the deformed theory is related to the undeformed one by %\textcolor{magenta}{\emph{Have we already set $k=2\pi$?}}
%{\color{ForestGreen}(we can absorb $\hat{k}$ by rescaling $\tilde{\lambda}=\lambda\sqrt{\hat{k}}$ and $\tilde{\mathcal{J}}_+=\frac{\mathcal{J}_+}{\sqrt{\hat{k}}}$)}
\be
\H_R = \frac{2}{\l^2 \hat{k}} \bigg(1-\l \J_+ - \sqrt{(1-\lambda\mathcal{J}_+)^2-\lambda^2  \hat{k}\mathcal{H}_R^{[0]}}\bigg) \;, \;\;\;\;\;\; \H_{L,R} \equiv  \frac{\H \pm \mathcal{P}}{2} \label{solhr}
\ee
%The total Hamiltonian density is simply given by $\mathcal{H}=\mathcal{H}_R+\mathcal{H}_L=2\mathcal{H}_R+\mathcal{P}$. The momentum density is given by $\mathcal{P}=\pi \phi'$. 
where we have reinstated an arbitrary level by rescaling $\l \r \l \sqrt{\hat k}$ and $\J_+ \r \J_+/\sqrt{\hat k}$ in the $\hat k=1$ result, while replacing $\H_R^{[0]}$ by the corresponding Hamiltonian density in the theory with level $\hat k$. %{\color{ForestGreen}(is it true that here we assume $H_R^{[0]}$ does not also rescale i.e. no sugawara?)}
Since \eqref{solhr} is symmetric in $\pi\leftrightarrow\phi'$,  the $J\bar{T}$ deformation defined using $J^{sh}$ leads to the same (classical) deformed theory as the one defined using $J^{top} $. %\textcolor{red}{True this is the reason? $\H_R$ need not be symmetric...} The same holds if we define the classical Hamiltonian deformation using the chiral $U(1)$ current $K_\a$. \emph{\textbf{Check!!}} 
From here on, we will be exclusively using the latter definition, as the flow operator  and the charges  we construct starting with the next subsection do depend on the choice.  %{\color{ForestGreen}(so I removed the tildes in this section)} % In the next section,  we will be using $\tilde{J}\bar{T}$ in order to compute the flow operator for the energy eigenstates. 

%[Since it will appear a lot, it is useful to introduce a notation for:
%\begin{align}
%F:=\sqrt{(1-\lambda\mathcal{J}_+)^2-\lambda^2\mathcal{H}_R^{(0)}}=1-\lambda\mathcal{J}_+-\frac{\lambda^2\mathcal{H}_R}{2}]
%\end{align}
Given this solution, we can compute all components of the stress tensor and remaining currents using \eqref{stresstham} and  \eqref{solhr} and re-express them in terms of $\J_\pm, \H_R, \mathcal{P}$.  For example, the components of the currents $K_\a, \bar K_\a$ introduced in \eqref{defK}, \eqref{defKb} are  %{\color{ForestGreen}(I think that for consistency here we should set $k=1$)} 
%
%\textcolor{red}{I think there should have been factors of $k$ in the Hamiltonian.}
%
\be \label{comprmcurrent}
K_t = K_\s =  \J_+ + \frac{\l  \hat{k}}{2} \H_R  \equiv \mathcal{K}_L \;, \;\;\;\;\;\;\; \bar K_t =  \J_- + \frac{\l  \hat{k}}{2} \H_R  \equiv \mathcal{K}_R\;, \;\;\;\;\; \bar K_\s =  K_\s - J^{top}_\s
\ee
where the new notation for the time components of the two currents has been introduced to harmonize with the notation for the Hamiltonians, and
\be \label{comptopcurrent}
J^{top}_{\sigma}=\mathcal{J}_+-\mathcal{J}_-+\frac{2\mathcal{J}_-+\lambda \hat{k}\mathcal{H}_R}{1-\lambda\mathcal{J}_+-\frac{\lambda^2  \hat{k}}{2}\mathcal{H}_R} \;, \;\;\;\;\;\; T_{\sigma V}=\frac{2\mathcal{H}_R}{1-\lambda\mathcal{J}_+-\frac{\lambda^2  \hat{k}}{2}\mathcal{H}_R}-\mathcal{H}_R
\ee
Since $\H_R$ is determined by the undeformed currents via \eqref{solhr}, all Poisson 
 brackets of the various deformed currents are entirely determined by the undeformed ones and are universal, even if somewhat cumbersome. See e.g. \cite{Guica:2020uhm} for a comprehensive list of commutators. 

Using the above-mentioned universal Poisson brackets, 
one may easily check that 

\be
 \{H, \H_L\} = - \p_\s \H_L\;, \;\;\;\;\; \{H,\mathcal{K}_L\}=-\partial_{\sigma} \mathcal{K}_L
  \ee 
which expresses the conservation of   these holomorphic currents in Hamiltonian language.  Here, $H = \int_0^R d\s \H$ is the total Hamiltonian. Using this, one can show that
%This continues to hold  also if we multiply them by functions of $U$. Thus
\begin{align}\label{leftchrg}
Q_f \equiv \int_0^R d\sigma f(U) \, \mathcal{H}_L \;, \hspace{1cm}P_{\eta}\equiv \int_0^R d\sigma \, \eta(U)\, \mathcal{K}_L  %\left(\J_+ + \frac{\l \hat{k} \H_R}{2}\right)
\end{align}
are conserved (left-moving) charges for any periodic $f, \eta$. On the other hand, the Poisson brackets of the Hamiltonian with the right-moving currents  are 
%{\color{ForestGreen}(same comment as before about k)}
%
\be
\{H,\mathcal{H}_R\}=\partial_{\sigma}\left( \mathcal{H}_R \frac{1+\lambda\mathcal{J}_++\frac{\lambda^2  \hat{k}}{2}\mathcal{H}_R}{1-\lambda\mathcal{J}_+-\frac{\lambda^2  \hat{k}}{2}\mathcal{H}_R}\right)\;, \;\;\;\;\;\; \{H, \mathcal{K}_R\}=\partial_{\sigma}\left(\mathcal{K}_R \frac{1+\lambda\mathcal{J}_++\frac{\lambda^2  \hat{k}}{2}\mathcal{H}_R}{1-\lambda\mathcal{J}_+-\frac{\lambda^2  \hat{k}}{2}\mathcal{H}_R}\right)
\ee
As in our Lagrangian analysis, it is possible to define the following charges\footnote{Note this notation is different from that of \cite{Guica:2020eab,Guica:2021pzy}, where $\bar Q_{\bar f}, \bar P_{\bar \eta}^{(KM)}$ denoted the (inconsistent) right-moving charges built from $v_{naive}$, while the  ones built from the correct field-dependent coordinate were denoted with calligraphic script: $\bar{\mathcal{Q}}, \bar{\mathcal{P}}$. %, with an additional tilde if these represented  flowed generators. 
 Here, we have decided to drop this cumbersome notation, as only the charges whose action on phase space is consistent will appear, as well as only currents constructed from $\bar K_\a$, to which the `KM' superscript referred.}:
\begin{align}
\bar{Q}_{\bar{f}}&=\int_0^R d\sigma \bar{f}(v)\mathcal{H}_R \hspace{1cm}\bar{P}_{\bar{\eta}}=\int_0^R d\sigma \, \bar{\eta}(v)\,  \mathcal{K}_R \label{RMcharges}
\end{align}
which are conserved for arbitrary periodic $\bar f(v),\bar{\eta}(v)$, provided the field-dependent coordinate $v$ satisfies 
\be
\frac{\p_t v - \{ H, v\}}{\p_\s v} =-\frac{1+\lambda\mathcal{J}_++\frac{\lambda^2  \hat{k}}{2}\mathcal{H}_R}{1-\lambda\mathcal{J}_+-\frac{\lambda^2  \hat{k}}{2}\mathcal{H}_R} =- \frac{1- \l \{ H, \phi\}}{1-\l\,  \p_\s\phi } \label{reqv}
\ee
As already discussed in the Lagrangian context, the obvious solution 
 $v_{naive} = V - \l \phi$ does not lead to consistent commutators. However, the solution to \eqref{reqv} above is ambiguous up to the addition of a zero mode 
 satisfying  $\p_t \varphi_0 - \{H,\varphi_0\} = \p_\s \varphi_0=0$, for which \cite{Guica:2020eab} showed there is a non-trivial solution. Appropriately subtracting this zero mode from $v_{naive}$ leads to consistent commutators for the generators \eqref{RMcharges} that, in particular, respect charge and momentum quantization. 
 %
% [ satisfies the above equation but,  as remarked in \cite{}, using $v_{naive}$ inside \eqref{RMcharges}  leads to charges whose commutators are not consistent with $U(1)$ charge and momentum quantization.  The problem stems from the presence of the zero mode of $\phi$,  $\varphi_0$ in the field-dependent coordinate.  For example, in the case of the commutation relation of $Q_0$ with \eqref{RMcharges}, the only non-zero contribution can come from this zero mode, which yields an answer that is proportional to $\l$ and thus incompatible with quantization.  As shown in the same article, it is possible to find a different solution to \eqref{reqv} where the zero mode is removed in a way consistent with the required commutators.]
%
%
%\be
%v_{imp} = \s - t - \l \phi + \frac{\l (\widetilde \phi_0 -\a t)}{R-\l Q_K}
%\ee
%where $\widetilde \phi_0 = \phi_0 + $ local corrections is the spectral flow generator in te $J\bar T$ - deformed CFT, satisfying \textbf{Properly motivate and define $\widetilde \phi_0$!}
%%
%\be
%\{\widetilde \phi_0, H\} = ? 
%\ee
%{\color{ForestGreen}(define also $\alpha$ here? \emph{yes, concisely})}.
%Using $v_{imp}$ instead of $v_{naive}$  in the commutation relations yields results consistent with charge quantization. 
%
%This presentation may leave the impression that the derivation of the expression of $v$ is somewhat arbitrary/the result of guesswork.
 In the next section, we give a first-principles derivation of these generators, whereby the correct expression for $v$ emerges from the definition we argue for.

\section{The Hamiltonian picture for the symmetries}\label{section3:Hamiltonian}

The Hamiltonian picture is the simplest way to understand how the extended symmetries of the theory are preserved under the flow. This section is mostly  a  review  of \cite{Guica:2020eab}, but in  which we streamline many of its arguments and computations, and minimize the set of assumptions being used.

%making use of a minimal set of assumptions.

More precisely, in \eqref{defnbp} we give a brief but global overview of the arguments that lead   to our final picture for the symmetries, simply quoting the needed technical results, which are derived in the subsequent subsections. In \ref{subsection32:classicalflowop}, we review and streamline the derivation of \cite{Guica:2020eab} of the classical limit of the flow operator. 
%\textcolor{red}{\emph{Does this terminology make sense? }} 
Finally, in \ref{section33:emergencefdcoord} we explicitly solve the flow equations for the conserved currents, and then use these expressions to   show the emergence of the correct field-dependent coordinate.

\subsection{Definition and  basic properties of the conserved charges \label{defnbp}}

\subsubsection*{Definition}

As explained in the previous section, the na\"{i}ve attempt to construct conserved charges associated to the right-moving field-dependent conformal and affine transformations fails because the resulting conserved generators do not have a proper action on phase space. To ensure the latter property, one may simply start by demanding it be true. That is, we demand that if two states in the undeformed CFT are related by the action of a Virasoro generator, $|n'\rangle = L_{-m} | n \rangle$, then the corresponding flowed eigenstates should be related via the action of a corresponding ``flowed'' Virasoro generator\footnote{Also   denoted as ``spectrum-generating operator'' in \cite{LeFloch:2019rut}, where it was first introduced.}. In equations, the states satisfy the flow equation
\vskip-3mm
\be \label{adiabatic}
\p_\l | n\rangle = \hat{\mathcal{X}}_{J\bar T} | n\rangle
\ee
where the flow operator is directly determined by the deforming operator $\O_{J\bar T}$ via the decomposition %{\color{ForestGreen}(hat on chi? this part is still quantum)}
\vskip-3mm
\be \label{extractflowop}
\p_\l H %= - \int d\s \O_{J\bar T}
 = D - [H,  \hat{\mathcal{X}}_{J\bar T} ]
\ee 
where  $\p_\l H$ is given by its definition \eqref{hamflow} and the operator  $D$ collects the diagonal elements of $\p_\l H$ in the energy eigenbasis. The \emph{flowed Virasoro} generators $\widetilde L_n$ are then taken to satisfy%, by definition 

\be
\p_\l \widetilde L_n = [\hat{\X}_{J\bar T}, \widetilde L_n] \label{defflgen}
\ee
%{\color{ForestGreen} Using the flow operator, one can define a covariant derivative along the flow and use it in order to define flowed generators as solutions of the differential equations $\mathcal{D}_{\lambda}(...)=0$,
with the initial condition that at $\lambda=0$, they reduce to the Virasoro generators of the undeformed CFT. We can obviously define such a flowed generator for each extended symmetry in the original CFT, which in the cases we consider herein will be (Virasoro $\times$ Kac-Moody)$^2$. The definition \eqref{defflgen} ensures that the commutation relations of the flowed generators  will be identical to those in the undeformed theory. %, by construction. Thus, Virasoro-KM. % (again we don't have time dependence here; could we introduce it from the beginning?). By looking at $\mathcal{D}_{\lambda}\frac{d(...)}{dt}=\{\partial_{\lambda}H,(...)\}$, we see that in order for the charges to be conserved, we need to have explicit time dependence for the charges. As long as we have $\{H,(...)\}=\alpha(...)$ with $\alpha$ that commutes with $H$ (this is the case in $J\bar{T}$), we can add the time dependence.] ???

\subsubsection*{Conservation}

In order for the flowed generators \eqref{defflgen} to correspond to symmetries, we need to show that they are conserved. For this, we need to define them on some arbitrary time slice, $t$, which in turn requires a definition of  the  operator  $\hat{\X}(t)$ that flows the states at time\footnote{We drop the $J\bar T$ label from $\hat{\X}$ in this paragraph, as the argument is entirely general. %, and applies  also to $T\bar T$.
}
 $t$. Given that the time dependence of energy eigenstates is simply $e^{- i E_n t}$, the flow operator at time $t$ is  related to that at $t=0$ by %\emph{\textcolor{red}{Reread Kruthoff and Parrikar} }

%It is trivial to show that {\color{ForestGreen}(the state has a factor of $e^{-i H t/\hbar}$, which brings an extra contribution prop to t; factors of $i/\hbar$?)}
{
\be \label{timedepflowop}
\hat{\X}(t) = \hat{\X}(0) - i D\, t
\ee
}
where, as above,  $D$ represents the diagonal elements of $\p_\l H$ in the energy eigenbasis.%, $D= \p_\l E = \D_\l H$ by definition.

We would now like to show that a current, $\widetilde \J$, defined to be ``covariantly constant'' along  the flow 
\vskip-3mm
\be
\hat{\mathcal{D}}_\l(t)  \widetilde{\J}_\a \equiv \p_\l \widetilde{\J}_\a - [\hat \X (t), \widetilde{\J}_\a]=0 \label{defflcur}
\ee
stays conserved, provided it was conserved in the initial theory. %{\color{ForestGreen} (If we call $\hat{\mathcal{D}}_\l(t)$ covariant derivative, then the flowed generators are the ones parallel transported. It would be nice to geometrize the construction.)} \textcolor{red}{Do you have something specific in mind?}
 To see this, 
we compute the change in the current conservation equation  as $\l$ is varied 
\be
\frac{d}{d\l} (\p_t \widetilde{\J}_t + i  [H,\widetilde{\J}_t] - \p_\s \widetilde{\J}_\s)%= [\p_t + H \pm \p_\s, [\X(t), \J]] + [\p_\l H, \J] 
= [\p_t \hat{\X}(t) +i  \hat{\mathcal{D}}_\l H, \widetilde{\J}_t] + [\hat \X(t), \p_t \widetilde{\J}_t + i [H, \widetilde{\J}_t] - \p_\s \widetilde{\J}_\s] 
\ee
The first term on the right-hand side vanishes, since $\p_t \hat \X (t)=-i D=-i \hat{\mathcal{D}}_\l H  $, where in the second equality we used \eqref{extractflowop}.
 Thus, if the current is initially conserved, it will stay conserved along the flow. %This argument ignores potential  contact terms.  \emph{\textcolor{red}{Think more?}}

Note that most currents we will be dealing with are (anti)chiral in the undeformed theory. Since the flow of zero is zero, they will continue to be (anti)chiral along the flow, so $\widetilde{\J}_\s = \pm \widetilde{\J}_t$. Of course, one is always free to add improvement terms to these currents, which will in general break the corresponding chirality condition.

%The first term vanishes via the definition of $\p_t \X= \mathcal{D}_\l H  = D$, while the second does by assumption of conservation at the previous step. \emph{\textbf{Careful!! }}

The operators $\widetilde L_n$ defined in \eqref{defflgen} correspond to the spatial Fourier modes of  current operators defined as in \eqref{defflcur}. The argument presented above shows their conservation follows automatically from their definition for \emph{any} adiabatic deformation, which  acts as \eqref{adiabatic} on the energy eigenstates. % {\color{ForestGreen}(actually this is our definition of adiabatic, right? I think that it's important that we do not have level crossing; in the single-trace case probably this is true only within twisted sectors (??))} \emph{\textcolor{red}{Are there any implicit constraints?}}  
This suggests the Virasoro symmetries can be transported along the flow for \emph{any} adiabatic irrelevant deformation, not just the  $T\bar T$ and $J\bar T$ ones. Of course, the special form of the commutator of the Hamiltonian with the charges highlighted in \cite{Guica:2021pzy} appears special to these deformations, and it would be very interesting to understand what happens in more  general situations.

%From this perspective, it is not entirely clear whether, and in which respects, the $T\bar T$ and $J\bar T$ deformations should be special. Of course, the fact that the commutator with $H$ is proportional to the charge is special to these types of deformations, and the resulting expressions for the generators are (almost) quasilocal. 

\subsubsection*{Exact solution for the classical currents}

In classical $J\bar T$ - deformed CFTs, just like in classical $T\bar T$ - deformed ones \cite{Guica:2022gts},  it is possible to find a closed-form solution for the classical limit of $\hat \X(t) \r i \X_{cls} (t)$. This  was originally found in \cite{Guica:2020eab}; in the next subsection, we streamline the derivation so that the steps are clearer and the set of assumptions used is minimised. 

Given $\X_{J\bar T}^{cls}$, one may simply solve

\be
\mathcal{D}_\l(t) \widetilde{\mathcal{J}}_\a =0 \label{defflcurq}
\ee
for the various chiral currents in the undeformed CFT, namely $\J_\pm, \H_{L,R}^{[0]}$, to find the deformed ones.   Here,  $\D_\l(t) = \p_\l + \{ \X_{cls}(t),\, \cdot \, \}$ is the classical limit of $\hat{\D}_\l$.  The solutions are denoted as $\widetilde{\mathcal{K}}_L, \widetilde{\H}_L, \widetilde{\mathscr{K}}_{R}$ and $ \widetilde{\mathscr{H}}_{R}$,  and are explicitly given in \eqref{solcurrent1},\eqref{solcurrent2},\eqref{solcurrent3} and \eqref{solcurrent4} %{\color{ForestGreen}(maybe we collect the expressions somewhere at the end of 3.3)} 
in the next section.  The difference in  notation between the left and right currents reflects the extra modifications needed in order to render the right-moving ones  anti-chiral.  

As discussed,  these currents are automatically conserved. The various conserved charges 
 correspond to their Fourier modes with respect to $t \pm \s$  %of these (anti)chiral currents
% \emph{Signs!}{\color{ForestGreen}(checked, $e^{inU}$ and $e^{-inv}$)}

\be \label{flowedleft}
\widetilde P_n = \int_0^R d\s \, e^{i n \frac{2\pi}{R} (t+\s)} \widetilde{\mathcal{K}}_L \;, \;\;\;\;\;\; \widetilde Q_n  = \int_0^R d\s \, e^{i n \frac{2\pi}{R}(t+\s)} \widetilde{\mathcal{H}}_L
\ee

\be \label{flowedright}
\; \widetilde{\!\!\bar P}_n = \int_0^R d\s \, e^{i n \frac{2\pi}{R}(t-\s)} \widetilde{{\mathscr{K}}}_R \;, \;\;\;\;\;\;\; \widetilde{\!\!\bar Q}_n  = \int_0^R d\s \, e^{i n \frac{2\pi}{R}(t-\s)} \widetilde{\mathscr{H}}_R
\ee
and their  algebra %of the generators \eqref{flowedleft},\eqref{flowedright} 
consists of two commuting copies of the  Witt  $\times \, U(1)$ Kac-Moody algebra, by construction. %\emph{Level?} % Note the zero modes $\tilde P_0, \tilde{\bar{P}}_0$ are precisely $J_0,\bar J_0$. The algebra implies they commute with all conserved charges, as expected. 

%{\color{ForestGreen} where we remind that $\sigma\sim\sigma +R$.}
The  expressions  for the right-moving charges \eqref{flowedright} can be significantly simplified by integrating by parts and resumming the resulting infinite set of terms, as in \eqref{resumexpp}.  As in the $T\bar T$ case, the correct field-dependent coordinate can be shown to unambiguously \emph{emerge} from this resummation. %\textcolor{violet}{which corresponds to a particular improvement of the corresponding current that renders it non-chiral, but with a much simpler expression }. {\color{ForestGreen}(ok, because we integrate by parts, but maybe we can say that the improvement depends on the mode, namely that for each $n$, the term is different)}
%{\color{ForestGreen}(They look local when expressed in terms of the field-dependent coordinate i.e. when we do the Fourier expansion in $v$. Is it ok to say that they are local in $v$?)}

% from the flow via  a simple integration by parts/ should then give us the deformed conserved current. We should see the field-dependent coordinate emerge from this in the classical limit. In the following, we'd like to recap the derivation of $\mathcal{X}_{J\bar T}^{cls}$ in a streamlined fashion, and then show that $v_{imp}$ emerges from the flow unambiguously, without any further input. \textcolor{magenta}{
%Jumping ahead, the solution for the currents is given in \eqref{} in section \ref{} below. They are automatically conserved.}

%The flow operator is inferred from its action on the states at $t=0$. We would like to have it for an arbitrary time, as we'd like to use it to flow arbitrary Heisenberg picture operators.   If we act on the states at time $t$, it receives an additional contribution proportional to $\p_\l E t$.

%The conserved charges 

\subsubsection*{Different operator bases}

 The explicit charges $\widetilde{P}_n, \, \widetilde{\!\bar{P}}_n, \widetilde{Q}_n, \, \widetilde{\!\bar{Q}}_n$ that result from the flow turn out to not be of the form \eqref{leftchrg}, \eqref{RMcharges} for the appropriate choice of $v$. Rather, the two sets of generators  differ  by certain non-linear shifts that resemble  an ``energy-dependent spectral flow'' %\emph{Or, the other way?}
%The generators $\bar{\mathcal{P}}_n,\bar{\mathcal{Q}}_n$ are the same as before but with $v_{imp}$ instead of $v$ and are related via
% energy-dependent spectral flow to the Virasro-KM generators:
%{\color{ForestGreen}(we wrote them in the intro 1.1,1.2 in different notations)}
\begin{align}\label{basisgenerators}
P_n&=\widetilde{P}_n+\frac{\lambda \hat{k} E_R}{2}\delta_{n,0}\;, \hspace{1.5cm} R Q_n= R\, \widetilde{Q}_n +\lambda E_R \widetilde{P}_n +\frac{\lambda^2 \hat{k} E_R^2}{4}\, \delta_{n,0}\\
\bar{P}_n&=\, \widetilde{\!\bar{{P}}}_n+\frac{\lambda\hat{k} E_R}{2}\delta_{n,0}\;, \hspace{1.5cm}
R_v \bar{Q}_n= R \; \widetilde{\!\bar{Q}}_n + \lambda E_R \, \widetilde{\!\bar{P}}_n+\frac{\lambda^2 \hat{k} E_R^2}{4} \, \delta_{n,0}
\end{align} 
where $R_v$, given in \eqref{notationch}, can be interpreted as the circumference of the emergent field-dependent coordinate, and $E_R$ stands for the right-moving Hamiltonian, i.e. it is a function on phase space. This is the classical counterpart of the relations \eqref{quasilocalgen1}, \eqref{quasilocalgen2} quoted in the introduction, where   $L_m, J_m$ and their right-moving analogues  denoted the \emph{quantum} operators (rescaled by factors of radii,  using the conventions of \cite{Guica:2021pzy}), whereas here $Q_m, P_m$ denote the corresponding \emph{classical} charges. 
%
%{\color{ForestGreen}(I think that we haven't defined $R_v$ yet. If we have in our conventions $R_v=R-\lambda w$ then it's not ok to call it field-dependent radius because it's the circumference)} 
%{\color{blue}The untilded generators above are more physical, in that they can be argued to act more locally on the fields in the theory, a criterion that can be made fully precise in the case of the left-moving generators.}
 As we will argue, the untilded generators above appear more physical, in that they  act more locally on the fields in the theory, a criterion that will be explained in section \ref{section41:symtr}. We will thus denote them as \emph{quasi-local} generators.  At the level of the associated conserved currents, it is easy to see that $\mathcal{K}_{L,R},\mathcal{H}_{L,R}$ are  quasi-local (in the sense that, to any finite order in the $\l$ expansion, they are sums of products of local fields),  while their tilded counterparts -  given in \eqref{solcurrent1}, \eqref{solcurrent2nc}, \eqref{solcurrent3} and  \eqref{solcurrent4nc},   which enter the expressions for the flowed charges   summarized in table \ref{table:generators}  -  are non-local, because they explicitly involve pieces that are integrated over all space. %\textcolor{blue}{It remains an interesting question how to derive  their form from ``first principles''.}  \textcolor{red}{You mean Noether procedure?}{\color{ForestGreen}(yes, but actually I think we could derive them by imposing that the computation is similar)}

Throughout this article, we will use the following notations%(but in the bulk section we will use $Q_{L/R}$)
\footnote{This proliferation of symbols represents our attempt to change as little as possible the notation in the previous literature, which we hope will not cause too much confusion in the current article.}  
\begin{equation}   \label{notationch}
   \boxed{
   \begin{aligned}
   Q_K&= P_0 = Q_L = J_0 + \frac{\l \hat{k}E_R}{2} \;, \;\;\;\;\;\;\;\; \;\;\bar{Q}_{\bar{K}} = \bar{P}_0=Q_R=\bar{J}_0+ \frac{\l \hat{k}E_R}{2} \nonumber\\[2pt]
&\;\;\;\;\; J_0=\widetilde{P}_0  \;, \;\;\;\;\;\;\;\bar J_0 = \widetilde{\bar{P}}_0 \;, \;\;\;\;\; 
Q_0 = J_0-\bar J_0 \;, \;\;\;\;\;w= J_0-\bar J_0  \nonumber \\[8pt] 
 &\hspace{2cm}R_v = R-\l w \;, \;\;\;\;\; R_Q\equiv R-\lambda Q_K    
 \end{aligned}   }
\end{equation}
The quasilocal generators can  be used to define analogues of CFT primary operators in the non-local $J\bar T$ - deformed CFT \cite{Guica:2021fkv}. Their interplay with the flowed generators can be used to determine the correlation functions  of these operators in the deformed theory in terms of the undeformed CFT correlators, both in double-trace \cite{Guica:2021fkv} and  single-trace \cite{Chakraborty:2023wel} $J\bar T$ - deformed CFTs.

The algebra of the quasilocal generators simply follows from \eqref{basisgenerators}. While the algebra of the left-moving generators remains Witt $\times $ Kac-Moody, that of the right-moving ones becomes a non-linear modification of it. Concretely %\textcolor{red}{\emph{You may want to replace $2\pi \hat k $ by $k$}} {\color{ForestGreen}(I would leave them like this to have the 2pi everywhere, since not all the terms have a k )}% \emph{Write them down!}
%{\color{ForestGreen}(we already need to introduce $Q_K$. Can we also introduce some notation for $R-\lambda Q_K$ ?)
\be \label{algebranonlinearjtbar}
\{ \bar P_m, \bar P_n\} = -\frac{i m k}{2} \delta_{m+n,0}-\frac{i m k\lambda }{2 R_Q} \bar{P}_m \delta_{n,0}+\frac{i n k\lambda}{2R_Q}\bar{P}_n \delta_{m,0}
\ee

\be
\{ \bar Q_m, \bar P_n\} = \frac{2\pi i n}{R_v}\bar{P}_{m+n}-\frac{i m k\lambda}{2 R_Q}\bar{Q}_m \delta_{n,0}+\frac{2\pi i n \lambda}{R_v R_Q}\bar{P}_m \bar{P}_n
\ee

\be
\{ \bar Q_m, \bar Q_n\} = -\frac{2\pi i(m-n)}{R_v}\bar{Q}_{m+n}-\frac{2\pi i m \lambda}{R_v R_Q}\bar{Q}_m \bar{P}_n+\frac{2\pi i n \lambda}{R_v R_Q}\bar{Q}_n \bar{P}_m 
\ee
%\textcolor{red}{Factors k?}
%
where we used $k=2\pi \hat k$. The commutation relations between the left and the right generators are 
\be
\{ Q_m, \bar Q_n\}= \frac{2\pi i n \lambda}{R R_Q}\bar{Q}_n P_m\hspace{2cm} \{ Q_m, \bar P_n\}= \frac{2\pi i n \lambda}{R R_Q}\bar{P}_n P_m
\ee
\be  \label{algebra:LMRMaffine}
\{P_m, \bar Q_n\} = \frac{i n k \lambda}{2 R_Q}\bar{Q}_n \delta_{m,0}\hspace{2cm}\{P_m, \bar P_n\} = \frac{i n k \lambda}{2 R_Q}\bar{P}_n \delta_{m,0}
\ee
This is just the classical Poisson bracket algebra. Its quantum generalization  is given by the usual replacement $\{\;,\:\} \r - i [\;, \;]$ to leading order in $\hbar$; \cite{Guica:2021pzy} discusses it to all orders.  
%
%Finally, let us fix the notation for the charges. The zero modes of the flowed $U(1)$ charges, $\widetilde{P}_0, \widetilde{\bar{P}}_0$, are identified with the chiral conserved charges in the undeformed CFT, $J_0$ and, respectively, $\bar J_0$. The relations \eqref{basisgenerators} imply that the zero modes of $P_0, \bar P_0$, denoted $Q_K, \bar Q_{\bar K}$ are related to them via
%
%\be
%Q_K = J_0 + \frac{\l E_R}{2} \;, \;\;\;\;\; \bar Q_{\bar K} = \bar J_0 + \frac{\l E_R}{2}
%\ee
%We will be using this notation throughout. 

\subsection{Classical limit of the flow operator in $J\bar T$ - deformed CFTs}
\label{subsection32:classicalflowop}

In this subsection, we explain how to obtain a closed-form expression for the classical flow operator in $J\bar T$ - deformed CFTs, $\X_{J\bar T}^{cls}$, significantly simplifying the presentation of \cite{Guica:2020eab}. 
%
%The operator can be obtained from first-order quantum-mechanical perturbation theory, which yields \eqref{}. Since we are mostly interested in the classical limit, we will be working with the classical analogue of it, which reads
In the classical limit, the decomposition \eqref{extractflowop} that defines the flow operator reduces to %{\color{ForestGreen}($\hbar=1$ in our conventions? should we put everywhere tilde on J in $\mathcal{O}$ or just mention again that at classical level it's the same?)}

\be \label{extractflowopcls}
\p_\l H = - \int_0^R d\s \, \O_{J\bar T}^{cls} = D + \{ H,  \mathcal{X}^{cls}_{J\bar T} \} 
\ee 
where $D$ is the expectation value of $\p_\l H$ in the eigenstate of interest, which is simply given by 
\be \label{diagonalpart}
D= \p_\l E = \frac{2 E_R Q_K }{R_Q}
\ee
We assume that the current which enters the deformation is the topological one, $J=J^{top}$.
Thus,  we simply need to understand how to write the integrated classical $J\bar T$ operator  in the form above. %\footnote{The original reference  \cite{} first approached this problem perturbatively, followed by an all-orders guess. It also introduced and then subtracted away a zero mode of $\phi$ whose action on the Hilbert space was not quite well defined. Here, we  shorten the derivation, along the lines of the $T\bar T$ one presented in \cite{}. }
Our new derivation is along the  lines of the $T\bar T$ one, presented in \cite{Guica:2022gts}.

The integrated Smirnov - Zamolodchikov operator constructed from any two conserved currents $J^{A,B}$  can be written as\footnote{This works at both the classical and the quantum level; here we work  classically  in order to be able to ignore operator ordering issues. %We are using the convention $\epsilon^{\sigma t}=1$.% {\color{ForestGreen}(this is consistent with $\epsilon^{UV}=-2$)}
 } \cite{Kruthoff:2020hsi} %{\color{ForestGreen}(ok with the total time derivative)}
 
\be
\int_0^R d\s \, \e^{\a\b } J^A_\a J^B_\b =  \frac{1}{R} \, \e^{\a\b} \int_0^R d\s J^A_\a  \int_0^R d\s J^B_\b -  \frac{d}{dt}  \int_0^R d\s d\tilde \s G(\s-\tilde \s)  J^A_t (\s) J^B_t (\tilde \s) \label{szburst}
\ee
where 
\be
G(\sigma, \tilde \s)=\frac{1}{2}sgn(\sigma-\tilde \s)-\frac{\sigma-\tilde \s}{R}
\ee is the Green's function on the cylinder, $\p_\s G(\s, \tilde \s) = \d (\s-\tilde \s)-\frac{1}{R}$. Applying this to the integrated $J\bar T$ operator with $J^A_\a= J_\a^{top}$ and $J^B_\a= T_{\a V}$, replacing the time derivative by a commutator with $H$ and the integrals over $ J^{top}_t, T_{t V}$ by the corresponding conserved charges ($w$ and, respectively,  $-E_R$), we find %\textcolor{red}{Notation!} %{\color{ForestGreen}(we use that the currents are conserved on-shell)}
%\be
%\int d\s \O_{JK} = \frac{1}{R} (Q_K \int J_\s - Q_J \int K_\s) \pm i [H, \int d\s d\tilde \s G(\s-\tilde \s)  J_t (\s) K_t (\tilde \s)  ] 
%\ee
%%, 
%
%\be
%\int d\s \O_{J\bar T} = \frac{1}{R }\e^{\a\b} \int J \int T - \p_t \int d\s d\tilde \s G(\s-\tilde \s)  J_t (\s) T_{t V} (\tilde \s)
%\ee
%{\color{ForestGreen}(we actually compute $\tilde{J}\bar{T}$ instead of $J\bar{T}$)}\emph{Explain steps \underline{concisely} for how to obtain the flow operator. Understand improvement with respect to symmvssp!}
%
{%\color{ForestGreen}
% Using the fact that $w$ is the zero mode of $\tilde{J}_t$ and $-E_R$ is the zero mode of $T_{tV}$:
\begin{align}\label{exprdefop}
\int_0^R d\sigma \, \mathcal{O}^{cls}_{{J}\bar{T}}=\frac{1}{R}\bigg(-E_R\int_0^R d\sigma J^{top}_{\sigma}-w \int_0^R d\sigma T_{\sigma V}\bigg)+ \left\{H, \int_0^R d\sigma d\tilde{\sigma} G(\sigma-\tilde{\sigma}){J}^{top}_t(\sigma)T_{tV}(\tilde{\sigma})\right\}
\end{align}}
%{\color{ForestGreen}(I think that the notation is very confusing because we have tilde everywhere for flowed quantities and here we use tilde for the topological current (even if for the flowed ones we use a wider tilde))}
The last term is already in the form we desire.  To put the remaining components in the form \eqref{extractflowopcls}, we use two tricks: first, we note the following identity 
\be
T_{\s V} = \H_R - \l \, \O^{cls}_{{J}\bar T}
\ee
where  $\mathcal{O}^{cls}_{J\bar{T}}=J^{top}_{\sigma}T_{tV}-J^{top}_{t}T_{\sigma V} %=\frac{2\mathcal{H}_R}{\lambda}\bigg(1-\frac{1}{1-\lambda\mathcal{J}_+-\frac{\lambda^2{\color{ForestGreen}k}}{2}\mathcal{H}_R}\bigg)
$, with the current components  given in \eqref{comptopcurrent}%{\color{ForestGreen}(I didn't find them elsewhere)}
. Plugging this into \eqref{exprdefop},  we can   pass the $\mathcal{O}^{cls}_{J\bar{T}}$ term to the left-hand side, which produces a factor of $R_v/R = 1 - \l w/R$. This takes care of the second term on the right-hand side of \eqref{exprdefop}. %{\color{ForestGreen}(notation for the classical limit? notation for that combination that will keep appearing?)}

Turning to the first term, we use a second trick to write $\int_0^R d\s  J^{top}_\s$ as a Poisson bracket commutator with $H$ plus some charge-dependent terms. For this, we relate the expression for $ J^{top}_\s$ to that for a putative classical  Smirnov-Zamolodchikov-type operator $\O_{ J\bar K}$% \equiv \e^{\a\b} \tilde{J}_\a \bar K_{\b}$%, whose classical expression turns out to be related to $\tilde J_\s$ as 

\be
\O_{J \bar K}^{cls}  \equiv    J^{top}_\s \bar K_t -  J^{top}_t \bar K_\s % = (\tilde J_\s - \tilde J_t) K_t = \frac{2 (\J_+ + \l {\color{ForestGreen}k\mathcal{H}_R}/2) (\J_- + \l {\color{ForestGreen}k\mathcal{H}_R}/2)}{1-\l \J_+ - \l^2{\color{ForestGreen}k}/2 \H_R}
 = \frac{ J^{top}_\s - \phi' - 2 \bar K_t}{\l} \label{ojtkb}
\ee
where we used  again the explicit expressions \eqref{comprmcurrent}. Applied to this classical operator, the formula \eqref{szburst} reads %\emph{Check sign!}{\color{ForestGreen}(checked)} 
%\textcolor{red}{Notation!}
\be
\int_0^R d\s \, \O^{cls}_{J \bar K} = \frac{1}{R} \left( \bar Q_{\bar K} \int_0^R d\s J^{top}_\s  - w \int_0^R d\s \bar K_\s \right)  +  \left\{H, \int_0^R d\s d\tilde \s G (\s-\tilde \s) J^{top}_t (\s) \bar K_t (\tilde \s)\right\}
\ee
%{\color{ForestGreen}where $\bar{Q}_{\bar{K}}=\bar{J}_0+\frac{\l\hat{k}E_R}{2}$ is the charge associated to $\bar{K}$. (I didn't find it introduced before)}. 
Writing $\bar K_\s = K_\s - J^{top}_\s = K_t - J^{top}_\s$,  using \eqref{ojtkb} and 
collecting the terms involving $\int_0^R d\s J^{top}_\s$, we immediately find an expression for it in terms of the conserved charges and a  Poisson bracket with $H$. 
%
%
% find %{\color{ForestGreen}(checked)}
%
%\begin{align}
%\int d\s \tilde J_\s  = \frac{1}{R-\l Q_K} {\color{ForestGreen}\left[ R_v Q_K + R\bar Q_{\bar K} +\{ H, \l R \int d\s d\tilde \s G (\s-\tilde \s) \tilde J_t (\s) \bar K_t (\tilde \s) \} \right]  }
%\end{align}
Putting everything together, the change in the Hamiltonian can be classically decomposed as %\emph{Re-check!!}{\color{ForestGreen}(checked, but I don't understand the tilde on $\mathcal{H}_R$; edit: now I understand, but I think that the notation is very confusing...)}
%\textcolor{red}{Should we intro a notation for $R-\l Q_K$?} {\color{ForestGreen}(yes, $R_Q$?)}

\begin{align}
\partial_{\lambda}H&=\frac{2 E_R Q_K}{R_Q}+\left\{H, \frac{R}{R_v}\int_0^R d\s d\tilde \s  G(\s-\tilde \s) \phi'(\s) \left[ {\mathcal{H}}_R (\tilde \s) + \frac{\lambda E_R}{R_Q} \bigg(\mathcal{J}_- (\tilde \s)+\frac{\lambda \hat{k}\mathcal{H}_R (\tilde \s)}{2} \bigg)\right] \right\}
\end{align}
which has precisely the expected form \eqref{extractflowopcls}. We may further simplify this expression by noting that the $\s$ integral yields   $ \int_0^R d\s G( \s- \tilde{\s}) \phi'(\sigma)=-\hat{\phi}_{nzm}$, which denotes  the scalar field with its winding and zero mode removed. % and it results from  performing the integral.
%\begin{align}
%\int d\s \tilde J_\s  &=\frac{R_v}{R-\lambda Q_K}\bigg[Q_K+\frac{R\bar{Q}_{\bar{K}}}{R_v} +\frac{\lambda R}{R_v}\bigg\{\int (\mathcal{J}_-+\frac{\lambda\mathcal{H}_R}{2})\hat{\phi}_{nzm},H\bigg\} \bigg]
%\end{align}
%
%Coming back to \ref{exprdefop} (I didn't replace now the commutator because it's long):
%\begin{align}
%R\int \mathcal{O}_{\tilde{J}\bar{T}}=-E_R R\{\phi_0,H\}-w(E_R-\lambda\int\mathcal{O}_{J\bar{T}})- R \partial_t\int d\sigma d\tilde{\sigma} G(\sigma-\tilde{\sigma})\tilde{J}_t(\sigma)T_{tV}(\tilde{\sigma})
%\end{align}
%we notice the appearance of the field-dependent radius $R_v=R-\lambda w$:
%\begin{align}
%R_v\int \mathcal{O}_{\tilde{J}\bar{T}}&=-w E_R - E_R R\{\phi_0,H\}- R \partial_t\int d\sigma d\tilde{\sigma} G(\sigma-\tilde{\sigma})\tilde{J}_t(\sigma)T_{tV}(\tilde{\sigma})
%\end{align}
%The last term can be rewritten as:
%Next, we rewrite the last term using the same manipulations as before:
%\begin{align}
%\partial_t\int d\sigma d\tilde{\sigma} G(\sigma-\tilde{\sigma})\tilde{J}_t(\sigma)T_{tV}(\tilde{\sigma})
%=\bigg\{\int \mathcal{H}_R \hat{\phi}_{nzm},H \bigg\}
%\end{align}
%This has precisely the expected form  \eqref{extractflowopcls} and allows us to  
We can therefore identify the classical flow operator to be %{\color{ForestGreen}(now we remove the tilde in $\mathcal{X}^{cls}_{J\bar T}$)}
\begin{align}
\boxed{\mathcal{X}^{cls}_{J\bar T}=-\frac{R}{R_v}\int_0^R d\s \, \mathcal{H}_R \, \hat{\phi}_{nzm} -\frac{\lambda R E_R}{R_v R_Q}\int_0^R  d\s \left(\mathcal{J}_-+\frac{\lambda \hat{k}\mathcal{H}_R}{2}\right)\hat{\phi}_{nzm}}
\end{align}
in perfect agreement with the expression found in \cite{Guica:2020eab}. Note this represents $\X^{cls}_{J\bar T} (0)$; to obtain  $\X^{cls}_{J\bar T} (t)$ we should add the appropriate  time dependence \eqref{timedepflowop}, with $D$ as in \eqref{diagonalpart}. 
%It might be useful to write it as:
%\begin{align}
%\chi_{J\bar{T}}=i\bigg(-\frac{R}{R_v}\frac{R-\lambda J_0}{R-\lambda Q_K}\int \mathcal{H}_R\hat{\phi}_{nzm} -\frac{\lambda R E_R}{R_v(R-\lambda Q_K)}\int \mathcal{J}_- \hat{\phi}_{nzm}\bigg)
%\end{align}

\subsection{Emergence of the field-dependent coordinate}
\label{section33:emergencefdcoord}

In this section, we explicitly solve the flow equation \eqref{defflcur} for each of the  conserved currents in the classical $J\bar T $ - deformed CFT, using the expression above for the flow operator. We then explicitly show the emergence of the field-dependent coordinate via integration by parts. While the method and ultimate solution are the same as in \cite{Guica:2020eab}, we find that our current derivation  is conceptually clearer  than that presented therein, who built the correct solution for the flowed charges as an expansion around the ones constructed from $v_{naive}$, which do not have a proper action on phase space.  %\textcolor{red}{\emph{Is this really a problem?}}

%As in the $T\bar T$ case, we will notice the emergence of the field-dependent coordinate by studying the flow of the symmetry currents. {\color{ForestGreen}We are looking for flowed generators, which by definition are covariantly constant along the flow. Equivalently, we require that their corresponding currents are covariantly constant.} 

Rather than solving the flow equation \eqref{defflcur} order by order in $\l$, which is possible in principle but hard in practice, we  instead try to guess as much of the solution as possible, directly at finite $\l$, using the Poisson brackets listed in the appendix of \cite{Guica:2020uhm}.  For the left-moving affine current, a natural first guess is the left current $\mathcal{K}_L = \J_+ + \l \hat{k} \H_R/2$, whose covariant  derivative along the flow is 

\be \label{flowLMcurrent}
\mathcal{D}_\l  \bigg( \mathcal{J}_+ + \frac{\lambda \hat{k}\mathcal{H}_R}{2} \bigg) =\frac{\hat{k}E_R}{2 R_Q}
\ee
Thus, our guess fails to be covariantly constant by just   a constant term. Remembering that % \emph{Check!}
\be \label{flowER}
\D_\l E_R = \frac{E_R Q_K}{R_Q}
\ee
it is not hard to show that the covariantly constant combination is given by
%\begin{align}
%\mathcal{D}_{\lambda}\bigg( \mathcal{J}_+ + \frac{\lambda\mathcal{H}_R}{2}- \frac{\lambda E_R}{2R}\bigg) =0
%\end{align}
%Hence, we have 
\be \label{solcurrent1}
\boxed{\widetilde{\mathcal{K}}_L= \mathcal{K}_L -\frac{\lambda \hat{k} E_R}{2R}= \mathcal{J}_+ + \frac{\lambda\hat{k}\mathcal{H}_R}{2}- \frac{\lambda \hat{k}E_R}{2R}}
\ee
i.e.,  the left-moving spectrally flowed current, as we anticipated. % We recover the fact that the spectrally flowed currents are solutions to the flow equation. 
 
For the right-moving affine current, we start from the natural guess $\mathcal{K}_R=\J_- + \frac{\l \hat{k}}{2} \H_R$ and compute 
\begin{align}\label{firstguessRM}
\mathcal{D}_\l(t) \bigg(\J_- + \frac{\l \hat{k}}{2} \H_R\bigg) &= \frac{E_R \hat{k} R}{2 R_Q}\bigg(\frac{1}{R}-\frac{\lambda\hat{\phi}'}{R_v}\bigg)-\hat{k}\partial_{\sigma}\bigg(\frac{\J_- + \frac{\l \hat{k} }{2} \H_R}{\Sigma} \, \Lambda \bigg)
\end{align}
where
\be \label{definitionsigma}
\Sigma \equiv 1-\l \J_+ - \frac{\l^2 \hat{k} \H_R}{2} \;, \;\;\;\;\;\; \hat{\phi}\equiv \phi-\frac{w\sigma}{R}
\ee
$\hat \phi$ is the scalar with its winding mode removed. The quantity $\Lambda$ stands for the expression 
\be
\Lambda \equiv R \frac{R-\lambda J_0}{R_Q} \bigg(\frac{\hat{\phi}}{R_v}-\frac{\widetilde{\phi}_0}{R_Q} \bigg)
+ \bigg(2 Q_K +\frac{\lambda R E_R}{R_Q}\bigg)\frac{t}{R_Q} 
\ee
where  the explicit time-dependent term originates in the linear time dependence of $\X^{cls}_{J\bar T}(t)$, written in \eqref{timedepflowop},
%{\color{ForestGreen}(I don't understand this part)} \textcolor{red}{ Why? We can quote eqn 3.4, if that makes it clearer. } 
which is indicated in the argument of $\D_\l$. We will henceforth drop this explicit notation from the covariant derivative, as $\D_\l$ will always stand for $\D_\l(t)$.

The quantity   $\widetilde \phi_0$ appearing in $\Lambda$ is related to the \emph{spatial} zero mode, $\phi_0$, of the scalar as  
\be
\widetilde \phi_0 =  \phi_0-\frac{\lambda R}{R_v}\int_0^R d\sigma \left(\mathcal{J}_-+\frac{\lambda \hat{k}\mathcal{H}_R}{2}\right)\hat{\phi}
\ee
and corresponds to the generator of spectral flow in the $J\bar T$ - deformed CFT \cite{Guica:2020eab}.  More precisely,  its commutation relations with the various currents are 
\begin{align}
&\{ \widetilde{\phi}_0, \mathcal{K}_L\} = \frac{\hat{k}}{2 R} \;, \;\;\;\;\; \{ \widetilde{\phi}_0, \H_L\} = \mathcal{K}_L\;, \;\;\;\;\; \{ \widetilde{\phi}_0, {\mathcal{K}_R\} =\frac{\hat{k}}{2R}-\frac{\lambda \hat{k}}{2R_v}\partial_{\sigma}\bigg(\frac{1-\lambda\phi'}{\Sigma}\hat{\phi}\bigg)}\;, \;\;\;\;\; \nonumber\\
&\{ \widetilde{\phi}_0, \mathcal{H}_R\} =\frac{\mathcal{J}_-+\frac{\lambda \hat{k}\mathcal{H}_R}{2}}{R_v}+\frac{\lambda}{R R_v}\partial_{\sigma}\bigg[\bigg(\mathcal{J}_-+\frac{\lambda \hat{k}\mathcal{H}_R}{2}\bigg)\hat{\phi}\bigg]
\end{align}
At the level of the conserved charges, the above brackets translate into an action that takes $Q_n$, $\bar Q_n$ into $P_n,\bar P_n$, and the latter two into the identity \cite{Guica:2021pzy}, fulfilling the same role that the scalar zero mode,  $\phi_0$,  plays in a standard CFT. %{\color{ForestGreen}(cut the sentence in two? It's a bit long)} 

 We may now try to guess whose covariant derivative reproduces the first term on the right-hand side of \eqref{firstguessRM}. Guided by the somewhat similar left-moving case, we compute
\begin{align}
\mathcal{D}_{\lambda}\bigg(\frac{\lambda E_R}{2}\bigg(\frac{1}{R}-\frac{\lambda\hat{\phi}'}{R_v}\bigg) \bigg)&=\frac{E_R R}{2 R_Q}\bigg(\frac{1}{R}-\frac{\lambda\hat{\phi}'}{R_v}\bigg)-\frac{\lambda E_R}{2 R_v}\partial_{\sigma}\bigg[\frac{1-\lambda\phi'}{\Sigma}\Lambda \bigg]
\end{align}
Thus,  the flow of  the following combination 
%\textcolor{red}{\emph{Notation!}} %{\color{ForestGreen}($\widetilde{\mathcal{K}}_-$ just like for the LM?)}
%
\begin{align}\label{solcurrent2nc}
\widetilde{\mathcal{K}}_R\equiv\mathcal{J}_-+\frac{\lambda \hat{k}\mathcal{H}_R}{2}-\frac{\lambda \hat{k} E_R}{2}\bigg(\frac{1}{R}-\frac{\lambda\hat{\phi}'}{R_v}\bigg) 
\end{align}
is a total spatial  derivative
\begin{align}\label{flowofKtime}
\mathcal{D}_{\lambda}\widetilde{\mathcal{K}}_R&=-\partial_{\sigma}\left( \widetilde{\mathcal{K}}_R \, \frac{\Lambda}{\Sigma}  \right)
\end{align} 
To proceed, it is also useful to note that the quantity $\frac{\Lambda}{\Sigma}$, without a derivative, appears on the right-hand side of the flow of the following combination
\begin{align}
&\mathcal{D}_{\lambda}\bigg[\lambda\bigg(\frac{\hat{\phi}}{R_v}-\frac{\widetilde{\phi}_0}{R_Q} +\frac{t}{R}\frac{2Q_K}{R_Q}\bigg)\bigg]\equiv \D_\l (\l \Phi)=\bigg(\frac{1}{R}-\frac{\lambda\hat{\phi}'}{R_v}\bigg)  \frac{\Lambda}{\Sigma}
\end{align}
where the explicitly time-dependent term on the left-hand side has been adjusted to reproduce  the time dependence of $\Lambda$ on the right-hand side. One may easily show that
%Denoting the argument of the covariant derivative {\color{ForestGreen}above} as $\l \Phi$, one may easily show that \emph{Check!} {\color{ForestGreen}(for me it doesn't work)} \textcolor{red}{You mean, b/c of the tilde? It seems it'd be correct with a tilde, no?}{\color{ForestGreen}(I rechecked now and it's ok, but I would keep 3.41 also)}

\be
\D_\l \left[ (\l \Phi)^n  \widetilde{\mathcal{K}}_R \right] = n (\l\Phi)^{n-1} \widetilde{\mathcal{K}}_R \, \frac{\Lambda }{R \Sigma} - \p_\s \left[(\l \Phi)^n \widetilde{\mathcal{K}}_R\,\frac{\Lambda}{\Sigma}  \right]
\ee
%
%\begin{align}
%&\mathcal{D}_{\lambda}\bigg(\frac{R^n}{n!}\partial_{\sigma}^{n}(\lambda^{n}\Phi^{n}\widetilde{\mathcal{K}}_R)\bigg)&=\partial_{\sigma}^{n} \bigg(\frac{\lambda^{n-1}\Phi^{n-1}\Lambda R^{n-1}\widetilde{\mathcal{K}}_R}{(n-1)! \Sigma}\bigg)-\partial_{\sigma}^{n+1}\bigg(\frac{\lambda^n\Phi^{n}R^n\Lambda\widetilde{\mathcal{K}}_R}{n! \Sigma} \bigg)
%\end{align} 
Using the fact that $\mathcal{D}_{\lambda}$ and $\partial_{\sigma}$ commute, it is then trivial  to see that the expression for the flowed current is 
\begin{align}\label{solcurrent2}
\boxed{\widetilde{\mathscr{K}}_R = \sum_{n=0}^{\infty} \frac{1}{n!} \partial_{\sigma}^n\big(\lambda^n\Phi^n R^n\widetilde{\mathcal{K}}_R\big) }
\end{align}
since $\D_\l  \widetilde{\mathscr{K}}_R =0$ and it  reduces to $\J_-$ at $\l=0$.
%which produces the desired covariantly constant chiral current. 
Such infinite formal sums also appeared in the solution for the analogous flowed currents in $T\bar T$ - deformed CFTs \cite{Guica:2022gts}.

The associated right-moving flowed affine charges  are the Fourier modes of this current 
%\textcolor{red}{Factors $2\pi$?}{\color{ForestGreen}(yes, I added below)}
%\be
%\tilde{\bar{\mathcal{P}}}_m = \int d\s e^{-i m (\s-t)} \bar{\mathscr{K}}
%\ee
%
\begin{align}\label{resumexpp}
\widetilde{\bar{P}}_m&=\int^R_0 d\sigma \,e^{2\pi im \frac{t-\sigma}{R}}\bigg[ \sum_{n=0}^{\infty} \frac{1}{n!} \partial_{\sigma}^n\big(\lambda^n\Phi^n R^n\widetilde{\mathcal{K}}_R\big) \bigg]=\int^R_0 d\sigma\, e^{ 2\pi i m \frac{t-\sigma}{R}} \bigg[ \sum_{n=0}^{\infty} \frac{(-im \lambda \Phi)^n}{n!}\bigg]\widetilde{\mathcal{K}}_R\nonumber\\
&=\int^R_0 d\sigma \,  \exp\left(-2\pi im \frac{\sigma-t-\lambda R\Phi}{R}\right)\, \widetilde{\mathcal{K}}_R
\end{align}
where in the second equality we have integrated by parts the $\s$ derivatives,  and in the third we found that the result resums nicely  into a shift of the  exponent, which effectively behaves as an \emph{emergent} (normalised)  right-moving coordinate

\be
\hat v =\frac{ \s - t - \l  R   \Phi }{R} = \frac{\s -t - \l (\phi-\varphi_0) }{R_v} \equiv \frac{v}{R_v} \label{emfdepcoord}
\ee
where 
\be \label{exprzeromode}
\varphi_0 \equiv \frac{R_v }{R_Q} \left[ \widetilde \phi_0 - \left(\frac{Q_K}{R} + \frac{\bar Q_{\bar K}}{R_v}\right) t \right]
\ee
 One may easily check that, interestingly, the  $\varphi_0$ so defined is the spatio-temporal zero mode of the field $\phi$, in that it is  $\s$ - independent (by construction) and  also it satisfies

\be
\p_t \varphi_0 - \{ H, \varphi_0\} =0
\ee
The  right-moving field-dependent coordinate $v$ thus contains $\phi$ - \emph{except} its zero mode - which makes it into a \emph{non-local} function of the fields. 
%{\color{ForestGreen}%Let us remark that the expression of $v$ is non-local, because of the extraction of the zero mode of $\phi$.
  Writing the generators \eqref{resumexpp} in terms of $\widetilde{\mathcal{K}}_R$, rather than  $\widetilde{\mathscr{K}}_R$, amounts to transferring part of the non-locality of the conserved current onto the field-dependent coordinate. The right-moving affine charge is conserved by construction.

Finally, let us  also consider the generators of the (pseudo)conformal transformations. For the left-movers, it is easy to show that 
%\emph{\textcolor{red}{Complete!}} 
\begin{align}
\mathcal{D}_{\lambda}\mathcal{H}_L&=\frac{E_R}{R_Q}\mathcal{K}_L
\end{align}
Using \eqref{flowLMcurrent} and \eqref{flowER}, we obtain
\begin{align}
\mathcal{D}_\l  \bigg(\H_L - \frac{\lambda E_R}{R}\mathcal{K}_L + \frac{\lambda^2\hat{k} E_R^2}{4 R^2} \bigg) =0
\end{align}
from which we read off:
\begin{align}\label{solcurrent3}
\boxed{\widetilde{\mathcal{H}}_L =\H_L - \frac{\lambda E_R}{R}\mathcal{K}_L + \frac{\lambda^2 \hat{k} E_R^2}{4 R^2}}
\end{align}
For the right-movers, we start from the flow of $\mathcal{H}_R$, which can be expressed, using the same notations as before, as 
%\emph{\textcolor{red}{Change to new notation! }} 
\begin{align}
\mathcal{D}_{\lambda}\mathcal{H}_R&=-\partial_{\sigma}\bigg[ \frac{\mathcal{H}_R}{\Sigma}\Lambda \bigg] +\frac{R E_R}{R_v R_Q}\mathcal{K}_R+\frac{w}{R_v}\mathcal{H}_R
\end{align}
This expression suggests considering the flow of
\begin{align}
\mathcal{D}_{\lambda}\bigg(\frac{R_v\mathcal{H}_R}{R}\bigg)=-\partial_{\sigma}\bigg[\frac{(R_v\mathcal{H}_R/R)\Lambda}{\Sigma}\bigg]+\frac{E_R}{R_Q }\mathcal{K}_R %\bigg(\mathcal{J}_-+\frac{\lambda \hat{k}\mathcal{H}_R}{2}\bigg)
\end{align}
One can also show that
\begin{align}
&\mathcal{D}_{\lambda}\bigg[\frac{\lambda E_R\widetilde{\mathcal{K}}_R}{R}+\frac{\lambda^2 \hat{k} E_R^2}{4R}\bigg(\frac{1}{R}-\frac{\lambda\hat{\phi}'}{R_v}\bigg)\bigg]=-\partial_{\sigma}\bigg[\frac{\Lambda}{\Sigma}\bigg(\frac{\lambda E_R\widetilde{\mathcal{K}}_R}{R}+\frac{\lambda^2 \hat{k}E_R^2}{4R}\bigg(\frac{1}{R}-\frac{\lambda\hat{\phi}'}{R_v}\bigg)\bigg)\bigg] +\frac{E_R}{R_Q}\, \mathcal{K}_R %\bigg(\mathcal{J}_-+\frac{\lambda \hat{k}\mathcal{H}_R}{2}\bigg)
\end{align}
Using these relations, it is 
 easy to find a combination that obeys the same type of flow equation as  $\widetilde{\mathcal{K}}_R$ did  before 
\begin{align}\label{solcurrent4nc}
\widetilde{\H}_R\equiv \frac{R_v \mathcal{H}_R}{R}-\frac{\lambda E_R}{R}\mathcal{K}_R+\frac{\lambda^2 \hat
k E_R^2}{4R}\bigg(\frac{1}{R}-\frac{\lambda\hat{\phi}'}{R_v}\bigg)
\end{align}
namely\be
\mathcal{D}_{\lambda} \widetilde{\H}_R =-\partial_{\sigma}\bigg(\frac{\widetilde{\H}_R\Lambda}{\Sigma}\bigg)
\ee
whose solution is, as before \be \label{solcurrent4}
\boxed{\widetilde{\mathscr{H}}_R = \sum_{n=0}^{\infty} \frac{1}{n!} \partial_{\sigma}^n\big(\lambda^n\Phi^n R^n\widetilde{\mathcal{H}}_R\big) }
\ee
As before, the expression for the conserved charges can be integrated by parts to exhibit the emergence of the field-dependent coordinate \eqref{emfdepcoord} 
%{\color{ForestGreen}(Did we rescale them by R? should we integrate from 0 to R or from 0 to 1?) \textcolor{red}{No, we just rescaled the $L$'s, and only in the intro and  for cosmetic purposes. I would always integrate to $R$}
\begin{align}
\widetilde{\bar{{Q}}}_m&=\int^R_0 d\sigma \,  e^{-\frac{2\pi i m\, V}{R} }\, \widetilde{\mathscr{H}}_R=\int^R_0 d\sigma \,  e^{- \frac{2\pi i m \, v}{R_v}}\, \widetilde{\mathcal{H}}_R
\end{align}
%}
%which  matches the expression \eqref{} for the spectrally flowed pseudoconformal charges.
Thus,  the field-dependent coordinate that we  previously  proposed  in order to be able to construct right-moving conserved charges  is a direct consequence of the flow equation.

The flowed charges we built in this section satisfy a (Witt-Kac-Moody)$^2$ algebra, by construction. In the table below, we collect the expressions for all the  flowed generators (after performing the recommended  integrations by parts for the right-moving ones), which we now write in terms of a set of arbitrary periodic functions of $\hat U=U/R$ and $\hat v =v/R_v$, rather than in the  
 Fourier basis
 %\textcolor{red}{I would remove the top row and last column, and write expl the $d\s$}
\begin{table}[H]
{\tabulinesep=1.35mm
\begin{tabu}{|c|c|}
\hline
\textbf{\small{affine LM}}    &  $\widetilde{P}_{\eta}=\int^R_0 d\sigma\;\eta(\hat{U})\left(\mathcal{J}_+ + \frac{\lambda\hat{k}}{2}\mathcal{H}_R-\frac{\lambda \hat{k} E_R}{2R}\right)$  \\ \hline
\textbf{\small{affine RM}}    &  $\widetilde{\bar{P}}_{\bar{\eta}}=\int^R_0  d\sigma\;\bar{\eta}(\hat{v})\left[\mathcal{J}_- + \frac{\lambda \hat{k}}{2}\mathcal{H}_R-\frac{\lambda \hat{k}E_R}{2}\left(\frac{1}{R}-\frac{\lambda\hat{\phi}'}{R_v}\right)\right]$              \\ \hline
\textbf{\small{conformal LM}} &   $\widetilde{Q}_{f}=\int^R_0 d\sigma f(\hat{U})\left[\mathcal{H}_L -\frac{\lambda E_R}{R}\left(\mathcal{J}_+ + \frac{\lambda \hat{k}}{2}\mathcal{H}_R\right)+\frac{\lambda^2 \hat{k}E_R^2}{4R^2}\right]$    \\ \hline
\textbf{\small{(pseudo)conformal RM}} & $\widetilde{\bar{Q}}_{\bar{f}}=\int^R_0 d\sigma \bar{f}(\hat{v})\left[\frac{R_v\mathcal{H}_R}{R} -\frac{\lambda E_R}{R}\left(\mathcal{J}_- + \frac{\lambda \hat{k}}{2}\mathcal{H}_R\right)+\frac{\lambda^2\hat{k} E_R^2}{4R}\left(\frac{1}{R}-\frac{\lambda\hat{\phi}'}{R_v}\right)\right]$             \\ \hline
\end{tabu}
\caption{List of all the flowed generators}
\label{table:generators}
}
\end{table}
%{\color{blue}for arbitrary \textbf{periodic} functions $\eta,\bar{\eta},f,\bar{f}$ of their arguments, which is required for consistency (if they are not periodic we cannot integrate on quantities on sigma).}

\noindent Given these explicit expressions,  it is all but natural to define a set of \emph{quasi-local} generators $P_{\eta}, \bar P_{\bar \eta}, Q_f$ and $\bar Q_{\bar f}$ as integrals of the quasi-local currents $\mathcal{K}_{L,R}$ and $\H_{L,R}$ against the various functions of $\hat U, \hat v$ above.  They will obviously be related via 
 \eqref{basisgenerators} to the flowed ones.

\section{ $J\bar T$ symmetries in the Lagrangian formalism}\label{section4:Lagrangian}

We would now like to understand the field-dependent symmetries in the Lagrangian language. We concentrate on the $J\bar T$ - deformed free boson, for simplicity.

In more detail, in \ref{section41:symtr} we translate the action of the extended symmetries from the Hamiltonian to the Lagrangian formalism, finding that all the field-dependent transformations must be accompanied by a large affine ``compensating'' one. In \ref{section42:noethercurrents}, we construct the associated Noether currents directly in the Lagrangian formalism, and discuss  which representative from each equivalence class of such conserved currents corresponds to the Hamiltonian result. Finally, in \ref{section43:commutators} we compute a set of charge commutators directly in the Lagrangian formalism, and show how one may recover the fact that the field-dependent coordinate needs to have its zero mode removed.  
%\textcolor{red}{Anything else we want to say?} {\color{ForestGreen}(maybe mention that we also look at the transformations associated to the flowed charges, if we explain what we mean by more non-local)}

\subsection{From Hamiltonian to Lagrangian transformations}
\label{section41:symtr}
In the previous section, we have found symmetry generators that are guaranteed to act correctly on the Hilbert space of the theory, or phase space in the classical limit.  The variation of the scalar field under a symmetry generated by a charge $Q_\e$ is given by %\textcolor{red}{Sign?} %the equal-time commutator/PB 
\be \label{variationundersymmtr}
\d_\e \phi = -\{ Q_\e, \phi \}
\ee
To find how the symmetries act in Lagrangian language, we simply translate the above variation  to the Lagrangian formalism, using  the fact that

\be
\dot \phi = -\{ H, \phi\} = 2 \frac{\J_- + \frac{\l \hat{k}}{2}\H_R}{\Sigma} + \phi'
\ee
where a dot denotes a time derivative, and a prime, a $\sigma$ derivative. 
This, in particular, implies that 
\be
\p_V \phi = -   \frac{\J_- + \frac{\l \hat{k}}{2}\H_R}{\Sigma}\;, \;\;\p_U \phi= \frac{\J_- + \frac{\l \hat{k}}{2}\H_R}{\Sigma} + \phi'
\ee
with $\Sigma$ defined in \eqref{definitionsigma}. In the following,   we will treat each type of symmetry in part,  concentrating on the quasi-local symmetry generators, which act more naturally on the fields of the theory than the flowed generators derived in the previous section. 

Let us explain what we mean. 
The distinction between the actions of the two sets of generators is easiest to draw for the left-moving ones. The quasi-local generators $P_\eta, Q_f$ act locally on the fields in the theory, as can be noted from \eqref{leftmovingtransf} below.  Meanwhile, the action \eqref{actionflgn}, \eqref{actionflgn2} of the flowed generators $\widetilde{P}_\eta, \widetilde{Q}_f$ is not local, because removing the zero mode of the symmetry parameter involves an integral over all space.

As for the right-moving generators, the distinction is less clear, since the associated transformations are always non-local. If we consider the ones \eqref{rightafftransf}, \eqref{rightpseudocftr} generated by the ``quasi-local'' charges, there are two sources of non-locality: the field-dependent coordinate $v$ (which is non-local, as explained in the previous section) and the coefficient of the compensating transformation, which is an integral over all space. However, the flowed right-moving generators, which lead to \eqref{rightaffineflowedaction}, \eqref{rmpseudoflowed} act even more non-locally on the fields, due to the missing  zero modes of the symmetry parameters. This hopefully justifies our choice of terminology, which effectively ignores the first two types of non-locality.

%More precisely, {\color{ForestGreen}
%We find it useful to summarize here the locality properties of the various transformations derived in this section. There are two sources of non-locality in the symmetry transformations: the different effect of constant transformations compared to the others and the explicit presence of conserved charges (integrated quantities). Nevertheless, the latter can be tracked back to the extraction of the zero mode in the field-dependent coordinate. The LM quasilocal affine and conformal generators act locally on $\phi$. Their flowed counterparts act differently depending on whether they are constant or not, thus being non-local. The action of all RM transformations is non-local. In the case of the quasilocal ones, if treating $v$ as a usual coordinate, the only source of non-locality is the presence of conserved charges. In the case of the flowed ones, additional non-locality comes from the different effect of constant ones with respect to the rest. 

\subsubsection*{Left affine and conformal symmetries}

The generators of these symmetries are 
%\emph{\textcolor{red}{Policy factor $k$!}} {\color{ForestGreen}(factors of $\frac{k}{4\pi}$ instead of $\frac{k}{2}$ if we keep the conventions from the previous section)} \textcolor{red}{One should add a hat, correct?} 

\be \label{leftmovingch}
P_\eta = \int_0^R d\s  \, \eta_p(U)  \left(\J_+ + \frac{\l \hat{k}}{2} \H_R \right) \;, \;\;\;\;\;\; Q_f = \int_0^R d\s \, f(U) \H_L
\ee
where we have added a subscript, '$p$', to the function labeling the affine charges, to emphasize the fact that it is periodic - a specification that will  be useful later in this section. However, we do not write this label on the $\eta$ appearing in $P_\eta$,  since for it there is no possibility of confusion: this function is always   periodic.  We compute  
\be \label{leftmovingtransf}
\d_\eta \phi =-\{P_\eta, \phi\} =  \frac{\hat{k}}{2}  (1-\l \p_V \phi)  \eta_p(U) \;, \;\;\;\;\;\;\;\d_f \phi=- \{ Q_f,\phi\} =   \p_U \phi f(U)
\ee
The action of $Q_f$ is noting but the standard action of a left diffeomorphism $\xi = - f(U) \p_U$. The action of $P_\eta$
can be understood as the  the standard affine shift by an arbitrary (periodic) function of the left-moving coordinate, accompanied by a right diffeomorphism  

\be
\phi \r \phi +\frac{\hat{k}}{2}\eta_p(U) \;, \;\;\;\;\;\; V \r V   +\frac{\l \hat k}{2} \eta_p (U) \label{actlmsymm}
\ee 
Note this leaves unchanged the combination   $ V - \l \phi$.  
%\textcolor{red}{I feel we should be able to change our definitions so that these ugly signs disappear.}
%{\color{ForestGreen}Since the zero mode of $\phi$ shifts with the zero mode of the function $\eta$, let's denote it $[\eta]_{zm}$, we obtain that the field-dependent coordinate shifts $v\rightarrow v+ \lambda [\eta]_{zm}$.(because $V$ shifts like this, while $(\phi-\varphi_0)$ does not shift) If we want to have the field dependent coordinate without zero mode after the transformation, we should do an extra isometry $V\rightarrow V-\lambda [\eta]_{zm}$, but this would bring an extra contribution to the charges proportional to $E_R$ which would amount for the spectral flow. (is there a problem to induce a zero mode to $v$ as long as it's not the zero mode of $\phi$ which is by definition ``field-dependent"? this should not affect the algebra in any way, right?)}  {\color{red}???} {\color{ForestGreen}(here I just wanted to say that not any zero mode in $v$ is problematic, but only the zero mode of $\phi$ in $v$. For ex, the transformation above clearly induces a zero mode in $v$, which is $[\eta]_{zm}$)}

%{\color{red} I think we should add a line with the explicit action of $\widetilde{P}_{\eta}$ and $\widetilde{Q}_{f}$. In fact, I thought we already had something like this - did it get erased?} {\color{ForestGreen}(I think we only had it for RM)}

Clearly, these transformations act locally on the scalar field. On the other hand, the action of the flowed left-moving affine and conformal charges is non-local, as we can see from the explicit expressions for the transformations they generate
\begin{align}\label{actionflgn}
\widetilde{\delta}_{\eta}\phi=-\{\widetilde{P}_{\eta},\phi\}=\frac{\hat{k}}{2}\big[\eta_p(U)-\lambda\partial_V\phi(\eta_p(U)-[\eta_p]_{zm})\big]
\end{align}
\begin{align}\label{actionflgn2}
\widetilde{\delta}_{f}\phi=-\{\widetilde{Q}_{f},\phi\}=\partial_U\phi f(U)-\frac{\lambda\hat{k} E_R}{2R}\big[f(U)-\lambda\partial_V\phi(f(U)-[f]_{zm})\big]+\frac{\lambda \partial_U\phi\partial_V\phi}{1-\lambda\partial_V\phi}[f]_{zm}
\end{align}
in which certain zero modes (denoted as $[\;]_{zm}$) of the functions involved are explicitly subtracted - a non-local procedure. Unlike \eqref{actlmsymm},  the flowed left-moving affine charges leave invariant 
%, the shift in $V$ is absent from \eqref{actlmsymm} when $\eta(U) = const.$, which leads to the fact that 
the field-dependent coordinate

%Thus, the action on the scalar field is not local,  {\color{red}Can we make this fully sharp?} but the field-dependent coordinate {\color{ForestGreen}(since in $v$ we remove the zero mode of $\phi$, does it make sense to say that $v$ itself is non-local? and then automatically the flowed generators which leave $v$ invariant are non-local)}

\be
v = V-\l (\phi - \varphi_0)
\ee
including when $\eta_p(U) = const.$. 
%is left invariant by all the \textcolor{red}{flowed} left affine transformations, including the constant ones. 
This will be important when we study the charge algebra in section \ref{section6:asg}.
 %
 %is required by having zero commutation relations between $Q_0$ and the field-dependent charges {\color{ForestGreen}(RM charges)}; clearly, the currents are invariant under constant shifts of $\phi$, so all we need is that $v$ be also invariant.}  Conversely, we could have inferred the above expression for the field-dependent coordinate from the requirement that it be invariant under constant shifts of $\phi$. 
%{\color{ForestGreen}(anytime we have a theory for which we can express the deformed action as the undeformed one in terms of field-dependent coordinates we can consider transformations which leave the (hatted) field-dependent coordinates fixed and then the symmetries should be those of the undeformed theory in the new coordinates, correct? these should be the flowed charges)}  {\color{red}Don't see why.} 
%
%{\color{ForestGreen} I was thinking that we can change coordinates $(u=U,v)$ and then $\partial_v\phi=\frac{\partial_V\phi}{1-\lambda\partial_V\phi}$ and $\partial_u\phi=\frac{\partial_U\phi}{1-\lambda\partial_V\phi}$ so the last term is a diffeo on $v$ with $\lambda[f]_{zm}\partial_U\phi$}
Note that for the flowed left-moving  conformal transformations, the non-locality is also manifest  in the explicit presence of $E_R$ in the symmetry transformation.

\subsubsection*{Right affine symmetries}

The expression for  the right-moving affine generators is 
%{\color{ForestGreen}(same comment as before if we want to keep $k$ arbitrary; we can introduce $\hat{k}$)} 

\be \label{rmaffch}
\bar P_{\bar \eta} = \int_0^R d\s \, \bar \eta_p \left(\hat v\right) \left( \J_- + \frac{\l \hat{k}}{2} \H_R \right) \;, \;\;\;\;\; \hat v \equiv \frac{v}{R_v} = \frac{V - \l \phi + \l \varphi_0}{R_v}
\ee
where the expression for $\varphi_0$ is given in \eqref{exprzeromode} and we have again added a subscript `$p$' to the function parametrizing the transformation, to emphasize it is a periodic function of its argument (lest $\bar P_{\bar \eta}$ is not conserved). %{\color{ForestGreen}(only in the case of RM affine we will write this p index, since all the other functions are periodic; hmm actually not true since $\eta$ can be nonperiodic but only when compensating.)}
The change of the scalar under this transformation is  

\be
\d_{\bar \eta} \phi = -\{\bar{P}_{\bar \eta} , \phi\} = \frac{ \hat{k}}{2}  (1-\l \p_V \phi) \bar \eta_p - \l \bar{P}_{\bar \eta'} \left\{ \frac{\varphi_0}{R_v} ,\phi\right\}
\ee
where the prime in $\bar{P}_{\bar{\eta}'}$ denotes a derivative with respect to $\hat{v}$. The first term is just the standard transformation of the scalar under a right affine transformation, similar to the one above. The second term is specifically the contribution of the zero mode of $\phi$ that consistency required to be   subtracted from the field-dependent coordinate. 
%{\color{ForestGreen}We notice that this term is charge-dependent. Having charge-dependent coefficients in the symmetry transformations leads to non-linear symmetry algebras.} \textcolor{red}{Is this a general statement?}  {\color{ForestGreen} (I think so because the Poisson bracket is defined as a variation of a charge. The charge depends on the field, so if the field variation has some charge-dependent coefficient it will go out of the integral, while the rest should integrate to another charge in the theory, assuming that the algebra closes. In this way, we get some QQ terms.)} {\color{red} I'm not sure we need to bring up this non-linear algebra stuff here.} {\color{ForestGreen}(I agree, but maybe we can say something about non-locality here since we have charge-dependent transformations. I would say we have two notions of non-locality, one is when the transformation acts differently on zero modes than on the rest and another one is when we have charge-dependent transformations)}

The corresponding commutator evaluates to 
\be \label{largeaffinetr}
% \left\{ \frac{\widetilde \phi_0 - \a t}{R-\l Q_K } ,\phi\right\} = 
 \left\{ \frac{\varphi_0}{R_v} , \phi \right\} 
 =  \frac{\hat{k}}{R_Q} (1-\l \p_V\phi) \, \hat{\mathcal{T}} \;, \;\;\;\;\;\; \hat{\mathcal{T}} \equiv \frac{1}{2} \left(\frac{U}{R}-\frac{v}{R_v}\right)
\ee
which may be interpreted as a large affine transformation that is being performed simultaneously on the left and on the right\footnote{Note that, due to the non-linearity, the coefficient of the $\eta, \bar \eta$ winding term that enters in the symmetry generator (subscript of $\d$ below) is different from the coefficient of the linear terms which shift $\phi$ %\textcolor{ForestGreen}{(checked)}
\be
 \d_{\bar \eta = \hat{v}} \phi - \d_{\eta = \hat{U}} \phi  =  \hat{k}  (1-\l \p_V \phi)\hat{\mathcal{T}} + \frac{\l Q_{\bar K}}{R_v} \{ \varphi_0, \phi\} = \hat{k}  (1-\l \p_V \phi) \hat{\mathcal{T}} \left(1 + \frac{\l Q_{\bar K}}{R_Q}\right)=\frac{\hat{k}R_v}{R_Q}(1-\l \p_V \phi) \hat{\mathcal{T}}
\ee
%{\color{blue}Note the coefficient of this transformation is $\l \bar P_{\bar \eta'}/R_v$, due to the non-linearity. } \textcolor{red}{Please mark if  you move  things around. } 
}, which is so that no additional net spatial winding is introduced in the scalar field. The field-dependent time coordinate introduced above reduces to the standard CFT time, $t$, in the $\l \r 0$ limit.

The full expression for the right-moving affine transformation is then given by  
%{\color{red} Signs!} {\color{ForestGreen}(? i think it's correct)}
\be \label{rightafftransf}
\d_{\bar \eta} \phi =  \frac{\hat{k}}{2}  (1-\l \p_V \phi) \left(\bar \eta_p  - \frac{2\l \bar P_{\bar \eta'}}{R_Q} \hat{\mathcal{T}} \right)%,\hspace{1.5cm}\hat{\textbf{t}}=\frac{1}{2}\left(\frac{U}{R}-\frac{v}{R_v}\right)
\ee
Note that, at the level of the action on $\phi$ (but not at the level of the functions that enter the generator, which do need to be periodic) this behaves as if the periodic function that parametrises the affine transformation, $\bar \eta_p$, has picked up a  winding term proportional to the affine charge associated with $\bar \eta_p'$, and similarly for the previously absent left-moving affine transformation, leading to the `total' affine transformations 

%{\color{red} Check signs!}{\color{ForestGreen}(plus in $\bar{\eta}$)}

\be  \label{comptransformations}
\bar \eta (\hat v) = \bar \eta_p  (\hat v) + \frac{\l \bar P_{\bar \eta'}}{R_Q}\,  \hat v \;\;\;\;\;\;\;\; \eta (U) =  -  \frac{\l \bar P_{\bar \eta'}}{R_Q} \, \hat U
\ee
This perspective will be particularly natural when we discuss holographic implementation of the symmetries. The fact that the field-dependent transformation needs to be accompanied by a  large compensating transformation with a charge-dependent coefficient is entirely analogous    to the way the field-dependent symmetries are implemented in $T\bar T$ - deformed CFTs  \cite{Guica:2022gts}.
% {\color{ForestGreen}A difference with respect to the case of $T\bar{T}$ is that here by construction the winding, thus also the field-dependent radius, will commute with everything.} \textcolor{red}{Later}. 

%Consequently, 
%
%\be
%\d_{\bar \eta} \phi =  - \frac{k}{2}  (1-\l \p_V \phi) \bar \eta + \frac{\l \bar{\mathcal{P}}_{\bar \eta'}}{R_v}   (\d_{\bar \eta = \frac{v_{imp}}{R_v}} \phi - \d_{\eta = \frac{U}{R}} \phi )
%\ee

 Thus, the transformation induced by $\bar{P}_{\bar \eta}$ consists of a field-dependent affine transformation, accompanied by a large shift in the value of the scalar field. %The latter is necessary in order to keep the $U(1)$ charge unchanged, as required by charge quantization. 
 To gain some physical insight into why this accompanying transformation is necessary, in appendix \ref{Appendixfreeboson} we discuss the action of the affine symmetries at the level of the $J\bar T$ - deformed free boson solution. As we explain there, the on-shell scalar field has a mode expansion in $\hat U, \hat v$ whose coefficients are precisely the conserved affine charges $P_n, \bar P_n$. In particular, the coefficients of the linearly growing modes with $\hat U $ and, respectively, $ \hat v$ are 
 
 \be \label{relationundefdef}
P_0= Q_K = J_0 + \frac{\l \hat{k} H_R}{2} \;, \;\;\;\;\;\;\;\;\bar P_0= \bar Q_{\bar K} = \bar J_0    + \frac{\l  \hat{k} H_R}{2}
 \ee
The necessity of the large affine compensating  transformation \eqref{largeaffinetr} can be understood as a consequence the fact that, under a right-moving affine transformation, $\d H_R \propto \bar P_{\bar \eta'}$, while we are requiring that $\d J_0 = \d \bar J_0 =0$ to respect shift charge quantization, which leads to a change in the linearly growing modes. % \textcolor{red}{Do we have a similar understanding for $T\bar T$?}
% 
 % Perhaps one way to make sense of this is to first have a large gauge transformation (of the rough form $\phi \r \phi - t H_R$) that removes the zero mode, followed by the usual afine transformation. 
% {\color{blue} One can also check that 
%
%\be
%\{ \varphi_0, \bar P_{\bar \eta}\} =0 \;\;\; \mbox{for} \;\;\;\; \eta \neq I
%\ee
%so the zero mode of the field is unchanged in all but the constant transformations.}

We may also compute the change in the scalar field under a flowed right affine transformation
\begin{align} \label{rightaffineflowedaction}
\widetilde{\delta}_{\bar{\eta}}\phi&=\frac{1}{2}\big[\bar{\eta}_p-\lambda\partial_V\phi(\bar{\eta}_p-[\bar{\eta}_p]_{zm})\big]-\frac{\lambda \bar{P}_{\bar{\eta}'}}{R_Q}(1-\lambda\partial_V \phi)\hat{\mathcal{T}}
\end{align}
Again, the result only differs from \eqref{rightafftransf} if $\bar \eta_p$ has a constant mode. If $\bar \eta_p$ is constant,  one does \emph{not} perform an accompanying $V$ translation when the affine transformation shifts the scalar field.
 This non-uniform behaviour with respect to the zero mode of the symmetry parameter makes the field variation  non-local, even 
 if we ignore the charge-dependent term and we treat $v$ as a usual (quasi-local) coordinate. %, the transformation is because of the subtraction of the zero mode.

%We can see that the effect is to subtract the zero mode of the function from the transformation of $V$
%\begin{align}
%\phi\r \phi+\eta + \bar \eta \;, \;\;\;\;\; V\rightarrow V + \lambda [\bar{\eta}(\hat{v})]_{nzm}
%\end{align}

\subsubsection*{Right pseudoconformal transformations}
Finally, we turn to  the transformation of $\phi$ under the right-moving pseudoconformal symmetries generated by 
\be
\bar{Q}_{\bar f} = \int_0^R d \s \bar f (\hat{v}) \H_R 
%\;, \;\;\;\;\;\; v_{imp} = \s -t - \l (\phi - \varphi_0)
\ee
An almost identical computation as that for the right-moving affine symmetries leads to 
\be
\d_{\bar f} \phi \equiv -\{ \bar{Q}_{\bar f} ,\phi\} = -\bar f (\hat{v}) \p_V \phi- \l \bar Q_{\bar f'} \left\{ \frac{\varphi_0 }{R_v} ,\phi\right\}
\ee
Using \eqref{largeaffinetr}, we obtain  
\begin{align}\label{rightpseudocftr}
\delta_{\bar{f}}\phi&=-\bar{f}(\hat v)\partial_V\phi -\frac{\lambda \hat{k} \bar{Q}_{\bar{f}'}}{R_Q}(1-\lambda\partial_V \phi)\hat{\mathcal{T}}
\end{align}
The first term is simply the transformation of the scalar under a field-dependent  change of the $V$ coordinate. The second term corresponds, as before, to an  accompanying large affine transformation whose coefficient is proportional to the change, $\bar Q_{\bar f'}$, in the right-moving energys and ensures, as before, that the first transformation does not affect the quantized $U(1)$ charge. 

%because the corresponding expansion coefficients depend on $H_R$, whose variation is precisely $\bar Q_{\bar f'}$, and not taking this into account would yield to a violation of $U(1) $ charge quantisation. 

%This corresponds to a RM field-dependent transformation, accompanied by a winding term and an affine transformations with  $\bar \eta$ and $\eta$ not  periodic, but have a winding term proportional to the charge {\color{ForestGreen}(mention that in 4.9 we already considered this?)}. This is more natural from the holographic perspective. In the appendix, we work out the action of the above generators on a free $J\bar T$ - deformed scalar, and try to understand there the physical interpretation of the compensating transformation. 
Let us also compute the corresponding flowed transformations 
\bea \label{rmpseudoflowed}
\widetilde{\delta}_{\bar{f}}\phi&=&-\frac{R_v}{R}\bigg(\bar{f}\partial_V\phi+\frac{\lambda \hat{k} \bar{Q}_{\bar{f}'}}{R_Q}(1-\lambda\partial_V \phi)\hat{\mathcal{T}}\bigg)+\frac{\lambda\partial_V\phi}{R}\bar{P}_{\bar{f}}- \nonumber\\&& \hspace{1.2cm}-\frac{\lambda E_R}{2R}\bigg[\bar{f}-\lambda \partial_V\phi(\bar{f}-[\bar{f}]_{zm})+(1-\lambda \partial_V\phi)\frac{\lambda\bar{P}_{\bar{f}'}}{R_Q}\hat{\mathcal{T}}\bigg]
\eea
which suffers from the same types of non-locality as  the right-moving affine case  discussed.

Equations \eqref{leftmovingtransf}, \eqref{rightafftransf} and \eqref{rightpseudocftr} tell us, in principle, 
how the field in the Lagrangian formalism should change under each type of transformation.  One should note, however, that Hamitonian and Lagrangian transformations only need to agree on-shell, whereas we are of course looking for the off-shell transformation of the fields. One concrete question that arises is whether the charge-dependent coefficients of the compensating large affine transformations, which are time-independent on-shell, should be mapped to the off-shell charge at some particular time $t$, which is time-dependent, or to the time average of the charge, which is by construction a constant. As in \cite{Guica:2022gts} for $T\bar T$, we will be choosing the latter option, which greatly simplifies the variation of the action. It would nonetheless be worthwhile to  properly understand the reason  behind this choice.

\subsection{The conserved Noether currents}
\label{section42:noethercurrents}

Given the explicit symmetry transformations in Lagrangian language, we can try to go full circle and construct the associated Noether currents directly in the Lagrangian formalism. Generally speaking, a continuous quasi-symmetry %\footnote{A quasi-symmetry is a transformation of the fields that leaves the Lagrangian invariant up to a total derivative term.}
 with parameter $\e$ is defined as a transformation that leaves the action invariant up to a boundary term, denoted $M_\e^\mu$ 
 %\emph{\textcolor{red}{Is this form sufficient to ensure the existence of useful Ward identities?}}{\color{ForestGreen}(it is sufficient to ensure a conserved current, so I think so)} \textcolor{red}{Should this be covariant?} {\color{ForestGreen}(yes)}

\be \label{variationactionbdte}
\d_\e S = \int d^d x \, \sqrt{g}\,  \nabla_\mu M^\mu_\e
\ee
At the same time, under a general variation, the action transforms as

\be
\d S = \int d^d x \, \sqrt{g}\, \bigg(E \d \phi + \nabla_\mu \Theta^\mu (\phi,\d\phi) \bigg) \label{genvaract}
\ee 
where $E$ are the equations of motion and $\Theta^{\mu}$ is the presymplectic potential, defined via the above equation. %  collects the boundary terms obtained in varying the action to reach the eom.
 It follows that the quantity 
%\textcolor{red}{Sign?} {\color{ForestGreen}($\Theta-M$ to get the hamiltonian left affine charges)}
 %
 \be \label{defnoethercurrent}
 J^\mu \equiv   M^\mu_\e - \Theta^\mu (\phi,\d_\e \phi)
 \ee
is conserved on-shell, since by construction it satisfies 
%{\color{ForestGreen}(covariant)}

\be \label{standarddivergence}
\nabla_\mu J^\mu = E \d_\e \phi
\ee 
This is the standard (canonical) Noether current associated to the given symmetry transformation. As is well known, this current is ambiguous up to the addition of exact terms  $J^{\mu}\rightarrow J^{\mu}+\nabla_{\nu}k^{[\mu\nu]}$, as well as terms proportional to the equations of motion.   Another type of ambiguity that will be relevant below is that any current whose divergence is proportional to the equations of motion (not necessarily with the above proportionality coefficient) is also conserved. These ambiguities are important, as they can affect the well-definiteness of the charges, as well as their algebra.% {\color{violet}[Of course, one should also check that $M_\e$ is such that these transformation correspond to invariances of the action by  understanding how they act directly in the Lagrangian formulation.]}

% {\color{blue} To be more precise, there are some ambiguities in the construction above, since the 1-to-1 correspondence is at the level of equivalence classes, namely between equivalence classes of transformations and equivalence classes of conserved currents. We can distinguish two cases. First, we notice that if we modify $J^{\mu}$ by an exact term and/or a term $t^{\mu}$ which is 0 on-shell $J^{\mu}\rightarrow J^{\mu}+\partial_{\nu}k^{[\mu\nu]}+t^{\mu}$, the  divergence $\partial_{\mu}J^{\mu}$ still vanishes on-shell. The associated conserved charges clearly do not see these ambiguities. \textcolor{red}{Yes, they see the k if the manifold has a boundary or if k has winding (our case).}{\color{ForestGreen}(yes, I agree)} (cite advanced gr lecture notes) Second, if we change the divergence itself with a term proportional to the eom, of course it is still 0 on-shell, but in this case we see this ambiguity appearing in the corresponding conserved charges, so it's not as harmless as the first case. It amounts for changing $J^{\mu}$ by the primitive of some quantity proportional to the eom, which in principle can have terms proportional to the coordinates/non-periodic terms; nothing guarantees that the corresponding charges and algebra are well-defined (very badly written, but I want to emphasize that we are doing some integrals which in principle can give currents that do not make sense)
%}

Let us now particularize the discussion to the $J\bar{T}$-deformed free boson, whose  action is given by %(but in the previous section we had a minus sign in the action)} 
%{\color{ForestGreen}(factor of $\hat{k}$?)}
%The action of the $J\bar T$ - deformed free boson  is given by 
%
\be \label{actiondefboson}
S=- \, \hat{k}^{-1}\int dU d V \frac{\p_U\phi \p_V \phi}{1-\l \p_V\phi}
\ee
where the prefactor is due to our having rescaled $\phi$ in order to generate an arbitrary level. The components of the presymplectic potential read

\be \label{generalpresymplectic}
\Theta^U (\d\phi) =-   \frac{ 2 \,  \hat k^{-1} \p_V \phi}{1-\l \p_V \phi} \, \d\phi \;, \;\;\;\;\;\; \Theta^V (\d \phi) =-  \frac{ 2\,   \hat k^{-1} \p_U \phi}{(1-\l \p_V \phi)^2} \d \phi
\ee
We will be taking the coordinates to be inert under the various transformations, so the variation of the action is computed as 
 
\be
\d S = -\, \hat{k}^{-1}\int dU dV \left( \frac{\p_U \d \phi \; \p_V \phi}{1-\l \p_V \phi} + \frac{\p_U \phi \p_V \d \phi}{(1-\l \p_V \phi)^2} \right) %{\color{blue}  \equiv \int dU dV \frac{1}{2} \d \mathcal{L}}
\ee
Finally, for future use, we also write the ``normalised equation of motion'', $E$, defined via the variation \eqref{genvaract} 
\be \label{definitionofE}
E = \frac{4}{\hat k (1-\l \p_V \phi)} \p_V \left(\frac{\p_U\phi}{1-\l \p_V\phi}\right)
\ee
For each type of transformation in part, we will show that  the variation of the action takes the form \eqref{variationactionbdte} for some $M_\e$, and then compute the associated  Noether current.

\subsubsection*{Left affine transformations}

 For left affine transformations, we have 
 %{\color{ForestGreen}(I switched signs and rechecked until 4.41, according to the sign of the action in the previous section)} \emph{\textcolor{red}{Maybe, let's discuss}.}
\begin{align}
\Theta^U&=- \, \eta(U) \, \partial_V\phi \hspace{1.5cm}\Theta^V=-\,\eta(U)\, \frac{\partial_U\phi}{1-\lambda\partial_V\phi}
\end{align}
The variation of the Lagrangian is a total derivative %\textcolor{red}{Is this also valid at the non-linear level?}
%(${\color{red} \times 2}$)
%
\be
\p_\eta \mathcal{L}=\, \l \p_V\left(\eta(U) \frac{\p_U\phi \p_V \phi}{1-\l \p_V\phi}\right) - \, \eta'(U) \p_V \phi
\ee
%which is clearly  for $\eta$ periodic
where the function $\eta(U)$ was left arbitrary for now. To find the associated Noether  current, we must write the above total derivatives in an intelligent way, such as 
\begin{align}
\delta_{\eta}\mathcal{L}&=-\,\partial_U(\eta\partial_V\phi)+\,\partial_V\bigg[\eta \partial_U\phi+\eta \frac{\lambda\partial_U\phi\partial_V\phi}{1-\lambda\partial_V\phi}\bigg]
\end{align}
For this choice 
\be
M^U_\eta =-\eta\partial_V\phi \;, \;\;\;\;\; M^V_\eta = \eta\bigg(\partial_U\phi+\frac{\lambda\partial_U\phi\partial_V\phi}{1-\lambda\partial_V\phi}\bigg)
\ee
and using \eqref{defnoethercurrent} and the fact that $\eta = \eta_p$,  we find the correct Noether current, denoted  $K_\eta$  %{\color{ForestGreen} (I'd keep the ref in order to be clear that this is computed as J above)}
%(with the definition $J=M-\Theta$) (${\color{red} \times 2}$)
%\textcolor{red}{Shouldn't we call it K?
\be
K^U_\eta =0 \;, \;\;\;\;\; K^V_\eta =2\eta_p \frac{\partial_U\phi}{1-\lambda\partial_V\phi}
\ee
which agrees with the previous expression \eqref{defK}. %, \eqref{leftmovingch}  with the Hamiltonian (see also \eqref{defK})
% (we have $J^t=-(...)$ so the charge computed from $J_t$ is positive). ({\color{red} Cand cobori indicele, $J_U = \p_U \phi = \J_+$ la $\l=0$. Deci, iti trebuie factorul de 2 in definitie, care vine din $\sqrt{g}$. Asta o sa-ti dea un curent identic cu cel pe care l-ai folosit in Hamiltonian formalism. })
 Different choices of how to split the derivatives among the two $M_\eta^\a$ terms will correspond to various improvements of the Noether current. 
%\emph{Correct?} (yes)

\subsubsection*{Right affine transformations}

For right affine transformations, corresponding to 
%\emph{Better notation?}{\color{ForestGreen}(I am ok with writing  $\frac{\bar{P}_{\bar{\eta}'}}{R_Q}$ and the factor of 2 in "time")}
%{\color{blue}
%\be \label{rightaffinevar}
%\d_{\bar \eta} \phi = - \frac{1}{2} (1-\l \p_V \phi) (\bar \eta - \l \bar{\mathcal{P}} \tilde t )\;, \;\;\;\;\; \bar{\mathcal{P}} = \frac{\bar P_{\bar \eta'}}{R-\l Q_K} \;, \;\;\;\; (2?) \tilde t = \frac{U}{R} - \frac{v}{R_v}
%\ee}
%{\color{ForestGreen}(if we want to write it also here)}
\begin{align}\label{rmafftr}
\d_{\bar \eta} \phi =  \frac{\hat{k}}{2}  (1-\l \p_V \phi) \left(\bar \eta_p  - \frac{2\l \bar P_{\bar \eta'}}{R_Q} \hat{\mathcal{T}} \right)%,\hspace{1.5cm}\hat{\textbf{t}}=\frac{1}{2}\left(\frac{U}{R}-\frac{v}{R_v}\right)
\end{align}
we find that
\be
\Theta^U = -\p_V \phi  \left(\bar \eta_p  - \frac{2\l \bar P_{\bar \eta'}}{R_Q} \hat{\mathcal{T}} \right)\;, \;\;\;\;\; \Theta^V = -\frac{\p_U \phi}{1-\l \p_V\phi}   \left(\bar \eta_p  - \frac{2\l \bar P_{\bar \eta'}}{R_Q} \hat{\mathcal{T}} \right)
\ee
The variation of the action takes the form  %{\color{ForestGreen}(index p always?)} %{\color{red} $\times 2$}
%\textcolor{red}{Is this non-linear?}
\begin{align}
\d_{\bar \eta} \mathcal{L} &=\partial_V\bigg[\lambda \left(\bar \eta_p  - \frac{2\l \bar P_{\bar \eta'}}{R_Q} \hat{\mathcal{T}} \right)\frac{\partial_U\phi\partial_V\phi}{1-\lambda\partial_V\phi}\bigg]+\partial_U\bigg(\frac{\bar{\eta}_p}{\lambda }\bigg)+\lambda \frac{\bar{P}_{\bar{\eta}'}}{R_Q}\bigg(\frac{\partial_V\phi}{R}-\frac{\partial_U\phi}{R_v}\bigg)
\end{align}
where we used $\p_U v = - \l \p_U \phi$ and $\p_V v = (1-\l \p_V \phi)$ and treated $\bar P_{\bar \eta'}, R_Q$ as  constants. 
While the variation of the action under this transformation is clearly a boundary term, this  term is not single-valued, due to the presence of winding in the scalar field. This implies, in particular, that the last two total derivative terms in the action variation above cannot be discarded (the first two boundary terms above are  periodic in $\s$, so their spatial derivatives can be dropped)
% In particular, while the first two boundary terms above are clearly  periodic in $\s$ and their spatial derivatives can be dropped, the last two terms are not,
  and we are left with a finite remainder in the Lagrangian. % {\color{red}
   Thus, strictly speaking, the action \eqref{actiondefboson} is not invariant under the right-moving affine transformations, due to the large affine compensating term. 
    
    Note, nonetheless, that in order for a conserved Noether current to exist,
 one does not need that the action variation be exactly zero, but only  that it be a total derivative.  The fact that the variations of all conserved charges, including the energy, under the right affine transformations are finite strongly suggests that their action on the phase space of the theory is well-defined. Thus, one should expect to find a boundary term - which does not affect the Noether current \eqref{defnoethercurrent}, by definition - that would render the variation of the action finite. We leave this interesting question to future work, as well as studying  its impact on the physical consequences of the symmetries, such as the Ward identities.

We will thus concentrate on finding the conserved Noether current associated with the right-moving affine transformations. For this we need, again, to write the boundary terms arising from the action variation in an intelligent manner. 
Concentrating on the part of the action variation that only depends on $\bar \eta_p$, we may write it as
\bea
\left. \d_{\bar \eta} \L \right|_{\bar \eta_p} &= &\l  \p_V\left(\bar \eta_p(\hat v) \frac{\p_U\phi \p_V \phi}{1-\l \p_V\phi}\right) -\frac{\bar \eta'_p(\hat v)}{R_v}  \p_U \phi = \p_V \left[\bar \eta_p \left(\frac{1}{1-\l \p_V \phi}-1\right) \p_U \phi\right] - \frac{\p_V \bar \eta_p \, \p_U \phi}{1-\l \p_V \phi} \nonumber\\
&=& \p_V \left[\bar \eta_p \left(\frac{1}{1-\l \p_V \phi}-2\right) \p_U \phi\right] +\bigg( \p_V (\bar \eta_p \p_U \phi) + \frac{1}{ \l} \p_U \bar \eta_p  \bigg)
\eea
The last two terms combine into $\p_U (\bar \eta_p \p_V \phi)$ after using the relation between $\p_V \bar{\eta}_p$ and $\p_U \bar{\eta}_p$ that follows from the definition of the field-dependent coordinate. %, which satisfies $\p_U v = - \l \p_U \phi$ and $\p_V v = (1-\l \p_V \phi)$.  It is very clear these boundary terms can be dropped. \emph{True? They may contribute constants at infinity. This seems even true for the left affine transformations.} 
The contribution of the terms proportional to $\hat{\mathcal{T}}$ can be written as %{\color{ForestGreen}(with $\tilde{t}=\hat{U}-\hat{v}$, we can rescale by 2)}
%{\color{blue}
%\be
%\left. \d_{\bar \eta} \L \right|_{\tilde t} = -\frac{\lambda\bar{P}_{\bar{\eta}'}}{R-\lambda Q_K} \partial_V\bigg(\frac{\lambda}{2}\frac{\partial_U\phi\partial_V\phi}{1-\lambda\partial_V\phi}\tilde t \bigg)+\frac{\lambda}{2}\frac{\bar{P}_{\bar{\eta}'}}{R-\lambda Q_K}\bigg(\frac{\partial_V\phi}{R}-\frac{\partial_U\phi}{R_v}\bigg) 
%\ee}
\be
\left. \d_{\bar \eta} \L \right|_{\hat{\mathcal{T}}} = \frac{2\l  \bar{P}_{\bar{\eta}'}}{ R_Q} \left[ \p_V \left( \frac{\p_U \phi}{1-\l \p_V \phi} \hat{\mathcal{T}} \right)  + \p_U (\p_V\phi \hat{\mathcal{T}}) - 2 \hat{\mathcal{T}} \p_V \left( \frac{\p_U \phi}{1-\l \p_V \phi} \right) \right] 
\ee
If one ignores the last term, which is proportional to the equations of motion, one finds that the $M_\e$ one reads from the above is such that the current \eqref{defnoethercurrent} precisely coincides with   \eqref{defKb} and the current  we obtained in the Hamiltonian formalism 
%\emph{\textcolor{red}{Notation! Did you update the conventions?}} {\color{ForestGreen}(yes)}
%{\color{ForestGreen}(again with $J=\Theta-M$)}

\be \label{compnoethercurrentrm}
\bar{K}^U_{\bar{\eta}}= 2 \bar\eta_p \p_V \phi \;, \;\;\;\;\;\; \bar{K}^V_{\bar{\eta}}= 2\bar{\eta}_p\frac{\l \p_U \phi \p_V \phi}{1-\l \p_V \phi}
\ee
%{\color{ForestGreen}($\mathcal{J}_-+\frac{\lambda\mathcal{H}_R}{2}=-\frac{\partial_V\phi}{1-\lambda\partial_V\phi}(1-\lambda\phi')$ using the expressions from section 2.1, while for the LM current we obtain the correct sign)}
This current does not satisfy  \eqref{standarddivergence}, with  $E$ given in \eqref{definitionofE}, but rather 
\be
\nabla_{\mu} \bar{K}_{\bar{\eta}}^\mu  = E\left(\d_{\bar \eta} \phi + (1-\l\partial_V\phi)\frac{\hat{k}\l \bar{P}_{\bar{\eta}'}}{R_Q} \hat{\mathcal{T}} \right)
\ee
%(when we take the variation of the action, the $E\delta\phi$ term is $\frac{4}{1-\lambda\partial_V\phi}\partial_V\bigg(\frac{\partial_U\phi}{1-\lambda\partial_V\phi}\bigg)\delta\phi$ and what we get here is $\partial_{\mu}J^{\mu}=2k\bar{\eta}\partial_V\bigg(\frac{\partial_U\phi}{1-\lambda\partial_V\phi}\bigg)$, as if we would have fully removed the $\tilde{t}$ part from $\bar{\eta}$ )
Note that the shift proportional to $\hat{\mathcal{T}}$ precisely cancels the corresponding large affine term in  $\d_{\bar \eta} \phi$. Even though this is not the standard Noether current \eqref{defnoethercurrent}, its divergence is proportional to the equations of motion, so it is equally conserved on-shell.

 If one nonetheless insists upon having a current that satisfies \eqref{standarddivergence} exactly, one is forced to include winding terms in the current components e.g. 
\be\label{extrawindterm}
\Delta \bar{K}_{\bar{\eta}}^U = \frac{2  \bar P_{\bar{\eta}'}}{R_Q} \hat{v} \;, \;\;\;\;\; \Delta \bar{K}_{\bar{\eta}}^V = -\frac{4\l \bar P_{\bar{\eta}'}}{ R_Q} \frac{\partial_U\phi}{1-\lambda\partial_V\phi}\hat{\mathcal{T}}  
\ee
which render the current  ill-defined, i.e. not integrable, since it is not single-valued with respect to  $\s$. We will therefore no longer consider these terms.

%{\color{blue}We note the Noether current constructed via the standard procedure is ill-defined. We therefore consider the alternate current \eqref{}, which satisfies 

%\be
%\p_\mu J^\mu = eom\, \d\phi + \tilde t \, eom
%\ee
%which is still conserved on-shell and has the advantage of yielding the same expression for the conserved charges as the Hamiltonian formalism. We can in fact see that in the Hamiltonian formalism, the divergence of the current does not equal the eom, but there is an extra term. It would be interesting to match the two. }

The expressions \eqref{compnoethercurrentrm} that we have obtained for the currents are identical to those reviewed in section \ref{section2:review}, which were first proposed in \cite{Guica:2020uhm}. The additional contribution of this section is to derive them from the action of the $J\bar T$ - deformed boson. A non-trivial feature is that, even though the transformation of the fields must contain the large affine compensating term, this term drops out at the level of the Noether currents \eqref{defKb}, \eqref{freebosonTv}. 
Note also that, without the large affine term proportional to $\hat{\mathcal{T}}$, the action would be perfectly invariant under the field-dependent affine transformations (the main criterion used in the early work \cite{Guica:2020uhm}), but the action of the symmetries on the phase space would be inconsistent.

 \subsubsection*{Left conformal transformations} 
For the left conformal transformations, from \eqref{leftmovingtransf}, we find
\be
\Theta^U =-   \frac{ 2 \,   \hat k^{-1} \p_V \phi \p_U \phi}{1-\l \p_V \phi}f(U) \, \;, \;\;\;\;\;\; \Theta^V =-  \frac{ 2\,   \hat k^{-1}(\p_U \phi)^2}{(1-\l \p_V \phi)^2}f(U) 
\ee
The variation of the Lagrangian takes the form 
\begin{align}
\delta_f\mathcal{L}&=\partial_U\bigg(-  \frac{2}{\hat{k}} \frac{\partial_U\phi\partial_V\phi}{1-\lambda\partial_V\phi}\;f\bigg)
\end{align}
from which we extract $M^U=\Theta^U$ and $M^V=0$. Hence, we reproduce  
\begin{align}
T_f^U{}_U&=0\hspace{2cm}T_f^V{}_U=\frac{2(\partial_U\phi)^2}{\hat{k}(1-\lambda\partial_V\phi)^2}f(U) = 2 \, T_{UU} f(U)
\end{align}
in agreement with \eqref{freebosonHL}, where we denoted the generator of left translations by $f(U)$ as $T^\a_f{}_U$.  
%
%\begin{align}
%M_t(\delta_f)-\Theta_t(\delta_f)&=\frac{(\partial_U\phi)^2}{(1-\lambda\partial_U \phi)^2}f(U)  {\color{red} \hat k^{-1}}
%\end{align}
%in agreement with the Hamiltonian result which can be readoff from \eqref{freebosonHL}  {\color{red} This is not the Hamilt result. We should explain it's rescaled}. {\color{ForestGreen}(why? I thought we have a $1/\hat{k}$ in the energy-mom tensor)}
%We can also perform the transformation at
%One can easily show the action \eqref{actiondefboson}  is invariant also under the non-linear transformation  $U\rightarrow U'=F(U)$,  which implies $\p_{U'} \phi = \p_U \phi/F'(U)$. 

%{\color{blue}Working at infinitesimal level, the transformation of the fields is, which  we plug into xxx, and of the Lagrangian. The Noether current is proportional to the left stress tensor, as expected.   This transformation can also be performed at non-linear level, with $U\r U'=f(U)$, $\phi$ invariant, which implies $\p_{U'} \phi = \p_U \phi/f'(U)$, which obviously leaves \eqref{} invariant.  }
 
 \subsubsection*{Right pseudoconformal transformations}

In this case, the field transformations are given by \eqref{rightpseudocftr}, which consists of a periodic piece, proportional to $\bar f$, and a large affine one, proportional to $\hat{\mathcal{T}}$. The contribution of the presymplectic potential to the current is simply obtained by plugging this expression into \eqref{generalpresymplectic}. As before, we are working in an active picture where the field transforms, but the coordinates are inert. %\emph{Is this active or passive?} {\color{ForestGreen}(active)}.
 We are thus only left to evaluate the transformation of the Lagrangian. % Note that here, unlike in the previous subsubsections, we will be working at the infinitesimal level. 
%Using the transformation of the field \eqref{rightpseudocftr}, we can compute the variation of the Lagrangian. 
%
Concentrating  on the part that only depends on  $\bar{f}$, we can write it as
 \begin{align}
\left. \d_{\bar f} \L \right|_{\bar f} =  \partial_V\bigg(\frac{2}{\hat{k}}\frac{\partial_U\phi\partial_V\phi}{1-\lambda\partial_V\phi}\bar{f} \bigg)
\end{align}
The contribution of the terms proportional to $\hat{\mathcal{T}}$ can be written as
 \begin{align}
\left. \d_{\bar f} \L \right|_{\hat{\mathcal{T}}} =
\frac{2\l \bar{Q}_{\bar{f}'}}{ R_Q} \left[ \p_V \left( \frac{\p_U \phi}{1-\l \p_V \phi} \hat{\mathcal{T}} \right)  + \p_U (\p_V\phi \hat{\mathcal{T}}) - 2 \hat{\mathcal{T}} \p_V \left( \frac{\p_U \phi}{1-\l \p_V \phi} \right) \right]  \label{delfbL}
\end{align}
Of course, this is exactly the same contribution as for the right affine transformations, upon  replacing $\bar{Q}_{\bar{f}'}$ by $\bar{P}_{\bar{\eta}'}$. As before, if we ignore the last term in \eqref{delfbL}, which is proportional to the equation of motion,  we obtain the following Noether current, denoted $T^\a_{\bar f}{}_V$ 
\begin{align}
T^U_{\bar{f}}{}_V&=-\frac{2\bar{f}(\partial_V\phi)^2}{\hat k(1-\lambda\partial_V\phi)}= -2 \bar f \, T_{VV}\;, \hspace{1cm}T^V_{\bar{f}}{}_V=-\frac{2 \lambda\bar{f}\partial_U\phi(\partial_V\phi)^2}{\hat k (1-\lambda\partial_V\phi)^2} = - 2 \bar f \, T_{UV}
\end{align}
as expected of a generator of right-moving translations, where the components of the right-moving stress tensor are given by \eqref{freebosonTv}, up to a factor of $\hat k$. This fully reproduces the charge \eqref{RMcharges} in the  Hamiltonian analysis.  %\textcolor{red}{\emph{Sign?}} {\color{ForestGreen}(I am confused about the sign, we get $T_{tV}=\frac{1}{\hat{k}}\frac{(\partial_V\phi)^2(1-\l \phi')}{(1-\lambda\partial_V\phi)^2}$) which should give the charge in our conventions, but we had $-E_R$ the zero mode of $T_{tV}$ in the previous section. are we doing the translations in the opposite way?)}
%
%which reproduces the charges $\hat{k}^{-1}\int d\sigma \bar{f}\frac{(\partial_V\phi)^2(1-\lambda\phi')}{(1-\lambda\partial_V\phi)^2}$ {\color{ForestGreen}(rescaling)}.
As in the affine case, the divergence of this current again differs from \eqref{standarddivergence} by a term proportional to the equations of motion 
\be
\nabla_\a T^\a_{\bar f}{}_V  =E\bigg(\delta_{\bar{f}}\phi+(1-\lambda\partial_V\phi)\frac{\hat{k}\lambda\bar{Q}_{\bar{f}'}}{R_Q}\hat{\mathcal{T}} \bigg) 
\ee 
which is such that the large affine term completely cancels from the variation on the right-hand side.  Since its divergence is proportional to the equations of motion, this current is obviously conserved on-shell. 

%\textcolor{red}{Include a short discussion about conservation. We also need some appropriate notation for these currents.}

\subsection{Select charge commutators in the Lagrangian formalism}\label{subsection:commutatorsfieldth}
\label{section43:commutators}
In this subsection, we explicitly compute a few select commutators of the transformations discussed above, directly in the Lagrangian formalism. Our main aim is 
 to illustrate the following  two facts: 
\bi
\item[ i)] that the compensating large affine transformations found above are necessary to ensure  a symmetry algebra that is consistent with $U(1)$ charge quantization %(which will be non-linear)
\item[ii)] that the antisymmetry of the charge variations  requires, directly  in the Lagrangian formalism, that the zero mode of $\phi$ in the field-dependent coordinate be removed
\ei
We will concentrate again on the quasi-local charges, whose action on the fields is simpler.

For the first task, we focus on the commutators with the global  shift charge  %{\color{ForestGreen} which we defined as}
 $Q_0$, whose expression in the $J\bar T$ - deformed free scalar theory is given by \eqref{defchzm}% {\color{ForestGreen}(it differs from the one derived from the action by a factor of $\hat{k}$ because we shift in units of $\hat{k}$; $Q_0$ defined as $J_0+\bar{J}_0$ is the normally quantized one)}

\be
Q_0 = \int_0^R d\s  \left( \frac{\p_U \phi}{(1-\l \p_V \phi)^2} - \frac{\p_V \phi}{1-\l \p_V \phi}\right)
\ee
As a warm-up, let us start by computing the variation of $Q_0$ under a left affine transformation, with parameter $\eta(U)$, assumed arbitrary for now 
\be \label{variationQ0left}
\d_\eta Q_0 = \frac{\hat{k}}{2} \int_0^R d\s \left[ \p_\s \left(\frac{\eta (U)}{1-\l \p_V \phi}\right)  - \frac{2 \l \eta}{(1-\l \p_V \phi)^2}\left(  \p_U \p_V \phi + \frac{\l \p_U \phi \p_V^2 \phi }{1-\l \p_V \phi}\right)\right]
\ee
The term in the second paranthesis is  proportional to the equation of motion, so we drop it. The first term integrates to zero if $\eta(U)$ is periodic, yielding $\d_\eta Q_0 =0$, as expected.

It is easy to see that we obtain the same result for the left conformal transformations:
\begin{align}\label{changeQ0leftcf}
\delta_f Q_0 &=\int_0^R d\sigma \bigg[ \frac{1}{(1-\lambda\partial_V\phi)^2}(f\partial_U^2\phi+f'\partial_U\phi)+\bigg(\frac{2\lambda \partial_U\phi}{(1-\lambda\partial_V\phi)^3}-\frac{1}{(1-\lambda\partial_V\phi)^2}\bigg)f\partial_U\partial_V\phi \bigg]\nonumber\\
&=\int_0^R d\sigma \bigg[ \partial_{\sigma}\bigg(\frac{f\partial_U\phi}{(1-\lambda\partial_V\phi)^2}\bigg)-\frac{2f}{(1-\lambda\partial_V\phi)^2}\bigg(\partial_U\partial_V\phi+ \frac{\lambda\partial_U\phi \partial_V^2\phi}{1-\lambda\partial_V\phi} \bigg)\bigg]%=\int_0^R d\sigma \partial_{\sigma}\bigg(\frac{f\partial_U\phi}{(1-\lambda\partial_V\phi)^2}\bigg)+0_{onshell}
\end{align}
For $f$ periodic, the total derivative term integrates to zero, while the second term vanishes on-shell,  so $Q_0$ is unchanged under a left conformal transformation, as expected.

\newpage

The analogous computation of the variation of $Q_0$ under a \emph{purely right-moving} transformation $\bar \eta(v)$, which may or may not have winding  yields, up to on-shell vanishing terms that we drop %\emph{\textcolor{red}{Careful notation!!}}
\bea \label{variationQ0right}
\d_{\bar \eta} Q_0 &= & -\frac{\hat{k}}{2} \int_0^R d\s\left[ \p_\s \left(\frac{\bar \eta (v)}{1-\l \p_V \phi}\right) % + \frac{2 \l \bar \eta}{(1-\l \p_V \phi)^2}\left(  \p_U \p_V \phi + \frac{\l \p_U \phi \p_V^2 \phi }{1-\l \p_V \phi}\right)
 -2 \bar \eta (v) \,  \p_\s  \frac{1}{1-\l \p_V \phi} \right] \\
& = &   \frac{\hat{k}}{2} \int_0^R d\s  \p_\s \left( \frac{\bar \eta }{1-\l \p_V \phi}\right) -  \hat{k} \int_0^R d\s \p_\s \bar \eta  + \frac{\hat{k}}{R_v} \int_0^R d\s \bar \eta' \p_\s v \frac{\l \p_V \phi}{1-\l \p_V\phi}  \nonumber
%\\&=& \frac{k}{2} \int d\s  \p_\s \left( \frac{\bar \eta (1-2\l \p_V \phi)}{1-\l \p_V \phi}\right) + \frac{k \l}{R_v} \int d\s \frac{\bar \eta ' \p_V \phi}{1-\l \p_V \phi} (1-\l\p_U \phi-\l \p_V \phi)\nonumber \\
%&=&k \int d\s \p_\s \bar \eta - \frac{k}{2} \int d\s  \p_\s \left( \frac{\bar \eta }{1-\l \p_V \phi}\right) - \frac{k \l}{R_v} \bar{\mathcal{P}}^{KM}_{\bar \eta'}
\eea  
If $\bar \eta$ were periodic, then the first two terms would drop out, and the last term would %{\color{blue}integrate to $\l \hat{k}\bar P_{\bar \eta'}$,} 
yield% \emph{\textcolor{red}{Notation!!}}
\be
\d^{naive}_{\bar \eta_p} Q_0 =  \frac{\l \hat{k}}{R_v} \bar{P}_{\bar \eta'}
\ee
where we used \eqref{rmaffch}.  However, this commutator, which is the same as that found in \cite{Guica:2020uhm}%{\color{ForestGreen}(now it agrees with commutators)}
, is inconsistent with the quantization of $Q_0$, as pointed out in \cite{Guica:2020eab}. On the other hand, if we do include the compensating transformation present in \eqref{comptransformations}, then the second term in \eqref{variationQ0right} brings in  a winding contribution proportional to $-\l\hat{k}\bar P_{\bar \eta'}/R_Q$, while the winding piece of $\bar \eta$ also contributes to the last term a constant piece $\frac{\l\hat{k}\bar{P}_{\bar{\eta}'}}{R_Q}\frac{\l \bar Q_{\bar K}}{R_v}$. The first term in \eqref{variationQ0right} combines with the corresponding left-moving winding contribution \eqref{variationQ0left} to yield the spatial derivative of a periodic function, which can be dropped. All in all, the variation of $Q_0$ under the full transformation \eqref{comptransformations} is given by % \textcolor{red}{\emph{Notation!}} {\color{ForestGreen}(I'm ok with it)}
\be \label{correctrezq01}
\d_{\bar \eta} Q_0 =\frac{\l \hat{k} \bar P_{\bar \eta'}}{R_v}-\frac{ \l \hat{k}\bar P_{\bar \eta'}}{R_Q} + \frac{\l^2 \hat{k}\bar Q_{\bar K} \bar P_{\bar \eta'}}{R_v R_Q}  = 0
\ee 
which is the expected consistent result that also  agrees with the one obtained in the Hamiltonian formalism.

The same mechanism applies to  the right-moving pseudoconformal transformations, for which ($\bar f$ is always assumed to be periodic)
\be
\d_{\bar f} Q_0 = \;\!\!\int_0^R \!\!d\s \left[ \p_\s \left( \frac{\bar f \p_V \phi}{(1-\l \p_V \phi)^2} \right) - \frac{2 \bar f }{(1-\l \p_V \phi)^2} \left(  \p_U \p_V \phi + \frac{\l \p_U \phi \p_V^2 \phi }{1-\l \p_V \phi}\right) - \frac{2 \l \bar f \p_V \phi}{(1-\l \p_V \phi)^3} \p_V \p_\s  \phi \right]
\ee
The first term is a total $\s$ derivative of a periodic quantity, the second term is proportional to the equations of motion, whereas the last term can be identified with $\bar{Q}_{\bar f'}$ by rewriting %{\color{ForestGreen}(old: the $\hat{k}$ is necessary since we need to rescale $\phi$ and $\lambda$ as explained in the previous section to put back the factors of $\hat{k}$ that were missing; correct? do we need to put these factors anywhere else? it makes sense because the contribution to $E_R$ for ex needs to be $\# \frac{1}{\hat{k}} Q_R^2$)}
\begin{align}
\bar{Q}_{\bar f'} &= \frac{1}{\hat{k}}\int_0^R d\s \bar f' \left( \frac{(\p_V \phi)^2}{1-\l \p_V \phi} - \frac{\l \p_U \phi (\p_V \phi)^2}{(1-\l \p_V \phi)^2}\right) = \frac{R_v}{\hat{k}} \int_0^R d\s \p_\s \bar f \frac{(\p_V \phi)^2}{(1-\l \p_V \phi)^2} =\nonumber\\
&= - \frac{2 R_v}{\hat{k}} \int_0^R d\s \bar f  \frac{\p_V \phi \p_V \p_\s \phi}{(1-\l \p_V \phi)^3}
\end{align}
where the overall $\hat k$ factor is due to the rescaling of $\phi$. 
We therefore find
\be
\d^{naive}_{\bar f} Q_0 = \frac{ \l \hat{k}}{R_v} \bar{Q}_{\bar f'}
\ee
This commutator, however, is not consistent with the quantization of $Q_0$. Just like in the case of right-moving affine transformations, the full transformation includes also a right-moving affine transformation with $\bar{\eta}=\frac{\lambda \bar{Q}_{\bar{f}'}}{R_Q}\hat{v}$ and a left-moving affine transformation with $\eta=-\frac{\lambda \bar{Q}_{\bar{f}'}}{R_Q}\hat{U}$, whose total contribution to the final result is  $-\frac{\lambda \bar{Q}_{\bar{f}'}}{R_v}$. Overall, we find 
\be \label{correctrezq02}
\d_{\bar f} Q_0 =0
\ee
which is the expected answer, consistent with charge quantization. 

%{\color{blue}in agreement with the Poisson bracket analysis. This commutator, however, is not xxx. \emph{Fix it!} 
%(I don't understand this, shouldn't this variation be 0 because the undeformed charges commute with the other generators? we can fix this in the same way as above, by adding compensating large affine transformations with charge-dependent coefficient) \emph{Exactly. Can you work it out, following the model above?}
%}
%
%{\color{blue}
%
%\bi
%\item is the commutator with the winding trivial? {\color{ForestGreen}(yes, by construction. The transformations leave $v$ invariant (up to constants), so $R_v$ is also invariant)}
%\item is it interesting to also compute the commutators with $P$?
%\ei
%}

The second issue we would like to discuss concerns the zero mode of the field-dependent coordinate. The specific question that we would like to address is whether  it is possible to see, directly in the Lagrangian formalism, that the zero mode of $\phi$  needs to be removed from $v$.  The answer is affirmative and  can be illustrated on a simple commutator, e.g. that of $H_L$ with the right affine charges. This commutator can be computed in two ways: by taking the variation of $ H_L$, which for the $J\bar T$ - deformed free boson is simply given by %{\color{blue}(same question as before in this part, are we missing a factor of $\frac{1}{\hat{k}}$?)}

\be
H_L = \frac{1}{\hat{k}} \int_0^R d\s\, \frac{(\p_U\phi)^2}{(1-\l \p_V\phi)^2} \label{HLexpl}
\ee
with respect to the transformation \eqref{rmafftr} generated by $\bar \eta$, or by computing the variation of $\bar P_{\bar \eta}$ under a constant left-moving translation. We expect that

\be
\d_{\bar \eta} H_L =  -\d_{\xi_L=\p_U} \bar{P}_{\bar \eta} \label{HLPbctc}
\ee
by antisymmetry of the charge algebra. Plugging in  \eqref{rmafftr}  into the variation of \eqref{HLexpl} and using the explicit expression for the field dependent coordinate,  the contribution from the terms proportional to $\bar{\eta}_p$ can be shown to vanish on-shell. The terms originating from the compensating transformation %in $\hat{\mathcal{T}}$ 
give %(rechecked with the correct sign from the algebra of modes)
\be
\d_{\bar \eta} H_L = - \frac{\l \bar P_{\bar{\eta}'} Q_K}{R R_Q}
\ee
in perfect agreement with \eqref{algebra:LMRMaffine}. To compute the right-hand side of \eqref{HLPbctc},  we plug $\d_{\xi_L =\p_U}\phi =  \p_U \phi$   into the variation of the expression for  $\bar P_{\bar \eta}$ %\textcolor{red}{\emph{Correct ? $v$ or $\hat v$?}} {\color{ForestGreen}(yes, v here)}
\be \label{naivevarcomm}
\d_{\xi_L=\p_U} \bar{P}_{\bar \eta} = - \int_0^R d\sigma\bigg[\partial_{\sigma}\bar{\eta}_p \frac{\partial_V\phi}{1-\lambda\partial_V\phi}\delta_{\xi_L=\p_U} v -\bar{\eta}_p\partial_{\sigma}\bigg(\frac{\lambda\partial_V\phi\partial_U\phi}{1-\lambda\partial_V\phi}\bigg)\bigg]
%=\int_0^R d\sigma\partial_{\sigma}\bigg(\bar{\eta}_p\frac{\lambda\partial_V\phi\partial_U\phi}{1-\lambda\partial_V\phi}\bigg)=0
\ee
%where in the second equality we used $\delta v=-\lambda\partial_U\phi$, which follows from the assumption that $v=V-\lambda\phi$, maybe up to a constant that is \emph{not} field-dependent (zero variation). 
 Notice that if we simply take $\d v = - \l \d\phi = -\l \p_U\phi$, the above expression integrates to zero, in disagreement with our expectation.  Since $v$ is fixed only up to a constant term, $c$, the most general form of the variation is $\d v = \d c - \l \d\phi$. Requiring that \eqref{HLPbctc} hold, we find that $\d c$ should equal  
%Certainly, this assumption is wrong, as the would-be Poisson bracket is not well-defined (not antisymmetric). In order to fix this issue, we note that the result we need to obtain for the consistency of the algebra,
 $\frac{\l  Q_K R_v}{R R_Q}$, which is proportional to the variation of the zero mode of $\phi$ under a left translation  %{\color{ForestGreen}(I rechecked the factors, the only thing I am not sure about is the sign, I thought $H_L$ generates $\partial_U\phi$, using for ex ec 1 and 2 from the beginning of the section, but for $U$ tr instead of time tr.)}
\begin{align}
\delta_{\xi_L=\partial_U}\varphi_0&=[\p_U \phi]_{zm}=\frac{Q_K R_v}{R R_Q}
\end{align}
Hence, subtracting  the zero mode of $\phi$ from the field-dependent coordinate
\begin{align}\label{fielddpcoordct}
v=V-\lambda(\phi-\varphi_0)
\end{align}
brings an extra contribution to $\delta v$ in \eqref{naivevarcomm} that yields the expected result
\begin{align}
\delta_{\xi_L=\partial_U}\bar{P}_{\bar{\eta}}&=-\frac{\lambda Q_K R_v}{R R_Q}\frac{1}{R_v}\int_0^R d\sigma \bar{\eta}'_p\partial_{\sigma}v \frac{\partial_V\phi}{1-\lambda\partial_V\phi}=\frac{\lambda Q_K \bar{P}_{\bar{\eta}'}}{R R_Q}
\end{align}
Of course, in order to fully convince ourselves that \eqref{fielddpcoordct} is the correct way in which the ambiguity in the zero mode of $v$ should be fixed, we should compute other similar commutators, such as $\{H_R,\bar{P}_{\bar{\eta}}\}$, and demand again antisymmetry in the  variations.  A particularly simple argument involves the commutators related by antisymmetry to \eqref{correctrezq01}, \eqref{correctrezq02} which imply

\be
\d_{\phi \r \phi +const. } \bar P_{\bar \eta} = \d_{\phi \r \phi +const. } \bar Q_{\bar f} =0
\ee
remembering that $Q_0$ generates constant shifts in $\phi$. Since the quasi-local currents from which $\bar P_{\bar \eta}, \bar Q_{\bar f}$ are built are invariant under such shifts, we conclude that the field-dependent coordinate entering the functions $\bar \eta, \bar f$ also should. However, this can only happen if the zero mode of $\phi$ is subtracted from $\phi$ inside $v$, yielding \eqref{fielddpcoordct}.

In conclusion, we showed that it is possible to  obtain, directly in the Lagrangian formalism,  the correct zero mode of the field-dependent coordinate, by requiring well-definiteness of the charge variations.

\section{The holographic dual of $J\bar T$ - deformed CFTs}\label{section5:holography}

Having discussed the extended symmetries of $J\bar T$ - deformed CFTs in both the Hamiltonian and Lagrangian formulations, we would next like to understand them also in the dual holographic description of these theories, where they correspond to asymptotic symmetries of the dual spacetime. Given that the asymptotic symmetry group calculation is entirely independent of what we did so far, finding a match would represent  a precision test of the proposed holographic dictionary for these theories.

The holographic dictionary for   $J\bar T$ - deformed holographic (large $c$, large gap) CFTs was derived in \cite{Bzowski:2018pcy}, for the specific  case when $J$ is  a chiral current. Note, however, that in the field-theory framework we used above, the right-moving component of the $U(1)$ current cannot be entirely set to zero, even if initially (at $\l=0)$ one may require it to vanish. Therefore, in this section we slightly generalise the proposal of \cite{Bzowski:2018pcy} to a non-chiral current, which corresponds to considering two Chern-Simons gauge fields on AdS$_3$, rather than just one. As we will see, this helps clarify  certain loose ends left by the analysis of \cite{Bzowski:2018pcy}. We then recheck the dictionary works as expected in this more general case. 
%
%This section builds up on and completes \cite{Bzowski:2018pcy}, who dealt with chiral $J$. However, in the field-theory analysis, it is more natural and simpler to consider a non-chiral one, as we have in the previous sections. We will therefore review and slightly generalise the results of \cite{Bzowski:2018pcy} for the case of a non-chiral current.
%
% {\color{ForestGreen}(We will comment on the chiral case as a particular case of our analysis.)} 
 %\emph{\textcolor{red}{References everywhere!}}

As reviewed at length in \cite{Bzowski:2018pcy}, finding the holographic dual to a double-trace deformation proceeds in two steps:

\bi
\item[i)] relating the sources and expectation values of operators in the deformed and undeformed theory using large $N$ field theory %- we will use the variational principle to implement this
\item[ii)] using the \emph{undeformed} AdS/CFT dictonary to relate the  sources and expectation values in the undeformed theory to bulk quantities, and then translating the results of the first step to mixed boundary conditions for these bulk fields
\ei
We perform the first step in subsection \ref{subsection51:deformedth},  closely following the derivation of \cite{Bzowski:2018pcy}, who used the variational principle, but relaxing the chirality condition on the current. In  \ref{subsection52:holographicdictionary} we  discuss the second step, a.k.a the holographic dictionary,  paying particular attention to the Chern-Simons action that we choose. Finally, in  \ref{subsection53:checks}, we check that the resulting spectrum and thermodynamics perfectly match those of $J\bar T$ - deformed CFTs, thus improving upon the results of \cite{Bzowski:2018pcy}, who only found a match by making a certain (\emph{ad hoc}) assumption.

\subsection{Relating the deformed and undeformed theories}\label{subsection51:deformedth}

\subsubsection*{The variational principle}

In a large $N$ field theory, one may use the Hubbard-Stratonovich trick and large $N$ factorization to relate the partition function (more precisely, the generating functional of correlators) in presence of a double-trace deformation to the undeformed partition function with sources that have been shifted by the deforming operator's expectation value \cite{Gubser:2002vv}. %\textcolor{red}{Check ref!}
 As nicely explained in {\cite{Papadimitriou:2007sj}, these manipulations on the large $N$ partition function are %described already at the level of the variational principle for the (classical limit) of the generating functional of correlation functions in the deformed theory/
captured by  the variational principle satisfied by the classical limit of the generating functional   (identified with the bulk on-shell action) in presence of the deformation.   

For the discussion of the variational principle, it is an efortless generalisation to consider $JT^a$ - deformed CFTs \cite{Anous:2019osb} instead of $J\bar T$ ones, which corresponds to taking the parameter of the deformation, $\l^a$, which is a vector in tangent space, to not necessarily be null. %\textcolor{red}{Were we the first to discuss this?} 
 We will focus on sources and expectation values of the stress tensor and current appearing in the deformation, which are the mainly affected operators. As explained in \cite{Bzowski:2018pcy}, for such Lorentz-breaking deformations, it is natural to couple the stress tensor to the vielbein, rather than the metric. Unlike \cite{Bzowski:2018pcy}, we will be working in Lorentzian signature\footnote{Note  there is a relative minus sign in the definition of the stress tensor with respect to the Euclidean case.} in this section. The variation of the on-shell action (identified with the classical limit of  the deformed generating functional) takes the form  
  
  \be\label{variationgenfun}
  \d S [e_\a{}^a, \mathrm{a}_\a] =  \int d^2 x \left(e \, T^\a{}_a \d e_\a{}^a  +  e \, J^\a \d \mathrm{a}_\a \right)
  \ee
  where all the quantities   are in the deformed theory at finite $\l^a$ and thus carry, in principle, a $\l^a$ label, which we will be ommitting to not clutter the notation, unless a confusion is possible. In the above, Greek letters denote spacetime indices, while Latin letters denote tangent space ones. 
  
   The variation of the Lorentzian  on-shell action  when $\l_a $ is infinitesimally changed, %[in the conventions of \cite{Anous:2019osb} \emph{Check sign def.!} \emph{\textcolor{red}{Check conventions!}} ] 
   is given by 
 \emph{minus} the double-trace deformation of the boundary theory action \eqref{flact} %{\color{ForestGreen} (as explained in \cite{Bzowski:2018pcy})} %of the deformed action, the relation between the generating functionals in the deformed theories (identified with the on-shell action in the bulk) is 
 
 \be
 \p_{\l^a} S^{[\l^a]} %[e_\a{}^a, \mathrm{a}_\a] 
 = - \int d^2 x \, e  \, \O_{JT^a}^{[\l^a]}
 \ee
In order to ensure the deformation is well-defined  also in presence of external sources, we assume  the current $J^\a$ is  exactly conserved, i.e. it is not anomalous.  
 The requirement that $S^{[\l^a+\Delta \l^a]}$ have a good variational principle translates into being able to write %\emph{\textcolor{red}{Sign!}}

\be
\d S^{[\l^a+\Delta \l^a]} =   \d S^{[\l^a]} - \Delta \l^a \int d^2 x \, \d \left(  e   \, \e_{\a\b} J^\a T^{\b}{}_a \right)
\ee
again in the form \eqref{variationgenfun}. Plugging in, we have 
%{\color{ForestGreen} (checked)}
%
\bea \label{newsourcevev}
\d S^{[\l^a+\Delta \l^a]}\!\!  &= & \!\! \int d^2 x \left[
%\left [e \, T^\a{}_a \d e_\a{}^a + e \, J^\a \d \mathrm{a}_\a - \Delta \l^a \d (e  \,\e_{\a\b} T^\a{}_a J^\b) \right] \\&= & 
e \, T^\a{}_a (\d e_\a{}^a + \Delta \l^a \e_{\a\b} \d J^\b)  +    e J^\a (\d \mathrm{a}_\a  -   \Delta \l^a  \e_{\a\b} \d T^\b{}_a)  + \Delta \l^a T^\a{}_a \d(e \e_{\a\b}) J^\b \right] \nonumber \\
&= & \!\! \int d^2 x \left[ e \, T^\a{}_a \d (e_\a{}^a +\Delta \l^a \e_{\a\b}  J^\b)  +   e J^\a \d ( \mathrm{a}_\a  -   \Delta \l^a  \e_{\a\b} T^\b{}_a) + \right. \nonumber \\[3pt]
&& \hspace{3cm} +  \left.  \Delta \l^a T^\a{}_a  \left( \d (e\e_{\a\b}) - 2e \d\e_{\a\b} \right) J^\b \right]
\eea
%
%{\color{blue}
%Let us fix the convention (to agree with 2.26 from the holographic JTbar paper):
%\begin{align}
%S&=S_{CFT}-\int \mathcal{O}_{J\bar{T}}
%\end{align}
%\begin{align}
%\delta W&=e \, T^\a{}_a \d e_\a{}^a + e \, J^\a \d \mathrm{a}_\a + \Delta \l^a \d (e  \,\e_{\a\b} T^\a{}_a J^\b)=\nonumber\\
%&= e \, T^\a{}_a (\d e_\a{}^a + \Delta %\l^a \e_{\a\b} \d J^\b) + e J^\a (\d \mathrm{a}_\a + \Delta \l^a  \e_{\a\b} \d T^\b{}_a)  + \Delta \l^a T^\a{}_a \d(e \e_{\a\b}) J^\b=\nonumber \\
%&= e \, T^\a{}_a \d (e_\a{}^a +\Delta \l^a \e_{\a\b}  J^\b) + e J^\a \d ( \mathrm{a}_\a + \Delta \l^a  \e_{\a\b} T^\b{}_a) + \Delta \l^a T^\a{}_a  \left( \d (e\e_{\a\b}) - 2e \d\e_{\a\b} \right) J^\b 
%\end{align}
%}
Using $\e_{\a\b} = e \hat \e_{\a\b}$, where $\hat \e_{\a\b}$ is the Levi-Civita symbol, one can show the term in the last paranthesis vanishes, and thus the variation of the action $\d S^{[\l^a + \Delta \l^a]}$ again takes the form \eqref{variationgenfun}, with slightly modified sources and vevs. From \eqref{newsourcevev}, we can easily read off the  change in these data with the flow parameter $\l$, defined by writing $\l^a = \l \hat \l^a$, where $\hat \l^a$ is a unit tangent space vector or its null counterpart %\emph{Ambiguities?} %which  are extremely simple
%
%\emph{\textcolor{red}{Recheck signs!!!}}{\color{ForestGreen}(checked)}
\be \label{defundefrel}
\p_\l e_\a{}^a = \hat \l^a \e_{\a\b} J^\b \;, \;\;\;\;\; \p_{\l} \mathrm{a}_\a = -   \hat \l^a \e_{\a\b} T^\b{}_a \;, \;\;\;\;\; \p_\l (e J^\a) = \p_\l (e T^\a{}_a) =0
\ee
%where $\hat{\l}^a=\l^a/\l$. \emph{How do you define $\l$?}(if the vector is not null it's its norm; for the null vector we  can bring it in the form $(\lambda,\lambda)$ up to signs, right?)
These flow equations are trivially solved by

\be\label{generalmapdefundef}
e^{[\l]\, a}_\a = e_\a^{[0] \, a} + \l^a \e_{\a\b} J^\b \;, \;\;\;\;\; \mathrm{a}_\a^{[\l]} =  \mathrm{a}_\a^{[0]}  -   \l^a \e_{\a\b} T^\b{}_a 
\ee

\be
(e\, J^\a)^{[\l]} = (e \, J^\a)^{[0]} \;, \;\;\;\;\; (e \, T^\a{}_a)^{[\l]} = (e\,  T^\a{}_a)^{[0]} \label{relvevs}
\ee
The shifts    by the expectation values  in \eqref{generalmapdefundef} can be considered in either the undeformed or deformed theory, as implied by \eqref{relvevs}. 

These solutions are rather similar to those obtained in \cite{Bzowski:2018pcy}, if one particularises $\l^a$ to be a null vector, additionally assumes that $J^\a = \e^{\a\b} J_\b$, %(which implies, in particular, that $e^{[\l]} = e^{[0]}$) %,  a condition that is easily checked to be preserved along the restricted flow, 
and considers  a coupling of the current to $\e^{\a\b}\mathrm{a}_\b$ instead. The only expression that looks rather different from \cite{Bzowski:2018pcy} is that  for $T^{\a}{}_a$, which in our current derivation is significantly simpler. It should though be noted that the positioning of the tangent space and spacetime indices in the non-linear expression obtained in \cite{Bzowski:2018pcy} is opposite from ours, which partly explains the difference.  %\footnote{\color{magenta}{To ease comparison with the formulae of \cite{Bzowski:2018pcy}, let us rewrite \eqref{generalmapdefundef}-\eqref{relvevs} with the same placement of the indices

\subsubsection*{Solution for vanishing deformed sources }

To compare the conserved charges of the spacetime with those in the $J\bar T $ - deformed CFTs studied in the previous sections, we are mostly interested in the deformed theory on two-dimensional flat space, with no external sources coupled to the deformed currents, namely $e_{\a}{}^{a [\l]} = \d^a_\a$ and $\mathrm{a}_\a^{[\l]} =0$. Our analysis above indicates the partition function of this theory is the same as 
that of the undeformed CFT on a two-dimensional space with a nontrivial metric/vielbein and an external source for the current, given by %\eqref{generalmapdefundef} 

\be
e_\a^{[0]\, a} = \d_\a^a - \l^a \e_{\a\b} J^\b \;, \;\;\;\;\; \mathrm{a}_\a^{[0]} =   \l^a \e_{\a\b} T^\b{}_a \label{sourcesflatsp}
\ee
Since we assume that the $U(1)$ current we use to define the deformation is not anomalous%(as its anomaly would be seen already at classical level in the bulk)
, it is exactly conserved even in presence of external gauge fields, and we can globally write 

\be
 J^\a = \e^{\a\b} \p_\b \phi \label{jbos}
\ee
for some scalar field $\phi$. Note this equation holds in the deformed theory (so, all quantities carry in principle the $[\l]$ label), where the metric is Minkowski. This is the \emph{definition} of $\phi$, which simply corresponds to the bosonisation of the non-anomalous current $J$ that enters the deformation.

Let us now 
particularise the discussion to the $J\bar T$ deformation, where $\l^a$ is a lightlike tangent space vector, with components $\l^+=0,\l^-=\l$.   To avoid confusions, the tangent space indices will be denoted as $\pm$, while  the spacetime coordinates are $U,V$, as before.
 The CFT vielbein \eqref{sourcesflatsp} then takes the form 

\be
e_\a^{[0]\, a} = \d_\a^a - \l^a \p_\a \phi = \begin{pmatrix}
1 & -\l \partial_U\phi\\
0 & 1-\l \partial_V\phi
\end{pmatrix}
\ee
%where we have particularised the discussion to $J\bar T$, where $\l^a$  and
  Note that 
the form of $e^{[0]}$ indicates the CFT background \eqref{sourcesflatsp} is simply a coordinate transformation $x^a \r x^a - \l^a \phi + const$ of two-dimensional Minkowski space. 
  The inverse vielbein and the metric are %\textcolor{red}{Check positioning of the indices!} {\color{ForestGreen}(it's ok, the tangent one labels rows; the product is clearly identity summing $(\alpha a)(a \beta)=(\alpha\beta)$)}
  
  %\emph{Check!} {\color{ForestGreen}(checked)}

\be
e^{[0]\,\alpha}_a=\begin{pmatrix}
1 & \frac{\l\partial_U\phi}{1-\l\partial_V\phi}\\
0 & \frac{1}{1-\l\partial_V\phi}
\end{pmatrix} \;, \;\;\;\;\; \g_{\a\b}^{[0]} = \left(\begin{array}{cc} - \l \p_U \phi & \frac{1-\l \p_V\phi}{2}  \\ \frac{1-\l \p_V\phi}{2}  &0  \end{array} \right)  \label{defmet}
\ee
Note that $J^\a$ in \eqref{jbos} corresponds to the \emph{topological} current studied in section \ref{section2:review}%(where it  was denoted $\tilde J$)
. Note also that, when working out the flow of the conserved currents in section \ref{section3:Hamiltonian}, we used precisely the $J\bar T$ defomation driven by this topological current, so the results can be easily compared. As in that section, we expect to have an additional conserved current, roughly corresponding to shifts in $\phi$. Since, in the undeformed theory, the topological and the shift current are Hodge dual to each other, we expect this additional current, denoted $J'$, to satisfy 

\be \label{definitionjprime}
J'^{[0]}{}_\a = \e_{\a\b}^{[0]} J^{[0]\, \b}
\ee 
even in the deformed metric  \eqref{defmet}. Note that, in the quantum theory, the current $J'$ will be anomalous, given that we decided to not place the anomaly in its dual current. 
% we assumed that its dual current does not have an anomaly. [there is also a $U(1)$ anomaly, which we have chosen to place entirely in the shift current] \textcolor{red}{\emph{How does one see how the  anomaly is moved from one current to the other?}} {\color{ForestGreen} See 9.10-9.21 \href{https://www.ipht.fr/Pisp/francois.gelis/Physics/2018-QFT.pdf}{https://www.ipht.fr/Pisp/francois.gelis/Physics/2018-QFT.pdf} we can add a footnote)} \emph{\textcolor{red}{Note that we are using a non-standard coupling to the gauge potential, do we want to change?}} 
This anomaly reads %\textcolor{red}{\emph{Do we want a reference here?}} {\color{ForestGreen}(I think we don't need)}
\be \label{anomalyshift}
\nabla_\a J'^{[0] \a} = -\frac{k}{2\pi} \e^{\a\b}_{[0]} \p_\a \mathrm{a}_\b^{[0]} 
\ee
%\emph{Is it true that the factor of $\epsilon$ is due to our non-standard coupling to the vector potential, or that choice is encoded in the choice of which current to bosonise (i.e., we can simply say that $J$ corresponds to the vector current)? } {\color{ForestGreen}(we choose that the topological current is the conserved one and this allows us to bosonise it i.e. to write globally $J=\epsilon\p\phi$ which makes it automatically conserved. Then, we couple the topological current to an external gauge field $a_{\alpha}^{[0]}$ and our coupling is standard $Ja$. Coupling to this source makes the other current related to it by hodge duality, in our case the shift one, to have an anomaly with the non-conservation law as above. The epsilon is general in the sense that if we couple the conserved current to a source, the one which is not conserved will have the anomaly given by the expression above with epsilon. The choice is already made in the bosonisation)}
The components of the currents $J^{[0]\, \a}$ and $J'^{[0]\, \a}$ in the undeformed CFT with the non-trivial metric \eqref{defmet} are 
%\textcolor{red}{Re-check!} {\color{ForestGreen}(rechecked)}
%
\be
J^{\a}_{[0]} = \frac{e^{[\l]}}{e^{[0]}} J^\a_{[\l]} = \left( - \frac{2\partial_V\phi}{1-\l\partial_V\phi},  \frac{2\partial_U\phi}{1-\l\partial_V\phi} \right) \;, \;\;\;\;\;J^{[0]}_\a = \left( \partial_U\phi\frac{1+\l\partial_V\phi}{1-\l\partial_V\phi} , -\p_V \phi\right)
\ee
%{\color{blue}
%\be
%J_{[0]}^U=\frac{2\partial_V\phi}{1-\l\partial_V\phi}\hspace{1cm}Y^V=-\frac{2\partial_U\phi}{1-\l\partial_V\phi}\hspace{1cm}Y_U=-\partial_U\phi\frac{1+\l\partial_V\phi}{1-\l\partial_V\phi}\hspace{1cm}Y_V=\partial_V\phi
%\ee}
and, respectively 
% {\color{ForestGreen}(checked)}
\be
J'^{\a}_{[0]} =\left(\frac{2\p_V\phi}{1-\l\partial_V\phi},\frac{2\p_U\phi(1+\l \p_V\phi)}{(1-\l\p_V\phi)^2}\right)  \;, \;\;\;\;\;\; J'^{[0]}_{\a} = (\p_U \phi, \p_V \phi) \label{exprjp}
\ee
%{\color{blue}
%\be
%X^+=\frac{2\partial_-\phi}{1-\mu\partial_-\phi}\hspace{1cm}X^-=\frac{2\partial_+\phi(1+\mu\partial_-\phi)}{(1- \mu\partial_-\phi)^2}\hspace{1cm}X_+=\partial_+\phi\hspace{1cm}X_-=\partial_-\phi
%\ee}
where we used the convention $\e^{-+} = 2$, which is the same as that of section \textcolor{red}{\ref{section2:review}}.  %{\color{ForestGreen}(in flat space, otherwise $\e^{VU} = \frac{2}{e}$, which corresponds to $\e_{UV} = \frac{e}{2}$)}. 
Of course, in the deformed theory, where the metric is Minkowski, we have $J_\a^{[\l]} = ( \p_U \phi, - \p_V \phi)$, as follows from \eqref{jbos}. % with the convention $\e_{UV} = \frac{1}{2}$. 

We naturally expect the conserved current $J'^{[0]}$ in the undeformed CFT with the metric \eqref{defmet}  to correspond to a conserved current, $J'^{[\l]}$, in the deformed one on flat space, given the equality of partition functions.  Mapping the conservation equations (and the associated conserved charges) corresponds to the requirement that 

\be
\star_{[\l]} J'^{[\l]} = \star_{[0]} J'^{[0]}  \label{relvevsjp}
\ee
which is the same condition as the one we had for the flow of the expectation value of the topological current, \eqref{relvevs}. This condition allows us to relate the expectation value of some conserved current $J'$ in the deformed theory to that of the shift current in the undeformed one. 
We find the following expression for the components of $J'$ in the deformed theory 

\be
J'^{[\l]}_U =  \partial_U\phi\frac{1+\l\partial_V\phi}{1-\l\partial_V\phi}\hspace{1cm}J'^{[\l]}_V=\partial_V\phi \label{jpcomp}
\ee
The currents \eqref{relvevsjp} above are covariantly conserved (with respect to the corresponding metric): for $J'^{{[\l]}}$  this is expected since the sources in the deformed theory are set to zero, while for $J'^{[0]}$, this follows from the fact that the anomalous term \eqref{anomalyshift} that breaks the conservation of $J'^{[0]}$ actually drops out for an external source of the special form \eqref{sourcesflatsp}, since it is proportional to the conservation equation for the stress tensor.   The $J'$ conservation  equation  implies that $\phi$ must satisfy%\footnote{We would of course have obtained the same equation by requiring conservation of $J'^{[0]}$ in the metric \eqref{defmet} \emph{Check!}  {\color{ForestGreen}(checked with mathematica)},  so the anomaly does not appear to play a role. The reason is that $\nabla_\a J'^{[0]\, \a} = - \frac{k}{2\pi} \l^a \nabla_\a T^{[0]\,\a}{}_a =0$, thanks to the conservation of the stress tensor in the undeformed theory in an arbitrary background metric. } 

\be
\p_V \left(\frac{\p_U \phi}{1-\l \p_V\phi}\right) =0
\ee
which is precisely the equation of motion of a $J\bar T$ -deformed free boson. The reason for obtaining such a decoupled equation is that, as explained in section \ref{section2:review}, we effectively assumed  to have a decoupled free boson sector in the undeformed CFT. Note that, even though we constructed $J'^{[\l]}$ to correspond to what might look like the undeformed shift current in the background metric \eqref{defmet}, it need not correspond to the Noether current associated with the shift symmetry of $\phi$ in the deformed theory. A quick comparison with the $J\bar T$ - deformed free boson formulae \eqref{freebosonTv}-\eqref{freebosoncurrents} shows this current maps instead to 

\be\label{shiftcurrentdefth}
J'_\a  \leftrightarrow J^{sh}_\a - \l T_{\a V}
\ee
Obviously, in a more general theory the expressions for $J^{sh}_\a$ and $\l T_{\a V}$ would change, but it is expected that their difference,  which should correspond to the sum of the zero modes of $\mathcal{K}_L$ and $\mathcal{K}_R$ from our general Hamiltonian analysis still takes the form \eqref{jpcomp}. One may be able to show this directly by translating the general results of the Hamiltonian analysis to the Lagrangian formalism.

It is interesting to identify the currents whose charge stays constant along the flow. Of course, the topological current $J$ is one such current, and the associated conserved charge is the winding, $w$, of $\phi$. The above expression for $J'$  shows its associated conserved charge is  $Q_L + Q_R = Q_0 + \l E_R$, 
so $J'$ cannot have this property. %, , if we simply use the $J\bar T$ - deformed  free boson formulae. {\color{ForestGreen}(I would mention in a footnote that the construction of flowed charges in section 3 was completely general and their sum gives the current $J'$ mentioned above up to maybe contributions which do not affect the charge)}  
On the other hand, the combination

\be \label{nonanomalouscurrent}
J'^{[0] \a} + \frac{k}{2\pi}  \e^{\a\b}_{[0]}  \mathrm{a}_\b^{[0]} 
\ee
which, according to  \eqref{anomalyshift},
 is exactly conserved, does have it. To see this, we plug in $\mathrm{a}^{[0]}$ from \eqref{sourcesflatsp} and find  that the shift in $J'$ equals $\frac{k}{2\pi} \l^a T^\a{}_a$,  
 %{\color{ForestGreen}(index a up? we defined $\lambda^-=\lambda$ earlier)}
  which, using  \eqref{shiftcurrentdefth}, is precisely the necessary shift  % in the free theory
  to give $J_\a^{sh}$. % and, if we use the boson formula, yields the correct charge. 
 %{\color{red} Recheck!} {\color{ForestGreen}(rechecked)} 
 This suggests a general way to identify a current whose associated conserved charge is constant along the flow: it should have vanishing anomaly.

We now turn to the holographic dictionary, and show how these expected field-theoretical features are realised in the bulk.

{\color{blue}

}

{\color{blue}
%It will be useful for later to write down the components of the shift current, the topological current and the external sources, which we denote by (better notations?)
%\begin{align}
%X_{\alpha}&=J_{\alpha}^{shift}\hspace{1cm}Y_{\alpha}=J_{\alpha}^{top}=\epsilon_{\alpha\beta}\gamma^{\beta\rho}\partial_{\rho}\phi\hspace{1cm}Z_{\alpha}=-\mu^a \epsilon_{\alpha\beta}T^{\beta}_{\;a}
%\end{align}
%and which are given by (here we assume that the shift current is of the form $J_{\alpha}=\partial_{\alpha}\phi$, thus related by hodge duality to the topological one)
%\begin{align}
%Z^+&=-\frac{2\mu T_{--}}{(1-\mu\partial_-\phi)^2}\hspace{0.5cm}Z^-=\frac{2\mu T_{+-}}{(1-\mu\partial_-\phi)^2}\hspace{0.5cm}Z_+=\frac{\mu T_{+-}}{1-\mu\partial_-\phi}+\frac{2\mu^2\partial_+\phi T_{--}}{(1-\mu\partial_-\phi)^2}\hspace{0.5cm}Z_-=-\frac{\mu T_{--}}{1-\mu\partial_-\phi}
%\end{align}
%We can express the source in terms of the tangent space energy-momentum tensor:
%\begin{align}
%T_{--}&=(1-\mu\partial_-\phi)^2 T_{--}^{tg}\hspace{1cm}T_{+-}=-\mu\partial_+\phi(1-\mu\partial_-\phi)T_{--}^{tg}
%\end{align}
%such that the expressions for the components of $Z$ simplify:
%\begin{align}
%Z^+&=-2\mu T_{--}^{tg}\hspace{0.5cm}Z^-=-\frac{2\mu^2\partial_+\phi T_{--}^{tg}}{1-\mu\partial_-\phi}\hspace{0.5cm}Z_+=\mu^2 \partial_+\phi T_{--}^{tg}\hspace{0.5cm}Z_-=-\mu(1-\mu\partial_-\phi)T_{--}^{tg}
%\end{align}
%\begin{align}
%Z_+^{tg}&=0\hspace{1cm}Z_-^{tg}=-\mu T_{--}^{tg}
%\end{align}
}

{\color{blue}

%We would like to build the most general background that has $e_{\a}{}^a = \d^a_\a$ and $\mathrm{a}_\a =0$. If we \emph{ define} a scalar $\phi$ via $\e_{\a\b} J^\b=  \p_\a \phi$, \emph{Sign!} then the undeformed background must have been

%\be
%e_\a^{(0)\, a} = \d_\a^a - \mu^a \p_\a \phi \;, \;\;\;\;\; \mathrm{a}_\a^{(0)} = - \mu^a \e_{\a\b} T^\b{}_a
%\ee
%The form of $e^{(0)}$ tells us that the new background is simply a coordinate transformation $x^a \r x^a - \mu^a \phi$ on the old one, where the argument is defined only up to a constant shift. Let us now particularise to $J\bar T$, where $\mu^a$ is lightlike. 
%Then, the boundary metric

%\be
%dx^+ dx^- \;\;\; \r \;\;\; dx^+ (dx^- - \mu \p_+ \phi dx^+ - \mu \p_- \phi dx^-) 
%\ee
%where $\mu^a = \mu \d^a_-$. The resulting gravitational background is simply obtained by taking a Ba\~{n}ados geometry and replacing everywhere $x^-$ by $x^--\mu \, \phi (x^+,x^-)$, including in the argument of $\bar{\mathcal{L}}$. The non-chiral current is modellel by two Chern-Simons gauge fields.  To relate $\phi$ to the expectation value of the currents, we need the holographic dictionary. 

}

\subsection{The holographic dictionary for $J\bar T$ - deformed CFTs}\label{subsection52:holographicdictionary}

As reviewed in the introduction to this section,  the holographic dictionary for a CFT deformed by a multitrace operator is obtained by translating the field-theory relation between deformed and undeformed sources and expectation values to bulk language, using the standard AdS/CFT dictionary.

\subsubsection*{Building the bulk solution}

The dual gravitational action that models the universal (conserved current) sector of a $J\bar T$ - deformed holographic CFT is three-dimensional Einstein gravity with a negative cosmological constant, coupled to two $U(1)$ Chern-Simons gauge fields $A_\mu, B_\mu$ that model the chiral and anti-chiral components of the currents. %In order to fully implement the mixed boundary conditions, we need the holographic dictionary, but let's see first how far we can go without explicitly it. 
On-shell, the metric is locally AdS$_3$ and the connections are simply flat. In the Fefferman-Graham gauge, the most general pure $AdS_3$ solution takes the form %\textcolor{red}{\emph{Careful! $(0)$ FG vs $[0]$ undeformed! }}
\begin{align}\label{FGexpads3}
ds^2&=\frac{\ell^2 dz^2}{z^2}+\frac{1}{z^2}\bigg(g^{(0)}_{\alpha\beta}+z^2 g^{(2)}_{\alpha\beta}+z^4 g^{(4)}_{\alpha\beta}\bigg)
\end{align}
The boundary is located at $z=0$, $g^{(4)}= \frac{1}{4}g^{(2)}(g^{(0)})^{-1}g^{(2)}$  and the trace and divergence of   $g^{(2)}$ are determined in terms of $g^{(0)}$ via the holographic Ward identities \cite{Henningson:1998gx}. In radial gauge, $A_z=B_z=0$, which implies the gauge connections are $z$ - independent. 

As discussed in the previous subsection, the $J\bar T$ - deformed CFT on flat space corresponds to the undeformed CFT in the non-trivial metric  \eqref{defmet}. Consequently, we should take\footnote{ 
Note that in this section, the only field-theory quantities  that appear are the ones marked with ``$[0]$'' in the previous one (undeformed CFT quantities in a non-trivial background), so we will mostly drop the $[0]$ label. }
\be \label{boundarymet}
g^{(0)}_{\a\b} = \g^{[0]}_{\a\b} = \left(\begin{array}{cc} - \l \p_U \phi & \frac{1-\l \p_V\phi}{2}  \\ \frac{1-\l \p_V\phi}{2}  &0  \end{array} \right) 
\ee
Noting this can be induced by a simple coordinate transformation on a Minkowski boundary metric $ds_{(0)}^2 = dU dV$

\be \label{coordtrdef}
U \r U\;, \;\;\;\;\; V \r  V - \l \phi + const.
\ee
the most general solution with these boundary conditions is easy to write down: it simply corresponds to applying the above coordinate transformation to a general Ba\~{n}ados metric, which is parametrised by two arbitrary functions $\L$ and $\bar \L$. We obtain 

%{\color{red}

%(I removed the  $\ell$ factors from  the $\mathcal{L}s$, more standard convention). }
\begin{align}\label{deformedmetric}
ds^2&=\frac{\ell^2 dz^2}{z^2}+\bigg(\mathcal{L}+\bar{\mathcal{L}}(\lambda\partial_U\phi)^2-\frac{1}{z^2}(1+ z^4\mathcal{L}\bar{\mathcal{L}})\lambda\partial_U\phi\bigg)dU^2+\bar{\mathcal{L}}(1-\lambda\partial_V\phi)^2dV^2 +\nonumber\\
& \hspace{1.2cm}+(1-\lambda \partial_V\phi)\bigg(\frac{1+ z^4\mathcal{L}\bar{\mathcal{L}}}{z^2}-2 \bar{\mathcal{L}}\lambda\partial_U\phi\bigg)dUdV
\end{align}
with $\mathcal{L}=\mathcal{L}(U)$ and $\bar{\mathcal{L}}=\bar{\mathcal{L}}(v)$, where 
\be \label{exprfielddepcoord}
v \equiv V - \l \phi + const. 
\ee
The constant ambiguity in the definition of $v$ will be discussed in the next section.

Thus, the bulk metric is fully fixed once we specify the expectation value of the  current, encoded in $\phi$. Naturally, the latter also determines the %values of the 
bulk gauge fields.  
%{\color{blue} How we read off this vev holographically will be explained soon.  Note $\phi \sim$ vev $\tilde J$ has not yet been related holographically to the components of the bulk CS fields.} 
To see how, we need the details of the Chern-Simons holographic dictionary, to which we now turn.

%{\color{blue}
%To fully build the bulk background, we need to model the currents via CS field, for which we need the dictionary. // We also need to model the currents which give the conserved charges in the $J\bar{T}$ deformed CFT. They are modeled as $U(1)$ CS gauge fields in the bulk, on this nontrivial 2d metric and coupled to external sources.]}

\subsubsection*{The Chern-Simons action}

The currents that we would like to model in holography are two abelian $U(1)$ currents that are Hodge dual to each other. One of these currents is anomalous, and the other one is  not. These properties can be easily modeled \cite{Banados:2006fe} using two three-dimensional Chern-Simons gauge fields, whose (bulk) anomaly reproduces the field theory one
%{\color{blue}We need to model two currents, one of them, the topological one, which is conserved, and one which is anomalous after we couple to an external source. We need to add (subtract depending on conventions) to the latter the anomaly in order to keep it conserved in presence of the deformation. The anomaly in field theory should match the anomaly in the bulk. Thus, we need two gauge fields in the bulk, one which is invariant under gauge transformations and one which is not. }
%Such a Chern-Simons action was already introduced in \cite{Banados:2006fe} in order to model the 2d chiral anomaly 
%{\color{blue}Since what we have here is basically a 2d chiral anomaly (via bosonisation) i.e. the topological current and the shift current are related here by exactly the same relation $J^{shift}=\epsilon J^{top}$ as the vector and axial currents in 2d, we will use the same action:}
\be \label{proposalactioncs}
S_{CS} = - \frac{k}{8\pi} \int d^3 x \sqrt{g} \, \e^{\mu\nu\rho} (A_\mu+B_\mu) \p_\nu (A_\rho-B_\rho)
\ee
This action is by construction invariant under equal and opposite gauge transformations of $A,B$, but picks up a boundary term under equal shifts of the gauge potential. Thus, half of the bulk gauge symmetry is preserved at the boundary, and one expects  the associated current to be non-anomalous. The perhaps unusual minus sign in front of the action is due to our unusual conventions for the $\e$ tensor. 

% {\color{ForestGreen}(yes, keep comment)} }
%{\color{blue}Roughly speaking, $A+B$ gives the anomalous part, so it should be identified with/related to the source for the shift current, while $A-B$ gives the invariant part, so it should be identified with/related to the source for the topological current. We will make this relation precise by using the variational principle.}

As is well-known, the action \eqref{proposalactioncs} does not have a well-defined variational principle. To make it so, we need to add a boundary term \cite{Kraus:2006wn}. A choice that yields - in Minkowski space -  a chiral current coupling to $A_-$ and an anti-chiral one coupling to $B_+$ is  %\textcolor{red}{Notation bnd?}
\begin{align}
S_{bnd}&=  \frac{k}{16\pi}\int_{\partial} d^2x \sqrt{-\gamma}\gamma^{\alpha\beta}(A+B)_{\alpha}(A+B)_{\beta}
\end{align}
Only this term will contribute to the stress tensor, since \eqref{proposalactioncs} is topological. The variation of the action is: %\emph{\textcolor{red}{You can just put it on-shell, directly. In addition, the first line is not defined, likely wrong}}
%{\color{ForestGreen}(checked)}
\begin{align}\label{varactcs}
\delta S&=  \delta S_{CS}+\delta S_{bnd}= -\frac{k}{8\pi}\int d^3x\sqrt{-g}\epsilon^{\mu\nu\rho}(F^A_{\mu\nu}\delta A_{\rho}-F^B_{\mu\nu} \delta B_{\rho})+ \nonumber\\
&+ \frac{k}{8\pi}\int_{\partial} d^2x \sqrt{-\gamma}\bigg[(A_{\alpha}+B_{\alpha})(\gamma^{\alpha\beta} -\epsilon^{\alpha\beta})\delta A_{\beta}+(A_{\alpha}+B_{\alpha})(\gamma^{\alpha\beta}+\epsilon^{\alpha\beta})\delta B_{\beta}\bigg]+ \nonumber\\
&+\frac{k}{16\pi}\int_{\partial}d^2x \sqrt{-\gamma}\bigg[(A+B)_{\alpha}(A+B)_{\beta}-\frac{\gamma_{\alpha\beta}}{2}\gamma^{\mu\nu}(A+B)_{\mu}(A+B)_{\nu}\bigg]\delta\gamma^{\alpha\beta}
\end{align}
where we have taken into account the fact that the outward-pointing unit normal to the AdS boundary is $- (z/ \ell)\, \p_z$. %{\color{ForestGreen}(I agree, the boundary is at $z=0$ and the normal is positive inward)} \emph{Correct? Factors $\ell$ ok?}{\color{ForestGreen} (yes, the unit vector is $-\frac{z}{\ell}\partial_z$ but we also get an $\ell$ when we go from $\sqrt{-g}$ to $\sqrt{-\g})$} 
 From the second line above, we immediately read off the components of the two chiral currents, which are  linear combinations %{\color{red}$\frac{1}{2}(J \pm \star J)$  \emph{Check!}  of the deforming current } 
  of the form $\frac{1}{2}(\star J\pm J)$. 
%
%{\color{ForestGreen} On-shell, we can write:
%\begin{align}
%\hspace{-1.5cm}\delta S=\frac{k}{8\pi}\int d^2x \sqrt{-\g}\epsilon^{\alpha\beta}(A+B)_{\beta}\bigg[\epsilon_{\alpha\rho}(\delta A+\delta B)^{\rho}-(\delta A-\delta B)_{\alpha}\bigg]-\frac{k}{16\pi}\int d^2x\sqrt{-\g}\bigg[(A+B)_{\alpha}(A+B)_{\beta}-\frac{\gamma_{\alpha\beta}}{2}(A+B)^2\bigg]\delta\gamma^{\alpha\beta}
%\end{align}
%(quick check: in flat space we get from the first part $\frac{k}{2\pi}(A_+\delta A_- - B_-\delta B_+)$)}
It is then natural to identify  

\be
J^{[0] \a} =  \frac{k}{4\pi} \e^{\a\b} (A_\b + B_\b)  \label{holocur}
\ee  
since the right-hand side is conserved, using the Chern-Simons equations of motion, for \emph{any} flat connections $A, B$, just like the left-hand side is conserved with no restriction on $\phi$. This identification implies  of course that 

\be
J'^{[0]}_\a =   \frac{k}{4\pi} (A_\a+B_\a) =  \p_\a \phi
\ee
%{\color{blue}
%This looks exactly like a linear combination of $\tilde J^{[0]}$ and $J^{[0]}$, which are related via \eqref{relationcurrentsundef}. It is natural to identify $\tilde J^{[0]}$ with $\epsilon (A+B)$, which implies
%
%\begin{align}
%A_{\alpha}+B_{\alpha}=\frac{4\pi}{k}\partial_{\alpha}\phi
%\end{align}
%and so $\tilde J^{[0]}$ is automatically conserved for any $\phi$ upon using  the bulk eom,  as it is in field theory {\color{ForestGreen}(we
where in the last equality we used \eqref{exprjp}. Thus $\phi$, which encodes the  expectation values of the two currents (which are considered as being given), determines the sum of the two Chern-Simons gauge fields. The tangent space components of the sum are 
%\emph{Check!}{\color{ForestGreen}(rechecked)}

\be \label{sumofcurrents}
A_+ + B_+ = \frac{4\pi}{k}\frac{\p_U \phi}{1-\l \p_V \phi} \;, \;\;\;\;\;\;  A_- + B_- =\frac{4\pi}{k} \frac{\p_V \phi}{1-\l \p_V \phi}
\ee
formulae that will be useful later.  To identify the field theory source that couples to the topological current \eqref{holocur}, we simply interpret the second  line in the action variation \eqref{varactcs} as $\int d^2 x \sqrt{-\g} J^{\a} \d \mathrm{a}_\a$, with $J^\a$ given in \eqref{holocur}, plus some extra contribution proportional to the variation of the projectors.  Concretely, on-shell the action variation \eqref{varactcs} becomes
\bea
\delta S^{on-shell} & = &\frac{1}{2}\int_{\partial} d^2 x \sqrt{-\gamma}\, J^{[0]\alpha}\delta[A_{\alpha}-B_{\alpha}-\epsilon_{\alpha\beta}(A^{\beta}+B^{\beta})] - \\
&&\hspace{1cm} -\frac{k}{16\pi}\int_{\partial} d^2 x \sqrt{-\gamma}\bigg[(A+B)_{\alpha}(A+B)_{\beta}-\frac{\gamma_{\alpha\beta}}{2}(A+B)^2\bigg]\delta\gamma^{\alpha\beta} \nonumber 
\eea
This immediately leads to the identification
{\color{red} 
}
\be
\mathrm{a}_\a = \frac{1}{2} [ A_\a - B_\a - \e_{\a\b} (A^\b+B^\b)] \label{holosource}
\ee
The variation of the action also allows us  to read off  the Chern-Simons contribution to the energy-momentum tensor\footnote{ Note that a na\"{i}ve computation using just the last term in \eqref{varactcs} would yield \emph{minus} the above answer, 
 as $T_{\alpha\beta}=-\frac{2}{\sqrt{-\gamma}}\frac{\delta S }{\delta \gamma^{\alpha\beta}}$ in Lorentzian signature. The contribution of the $\e$ tensor variation is \emph{essential} for obtaining the correct relative sign between the stress tensor and the $U(1)$ charge, to which the $J\bar T$-deformed spectrum is highly sensitive. %\emph{Correct?}} {\color{ForestGreen}(yes, keep the comment. we got the correct signs only after taking into account this contribution from the projectors)}\emph{Comment literature!}
 }
\begin{align}\label{energymomtensor}
T^{CS}_{\alpha\beta}&=\frac{k}{8\pi}\bigg[ (A_\a+B_\a)(A_\b+B_\b) - \frac{1}{2} \g_{\a\b} (A+B)^2\bigg]
\end{align}
Note that, since it only depends on the sum $A+B$, the Chern-Simons contribution to the stress tensor is also fully determined by $\phi$, i.e. t only involves the expectation values of the currents. The full stress tensor is given by the  sum of the gravitational and the Chern-Simons contributions

\be
T_{\a\b} = T_{\a\b}^{g} + T_{\a\b}^{CS} \;, \;\;\;\;\; \;\; T_{\a\b}^{g}  = \frac{1}{8\pi G \ell} g^{(2)}_{\a\b}
\ee
where $g^{(2)}$ is the first subleading Fefferman-Graham coefficient in \eqref{FGexpads3}, and we have used the fact that the boundary metric \eqref{boundarymet} is flat to drop an additional potential contribution to $T^g$ proportional to its trace. The expression for the stress tensor simplifies if we write it in tangent space where, in  components, we have 

\be \label{energymomtg}
T_{++} = \frac{\mathcal{L} }{8\pi G \ell}+\frac{2\pi}{k}\frac{(\partial_U\phi)^2}{(1-\lambda\partial_V\phi)^2} \;, \;\;\;\; T_{--} = \frac{\bar{\mathcal{L}}}{8\pi G \ell\,}+\frac{2\pi}{k}\frac{(\partial_V\phi)^2}{(1-\lambda\partial_V\phi)^2}
\ee 
and $T_{+-}=T_{-+} =0$.
The spacetime components of the holographic stress tensor can be obtained by simply multiplying with $e^{[0]\, a}_\a$, given in \eqref{defmet}.

%\textcolor{red}{Check whether the "chiral" currents (5.42) are separately conserved.}
{\color{blue}

}

According to   \eqref{sourcesflatsp}, the gauge field source \eqref{holosource} should be identified with the stress tensor. In tangent space (where the calculation is simpler), we have

\be \label{sourcestgspace}
\mathrm{a}_+ = - B_+= 0 \;, \;\;\;\;\;\; \mathrm{a}_- = A_- =  -\l T_{--}% = {\color{red} -} \l \left( \frac{\ell \, \bar \L}{8\pi G}  + \frac{2\pi}{k}\bigg(\frac{\partial_V\phi}{1-\lambda\partial_V\phi}\bigg)^2 \right)
\ee
with $T_{--}$ given in \eqref{energymomtg}. Together with \eqref{sumofcurrents}, these  determine $A_{\pm}, B_{\pm}$ individually in terms of the presumably known functions $\bar \L, \phi$. The spacetime components are obtained by multiplying with the vielbein \eqref{defmet}, and we obtain   %{\color{red} Redo/check signs!} 

%{\color{ForestGreen}(checked)}

\be \label{gaugefieldA}
A_U = \frac{4\pi}{k}\frac{\p_U\phi}{1-\l \p_V\phi} + \l^2 \p_U\phi \,T_{--} \;, \;\;\;\;\; A_V = -\l (1-\l \p_V\phi) T_{--} 
\ee
\be  \label{gaugefieldB}
B_U = \l \p_U\phi \left(-\frac{4\pi}{k}\frac{\p_V\phi}{1-\l \p_V\phi} -\l T_{--}\right)  \;, \;\;\;\;\;\; B_V = \frac{4\pi}{k}\p_V \phi +\l (1-\l \p_V\phi) T_{--} 
\ee
Note that so far, in our holographic analysis, the field $\phi$ could be arbitrary. 
The flatness conditions for the above gauge fields result in 
an  equation of motion for it. Even though there are two gauge fields, one obtains a single condition on $\phi$ since, as we already discussed, the flatness condition for $A+B$ does not yield a constraint. The flatness condition for $A-B$ can be written, using 
\eqref{holosource}, as
%
%\be
% \e^{\a\b} \p_\a (A_\b -B_\b) = 2 \nabla_\a (\e^{\a\b} \mathrm{a}_\b) + \frac{4\pi}{k} \nabla_\a (A^\a+B^\a)=  -2 \l^a \nabla_\a T^\a{}_a + \frac{4\pi}{k} \Box \phi =0
%\ee
\be
 \e^{\a\b} \p_\a (A_\b -B_\b) = 2 \nabla_\a (\e^{\a\b} \mathrm{a}_\b) +  \nabla_\a (A^\a+B^\a)=  2 \l^a \nabla_\a T^\a{}_a + \frac{4\pi}{k} \Box \phi =0
\ee
Since the stress tensor is conserved, the first term may be dropped, and we simply obtain a free wave equation for $\phi$ in the non-trivial metric \eqref{defmet}, which takes precisely the form of the  equation of motion for a $J\bar{T}$-deformed free boson 
%{\color{ForestGreen}(rechecked)}
%
\begin{align}\label{eomdfboson}
\partial_V\left(\frac{\partial_U\phi}{1-\l\partial_V\phi}\right)=0
\end{align}
%{\color{blue}[Thus,  in the absence of the source, we obtain that $\phi$ satisfies the eom of a free scalar in a $J\bar{T}$-deformed CFT. When the sources are turned on, it is not obvious that this is still true.] } 
The general solution takes the form 
\be  \label{scalarfieldonshell}
\phi(U,V) = f(U) + g (v) 
\ee
from which it follows that the quantities

\be\label{tangtspparam}
 \frac{\p_U\phi}{1-\l \p_V\phi}= f'\equiv  \J(U)     \;, \;\; \;\;\; \frac{\p_V\phi}{1-\l \p_V\phi} =g'  \equiv   \bar \J(v)
\ee
are only functions of $U$ and, respectively, the field-dependent coordinate $v$. Note these quantities correspond precisely to the tangent space components of the ``chiral" currents $\frac{J'\pm J}{2}$, which should be identified with  $K_\a, \bar K_\a$  in section \ref{fbanalysislagr} for the  the $J\bar T$ - deformed free boson case, and the more general currents \eqref{comprmcurrent} discussed in  \ref{hamformjtb} in Hamiltonian language. 

Note  that the gauge field solution \eqref{gaugefieldA} - \eqref{gaugefieldB} can be generated via a coordinate transformation of the form \eqref{coordtrdef} on an AdS$_3$ background with non-trivial Chern-Simons fields turned on 
%\textcolor{red}{Rephrase slightly and notation!}  
%
\begin{align}
A_u=\frac{4\pi}{k}\partial_u\phi\hspace{1cm}A_v=0\hspace{1cm}B_u=0\hspace{1cm}B_v=\frac{4\pi}{k}\partial_v\phi
\end{align}
with $u=U$ and $v$ given in \eqref{exprfielddepcoord}, which simply correspond to generic expectation values for a holomorphic and an antiholomorphic $U(1)$ current in the dual theory and no sources turned on. The $T_{--}$ - dependent piece of the gauge fields \eqref{gaugefieldA}, \eqref{gaugefieldB}, which accounts for the external source, can be induced by a field-dependent gauge transformation 
\be \label{fielddepgaugetr2}
A_\a \r A_\a - \p_\a \Lambda \;, \;\;\;\;\; B_\a \r B_\a + \p_\a  \Lambda\;, \;\;\;\;\;\; \Lambda = \l \int^v dv' T_{--} (v')
\ee
where we used the fact, implied by \eqref{tangtspparam} that $T_{--}$ in \eqref{sourcestgspace} is purely a function of $v$. 

 To summarize, the most general solution to the gravity - Chern-Simons equations of motion  that satisfies the $J\bar T$ - deformed 
boundary conditions \eqref{sourcesflatsp} is parametrized by four arbitrary functions $\L(U), \bar \L(v), \J(U)$ and $\bar \J(v)$ of the left-moving  coordinate $U$ and  field-dependent  ``right-moving'' coordinate $v$, which is also determined by them up to a certain constant mode that we will discuss more thoroughly in the next section. In terms of these, the most general  metric is \eqref{deformedmetric}, and gauge fields \eqref{gaugefieldA}, \eqref{gaugefieldB}.  The holographic dictionary \eqref{relvevs} relates the expectation values in the deformed theory to the holographic expectation values encoded in this bulk solution as

\be
T^{[\lambda]}_{UU}%=T_{++}
 =  \frac{\mathcal{L} }{8\pi G \ell}+\frac{2\pi}{k}\J^2 \;, \;\;\;\;\;\;\; T^{[\lambda]}_{VU}=0 \nonumber 
\ee

\be
T^{[\lambda]}_{UV}=%\lambda\partial_U\phi T_{--} =
\frac{\l \J}{1+\l \bar{\J}}\bigg(\frac{\bar{\mathcal{L}}}{8\pi G \ell\,}+\frac{2\pi}{k}\bar{\mathcal{J}}^2\bigg)  \;, \;\;\;\;\;\; T^{[\lambda]}_{VV}=%(1-\lambda\partial_V\phi)T_{--}=
\frac{1}{1+\l \bar{\J}}\bigg(\frac{\bar{\mathcal{L}}}{8\pi G \ell\,}+\frac{2\pi}{k}\bar{\mathcal{J}}^2\bigg) \label{energymomdefth}
\ee
\begin{align}
&J^{U}_{top}=-2\J \frac{1-\lambda\J}{1+\l \bar{\J}}\hspace{1cm}J^{V}_{top}=\frac{2\bar{\J}}{1+\l \bar{\J}} \nonumber\\[4pt]
J_{U}^{shift}&=\frac{\J(1+2\l\bar{\J})}{1+\l\bar{\J}}+\lambda T^{[\lambda]}_{U V}\hspace{1cm}J_{V}^{shift}=\frac{\bar{\J}}{1+\l\bar{\J}}+\lambda T^{[\lambda]}_{V V}
\end{align}
It is not difficult to check that the above currents are conserved with respect to the Minkowski metric of the $J\bar T$ - deformed theory. The conserved charges are constructed by integrating these conserved currents against appropriate vectors. For the global conserved charges, these vectors are simply constant, and we will check in the next subsection that the resulting expectation values of the global conserved charges precisely reproduce the correct $J\bar T$ - deformed spectrum of energies and $U(1)$ charges.

Before we end this section, let us comment on the similarities and differences 
 with the analysis of \cite{Bzowski:2018pcy}, who were assuming the current to be chiral and thus modeling the system with a single Chern-Simons field. We can reach this case by setting $B_\a =0$ above, which  corresponds to restricting our parameters as  %$\bar \J = - \l \bar \L$ in our formulae. \emph{Correct?} {\color{ForestGreen}(I get $\bar{\mathcal{J}}=-\lambda\bigg(\frac{\bar{\mathcal{L}}k}{32\pi^2 G\ell}+\frac{\bar{\J}^2}{2}\bigg)$ which is a quadratic eq for $\bar{\J}$, maybe better write
  $\bar{\J}=-\frac{\l k}{4\pi}T_{--}$, with $T_{--}$ given in \eqref{energymomtg}.  Given that  the zero mode of $\bar \J$ is % $- Q_R$ \emph{Correct?}{\color{ForestGreen}
  $- Q_R/R_v$,  this constraint is consistent with the condition $\bar J_0 =0$ that \cite{Bzowski:2018pcy} effectively imposed. % {\color{ForestGreen}(checked, we get exactly $Q_R=\frac{\lambda k}{4\pi}E_R$, which is obvious because we solved for setting $B=0$ and $\bar{J}_0$ is the holonomy of $B$)} 
  We then observe that the  way in which the source $\mathcal{A}$ of \cite{Bzowski:2018pcy} appears in their holographic formulae (e.g. in the Chern-Simons stress tensor) is  identical  with how  $-\frac{k}{4\pi}\bar \J$ appears in our formulae,  so naturally the way that  spectrum check proceeds is the same. 
%  
%  
%  In particular, their constraint $\mathcal{A} = \l \bigg(\frac{\bar \L}{8\pi G\ell } + \frac{k}{8\pi}\mathcal{A}^2\bigg)$ associated to the boundary condition \eqref{} simply maps to the way $E_R$ is related to $E_R^{(0)}$. \emph{Does this mean $\phi$ should have the full stress tensor contribution in it in the chiral case?} {\color{ForestGreen}(we would expect that the holonomy of $A$ in \cite{Bzowski:2018pcy} is $J_0$, but one can easily see that this is not the case: we get $Q_R+\frac{\lambda k}{4\pi}E_R$ instead of $Q_R-\frac{\lambda k}{4\pi}E_R$. The issue probably comes from not switching signs in the energy-momentum tensor contribution, which determines the other signs by the requirement that $T_{\mu\nu}$ is positive.)}
 
As we already mentioned, the analysis of \cite{Bzowski:2018pcy} left certain ends loose. For example, it was found that the charge associated with the current entering the deformed metric was not the same as the chiral charge. We find it rather likely that this discrepancy is due to the fact that anomalies %- which cannot be avoided in the chiral case - 
may contribute to the variational principle analysis  of \cite{Bzowski:2018pcy}. In other words, in considering the Smirnov-Zamolodchikov deformation in presence of arbitrary gauge field sources, one needs to worry about potential anomalies in the conservation of the current. In our present analysis, we purposefully avoided this issue by using a non-anomalous current to build the $J\bar T$ operator. However, the current of \cite{Bzowski:2018pcy} is necessarily anomalous, case in which we may need to specifically subtract the anomalous piece from it as in \eqref{nonanomalouscurrent} when building the Smirnov-Zamolodchikov deformation in an arbitrary background field. %{\color{ForestGreen}(ok)} 
The resulting exactly conserved current, which drives the deformation of the metric,  would no longer be  chiral, which would affect the general form of the allowed metrics. However, it will have the correct winding mode, which would fix one of the problems that \cite{Bzowski:2018pcy} could not solve. We have not checked in detail whether this proposed  correction to the holographic dictionary for chiral $J$ is  fully consistent.

\subsubsection*{More general solution}

Finally, let us briefly comment on how our bulk solution is expected to change  if we turn on more general sources  in the deformed theory. This should allow us to eventually compute arbitrary holographic correlation functions of the conserved currents in this theory.

Turning on a source $\mathrm{a}^{[\l]}$ for the current is trivial, as it only contributes to the difference of bulk fields through \eqref{holosource}. A non-trivial deformed vielbein enters the boundary metric of AdS$_3$ via \eqref{generalmapdefundef}. It is clear that the bulk metric will depend only on $e, \phi$ and the  analogues of $\L, \bar \L$ for this new background. To fully determine the solution, we need to relate $\phi$ to the holographic expectation value of the currents,  captured by the sum of the two gauge fields as in \eqref{holocur}. %\textcolor{red}{\emph{It looks to me that the relation 5.32 does not change. Agree? If so, maybe add a few more details (below too much).} } {\color{ForestGreen}(I agree. The way we model the currents does not change. The CS contribution to the energy-momentum tensor will be the same in tangent space. The metric will look differently, so probably the metric contribution will have the analogues of $\L,\bar{\L}$. Of course the spacetime components will look differently because of the vielbein. 
The flatness condition for the gauge fields is expected to become
\begin{align}
\Box\phi&=-\frac{k}{4\pi}\epsilon^{\alpha\beta}f_{\alpha\beta}^{[\lambda]}
\end{align}
where $f_{\alpha\beta}^{[\lambda]}$ is the field strength associated to the non-trivial source in the deformed theory. This equation is expected to have  some more complicated solutions analogous to $\J,\bar{\J}$, in terms of which  one would be  parametrizing  the background. 

%Most of the expressions don't change if we add a source in the deformed theory:
%\be
%e_\a^{[0]\, a} = \d_\a^a - \l^a \e_{\a\b} J^\b \;, \;\;\;\;\; \mathrm{a}_\a^{[0]} =  \mathrm{a}_\a^{[\l]}+ \l^a \e_{\a\b} T^\b{}_a
%\ee
%The topological current is still conserved, we can introduce $\phi$ just like before, the combination \eqref{nonanomalouscurrent} is still conserved, but now with a different shift that includes the extra source. The components of the gauge fields get an extra contribution:
%\begin{align}
%A_U&=\frac{4\pi}{k}\frac{\partial_U\phi}{1-\lambda\partial_V\phi}+\lambda^2\partial_U\phi T_{--}+(\mathrm{a}_+^{[\l]}-\lambda\partial_U\phi \mathrm{a}_-^{[\l]})\hspace{1cm}A_V=(1-\lambda\partial_V\phi)(\mathrm{a}_-^{[\l]}-\lambda T_{--})\\
%B_U &= -\frac{4\pi}{k} \frac{\l \p_U\phi\p_V\phi}{1-\l \p_V\phi} -\l^2 \p_U\phi T_{--} -(\mathrm{a}_+^{[\l]}-\lambda\partial_U\phi \mathrm{a}_-^{[\l]})  \\
%B_V &= \frac{4\pi}{k}\p_V \phi +\l (1-\l \p_V\phi) T_{--}  - (1-\lambda\partial_V\phi)\mathrm{a}_-^{[\l]}
%\end{align}
%and of course in the deformed theory tangent space indices are the same as spacetime indices. We can impose that the gauge fields vanish at the horizon and try to solve for the sources, but we get no solution.

\subsection{Match to the $J\bar T$ - deformed  spectrum and thermodynamics}\label{subsection53:checks}

The aim of this subsection  is to check the holographic dictionary   we proposed, by showing that it  reproduces the correct $J\bar{T}$- deformed spectrum 
and thermodynamics for the full set of allowed global charges. Our methodology is a direct generalisation of that of \cite{Bzowski:2018pcy}. 

There are several ways in which  one may perform these consistency checks. The simplest one is to verify, directly in the deformed theory,  that the relationship between the entropy and the conserved charges
 $E_{L,R}, Q_{L,R}$ is given by the correct $J\bar T$ relation. % We can perform this check using either the Kraus formalism, or the covariant phase space one (though there may be a repetition with the next section). 
 A slightly different check  is that the relation between deformed and undeformed observables (energy, $U(1)$ charges) is the same as in $J\bar T$ - deformed CFTs. Finally, we check that  the thermodynamics is the same as that of $J\bar T$. 
 %\textcolor{blue}{Following \cite{},  
%we devise a simple procedure to obtain the deformed temperatures from the undeformed ones, and generalise it to obtain also the chemical potentials.} \textcolor{red}{ \emph{Comments first law?}}

\subsubsection*{The entropy formula in the deformed theory}

The simplest check is to compute the entropy-to-energy relation for constant backgrounds, parametrised by constant $\L, \bar \L, \J, \bar \J$. Integrating the expressions \eqref{energymomdefth} over $\s$, we obtain 
%{\color{red}\emph{Work out Factors!!}}
%
\be \label{spectrumbulk}
E_L = \left(\frac{\L}{8\pi G\ell} +\frac{2\pi}{k}\J^2 \right)R \;, \;\;\;\;E_R = \left(\frac{\bar \L}{8\pi G\ell} +\frac{2\pi}{k} \bar \J^2\right) R_v \;, \;\;\;\; Q_L = \J R \;, \;\;\;\;  Q_R = - \bar \J R_v
\ee
with $R_v = R-\l w$, $w=Q_L-Q_R$. Here, we have simply defined $Q_{L,R}$ as the conserved charges associated with the chiral combinations of the currents.  
The same expression for the energies and conserved charges can also be obtained using the covariant phase space formalism, which we present in  section  \ref{section62:conservedcharges}. The fact that two entirely different formalisms yield exactly the same expressions is a non-trivial check of the formulae we obtained.

Let us assume that $\L$ and $\bar \L$ are positive, so the corresponding background  is a black hole. Its horizon is located at $z = (\ell^2 \L \bar \L)^{-1/4}$  %\textcolor{red}{Correct?}{\color{ForestGreen}(yes)} 
and its Bekenstein-Hawking entropy is %{\color{red}%\emph{Factors!!}}
\be \label{entropyform}
S = \frac{1}{4 G}(\sqrt{\L}R + \sqrt{\bar \L} R_v ) = \sqrt{\frac{\pi\ell}{2 G}}\bigg( \sqrt{R E_L-\frac{2\pi}{k}Q_L^2} + \sqrt{E_R R_v -\frac{2\pi}{k} Q_R^2} \bigg)
\ee
where in the second step we have rewritten the answer in terms of the conserved charges \eqref{spectrumbulk}. Trading $\ell$ in the prefactor for the   Brown-Henneaux central charge $c=\frac{3\ell}{2G}$, we find  precisely the universal entropy formula for a $J\bar T$-deformed CFT as a function of the (deformed) conserved charges, which we review in appendix \ref{Appendix:thermoJTbar}.  Note this formula makes no use of, nor reference to,  which charges are quantized.

\subsubsection*{The deformed energy spectrum in terms of the undeformed one}

While the above match is satisfactory, one would also like to  obtain the relation between the deformed and undeformed observables, such as the energies and the charges. For this, one may use the fact that
the entropy, the angular momentum and the winding charge must not vary with $\l$: the first, because the deformation is adiabatic and does not change the number of states, while the latter two, because the corresponding charges are quantized. To fully determine the deformed spectrum, which depends on four parameters,  we need  one more $\l$ - independent quantity.  Our proposal is  that this is the charge associated to the non-anomalous current \eqref{nonanomalouscurrent}, which we expect to stay constant along the flow, and equal the quantized shift charge. %\textcolor{red}{\emph{Argument more?}}{\color{ForestGreen}(I think we have comments about this in section 5.1, I wrote something in green)}

Denoting the conserved charges in the deformed theory as $Q_L, Q_R$ and in the undeformed one as $J_0, \bar J_0$, the relations below follow %\emph{Factors!}
\be
Q_L-Q_R = w = J_0 -\bar J_0 \;, \;\;\;\; Q_L+Q_R - \frac{\l k}{2\pi} E_R = J_0 +\bar J_0
\ee
which allow us to derive  the known expressions for $Q_{L,R}$ in the deformed theory in terms of the quantized charges $J_0,\bar J_0$. Note that, in the bulk, this amounts to requiring that the holonomies of the gauge fields $A, B$ around the $\s$ direction 
\begin{align}\label{holonomies}
\frac{k}{4\pi}\oint A_{\sigma}&=Q_L-\frac{\lambda k}{4\pi} E_R=J_0\hspace{1cm}\frac{k}{4\pi}\oint B_{\sigma}=-Q_R + \frac{\lambda k}{4\pi}=-\bar{J}_0
\end{align}
not change with $\l$. 
% \textcolor{red}{Remove, I guess: }
%{\color{ForestGreen}(It's obvious that in the undeformed theory (setting $\lambda=0$ everywhere) the gauge transformations corresponds to shift in $\phi$. It is also clear that $J_0+\bar{J}_0$ corresponds in the deformed theory to shifting $\phi$ while keeping the deformed coordinates fixed namely accompanying the shift in $\phi$ by a RM diffeo that brings the $\lambda E_R$ term in the charges. Since the holonomies are still the undeformed charges, the gauge transformations should correspond to pure shift in $\phi$, even in the deformed theory. It's not obvious to me why the field-dependent gauge transformation that brings the $T_{--}$ pieces in the gauge fields has the effect of precisely canceling the diffeo part such that we are left with the pure shift in $\phi$ piece. Maybe it's related to the fact that what enters in the CS charges is always the combination $\xi\cdot A+\Lambda$ the diffeo can be rewritten as a field-dep gauge transformation?)}
 
 Further using the constancy of the (angular) momentum and the entropy  

\be
E_L - E_R = E_L^{[0]} - E_R^{[0]} \;, \;\;\;\;\; S = S^{[0]}
\ee
allows us to show that 
\begin{align}
RE_L-\frac{2\pi}{k}Q_L^2&=R E_L^{[0]}-\frac{2\pi}{k}J_0^2\hspace{1.2cm}R_v E_R-\frac{2\pi}{k}Q_R^2=R E_R^{[0]}-\frac{2\pi}{k}\bar{J}_0^2
\end{align}
which fully determines the deformed spectrum in  terms of the undeformed one, as in \eqref{defspec}.

\subsubsection*{The relation between the deformed and undeformed thermodynamics}

Since - as we just showed - the bulk spectrum perfectly matches that of a $J\bar T$ - deformed CFT, obviously the thermodynamics should also match. The relationship between the thermodynamic quantities of the deformed theory - which are to be read off from the smoothness conditions for the analytically continued deformed  spacetime - and those of the undeformed one - read off from the smoothness of the euclideanised charged BTZ solution - can be  obtained by noting the two solutions are  simply related by the coordinate transformation \eqref{coordtrdef}, accompanied by the gauge transformation \eqref{fielddepgaugetr2}. 

%In the following, we exhibit, following \cite{}, a very simple way to generate the thermodynamic quantities of this background from those of the undeformed CFT. The simple way the deformed thermodynamic quantities are generated is that we start from the undeformed euclidean BTZ solution, with coordinates identified according to the undeformed temperatures $T_{L,R}^{[0]}$, and we perform a simple coordinate transformation  $ V \r V -\l \phi$, with $\phi$ given by the constant solution \eqref{constantsolparam}. This entirely reproduces the correct relation between the deformed and undeformed temperatures. 

 In the bulk, the fact that the deformed solution written in terms of the field-dependent coordinates is identical to the undeformed one with some parameters $\L, \bar \L$ implies that the identifications of the field-dependent rescaled coordinates  $\hat{u}=\frac{U}{R},\hat{v}=\frac{v}{R_v}$ are identical to the identifications of the corresponding rescaled coordinates in the undeformed geometry 
 %\textcolor{red}{Careful factors of $2\pi$, probably best to divide by them below}
\begin{align}
\hat{u}\sim \hat{u}+ m +\frac{i n}{R T_L^{[0]}}\;, \hspace{1cm}\hat{v}\sim \hat{v}+ m -\frac{i n}{R T_R^{[0]}} \;, \;\;\;\;\;\; m, n \in \mathbb{Z}
\end{align}
where  $T^{[0]}_{L,R}$ are related to the parameters of the background in the standard way, fixed by smoothness 
\be
\frac{1}{T_L^{[0]}}=\frac{\pi\ell}{\sqrt{\mathcal{L}}}\;, \;\;\;\;\; \frac{1}{T_R^{[0]}}=\frac{\pi\ell R }{R_v\sqrt{\bar{\mathcal{L}}}}
\ee
Remember $v$ is related to $V$ via the shift $v = V -\l \phi + const$, where for constant solutions 
\begin{align}\label{constantsolparam}
\phi(U,V)&=\frac{\frac{Q_L R_v}{R}U -Q_R V}{R-\lambda Q_L}
\end{align}
which simply follows from requiring that the charges of the conserved currents associated with $\phi$ be given by $Q_{L,R}$. % \emph{Correct?}} {\color{ForestGreen}(yes, we get $\frac{Q_L}{R}$ for $\frac{\partial_U\phi}{1-\lambda\partial_V\phi}$ and $-\frac{Q_R}{R_v}$ for $\frac{\partial_V\phi}{1-\lambda\partial_V\phi}$)}
%
%{\color{blue}
%In the bulk, we can extract the temperatures from the identifications of the euclidean geometry required for smoothness. Expressed in the coordinates $\hat{u}=\frac{U}{R},\hat{v}=\frac{v}{R_v}$, \textcolor{red}{Notation!!!!} the metric is just BTZ, for which the smoothness of the euclidean geometry requires:
%Using these notations (better place to move it?):
%For the stationary solution we can simply write:  \textcolor{red}{Notation!!!}
%\begin{align}
%\hat{u}=\hat{U}=\frac{U}{R}\hspace{1cm}\hat{v}=\frac{V-\frac{\lambda Q_L}{R} U}{R-\lambda Q_L}
%\end{align}}
Then, using the relation between the two sets of coordinates: $\{\hat{u},\hat{v}\}$ and $\{U,V\}$, which we can write conveniently as 
\begin{align}
V&=(R-\lambda Q_L) \hat{v}+\lambda Q_L \hat{u}\hspace{1cm}U=\hat{u} R
\end{align}
we can read immediately off the deformed temperatures, associated to the periodicities of the  latter. More precisely,  the periodicity of the $U$ coordinate yields
\begin{align}
T_L=T_L^{[0]}=\frac{\sqrt{\mathcal{L}}}{\pi\ell}
\end{align}
From the periodicity of $V$ we obtain: 
\begin{align}
V+  m(R-\lambda Q_L +\lambda Q_L)-in\bigg[ \frac{R-\lambda Q_L}{R T_R^{[0]}}-\frac{\lambda Q_L}{R T_L^{[0]}}\bigg]=V+ m R-\frac{i n }{T_R}
\end{align}
The second terms trivially agree, while the identification of the last terms gives: 
\begin{align}
\frac{1}{T_R}&=\frac{R-\lambda Q_L}{R T_R^{[0]}}-\frac{\lambda Q_L}{R T_L^{[0]}}
\end{align}
which is precisely the relation \eqref{fieldtheorypotentialapp} between deformed and undeformed temperatures in the boundary theory.

Next, we would like to relate the chemical potentials in the bulk solution to those in the deformed CFT, summarized in appendix \ref{Appendix:thermoJTbar}. To derive the chemical potentials, we will study the first law of thermodynamics
\be \label{firstlawft1}
\d S = \frac{1}{T_L}\delta E_L +\frac{1}{T_R}\delta E_R + \mu_L \delta J_0 + \mu_R \delta \bar{J}_0
\ee
from  the bulk perspective, which will in particular justify why it is natural to take the variation with respect to $J_0, \bar J_0$, rather than e.g. $Q_{L,R}$. 
To check the first law, we will find it most convenient to use the covariant phase space formalism (see appendix \ref{appendixcovph}). However, if we use the Chern-Simons action \eqref{proposalactioncs} in this formalism, we cannot obtain  two independent conserved $U(1)$ charges associated to the gauge transformations of $A,B$ -  see \eqref{pbch} - which we nonetheless need for the first law. This issue can easily be fixed by slightly modifying the action by a boundary term that leads to the usual Chern-Simons  bulk 
\be
S'_{CS}=  - \frac{k}{8\pi} \int d^3 x \sqrt{g} \, \e^{\mu\nu\rho} (A_\mu \p_\nu A_\rho-B_\mu \p_\nu B_\rho) \label{newcsact}
\ee
  In this formalism, the entropy of the black hole is identified with the Noether charge associated with the Killing vector that generates the horizon \cite{Wald:1993nt}, which for a rotating black hole is %{\color{ForestGreen}(I wouldn't write the expression, we do not use it explicitly)}
\be
\xi_{hor} = \p_t + \Omega \p_\s = \frac{T}{T_L}\partial_U - \frac{T}{T_R}\partial_V
\ee
where $T=\frac{2 T_L T_R}{T_L+T_R}$ is the Hawking temperature. The associated conserved charge  is given by an integral over a codimension two surface of a form, $k_\xi$, which receives both gravitational contributions of the form \eqref{metriccontribph},  and Chern-Simons  ones \eqref{csdiffcontribution}. Since this charge is independent of where it is evaluated, we deduce that 

\be
\int_{S^{\infty}} k_\xi = \int_{hor} k_\xi = \int_{hor} k^{grav}_\xi + \int_{hor} k^{CS}_\xi
\ee 
The left-hand side yields the conserved charge associated with $\xi_{hor}$, whereas the first term on the rightmost side yields $T \d S$, as shown by \cite{Iyer:1994ys}.  Dividing by $T$, we obtain 

\be
\frac{1}{T_L} \d E_L + \frac{1}{T_R} \d E_R = \d S + \frac{k}{4\pi T}  \int \xi_{hor}^\l \, (A_\l \d A_\s - B_\l \d B_\s) 
\ee
Comparison with \eqref{firstlawft1} allows us to identify the charges as \eqref{holonomies}, explaining why it is the  $\d J_0, \d J_0$ variations that appear naturally in the first law. The chemical potentials are given by the standard formulae 

\be
\mu_L=%\frac{\Phi^A}{T} =
 -\frac{\xi_{hor}^\l A_\l}{T} = \pi \ell\bigg( -\frac{4\pi}{k}\frac{Q_L}{R\sqrt{\mathcal{L}}}-\frac{\lambda E_R}{R_v\sqrt{\bar{\mathcal{L}}}} \bigg) \;, \;\;\;\;\; \mu_R=%\frac{\Phi^B}{T} = 
 -\frac{\xi_{hor}^\l B_\l}{T} = \pi \ell\bigg( -\frac{4\pi}{k}\frac{Q_R}{R_v\sqrt{\bar{\mathcal{L}}}}+\frac{\lambda E_R}{R_v\sqrt{\bar{\mathcal{L}}}} \bigg)
\ee
We note these are identical to the field theory formulae \eqref{fieldtheorypotentialapp} from appendix \ref{Appendix:thermoJTbar}. %\textcolor{red}{Can one argue for the form of the result from the known modification of $A$ and $\xi$? I guess only the gauge transformations should be relevant...} {\color{ForestGreen}(I didn't manage to write any useful expression. Clearly the gauge fields change, but also the temperatures that enter in a nontrivial way)}

\section{Extended symmetries of the holographic dual}\label{section6:asg}

In the previous section, we performed several basic checks of the proposed holographic dictionary for  $J\bar{T}$-deformed holographic CFTs, namely that $AdS_3$ gravity coupled to the two $U(1)$ Chern-Simons gauge fields \eqref{proposalactioncs} via the mixed boundary conditions \eqref{sourcesflatsp} does yield the correct $J\bar T$ - deformed spectrum and thermodynamics, for a general non-chiral $U(1)$ deforming current. The aim of this section is to perform an even more detailed precision test of this dictionary, namely to show that the asymptotic symmetries of this bulk theory precisely reproduce the infinite-dimensional modified (Virasoro-Kac-Moody)${}^2$ symmetry algebra of  $J\bar{T}$-deformed CFTs. 

For the computation, we use the covariant phase space formalism \cite{Lee:1990nz,Iyer:1994ys} (see \cite{Compere:2018aar} for a comprehensive review). The reason for this choice is that the specification of the phase space of the theory - i.e., of the most general background allowed by the boundary conditions - is in principle sufficient %\textcolor{red}{Should we qualify?} 
to completely determine all the conserved charges and their algebra\footnote{By contrast, the formalism used in the previous subsection only provides a way to compute the expectation values of the conserved currents, but the symmetry parameters against which they are to be integrated to yield the conserved charges are not \emph{a priori} provided. More precisely, it is currently not clear why only  parts of the asymptotic diffeomorphism \eqref{solutiondiffeo} enter the charge. \label{ctsubft}}}. % \emph{Is it true that this is not fully determined in the Kraus formalism?} {\color{ForestGreen}(let's discuss this in the case of $T\bar{T}$, where we didn't have to throw away any terms. In the covariant formalism we obtained automatically that the charges depend only on the periodic part of the functions. If we just use $T\xi$, how can we get rid of the part in $\xi$ which comes from the primitives that are at the same order in the $r$ expansion as the functions we want?)} 
The calculation proceeds in a few algorithmic steps: 

\begin{enumerate}
\item \textbf{specifying the phase space},  identified with the space of solutions to the equations of motion that satisfy the given set of boundary conditions.   
 As explained in the previous section, the most general such solution is given by the metric \eqref{deformedmetric} and  gauge fields \eqref{gaugefieldA} - \eqref{gaugefieldB} and can be conveniently parametrized, in radial gauge,  by four periodic functions $\mathcal{L}(U),\bar{\mathcal{L}}(v),\mathcal{J}(U),\bar{\mathcal{J}}(v)$. Due to the non-dynamical nature of three-dimensional gravity and Chern-Simons theory, these solutions are locally gauge-equivalent to the vacuum solution. 
 \vskip2mm
 
\item \textbf{finding the most general allowed} (non-trivial) \textbf{diffeomorphisms  and gauge transformations}. Since 
all solutions belonging to the above phase space are related by combinations of (radial-gauge-preserving) diffeomorphisms and gauge transformations, the allowed transformations  are those whose effect is only to change the functions parametrizing the solutions. The allowed symmetry parameters are thus in one-to-one correspondence with infinitesimal changes in these functions. Their form turns out to be identical to the transformations \eqref{leftmovingtransf}, \eqref{rightafftransf} and \eqref{rightpseudocftr} we found in the Lagrangian formalism, including the large affine compensating terms. 

\vskip2mm

\item \textbf{computing the associated conserved charge differences}.  To each allowed transformation $(\xi, \Lambda)$ and to each pair of backgrounds that differ via an infinitesimal change in the values of the background fields  $\Phi \rightarrow\Phi+\delta\Phi$, we associate a charge difference $\slashed{\delta}Q_{\xi,\Lambda}$ expressed as an integral at spatial infinity over a codimension two  form
$k_{\xi,\Lambda}$ that is algorithmically constructed from the action, up to known ambiguities
\be \label{kformchrg}
\slashed{\delta}Q_{\xi,\Lambda} = \int_{S^{\infty}} k_{\xi,\Lambda} (\d \Phi,\Phi)
\ee
% which are closed asymptotically for all allowed transformations. These closed forms receive contributions from the various fields in the solutions. (In this section we use the standard notation for these forms, $k_{\xi,\Lambda}$) 
Assuming integrability, these charge differences can be further integrated in phase space to obtain the full conserved charges. As in the previous section, in order to reproduce the correct field-theoretical conserved charges using this formalism, we need to use the    action \eqref{newcsact} for the Chern-Simons  fields, which differs from  \eqref{proposalactioncs} by a boundary term.

\vskip2mm

\item \textbf{computing the charge algebra}.  In general,  the conserved charges span  a Poisson algebra, which is defined by the bracket 
\begin{align}
\{Q_{\xi,\Lambda},Q_{\chi,\Sigma}\}&\equiv \delta_{(\chi,\Sigma)}Q_{\xi,\Lambda}
\end{align}
As in the field theory analysis, consistency (i.e., antisymmetry) of this bracket fixes the field-dependence of the previously ambiguous zero mode  of the field-dependent coordinate $v$. % coordinate, just like in field theory
 %\textcolor{ForestGreen}{(I would say remove the zero mode of $\phi$ in $v$, otherwise one might think we need to remove any constant from $v$, which is not the case)}
\end{enumerate}

\noindent Thus, the conserved charges and their algebra are entirely determined by the boundary conditions. In this section, we work out the (rather technical) details of these steps. More precisely, in section \ref{section61:allowedtr} we review the parametrisation of the background and discuss the most general asymptotic diffeomorphisms and gauge transformations that preserve the boundary conditions. In section \ref{section62:conservedcharges}, we compute the conserved charges associated with each type of transformation, while in \ref{section63:asysymalgebra} we compute the asymptotic symmetry algebra, finding precisely the structure \eqref{algebranonlinearjtbar} - \eqref{algebra:LMRMaffine}, this time with an additional central extension. % In the last subsection, we comment on the representation theorem of \cite{Barnich:2010eb}, \textcolor{red}{Still to review: }which is not obeyed in this background. {\color{ForestGreen} (from explicit computations we see that the reason is the charge-dependence of the symmetry transformations, but it would be interesting to understand more generally how we need to extra modify the bracket in presence of such charge-dependent terms)}.
 We also comment on similarities and differences with the $T\bar T$ case  \cite{Guica:2019nzm,Georgescu:2022iyx}  and  the analysis of \cite{Compere:2013bya} in appendix \ref{appendixCSS}.

\subsection{The bulk phase space and the allowed symmetry generators}
\label{section61:allowedtr}

The bulk phase space, which we derived in the previous section, is viewed as the space of solutions to the equations of motion with the prescribed boundary conditions.   Each point in phase space consists of an $AdS_3$  metric and two flat $U(1)$ gauge fields. We partly fix the gauge redundancies by choosing all fields to be in radial gauge. As explained in the previous section, % the most general such metric is parametrised by two arbitrary periodic  functions $\L(U)$  and $\bar{\L} (v)$, as well as a scalar field $\phi$, with $v = V - \l \phi$.  On-shell, $\phi(U,V)$ is a sum of a function of $U$ and a function of $v$, and is specified via the left/right-moving currents 
 the most general such  background is  conveniently  parametrized in terms of four arbitrary \emph{periodic} functions $\mathcal{L}(U),\bar{\mathcal{L}} (v),\mathcal{J}(U),\bar{\mathcal{J}}(v)$, where $v$ is a field-dependent coordinate  that satisfies

\be
\p_U v  = - \l \p_U \phi %= - \frac{\l \mathcal{J}(U)}{1+\l\bar{\mathcal{J}}(v)}
 \;,  \;\;\;\;\;\p_V v = 1-\l \partial_V\phi%=\frac{1}{1+\l\bar{\mathcal{J}}(v)}
\ee
The scalar field $\phi$ is the bosonisation \eqref{jbos} of the field-theory $U(1)$ current, which on-shell consists of a left-moving piece parametrised by $\J(U)$ and a right-moving one, encoded in $\bar \J (v)$, both introduced in \eqref{tangtspparam}.
The derivatives of the scalar are related to the currents as 
\be
\p_U \phi = \frac{\mathcal{J}(U)}{1+\l\bar{\mathcal{J}}(v)} \; , \;\;\;\;\;  \partial_V\phi=\frac{\mathcal{J}(v)}{1+\l\bar{\mathcal{J}}(v)}
\ee
One may formally integrate the above equations and 
%
%We start by parametrizing our background conveniently. The metric is specified, as explained, by $\mathcal{L}(U),\bar{\mathcal{L}}(v)$ and $\phi(U,V)$. On-shell, $\phi(U,V)$ is a sum of a function of $U$ and a function of $v$, which can be
 express $v$ in terms of the primitives of $\mathcal{J}(U),\bar{\mathcal{J}}(v)$  as
 %
%\textcolor{blue}{Thus, we can write the components of the metric in terms of 4 functions $\mathcal{L}(U),\bar{\mathcal{L}}(v),\mathcal{J}(U),\bar{\mathcal{J}}(v)$, \textbf{periodic} in their variables. These relations do not fix the zero mode of $\phi$. In the field theory analysis, we obtained that the consistency of the symmetry algebra depends on extracting the zero mode of $\phi$ from the field-dependent coordinate. In other words, we do not have field-dependent constants in $v$. We will reach the same conclusion from the bulk analysis of the asymptotic symmetry algebra. For now, we can write, generally:} 
\begin{align} \label{fielddepcoordinitial}
v&=V-\l\left(\int^U dU'\mathcal{J}(U') + \int^v dv'\bar{\mathcal{J}}(v') + c_v\right)
\end{align}
where  the integrals are defined to not contain any constant mode,  placing instead this ambiguity in the integration constant $c_v$ which, \emph{a priori}, can be field-dependent% (with a non-zero variation in field space)
. This field-dependence is in principle fixed by the definition of $v$, even though we will need to wait until section \ref{section63:asysymalgebra} to access this information.  Note that the Fourier zero modes of  $\mathcal{J},\bar{\mathcal{J}}$ will lead to terms  linear in $U,v$, while the remaining contributions are periodic. 

%, very similar to \cite{Guica:2019nzm,Georgescu:2022iyx}, the bracket contains terms linear in $U,v$, corresponding to the zero modes of $\mathcal{J},\bar{\mathcal{J}}$. {\color{blue} The functions $\mathcal{L},\bar{\mathcal{L}}$ are also periodic and can have zero modes, since this was the case also in the undeformed metric. We will use the notation $R_v=2\pi r_v$, with $r_v$ the radius of the field-dependent coordinate. 

%Apart from the derivatives of $\phi$, the  gauge fields also depend on $T_{--}^{tg}$, which we need to express in terms of the functions introduced above. The CS contribution can be written directly from \eqref{energymomtensor}: {\color{ForestGreen}(factors of $2\pi$ probably removed)}
%\begin{align}
%T^{tg(CS)}_{--}&=\frac{2\pi}{k}\left(\frac{\partial_V\phi}{1-\l\partial_V\phi}\right)^2=\frac{2\pi}{k}\bar{\mathcal{J}}^2
%\end{align}
%Quick check, for constant solutions we have $T^{tg}_{--}=\frac{E_R}{2\pi R_v}=\frac{Q_R^2}{2\pi k R_v^2}$ (CS contribution only), which is exactly what we get by plugging in \eqref{scalarphipar} above. The metric contribution gives $\frac{\bar{\mathcal{L}}}{2\pi}$.  {\color{ForestGreen}(we will have this in the previous section already)}. We will rescale $\bar{\mathcal{L}}$ and also $\mathcal{L}$ by a factor of $\frac{k}{4\pi^2}$ in order for the gauge fields to look nicely. In the metric, we can absorb this into a rescaling of $\ell$. 
%}

We transcribe herein the form of the allowed metric \eqref{deformedmetric}  in terms of these functions 
\begin{align}\label{deformedmetricsec6}
ds^2&=\ell^2 \, \frac{ dz^2}{z^2}+\bigg(\mathcal{L}+\bar{\mathcal{L}}\frac{\l^2\mathcal{J}^2}{(1+\l\bar{\mathcal{J}})^2}-\frac{1}{z^2}(1+ z^4\mathcal{L}\bar{\mathcal{L}})\frac{\l\mathcal{J}}{1+\l\bar{\mathcal{J}}}\bigg)dU^2+\nonumber\\
&+\frac{1}{1+\l\bar{\mathcal{J}}}\bigg(\frac{1+ z^4\mathcal{L}\bar{\mathcal{L}}}{z^2}-\frac{2\l\bar{\mathcal{L}}\mathcal{J}}{1+\l\bar{\mathcal{J}}}\bigg)dUdV+\frac{\bar{\mathcal{L}}}{(1+\l\bar{\mathcal{J}})^2}dV^2
\end{align}
and of the gauge fields \eqref{gaugefieldA}, \eqref{gaugefieldB} 
%\textcolor{red}{Factors $2\pi$ and $k$, maybe also $\ell$?}
\begin{align} 
A_U &= \mathcal{J} + \frac{\l^2\mathcal{J}(\kappa\bar{\mathcal{L}}+\bar{\mathcal{J}}^2)}{2(1+\l\bar{\mathcal{J}})}\;, \;\;\;\;\; A_V = -\frac{\l(\kappa\bar{\mathcal{L}}+\bar{\mathcal{J}}^2)}{2(1+\l\bar{\mathcal{J}})}\;, \;\;\;\;\;\;\;\kappa\equiv \frac{k}{16\pi^2 G\ell} \nonumber \\
B_U &= -\frac{\l\mathcal{J}}{1+\l\bar{\mathcal{J}}}\bigg[\bar{\mathcal{J}}+\frac{\l(\kappa\bar{\mathcal{L}}+\bar{\mathcal{J}}^2)}{2}\bigg] \;, \;\;\;\;\;\; B_V = \frac{\bar{\mathcal{J}}}{1+\l\bar{\mathcal{J}}}+ \frac{\l(\kappa\bar{\mathcal{L}}+\bar{\mathcal{J}}^2)}{2(1+\l\bar{\mathcal{J}})} \label{paramgaugefields}
\end{align}
which have been rescaled  by an overall factor of $\frac{k}{4\pi}$ with  respect to the previous section ($(A_U)_{above}=\frac{k}{4\pi}(A_U)_{section5}$) to simplify the notation.
%(The factors of $k/4\pi$ are annoying to carry around, in the following I ignored them, meaning that I rescaled the gauge fields.) 
%\textcolor{red}{That's fine (if you mention it), above I mean relative factors.}

To summarize, the bulk phase space can be  parametrised  using four periodic functions % of their arguments
 $\mathcal{L}(U),\bar{\mathcal{L}}(v),\mathcal{J}(U),\bar{\mathcal{J}}(v)$ and an additional, possibly field-dependent constant $c_v$,  which only enters  in the field-dependent coordinates via \eqref{fielddepcoordinitial}. Note all the above metrics are diffeomorphic to AdS$_3$, and all gauge connections are flat; nonetheless, they all carry non-trivial conserved charges with respect to large diffeomorphisms, to which we now turn.

\subsubsection*{Allowed transformations in phase space}

We would like to characterise the most general gauge transformations (consisting of diffeomorphisms and two $U(1)$ gauge transformations) that keep the metric and Chern-Simons gauge fields within the prescribed phase space \eqref{deformedmetricsec6},   \eqref{paramgaugefields}. % [We expect that the asymptotic symmetries are combinations of diffeomorphisms, acting on the metric and gauge fields, and gauge transformations of the gauge fields.] 
Since the solutions for the bulk fields are all in radial gauge 
 \begin{align}
g_{zz}&=\frac{1}{z^2}\hspace{1cm}g_{zU,zV}=0\hspace{1cm}A_z=B_z=0
\end{align}
this condition must be preserved by the gauge transformations.
The form of the most general diffeomorphism that respects radial gauge is 
%{\color{blue}This fixes the  $r$ dependence as follows:}
\begin{align}
\xi^z&=z F_z(U,V)\nonumber \\
\xi^U&=F_U(U,V)+ \ell^2 \, \frac{\big[\l\mathcal{J}-z^2\mathcal{L}(1+\l\bar{\mathcal{J}})\big]\partial_V F_z(U,V)+\partial_U F_z(U,V)}{\mathcal{L}(1-z^4\mathcal{L}\bar{\mathcal{L}})} \nonumber\\
\xi^V&=F_V(U,V)+\ell^2 \, \partial_V F_z(U,V)\bigg[\frac{\l^2\mathcal{J}^2\bar{\mathcal{L}}+(1+\l\bar{\mathcal{J}})\mathcal{L}(1+\l\bar{\mathcal{J}}-2z^2\l\mathcal{J}\bar{\mathcal{L}})}{\mathcal{L}\bar{\mathcal{L}}(1-z^4\mathcal{L}\bar{\mathcal{L}})}\bigg]+ \nonumber\\
& \hspace{1.2cm} + \ell^2 \, \partial_U F_z(U,V)\bigg[\frac{\l\mathcal{J}-z^2\mathcal{L}(1+\l\bar{\mathcal{J}})}{\mathcal{L}(1-z^4\mathcal{L}\bar{\mathcal{L}})}\bigg] \label{diffeorad}
\end{align}
where the functions $F_{U,V,z} (U,V)$ are  arbitrary, so far. Acting with such a diffeomorphism on the gauge fields \eqref{paramgaugefields} takes them out of radial gauge. The allowed $U(1)$ gauge transformations will therefore contain a  compensating term, which is necessary in order to
%%
%We now consider the Lie derivative of the gauge fields with respect to vector fields of this form. Clearly $\mathcal{L}_{\xi}A_{z}\neq 0$, $\mathcal{L}_{\xi}B_{z}\neq 0$, so we need to compensate with a gauge transformation in order to
 bring  the gauge fields back to the radial gauge:
\begin{align}\label{eqfortrphsp}
\partial_z \Lambda_A(z,U,V)&= -(\mathcal{L}_{\xi}A)_z\hspace{1cm}\partial_z \Lambda_B(z,U,V)= -(\mathcal{L}_{\xi}B)_z
\end{align}
The solutions take the form 
\vskip-2mm
\be
\Lambda_A(z,U,V)= F_A(z,U,V)+G_A(U,V) \;, \;\;\;\;\; \Lambda_B(z,U,V)=F_B(z,U,V)+G_B(U,V) \label{defGAB}
\ee
where $G_{A,B} (U,V)$ are - so far - arbitrary $U(1)$ gauge transformations that preserve radial gauge, while $F_{A,B}(U,V,z)$ are the particular solutions to \eqref{eqfortrphsp} 
%This fixes the radial dependence of the gauge transformations:
\begin{align}\label{gaugerad}
F_A(z,U,V)&=-\ell^2\frac{2\l\mathcal{J}^2\bar{\mathcal{L}}-\mathcal{L}(1+\l\bar{\mathcal{J}})(2z^2\mathcal{J}\bar{\mathcal{L}}+\l(\kappa\bar{\mathcal{L}}+\bar{\mathcal{J}}^2))+\l^2 z^2\mathcal{J}\mathcal{L}\bar{\mathcal{L}}(\kappa\bar{\mathcal{L}}+\bar{\mathcal{J}}^2)}{2\mathcal{L}\bar{\mathcal{L}}(1-z^4\bar{\mathcal{L}}\mathcal{L})}\partial_V F_z-\nonumber\\
&-\ell^2\frac{2\mathcal{J}+z^2\l\mathcal{L}(\kappa\bar{\mathcal{L}}+\bar{\mathcal{J}}^2)}{2\mathcal{L}(1-z^4\mathcal{L}\bar{\mathcal{L}})}\partial_U F_z\\
F_B(z,U,V)&=-\ell^2\frac{(2\bar{\mathcal{J}}+\l(\kappa\bar{\mathcal{L}}+\bar{\mathcal{J}}^2))((1+\l\bar{\mathcal{J}}-z^2\l\mathcal{J}\bar{\mathcal{L}})\partial_V F_z -z^2 \bar{\mathcal{L}}\partial_U F_z)}{2\bar{\mathcal{L}}(1-z^4\mathcal{L}\bar{\mathcal{L}})}
\end{align}
%where we denoted by $F_{A,B}(z,U,V)$ the part which depends on $z$ and is completely fixed in terms of the radial function $F_z$ and by $G_{A,B}(U,V)$ the rest, namely arbitrary functions of $U,V$ only. These expressions are not important for now, but the point is that under diffeomorphisms \eqref{diffeorad} and gauge transformations
%\begin{align}
%A_{\mu}\rightarrow A_{\mu}+\partial_{\mu}\Lambda_A(z,U,V)\hspace{1cm}B_{\mu}\rightarrow B_{\mu}+\partial_{\mu}\Lambda_B(z,U,V)
%\end{align}
%given in \eqref{gaugerad}, the gauge fields remain $z$-independent. 
%
In order to determine the allowed form of the five functions $F_{U,V,z} (U,V)$ and $G_{A,B} (U,V)$ above, we require that the above gauge transformations map a point in phase space to another point in phase space. Since, locally, the backgrounds \eqref{deformedmetricsec6}, \eqref{paramgaugefields} are pure gauge, we expect a one-to-one map between the allowed gauge transformations in radial gauge and infinitesimal changes in the functions parametrising the backgrounds\footnote{
Our notation  $\d \bar \L(v),  \d \bar\J(v)$ stands for $ \d [\bar \L(v)],\d [\bar \J(v)]$, which in general are not just  functions of $v$, see %We chose to write $\d\bar{\L}(v),\d\bar{\J}(v)$, although they also depend on $U$, since the $U$ dependence is present only in the change of $v$, see 
\eqref{splittingvar}. }
%However, since the full solution is locally pure gauge, the most general such transformations will clearly simply correspond (in a one-to-one fashion) to an infinitesimal  change
%
\be
\L(U) \r \L(U) +\d \L(U) \;, \;\;\; \bar\L(v) \r \bar\L(v) +\d \bar\L(v)\;, \;\;\;\J(U) \r \J(U) +\d \J(U) \nonumber
\ee
\vskip-2mm
\be\bar\J(v) \r \bar\J(v) +\d \bar\J(v)\;, \;\;\;\; c_v \r c_v +\d c_v
\ee
%\textcolor{violet}{where the coordinate dependence is essential}.
 To find this map, we simply write
%
%\subsubsection*{Solving for the diffeomorphisms and gauge transformations}
%We now impose that the diffeomorphisms map a point in phase space to another point in phase space, namely that their effect on the metric is only to change the functions that parametrize it
\begin{align}\label{variationseqmet}
\mathcal{L}_{\xi}g_{\mu\nu}&=\frac{\partial g_{\mu\nu}}{\partial \mathcal{L}}\delta \mathcal{L}+\frac{\partial g_{\mu\nu}}{\partial \bar{\mathcal{L}}}\delta \bar{\mathcal{L}}+\frac{\partial g_{\mu\nu}}{\partial \mathcal{J}}\delta \mathcal{J}+\frac{\partial g_{\mu\nu}}{\partial \bar{\mathcal{J}}}\delta \bar{\mathcal{J}}
\end{align}
and similarly for the gauge fields  
\begin{align}\label{variatigaugetr}
(\mathcal{L}_{\xi}A)_\mu + \partial_{\mu} \Lambda_A &=\frac{\partial A_{\mu}}{\partial \mathcal{L}}\delta \mathcal{L}+\frac{\partial A_{\mu}}{\partial \bar{\mathcal{L}}}\delta \bar{\mathcal{L}}+\frac{\partial A_{\mu}}{\partial \mathcal{J}}\delta \mathcal{J}+\frac{\partial A_{\mu}}{\partial \bar{\mathcal{J}}}\delta \bar{\mathcal{J}}\\
(\mathcal{L}_{\xi}B)_{\mu} + \partial_{\mu} \Lambda_B &=\frac{\partial B_{\mu}}{\partial \mathcal{L}}\delta \mathcal{L}+\frac{\partial B_{\mu}}{\partial \bar{\mathcal{L}}}\delta \bar{\mathcal{L}}+\frac{\partial B_{\mu}}{\partial \mathcal{J}}\delta \mathcal{J}+\frac{\partial B_{\mu}}{\partial \bar{\mathcal{J}}}\delta \bar{\mathcal{J}} \nonumber
\end{align}
Since these equalities must hold order by order in the $z$ expansion, and respect the particular coordinate dependence of the functions involved,  \eqref{variationseqmet} fully determines the functions $F_{U,V,z}$ in terms of the change in the background parameters, while \eqref{variatigaugetr} additionally determine
%
% {\color{ForestGreen}Once we solve for $\xi$, we need to impose also that the gauge fields are mapped to gauge fields of the same form
%
%%These equations fix the functions
%
 $G_A(U,V), G_B(U,V)$. % from the gauge transformations \ref{gaugerad}. 

In order to solve \eqref{variationseqmet}, it is important to first understand the coordinates on which  the right-hand-side of this equation depends. Clearly,   $\delta\mathcal{L}$ and $\delta\mathcal{J}$ are functions of ${U}$ only. As for $\delta\bar{\mathcal{L}}$ and $\delta\bar{\mathcal{J}}$, we need to take into account the fact that, if $\mathcal{J},\bar{\mathcal{J}}$ change, then so will the field-dependent coordinate, $v$. Consequently, these variations can be written as 
\begin{align}\label{splittingvar}
\delta\bar{\mathcal{L}}&=\delta\bar{\mathcal{L}}_{int} ({v})+\bar{\mathcal{L}}'(v) \delta {v}\hspace{1cm}\delta\bar{\mathcal{J}}=\delta\bar{\mathcal{J}}_{int}({v})+\bar{\mathcal{J}}'(v) \delta {v}
\end{align}
where the first, ``intrinsic'' part corresponds to a variation of the Fourier coefficients in a would-be mode expansion of the corresponding function, while the second term originates from the variation of the field-dependent  coordinate. The intrinsic parts are functions of the field-dependent coordinate only,  while \eqref{fielddepcoordinitial} implies that  
\begin{align}\label{variationfielddepen}
\delta v&=-\frac{\l}{1+\l\bar{\mathcal{J}}}\left(\delta c_v+\int^U dU' \, \delta \mathcal{J}(U')+\int^v dv' \,\delta\bar{\mathcal{J}}_{int}(v')\right)% :=-\frac{\l}{1+\l\bar{\mathcal{J}}}\left(\delta c_v+[\eta(\hat{U})]_{nzm}+[\bar{\eta}(\hat{v})]_{nzm}\right)
\end{align}
where the primitives are defined to not contain any zero modes  and the constant $\d c_v$ should be considered as given.  \eqref{variationfielddepen}  in  turn dictates the form of the $U$ and $v$ - dependence of the full variations \eqref{splittingvar}.% \st{ It is important to notice that there exist ambiguities in this splitting, to which we will return later}{}.

Having determined the coordinate dependence of the right-hand-side of \eqref{variationseqmet}, all we have left to do is to solve this equation and determine the allowed dependence of the functions $F_{U,V,z}$ in terms of the functions $\d \L(U), \d \J(U), \d \bar \L_{int} (v)$ and $\d \bar \J_{int} (v)$, which are assumed to be known. Note, however, that we are still left with significant freedom in how to parametrize the functions of the coordinates $U$, $v$ that we will find, which will result in different relations between these functions and the variations $\d \L$, etc. of the background parameters that determine them.  We will choose our parametrisation of the various functions so that their relation to the change in the background: $\d \L$, etc. resembles as closely as possible the analogous relation %between the functions parametrising the diffeomorphisms and the change in the background
 in undeformed AdS$_3$, see  e.g. \eqref{variationofL}. This ensures, in particular, a good $\l \r 0$ limit of our analysis.

%{\color{magenta} Each choice of diffeo/gauge will correspond to some particular $\d L$ etc. Conversely, each $\d L$ etc will be linear in the diffeo function and can be thought as determining it. We will choose to parametrize the $4$ functions that change $\d \L$ etc as in BTZ, namely $\d \L = 2f' \L + f \L' +\ldots$. Of course, in the undeformed theory these simply parametrize the leading diffeos and gauge transformations. Here, we define the functions in this fashion and work out the $\l$ corrections to the diffeos + gauge that actually implement them. These diffeos + gauge are naturally in radial gauge (write 6.12-6.17). Imposing that the Lie derivative of the metric along them equals 6.19 with this form of $\d \L$  etc. plugged in yields the relations 6.32.}

We are now ready to solve \eqref{variationseqmet}.  The $VV$ component of this equation %at order $z^{\pm 2}$
 implies that  
%\textcolor{blue}{The $VV$ equation implies immediately that} 
$F_U(U,V)$ is a function of $U$ only, 
%\textcolor{blue}{ in order to have that $\delta\mathcal{L}$ is a function of $U$ only.} 
while  $F_z$ is the sum of a function of $U$ and one of $v$.   We find it convenient to parametrize them as follows
\be
F_U = f(U) - \frac{\ell^2 f_{zU}'(U)}{\L(U)}\;, \;\;\;\;\;F_z = f_{zU} (U) + f_{zV} (v) \label{Fufunction}
\ee
in view of our comments in the preceding paragraph. The fact that $\delta\mathcal{J}(U)$ is a function of $U$ only implies that $F_V$ is also the sum of a function of $U$ and a function of $v$
\be \label{functionVdiffeo}
F_V = -\l g_{VU} (U) + g_{Vv}(v)
\ee
where the explicit inclusion of the $\l$ coefficient in the parametrisation of the first function indicates that such a term is absent at $\l=0$. This function is determined by $\d \J(U)$ as %\textcolor{red}{Is $g_{VU}$ determined only up to a constant?}{\color{ForestGreen}(in this expression yes, but overall in $F_V$ the integration constants will just distribute as zero modes of the functions. Maybe we can add a footnote)} \textcolor{red}{Shall we make a comment about this? e.g. in principle we should keep track of all int const.}.
\be
g_{VU}(U) = \int^U \!\! \d \J(U) + \frac{l^2 \J(U) f''(U)}{2 \L(U)}-f(U) \J(U)
\ee
Finally, requiring that the $UV$ component of the metric at order $z^{-2}$ be proportional to $\d \bar{\J}$, which has the very specific $U$ dependence given in \eqref{splittingvar} - \eqref{variationfielddepen} fixes the relation between $f_{zU}$ and $f$ to be the standard AdS one
\be
f_{zU} = \frac{1}{2} f'(U)
\ee
Solving \eqref{variationseqmet} for $\delta\mathcal{L}$, we now find
\begin{align} \label{variationofL}
\d \L=2 \L(U) f'(U)+f(U)\L'(U) -\frac{\ell^2}{2}  f'''(U)
\end{align}
exactly as in undeformed AdS$_3$. Thus, 
 $f$ parametrizes (indeed) the infinitesimal left-moving conformal transformations $U\rightarrow U+ f(U)$ which are preserved by the deformation. Under such transformations $\J$, which is the expectation value of the chiral left-moving current, is expected to  change as $\delta\J=\partial_U(\J f)$. Note that  $\J$ is also expected to transform under  left affine transformations, parametrized by a function $\eta(U)$, as  $\d \J=\p_U\eta$. A useful parametrisation of $\d \J$ is, therefore
\begin{align} \label{variationofJ}
\d \J&=\partial_U \left[\eta(U)+f(U)\J(U)\right]
\end{align}
which we should view as a definition of the function $\eta(U)$. The function $g_{VU}(U)$ in \eqref{functionVdiffeo} is  then automatically fixed in terms of it. Similarly, we choose to write
\begin{align} \label{variationofJbarLbar}
\delta\bar{\J}_{int}(v)&=\partial_v [\bar{\eta}(v)+\bar{f}(v)\bar{\J}(v)]\hspace{1cm}\delta\bar{\L}_{int}(v)=2\bar{\L}(v)\bar{f}'(v)+\bar{f}(v)\bar{\L}'(v)-\frac{\ell^2}{2}\bar{f}'''(v)
\end{align}
as this reproduces the correct affine and conformal transformations at $\l=0$. The form of $\delta\bar{\L}_{int}$ automatically implies $f_{zV}(v)=\frac{\bar{f}'(v)}{2}$ and overall the form of $\delta\bar{\mathcal{L}}$ fixes  
\begin{align}
g_{Vv}(v)&=\bar{f}(v)-\l\bar{\eta}(v)-\frac{\ell^2(1+\l\bar{\J})\bar{f}''}{2\bar{\L}}
\end{align}
All together, we obtain the following final form for the functions that parametrize the radial gauge  diffeomorphisms
\be
F_U(U,V)=f-\frac{\ell^2 f''}{2\mathcal{L}}\;, \;\;\;\;\;
 F_V(U,V)=\bar{f}-\l \eta-\l \bar{\eta}-\frac{\l \ell^2\mathcal{J}f''}{2\mathcal{L}}-\frac{(1+\l\bar{\mathcal{J}})\ell^2\bar{f}''}{2\bar{\mathcal{L}}} \nonumber
\ee
\be
F_z(U,V)=\frac{f'+\bar{f}'}{2} \label{solutiondiffeo}
\ee
Using  \eqref{variationfielddepen}, the change in the field-dependent coordinate takes the following simple form 

\be\label{deltavfunctions}
\delta v=-\frac{\l}{1+\l\bar{\mathcal{J}}}(\eta+\bar{\eta}+f\mathcal{J}+\bar{f}\bar{\mathcal{J}}) 
\ee
The astute reader may nonetheless notice a problem with the above expression for $\d v$, namely that it only agrees with \eqref{variationfielddepen} for a very specific choice of $\d c_v$, equal to the sum of zero modes of the functions that appear above. As we already discussed, $\d c_v$ was supposed to be a given, which should have in principle also contributed to the form  of the asymptotic diffeomorphisms. The reason that our solution \eqref{solutiondiffeo} does not reflect this is that we have been exceedingly cavalier regarding the integration constants appearing in the functions above: a careful analysis does, indeed, show explicit factors of $\d c_v$ appearing. However, the solution is quite unelegant so, rather than presenting it, we will  proceed via a `trick'. 

We would like to bring the value of $c_v$ to the correct one, without modifying the functions $\d \L, \d \J, \d \bar\L$ and $\d \bar \J$ that parametrize the modified background. From \eqref{splittingvar}, this can be achieved by a simultaneous change in the `intrinsic' variations $\d \bar \J_{int}, \d \bar \L_{int}$ as
\begin{align}\label{ambiguitysplit}
\delta c_v\rightarrow \delta c_v -\frac{\mathcal{C}}{\lambda}\;, \;\;\;\;\;\; \delta\bar{\mathcal{J}}_{int}\rightarrow\delta\bar{\mathcal{J}}_{int}-\mathcal{C}\bar{\mathcal{J}}'\;, \;\;\;\;\;\;\delta\bar{\L}_{int}\rightarrow \delta\bar{\L}_{int} - \mathcal{C}\bar{\L}'
\end{align}
which additionally shift $\delta v\rightarrow \delta v+\mathcal{C}$. The latter two shifts modify the relation between the `intrinsic' variations and the functions $\bar \eta, \bar f$. The constant $\mathcal{C}$ is the difference between the variation of the true zero mode of the field-dependent coordinate under the given transformations, and the fiducial variation assumed in \eqref{deltavfunctions}. This `trick' allows for significant simplifications to the conserved charge analysis. The value of $\mathcal{C}$ only enters the charge algebra, which we study in section \ref{section63:asysymalgebra}.

%Given the above definitions of the functions $\bar \eta$ etc.  and $\d c_v$, which is in principle known/ an input, but not (yet) in practice, we can solve for $\d v$ using \eqref{}  and then the bulk diffeomoerphisms and gauge transformations. The solutions turn out to have an unpalatable dependence on the zero modes of the various functions. To avoid this issue, we will be using the following trick: we will be replacing $\d c_v$, which should be a given we cannot freely influence, by a particular combination of the zero modes of the functions, such that $\d v$ takes the very simple  form  
%
%
%where $'f'$ stands for fiducial, only in reference to the zero mode. The true $\d v$ is related to it via 
%%
%\be
%\d v^{true} = \d v^f +\# \mathcal{C}
%\ee
%\textcolor{red}{Using the expr \eqref{} for $\d v$, we obtain the following nice form for the functions}

%\textcolor{red}{One should note, however, that since we shifted $\d v$ from its true value, the $\d \J_{int}$ etc will also be shifted as xxx w.r.t \eqref{}, a transft one can easily check leaves $\d \J$ invar. }

%
{\color{blue}
%Now that we found the diffeomorphisms, we  act with them on the gauge fields. It is clear that they will not preserve the form of the gauge fields and we need some compensating gauge transformations to do so. Thus, we need to solve the following equations {\color{ForestGreen}(they were moved above)}:
%\begin{align}
%\mathcal{L}_{\xi}A_{U,V} + \partial_{U,V} \Lambda_A &=\frac{\partial A_{U,V}}{\partial \mathcal{L}}\delta \mathcal{L}+\frac{\partial A_{U,V}}{\partial \bar{\mathcal{L}}}\delta \bar{\mathcal{L}}+\frac{\partial A_{U,V}}{\partial \mathcal{J}}\delta \mathcal{J}+\frac{\partial A_{U,V}}{\partial \bar{\mathcal{J}}}\delta \bar{\mathcal{J}}\\
%\mathcal{L}_{\xi}B_{U,V} + \partial_{U,V} \Lambda_B &=\frac{\partial B_{U,V}}{\partial \mathcal{L}}\delta \mathcal{L}+\frac{\partial B_{U,V}}{\partial \bar{\mathcal{L}}}\delta \bar{\mathcal{L}}+\frac{\partial B_{U,V}}{\partial \mathcal{J}}\delta \mathcal{J}+\frac{\partial B_{U,V}}{\partial \bar{\mathcal{J}}}\delta \bar{\mathcal{J}}
%\end{align}
%which are equations for the functions $G_A(U,V),G_B(U,V)$ from the gauge transformations.
}
Finally, using the results we obtained for the allowed diffeomorphisms, let us solve  \eqref{variatigaugetr} for the gauge transformations $G_{A,B} (U,V)$ defined in \eqref{defGAB}. First, we evaluate the expressions for the radial-gauge-restoring gauge transformations \eqref{gaugerad}  on the explicit solution \eqref{solutiondiffeo} for the diffeomorphisms, finding
\begin{align}
F_A(z,U,V)&=-\ell^2\frac{2\mathcal{J}\bar{\mathcal{L}}(f''-z^2\mathcal{L}\bar{f}'')+\l(\kappa\bar{\mathcal{L}}+\bar{\mathcal{J}}^2)\mathcal{L}(z^2\bar{\mathcal{L}}f''-\bar{f}'')}{4\mathcal{L}\bar{\mathcal{L}}(1-z^4\mathcal{L}\bar{\mathcal{L}})} \nonumber\\
F_B(z,U,V)&=\ell^2\frac{\big(2\bar{\mathcal{J}}+\l(\kappa\bar{\mathcal{L}}+\bar{\mathcal{J}}^2)\big)\big(z^2\bar{\mathcal{L}}f''-\bar{f}''\big)}{4\bar{\mathcal{L}}(1-z^4\mathcal{L}\bar{\mathcal{L}})} \label{solFAB}
\end{align}
The solution we obtain for $G_{A,B}$ is then given by 
\begin{align}\label{solutiongauge}
G_A(U,V)&=\frac{\mathcal{J}\ell^2 f''}{2\mathcal{L}}-\frac{\l \bar{\mathcal{J}}^2 \ell^2\bar{f}''}{4\bar{\mathcal{L}}}+\eta-\l\int^{v} \bar{\mathcal{J}}\bar{\eta}'-\frac{\l}{2}\int^{v} (\kappa\bar{\mathcal{L}}+\bar{\mathcal{J}}^2)\bar{f}'\\
G_B(U,V)&=\frac{\bar{\mathcal{J}}(2+\l\bar{\mathcal{J}})\ell^2\bar{f}''}{4\bar{\mathcal{L}}}+\bar{\eta}+\l\int^{v} \bar{\mathcal{J}}\bar{\eta}'+\frac{\l}{2}\int^{v} (\kappa\bar{\mathcal{L}}+\bar{\mathcal{J}}^2)\bar{f}'
\end{align}
%\textcolor{violet}{The a priori independent integration constants $c_A,c_B$ can be absorbed in the zero modes of $\eta,\bar{\eta}$ respectively. This will introduce a zero mode $-\lambda(c_A+c_B)$ in the $V$ component of the diffeomorphisms, which can be nevertheless absorbed in the zero mode of $\bar{f}$. Thus, the constants $c_A,c_B$ do not parametrize new, independent transformations. Nevertheless, they are interesting and we will comment on them separately later.} {\color{ForestGreen}(let's keep this for now)} \textcolor{red}{I don't know, I can certainly do a gauge transf without doing a diffeo. I guess they can just be absorbed into $\eta, \bar \eta$, the integration constant is the zm of these functions. }
%
To summarize,  the most general allowed transformations that act on the phase space are combinations of the diffeomorphisms  \eqref{solutiondiffeo} and the gauge transformations \eqref{defGAB}, into which one should plug \eqref{solFAB} and  \eqref{solutiongauge}. They are parametrized by four arbitrary functions $f(U),\bar{f}(v),\eta(U),\bar{\eta}(v)$, which are related to the change in the periodic functions characterising the backgrounds via \eqref{variationofL}, \eqref{variationofJ} and a modification of the form \eqref{ambiguitysplit} acting on \eqref{variationofJbarLbar}, where the constant $\mathcal{C}$ is responsible for readjusting the solution so that it reproduces the correct variation of the zero mode of the field-dependent coordinate. This constant is currently treated as a free parameter, but will be fixed by requiring consistency of 
%
%  In addition, there is a constant mode ambiguity $c_v$ in the definition of $v$, whose variation results in an ambiguity in defining the `intrinsic' variations $\delta\bar{\mathcal{L}}_{int},\delta\bar{\mathcal{J}}_{int}$ via the split \eqref{splittingvar}. The latter will be resolved when discussing
   the charge algebra in section \ref{section63:asysymalgebra}. The radial symplectic form in phase space, evaluated on these variations, vanishes automatically,  which ensures that the associated charges are finite and conserved.

\subsubsection*{Winding terms}
Let us now discuss the periodicity properties of the functions parametrizing the transformations. The main constraint comes from the requirement that the gauge transformation parameters $\Lambda_{A,B}$ be periodic. This constraint follows from the fact \eqref{holonomies} that  the  integrals over $\s$ of the corresponding gauge fields are related to the winding and shift charge of the configuration %{\color{ForestGreen}(ok because we mentioned we rescaled the gauge fields)}
\be \label{invchargessection6}
w = J_0 -\bar J_0 = \int_0^R d\s (A_\s + B_\s) \;, \;\;\;\;\;\;\; Q_0 = J_0 + \bar J_0 = \int_0^R d\s (A_\s - B_\s)
\ee
where certain factors of $k/4\pi$ have been absorbed into a  rescaling of the gauge fields. 
Since we work in a winding superselection sector, any gauge transformation acting on $A+B$ needs to be periodic in $\s$. Since $F_{A,B}$, given explicitly in \eqref{gaugerad}, are periodic, we are left with the requirement that $G_A + G_B$ be periodic, which implies that
%
% $A_{\sigma}$ and $B_{\sigma}$ respectively. In order for them to stay fixed under gauge transformations, we need that $\Lambda_{A,B}$ are, respectively, periodic functions of $\sigma$. Since the $z$-dependent part is periodic, we need $G_{A,B}$ to be periodic. The requirement that $G_A+G_B$ is periodic imposes that:
\begin{align}\label{nowindetacomb}
\eta+\bar{\eta} \hspace{0.3cm}\text{is periodic}
\end{align}
On the other hand, the periodicity of the gauge transformations acting on $A-B$ is directly related to the requirement that the shift charge, $Q_0$, does not change under the corresponding transformation. In the field theory analysis, this condition was directly responsible for the appearance of the large affine compensating transformation. We will now show the same happens in the holographic analysis. Periodicity in $\s$ of the combination $G_A - G_B$ implies that
\begin{align} \label{nowindcond2}
\eta-\bar{\eta}-2\l \int^v \bar{\mathcal{J}}\bar{\eta}'-\l \int^v(\kappa\bar{\mathcal{L}}+\bar{\mathcal{J}}^2)\bar{f}' \hspace{0.6cm}\text{has no winding}
\end{align}
Before we further massage this expression, let us note that invariance of the periodicities of the spacetime coordinates $U,V$  under the allowed diffeomorphisms \eqref{solutiondiffeo} immediately imply that $f$ and the combination
%. For $U$, it is clear that this implies that $f(U)$ is a periodic function of $U$. For $V$, it implies that the following function has no winding:
\begin{align}
\bar{f} -\l (\eta +\bar{\eta})
\end{align}
have no winding, and  hence  both $f$ and $\bar{f}$ are periodic. The same could have been deduced from the relation between $f, \bar f$ and the periodic functions $\d \L, \d \bar{\L}_{int}$. The relation between $\eta, \bar \eta$ and the periodic functions $\d \J, \d \bar \J_{int}$ shows that the affine parameters can have at most a constant winding non-periodic term.  Note that the fact that $\eta+\bar{\eta}$ does not have winding implies that $\delta v$,  and thus also $\delta\bar{\mathcal{J}}$, $\d \bar \L$ do not.

In the following, we  denote by $w_{\eta}$ the winding of $\eta$ which, by \eqref{nowindetacomb},  is minus that of $\bar{\eta}$
\begin{align}
\eta&=\eta_p(\hat{U}) + w_{\eta}\, \hat{U}\hspace{1cm}\bar{\eta}=\bar{\eta}_p(\hat{v})-w_{\eta}\, \hat{v}
\end{align} 
where $\hat{U}=U/R,\hat{v}=v/R_v$ and $\eta_p, \bar \eta_p$ are periodic. % (here we need to write in terms of hatted variables in order for the overall combination to have the correct periodicity). Finally, from $G_A-G_B$ we find that: \textcolor{red}{Where does the denominator come from?}
Plugging this decomposition into \eqref{nowindcond2}, the no-winding condition becomes %{\color{ForestGreen}(we can delete later, here derivatives are wrt v. We should specify because in the field theory section they were mostly wrt $\hat{v}$)}
\begin{align}
2w_{\eta}-2\l ([\bar{\mathcal{J}}\bar{\eta}'_p]_{zm}-w_{\eta}[\bar{\mathcal{J}}]_{zm})-\l [(\kappa\bar{\mathcal{L}}+\bar{\mathcal{J}}^2)\bar{f}']_{zm}=0
\end{align}
where `$[\; ]_{zm}$' stands for Fourier `zero mode' and   the primes denote derivatives with respect to $\hat{v}$.
We obtain %{\color{ForestGreen}(here the charges are just a notation, they match the result we obtain for the charges later in this section. The signs differ from the field theory analysis because we are doing the diffeos differently (we can compare the way the functions shift V for example here and in the field theory sections). It's ok for me as long as we mention it.)} {\color{ForestGreen}(The factor of $\hat{k}$ we can see from field theory where we had $\frac{1}{\hat{k}}$ in front of pseudoconformal charges, here we will have the same factor (see later the result for pseudoconformal charges).)}
\begin{align}\label{winding}
w_{\eta}&=\frac{1}{[1+\l\bar{\mathcal{J}}]_{zm}}\left(\frac{\lambda}{2}[(\kappa\bar{\mathcal{L}}+\bar{\mathcal{J}}^2)\bar{f}']_{zm}+\l[\bar{\mathcal{J}}\bar{\eta}'_p]_{zm}\right)=-\frac{R_v}{R_Q}\left(\frac{\lambda \hat{k}}{2}\bar{Q}_{\bar{f}'}+\l\bar{P}_{\bar{\eta}'}\right)
\end{align}
where in the second equality we used the expressions for the conserved  charges that  will be derived in the next subsection, which can be viewed just as a notation, for now.

\subsection{The conserved charges}\label{section62:conservedcharges}

The goal of this subsection is to compute the conserved charges associated with the allowed symmetry transformations derived in the previous subsection.  Since we work in radial gauge, it turns out that all the symmetries parametrised by $f, \bar f, \eta, \bar \eta$ will act non-trivially on the phase space, and thus correspond to  asymptotic symmetries. Since the functions that parametrize the symmetries are independent, we will compute each of the associated  charges separately. To be more precise, we will compute the charges labeled by the periodic functions $f, \bar f$ and the periodic part $\eta_p, \bar \eta_p$ of the functions parametrising the affine transformations, the winding part of $\eta, \bar \eta$ being entirely fixed via \eqref{winding} by $\bar f$ and $\bar \eta_p$.

%winding terms of the transformations are fixed in terms of the periodic part of $\bar{\eta}$ and $\bar{f}$, so the 4 sets of independent transformations can be labeled by $\eta_p,f,\bar{\eta}_p,\bar{f}$.

 As usual in the covariant phase space formalism, we compute the difference in charges between two backgrounds that differ slightly, as explained in the beginning of this section. %\ref{section6:asg}. 
 The one-form $k_{\xi,\Lambda}$ appearing  in \eqref{kformchrg} receives, in our case, contributions from the metric  and the two $U(1)$ Chern-Simons gauge fields 
\begin{align}
k_{\xi,\Lambda_A,\Lambda_B}&=-\frac{1}{2}\epsilon_{\mu\nu\alpha}(K_g^{\mu\nu}+K_{CS}^{\mu\nu})dx^{\alpha}
\end{align} 
The metric contribution is 
\begin{align}
K^{\mu\nu}_{g,\xi}=- \frac{1}{8\pi G} \left(\xi^\nu  \nabla^\mu h - \xi^\nu \nabla_\s h^{\mu\s} + \xi_\s  \nabla^\nu h^{\mu\s} + \frac{1}{2} h  \nabla^\nu \xi^\mu - h^{\rho \nu} \nabla_\rho \xi^\mu\right)  \label{metriccontribph}
\end{align}
where $h_{\mu\nu} = \d g_{\mu\nu}$, while the Chern-Simons one, computed from the action \eqref{newcsact}, is 
\begin{align} \label{cscontribph}
K^{\mu\nu}_{CS;\xi,\Lambda_A,\Lambda_B}&=-\frac{4\pi}{k}\epsilon^{\mu\nu\rho}\bigg((\xi^{\lambda}A_{\lambda}+\Lambda_A)\delta A_{\rho}-(\xi^{\lambda}B_{\lambda}+\Lambda_B)\delta B_{\rho}\bigg)
\end{align}
The details are presented in appendix \ref{appendixcovph}. %As we mentioned also in the previous section, in order to have independent $U(1)$ charges associated to the two different gauge transformations, we are compelled to drop the off-diagonal terms in the Chern-Simons action \eqref{proposalactioncs}.
 The unusual factor in front of the Chern-Simons contribution is due to our rescaling of the gauge fields.

Note that in this section, the background field variations $\d A_\mu, \d g_{\mu\nu}$ can be arbitrary, in particular they can correspond to variations of the integer-valued charges $J_0, \bar J_0$, since the only way to construct a background with (ultimately quantized) such charges in this formalism is by adding up infinitesimal variations. One the other hand, when we consider variations of the background fields that are induced by an allowed diffeomorphism or gauge transformation, as in section \ref{section63:asysymalgebra},  then  the variation should be restricted to the superselection sector of fixed $J_0,\bar J_0$.

%{\color{blue} 
%We would like to emphasize that, as explained after \ref{invchargessection6}, we work in a particular superselection sector in which variations of the quantised charges $J_0$, $\bar{J}_0$ are not allowed. }\textcolor{blue}{So, how do you compute the charges of the background? I think we should allow background variations for $\d \Phi$, but not in those associated to the diffeos. }

\subsubsection*{Left affine charges}
We will start with the left-moving affine charges, which correspond to turning on $\eta_p$ and setting $f,\bar{f},\bar{\eta}_p$ to zero. The associated transformations are thus very simple, consisting of a diffeomorphism and gauge transformations given by 
\begin{align}
\xi_{\eta}&=-\frac{\l \hat{k}}{2}\eta_p(U) \partial_V\hspace{1cm}\Lambda_A=\frac{\hat{k}}{2}\eta_p(U)\hspace{1cm}\Lambda_B=0 \label{accshift}
\end{align}
where we rescaled $\eta_p$ from the previous subsection in order to match with the field theory convention \eqref{actlmsymm}, which  differed by a factor of $\hat{k}/2$.
%{\color{ForestGreen}(factors rechecked: the metric contribution has an overall $\frac{1}{8\pi G}$ factor in front, while the CS one has $\frac{k}{8\pi}$, but we have an extra $(\frac{4\pi}{k})^2$ because we rescaled the gauge fields. Plugging in, together with the expression for $\kappa$, most of the terms cancel and we are left with $\frac{4\pi}{k}\eta\delta \mathcal{J}$. I would say we need to multiply $\eta$ by a factor of $\frac{\hat{k}}{2}=\frac{k}{4\pi}$ because it enters as $-\l\eta$ in the diffeo on $V$ while in the field theory section $V$ was shifted with $\lambda \hat{k}/2$)}
We find that the corresponding charge variations are given by: 
 %{\color{ForestGreen}(now I am confused, in field theory the shift in $V$ was with + here is with -, I don't understand why the sign overall is the same)}
\begin{align}
\slash{\!\!\!\delta} P_{\eta}&=\int_0^R d\sigma k_{\eta_p}=\int_0^R d\sigma \;\eta_p(U) \delta\mathcal{J}(U)
\end{align}
The result is trivially integrable in phase space and we obtain the following set of conserved charges 
\be\label{resultleftaffineph}
P_{\eta}=\int_0^R d\sigma \; \eta_p(U)\mathcal{J}(U)
\ee
in agreement with the field theory result \eqref{leftmovingch}, using the fact that $\mathcal{J}=\frac{\partial_U\phi}{1-\l\partial_V\phi}$.  
 
 %However, in the field theory section we concluded that the field-dependent coordinate does not have a (field-dependent) zero mode. A left affine transformation generated by  $\eta_p$ induces a zero mode for the field-dependent coordinate, which is the zero mode of:
%\begin{align}
%\delta v&=-\frac{\l}{1+\l\bar{\mathcal{J}}}\eta(U)
%\end{align}
%Nevertheless, any constant induced in $\delta v$ can be reinterpreted as a change in $\delta \bar{\mathcal{J}}_{int}$, as we can see from \eqref{splittingvar}. While this reinterpretation/redistribution is innocuous for now, since it does not affect the result for the left affine charges, it will be important when we compute the charge algebra. (We made the same remark in the field-theory analysis, the problem with the zero mode of the field-dependent coordinate was not visible at the level of the conserved charges, but only when computing Poisson brackets.)} {\color{ForestGreen}edit: do we actually need this?}

\subsubsection*{Left conformal charges}
We now compute the charges associated to the left conformal symmetry preserved by the deformation. In our notation, these symmetry transformations correspond to turning on $f$, which is a periodic function of $U$, and setting $\bar{f},\eta_p,\bar{\eta}_p$ to zero. The diffeomorphisms  we consider are obtained by plugging %\textcolor{red}{Now the derivatives are wrt $U$ or $\hat U$? When did you switch notation? The formulae below look ugly, so  you should probably not have switched. Also, please use same notation for the functions turned on as previously.}
\begin{align}
F_z&=\frac{f'}{2}\hspace{1cm}F_U=f-\frac{\ell^2 f''}{2\mathcal{L}}\hspace{1cm}F_V=-\l \mathcal{J}\frac{\ell^2 f''}{2\mathcal{L}}
\end{align}
into \eqref{diffeorad}, and the gauge transformations are:
\be
\Lambda_A=-\frac{2\mathcal{J}+z^2\l(\kappa \bar{\mathcal{L}}+\bar{\mathcal{J}}^2)\mathcal{L}}{4\mathcal{L}(1-z^4\mathcal{L}\bar{\mathcal{L}})}\ell^2f''+\frac{\mathcal{J}\ell^2f''}{2\mathcal{L}}\;, \hspace{0.7cm}
\Lambda_B=\frac{\big(2\bar{\mathcal{J}}+\l(\kappa\bar{\mathcal{L}}+\bar{\mathcal{J}}^2)\big)z^2}{4(1-z^4\mathcal{L}\bar{\mathcal{L}})}\ell^2f''
\ee
We obtain the following charge variations %\textcolor{red}{Prefactors!}
\begin{align}
\slash{\!\!\!\delta} Q_f&=\int_0^R d\sigma k_f=\int_0^R d\sigma\bigg[\frac{1}{8\pi G\ell}f\delta\mathcal{L}+\frac{4\pi}{k}f\mathcal{J}\delta\mathcal{J}-\nonumber\\
&-\frac{\l \ell}{32\pi G}\bigg[\frac{2\l \bar{\mathcal{J}}'f'}{(1+\l\bar{\mathcal{J}})^2}\delta\mathcal{J}-\partial_{\sigma}\bigg(\delta\bar{\mathcal{J}}\frac{f'}{1+\l\bar{\mathcal{J}}}\bigg)+\frac{2f'}{1+\l\bar{\mathcal{J}}}(\l\mathcal{J}\partial_V +\partial_{U})\delta\bar{\mathcal{J}})\bigg]\bigg]
\end{align}
The first two terms are trivially integrable in phase space. The total $\s$ derivative on the second line can be dropped, as all the fields inside it are periodic. As for the last term, the differential operator in paranthesis annihilates all functions of $v$ only inside $\d \bar \J$, which itself takes the form \eqref{splittingvar}. Using the expression \eqref{variationfielddepen} for $\d v$, we note that the only contribution will be proportional to $\d \J$, which turns out to precisely cancel the first term on the second line. Consequently, the integrated conserved charge is, simply
%
%Since the function inside the sigma derivative does not have winding, we can just drop that term.  Next, we note that the combination $\l\mathcal{J}\partial_V+\partial_U$ kills all functions of $v$ only, which implies that only a part of $\delta v$ contributes from the last term. Overall, this contribution precisely cancels the first term on the second line. \textcolor{red}{You should make this clearer: only part of $\d v$ term survives.} Hence, we are just left with the result (we can change the sign of $f$): \textcolor{red}{Reinstate prefactors!}
\begin{align}\label{resultleftconformal}
Q_f&=\int_0^R d\sigma f(U)\bigg(\frac{\mathcal{L}}{8\pi G \ell}+\frac{2\pi}{k}\mathcal{J}^2\bigg)
\end{align}
which perfectly matches the field theory result. In particular, the choice $f=1$, which corresponds to $\xi=\partial_U$, reproduces the correct left-moving energy.

\subsubsection*{Right affine charges}
Next, we compute the conserved charges associated to right affine transformations, which are parametrized by  a periodic function $\bar{\eta}_p$, setting $\bar{f},f,\eta_p$ to zero. According to \eqref{winding}, this choice will lead to winding terms in both $\eta$ and $\bar \eta$
\begin{align}\label{transformationsrmaff}
&\hspace{0.6cm}F_z=0\hspace{1cm}F_U=0\hspace{1cm}F_V=-\frac{\lambda\hat{k}}{2}\bigg(\bar{\eta}_p+2 w_{\eta}\hat{\mathcal{T}}\bigg)\hspace{1cm}w_{\eta}=-\frac{\lambda R_v}{R_Q}\bar{P}_{\bar{\eta}'}\\
\Lambda_A&=\frac{\hat{k}}{2}\bigg(w_{\eta}\hat{U}-\l\int^v \bar{\J}(\bar{\eta}_p'-\frac{w_{\eta}}{R_v})\bigg)\hspace{1cm}\Lambda_B=\frac{\hat{k}}{2}\bigg(\bar{\eta}_p-w_{\eta}\hat{v}+\l\int^v \bar{\J}(\bar{\eta}_p'-\frac{w_{\eta}}{R_v})\bigg)
\end{align}
where we used again the notation introduced in the field theory analysis \eqref{largeaffinetr} and we rescaled $\bar{\eta}_p$ by $\frac{\hat{k}}{2}$ in order to consider the same transformations as in the field theory section. %{\color{red}(the derivatives are wrt v for $\bar{\eta}_p$, while in the notation $\bar{P}$ they are wrt $\hat{v}$)}

We then plug in the corresponding diffeomorphisms and gauge into the expression \eqref{kformchrg} for the charges, and carefully keep track of all periodic and non-periodic contributions. Since the calculation is slightly involved, we relegate the details to the appendix \ref{appendixcovph}, and simply quote the end result

\be \label{totalvariationinitial}
\slash{\!\!\!\delta} \bar{P}_{\bar \eta} = - \int_0^R d\s \bar \eta_p \p_\s v \d \bar \J_{int} +  \frac{w_\eta}{R_v} \int_0^R d\s \d \left(2 \hat{\mathcal{T}}  \J + \frac{\hat v}{\l}\right)
\ee
 The first term can be easily shown to correspond to a total variation in field space 

\be \label{variationRMaff1}
-\int_0^R d\sigma \bar{\eta}_p \partial_{\sigma}v \delta\bar{\mathcal{J}}_{int} =  
\delta\left(-\int_0^R d\sigma \bar{\eta}_p \bar{\mathcal{J}}\partial_{\sigma}v\right)
\ee
which is nothing but the field-theory expression \eqref{compnoethercurrentrm}  for the right-moving affine charge, if we use the identification \eqref{tangtspparam} for $\bar \J$.

Regarding the second term in \eqref{totalvariationinitial}, one may naively think that it is not integrable because of the charge-dependent coefficient $w_{\eta}$. However, as explained in appendix A of \cite{Compere:2015knw}, the integrability condition is modified in the case of field-dependent diffeomorphism and gauge transformations. Taking this modification into account, we obtain %, overall (I am ok also with saying directly that it is integrable and adding a footnote. We can also mention that we checked explicitly that this term comes from the $\delta Q$ part so it's automatically integrable)
\begin{align}
\bar{P}_{\bar{\eta}}&=-\int_0^R d\s \bar{\eta}_p \bar{\J}\partial_{\sigma}v+\frac{w_{\eta}}{R_v}\int_0^R d\s \bigg(2\hat{\mathcal{T}}\J+\frac{\hat{v}}{\lambda}\bigg) \label{incrmch}
\end{align}
The second term in this expression is problematic, most notably because the integrand is not periodic. The reader may remember, however, that the canonical Noether current \eqref{defnoethercurrent} constructed directly in field theory also suffered from lack of single-valuedness, and we needed to choose a different current representative in order to have a well-defined charge. In fact, the time component of the difference \eqref{extrawindterm} between the canonical and the well-defined current \eqref{compnoethercurrentrm}
%
%{\color{blue}
 %One may naively think that this term \textcolor{red}{You mean, the second one in (6.52)? }is not integrable. However, we need to take into account the fact that both the diffeomorphisms and the gauge transformations that we consider are field-dependent. As it was explained in appendix A in \cite{Compere:2015knw}, this field-dependence does not enter the charge variations. However, the integrability condition is modified when the vector fields and gauge parameters are field-dependent, and using the modified condition we see that this term is integrable. \textcolor{red}{I do not understand this argument.} Then, we obtain the following contribution to the charge:
%\begin{align}
%\frac{4\pi}{k}w_{\eta}\int d\sigma \bigg( \tilde{t}\mathcal{J}+\frac{\hat{v}}{\lambda}\bigg)
%\end{align} 
%which is problematic, most notably because the integrand is not periodic.}
%We compare it with the extra contribution to the current, which we discarded in the field theory analysis \eqref{extrawindterm}, for consistency reasons
\begin{align}
\Delta \bar{K}_{\bar{\eta}}^t&=\frac{\lambda \bar{P}_{\bar{\eta}'}}{R_Q}\left(2\hat{\mathcal{T}}\mathcal{J}+\frac{\hat{v}}{\lambda }\right)
\end{align}
precisely agrees with the second term in \eqref{incrmch}, upon
plugging in the expression \eqref{winding} for the winding. This suggests that we may similarly exploit an ambiguity in the definition of the conserved codimension-two form $k_{\xi,\Lambda}$ in order  to discard this inconsistent contribution.
 A natural source for this ambiguity in the bulk gauge theory is the addition of an extra  on-shell vanishing Noether current $\Delta S^\mu$ - proportional to the large affine term - whose effect on the boundary Wald charge $\mathcal{Q}^{\mu\nu}$ that enters the definition of $k_{\xi, \L}$  would be $\Delta S^\mu = \nabla_\nu ( \Delta \mathcal{Q}^{\mu\nu})$.  Choosing $\Delta S^\mu$ such that its $z$ component equals the divergence  of the `inconsistent' current \eqref{extrawindterm}, this immediately implies that adding this contribution to the bulk Noether current will allow us to discard the unwanted second term in \eqref{incrmch}. 

{\color{blue}
}

To conclude,  the standard, canonical procedure for constructing conserved currents in the covariant phase space formalism  yields currents that are inconsistent, both in field theory (see section \ref{section4:Lagrangian}) and in holography. Quite remarkably, we find that the inconsistent terms match perfectly between the boundary and the bulk. In both cases, we can discard these terms using an ambiguity in the Noether current that affects the proportionality coefficient between its divergence and the equations of motion.  The final result for the right-moving affine charge is 
\begin{align}
\bar{P}_{\bar{\eta}}&=-\int_0^R d\sigma \bar{\eta}_p \bar{\mathcal{J}}\partial_{\sigma}v
\end{align}
which agrees perfectly with the corresponding expression in field theory. For $\bar{\eta}_p=1$,  the conserved charge is $-Q_R$.

\subsubsection*{Right pseudoconformal charges}
Finally, we consider the conserved charges associated to right pseudoconformal transformations, which correspond to setting all the functions to zero except for $\bar{f}$, which  is periodic. In this case, using \eqref{winding}
\begin{align} \label{rmpseudocftr}
F_z&=\frac{\bar{f}'}{2}\hspace{0.7cm}F_U=0\hspace{0.7cm}F_V=\bar{f}-2\l w_{\bar{f}}\hat{\mathcal{T}}-\frac{(1+\l\bar{\mathcal{J}})\ell^2\bar{f}''}{2\bar{\mathcal{L}}}\hspace{0.7cm}w_{\bar{f}}=-\frac{\hat{k}R_v}{R_Q}\frac{\lambda}{2}\bar{Q}_{\bar{f}'}\nonumber\\
\Lambda_A&=\frac{2 z^2 \J \bar{\L}+\l(\kappa \bar{\mathcal{L}}+\bar{\mathcal{J}}^2)}{4\bar{\mathcal{L}}(1-z^4\mathcal{L}\bar{\mathcal{L}})}\ell^2\bar{f}''-\frac{\l \bar{\mathcal{J}}^2 \ell^2\bar{f}''}{4\bar{\mathcal{L}}}+w_{\bar{f}}\hat{U}+\l \frac{w_{\bar{f}}}{R_v}\int^{v} \bar{\mathcal{J}}-\frac{\l}{2}\int^{v} (\kappa\bar{\mathcal{L}}+\bar{\mathcal{J}}^2)\bar{f}'\nonumber\\
\Lambda_B&=-\frac{2\bar{\J} +\l(\kappa \bar{\mathcal{L}}+\bar{\mathcal{J}}^2)}{4\bar{\mathcal{L}}(1-z^4\mathcal{L}\bar{\mathcal{L}})}\ell^2\bar{f}''+\frac{\bar{\mathcal{J}}(2+\l\bar{\mathcal{J}})\ell^2\bar{f}''}{4\bar{\mathcal{L}}}-w_{\bar{f}}\hat{v}-\l \frac{w_{\bar{f}}}{R_v}\int^{v} \bar{\mathcal{J}}+\frac{\l}{2}\int^{v} (\kappa\bar{\mathcal{L}}+\bar{\mathcal{J}}^2)\bar{f}'
\end{align}
The result that we obtain by plugging in these transformations is
\begin{align}
\slash{\!\!\!\delta} \bar{Q}_{\bar{f}}=\int_0^R d\sigma
\bigg[ -\delta\bigg(\partial_{\sigma}v \bar{f} \bigg(\frac{\bar{\mathcal{L}}}{8\pi G\ell}+\frac{2\pi}{k}\bar{\mathcal{J}}^2\bigg)\bigg)+\frac{2}{\hat{k}} \frac{w_{\bar{f}}}{R_v} \delta\bigg(2\hat{\mathcal{T}}\mathcal{J}+\frac{\hat{v}}{\l}\bigg)\bigg]
\end{align}
The last term is the same as the one we found in field theory. The same discussion as for the right affine transformations applies here, as it originates from the same right affine winding contribution, just with a different coefficient. We can drop this term, as we did in the field theory analysis, by modifying the on-shell vanishing Noether current. Our final answer for  the pseudoconformal charges is
\begin{align}
\bar{Q}_{\bar{f}}&=-\int_0^R d\sigma \partial_{\sigma} v \bar{f}\bigg(\frac{\bar{\mathcal{L}}}{8\pi G \ell}+\frac{2\pi}{k}\bar{\mathcal{J}}^2\bigg)
\end{align}
in agreement with the Hamiltonian analysis. In particular, $\bar{f}=-1$ reproduces the correct right-moving energy.

\subsubsection*{The constant gauge transformations}

Finally, let us comment on constant gauge transformations, labeled by $c_A,c_B$. As we already mentioned, they are not independent transformations, but combinations of constant $\eta,\bar{\eta},\bar{f}$. Their associated conserved charges are given by:
\begin{align}
Q_{gauge}&=-\frac{2c_A}{\hat{k}}\int_0^R d\sigma \partial_{\sigma}v\bigg(\l (\kappa\bar{\mathcal{L}}+\bar{\mathcal{J}}^2)+\frac{2(1+\l \bar{\mathcal{J}})}{\l}\bigg)-\frac{2c_B }{\hat{k}}\int_0^R d\sigma \partial_{\sigma}v\big(2\bar{\mathcal{J}}+\l(\kappa\bar{\mathcal{L}}+ \bar{\mathcal{J}}^2)\big)=\nonumber\\
&=\frac{c_A}{\hat{k}}\bigg(Q_K-\frac{\l k}{4\pi}E_R\bigg)+\frac{c_B}{\hat{k}} \bigg(\bar{Q}_{\bar{K}}-\frac{\l k}{4\pi}E_R\bigg)=\frac{c_A}{\hat{k}}J_0+\frac{c_B}{\hat{k}} \bar{J}_0
\end{align}
Hence, the undeformed charges can be interpreted in the bulk as charges associated to the pure gauge transformations. In particular, for $c_B=-c_A$ (equal and opposite transformations of $A$ and $B$) we recover the winding charge.

To be more precise, these constant gauge transformations correspond to setting
\begin{align}
[\bar{f}]_{zm}=-\lambda([\eta]_{zm}+[\bar{\eta}]_{zm})
\end{align}
and thus, from \eqref{deltavfunctions}, they do not change $\delta v$ (the fact that the constant pieces do not change $\delta v$ was the feature of affine flowed transformations). It is then trivial to see that the flowed affine generators can be obtained from the flowed ones by turning on an extra constant $\bar{f}$ as above. (this would just bring the $E_R$ contribution)

More generally, in order to obtain the conformal/pseudoconformal flowed generators, we need to consider linear combinations of the transformations discussed, with $E_R$-dependent coefficients, in order to implement the spectral flow. It would be interesting to see if there exists another way to obtain directly the flowed generators, similar to the analysis of \cite{Georgescu:2022iyx} for $T\bar{T}$.

\subsection{The charge algebra}\label{section63:asysymalgebra}
Now that we obtained the conserved charges in the bulk, we proceed to computing their  algebra. The Poisson bracket between conserved charges is written generically as:
\begin{align} \label{definitionpoissonbr}
\{Q_{\xi,\Lambda},Q_{\chi,\Sigma}\}&\equiv \delta_{\chi,\Sigma}Q_{\xi,\Lambda}
\end{align}
where $\xi, \chi$ stand for diffeomorphisms and $\Lambda,\Sigma$ for the associated gauge transformations. As before, we will find it useful to work with the basis of conserved charges parametrised by the periodic functions $f,\bar f, \eta_p$ and $\bar \eta_p$. We compute each type of commutator in part and expand the corresponding functions in Fourier modes afterwards in order to compare with the field theory results. We will check the consistency, more precisely antisymmetry, or the commutators, which will fix the constant $\mathcal{C}$ introduced in section \ref{section62:conservedcharges}.

\subsubsection*{Left affine commutator $\{P_{\eta},P_{\beta}\}$}
We start with the simple case of the Poisson bracket between two left moving affine charges, parametrized by periodic functions $\eta_p(U),\beta_p(U)$, for which we obtain, using \eqref{variationofJ}:
\begin{align}
\{P_{\eta},P_{\beta}\}&=\delta_{\beta}P_{\eta}=\int_0^R d\sigma \eta_p \delta_{\beta}\mathcal{J}=\frac{\hat{k}}{2}\int_0^R d\sigma \eta_p \;\partial_{\sigma}\beta_p
\end{align}
The result is clearly a central term and is antisymmetric, since we can integrate by parts and drop the total sigma derivative. Expanding the functions in Fourier modes, we can particularize $\eta_p=e^{2\pi i m \hat{U}}$ and $\beta_p=e^{2\pi i n \hat{U}}$ for which we obtain
\begin{align}
\{P_m,P_n\}&=-\frac{ i m k}{2}  \delta_{n+m,0}
\end{align}
which is the expected $U(1)$ Kac-Moody result, for level $k$.

\subsubsection*{Right affine commutator $\{\bar{P}_{\bar{\eta}},\bar{P}_{\bar{\beta}}\}$}
We compute the Poisson bracket between two right-moving affine charges, parametrized by periodic functions $\bar{\eta}_p(v),\bar{\beta}_p(v)$:
\begin{align} \label{commutatorrmaff}
\{\bar{P}_{\bar{\eta}},\bar{P}_{\bar{\beta}}\}&=\delta_{\bar\beta}\bar{P}_{\bar{\eta}}=-\frac{\hat{k}}{2}\int_0^R d\sigma \bar{\eta}_p \partial_{\sigma}v \delta_{\bar\beta}\bar{\mathcal{J}}_{int}
\end{align}
which was derived in \eqref{variationRMaff1}. We already took into account the rescaling of both $\bar{\eta}_p$ and $\bar{\beta}_p$ by $\frac{\hat{k}}{2}$, in order not to carry this multiplicative factor inside $\delta_{\bar{\beta}}\bar{\J}_{int}$. Taking into account the constant $\mathcal{C}$ that can enter into $\delta\bar{\J}_{int}$, as explained in \ref{section61:allowedtr}, the variation of $\bar \J_{int}$ under this transformation takes the form
\begin{align}
\delta_{\bar\beta}\bar{\mathcal{J}}_{int}&=\bar{\beta}'-\mathcal{C}_{\bar\beta}\bar{\mathcal{J}}' = \bar{\beta}'_p-\frac{w_{\beta}}{R_v}-\mathcal{C}_{\bar\beta}\bar{\mathcal{J}}'
\end{align}
where we remind that the derivatives are with respect to $v$ and we put a label on the constant to emphasize that corresponds to this particular transformation. Plugging into \eqref{commutatorrmaff}, we find 
\begin{align}
&\frac{2}{\hat{k}}\{\bar{P}_{\bar{\eta}},\bar{P}_{\bar{\beta}}\}=-\int_0^R d\sigma \bar{\eta}_p\bar{\beta}_p' \partial_{\sigma}v +\frac{w_{\beta}}{R_v}\int_0^R d\sigma \bar{\eta}_p\partial_{\sigma}v +\mathcal{C}_{\bar{\beta}}\int_0^R d\sigma \bar{\eta}_p\partial_{\sigma}v \bar{\mathcal{J}}'=\nonumber\\
&=-\int_0^R d\sigma \bar{\eta}_p \partial_{\sigma}\bar{\beta}_p  +w_{\beta}[\bar{\eta}_p]_{zm} +\mathcal{C}_{\bar{\beta}}\frac{\bar{P}_{\bar{\eta}'}}{R_v}=-\int_0^R d\sigma \bar{\eta}_p \partial_{\sigma}\bar{\beta}_p  +w_{\beta}[\bar{\eta}_p]_{zm} -\mathcal{C}_{\bar{\beta}}\frac{w_{\eta} R_Q}{\l R_v^2}
\end{align}
The first term is clearly antisymmetric using integration by parts and it gives a central term. In order for the full result to be antisymmetric, which is of course needed for the consistency of the charge algebra, we need to fix:
\begin{align}\label{consistenalg}
\mathcal{C}_{\bar\beta}=\frac{\l R_v^2}{R_Q}[\bar{\beta}_p]_{zm}
\end{align}
%{\color{blue}Let us understand what this means from the point of view of \eqref{ambiguitysplit}. Under a right affine transformation, a zero mode is induced to the field-dependent coordinate. Nevertheless, this constant can be reinterpreted as a change in the intrinsic variation of $\bar{\mathcal{J}}$, that we need to take into account in order to obtain a well-defined Poisson bracket. Going back to the field theory analysis, we see that $\mathcal{C}$ is nothing but the change of the zero mode of $\phi$. (I'll think about it, factors don't seem to work)}

All together, the final result is:
\begin{align}
\{\bar{P}_{\bar{\eta}},\bar{P}_{\bar{\beta}}\}&=-\frac{\hat{k}}{2}\int_0^R d\sigma \bar{\eta}_p \partial_{\sigma}\bar{\beta}_p  -\frac{\hat{k}}{2}\frac{\lambda R_v}{R_Q} \bar{P}_{\bar{\beta}'}[\bar{\eta}_p]_{zm} +\frac{\hat{k}}{2}\frac{\lambda R_v}{R_Q} \bar{P}_{\bar{\eta}'}[\bar{\beta}_p]_{zm}
\end{align}
After expanding the functions in Fourier modes $\bar{\eta}_p=e^{-2\pi i m \hat{v}}$ and $\bar{\beta}_p=e^{-2\pi i n \hat{v}}$, we obtain 
\begin{align}
\{ \bar P_m, \bar P_n\} = -\frac{i m k}{2} \delta_{m+n,0}-\frac{i m k\lambda }{2 R_Q} \bar{P}_m \delta_{n,0}+\frac{i n k\lambda}{2R_Q}\bar{P}_n \delta_{m,0}
\end{align}
which matches  \eqref{algebranonlinearjtbar}.

Hence, although initially we considered $\mathcal{C}$, or equivalently $\delta c_v$, to be arbitrary, the requirement that the Poisson bracket defined as \eqref{definitionpoissonbr} is antisymmetric fixed this constant to a particular value, in the case of a right-moving affine transformation. Fixing the constant entering the variation of the field-dependent coordinate using the consistency of the charge algebra is precisely what we found in the field theory analysis of section \ref{section43:commutators}.

\subsubsection*{Left affine - right affine commutator $\{P_{\eta},\bar{P}_{\bar{\beta}}\}$}
Next, we compute the commutator between a left-moving affine charge parameterized by a function $\eta_p(U)$ and a right-moving affine charge parametrized by a function $\bar{\beta}(v)$:
\begin{align} \label{leftrightphsp}
\{P_{\eta},\bar{P}_{\bar{\beta}}\}&=\frac{\hat{k}}{2}\int_0^R d\sigma \;\eta_p(U)\delta_{\bar{\beta}}\mathcal{J}=\frac{\hat{k}}{2}w_{\beta}[\eta_p]_{zm}=-\frac{\hat{k}}{2}\frac{\lambda R_v}{R_Q}\bar{P}_{\bar{\beta}'}[\eta_p]_{zm}
\end{align}
where we used the fact that turning on $\bar{\beta}_p$ requires an extra large left-moving affine transformation $w_{\beta}\hat{U}$ and under such transformation $\mathcal{J}$ changes, as specified in \eqref{variationofJ}. This result is unambiguous and agrees with the field theory result \eqref{algebra:LMRMaffine}. On the other hand, we should be able to compute  the same commutator  via the variation
\begin{align}
%\{\bar{P}_{\bar{\beta}},P_{\eta}\}
-\d_\eta \bar{P}_{\bar \b}
&=\frac{\hat{k}}{2}\int_0^R d\sigma \bar{\beta}_p \partial_{\sigma}v \delta_{\eta}\bar{\mathcal{J}}_{int}
\end{align}
Under a left-moving affine transformation
\begin{align}
\delta_{\eta}\bar{\mathcal{J}}_{int}&=-\mathcal{C}_{\eta}\bar{\J}'
\end{align}
In order to obtain the correct result consistent with \eqref{leftrightphsp}, we need to set, in this case:
\begin{align} \label{fixedvalueceta}
\mathcal{C}_{\eta}&=\frac{\lambda R_v^2}{R_Q}[\eta_p]_{zm}
\end{align}
%In this case we obtain:
%\begin{align}
%\{\bar{P}_{\bar{\beta}},P_{\eta}\}&=\mathcal{C}\int d\sigma \bar{\beta}_p \partial_{\sigma}v \bar{\mathcal{J}}'=-\frac{\lambda R_v}{R_Q}[\eta]_{zm}\bar{P}_{\bar{\beta}'}=-w_{\beta}[\eta]_{zm}
%\end{align}
which makes the Poisson bracket well-defined{\footnote{To be more precise, we check here specifically the antisymmetry of the Poisson bracket. Nevertheless, since our results agree with the field theory ones for which the algebra is $(Virasoro\times Kac-Moody)^2$ in a more complicated basis, all the required conditions (for ex Jacobi identity) are automatically satisfied}}. 
Finally, let us particularize $\eta_p=e^{2\pi i m \hat{U}}$ and $\bar{\beta}_p=e^{-2\pi i n \hat{v}}$ for which we obtain
\begin{align}
\{P_m,\bar{P}_n\}&=\frac{i n k\lambda}{2 R_Q}\bar{P}_n\delta_{m,0}
\end{align}
which matches perfectly the field theory result \eqref{algebra:LMRMaffine}.

Just like for the right-affine commutator, in order to have a well-defined Poisson algebra, we need to set $\mathcal{C}_{\eta}$ to a specific value, which fixes the variation of the zero mode of the field-dependent coordinate, matching the field theory analysis of section \ref{section43:commutators}. Both in field theory and in the holographic analysis, the conserved charges do not depend on the zero mode of $\phi$ (or $\delta v$) and it is only at the level of the charge algebra that the apriori ambiguity gets fixed. The same mechanism of fixing additional constants in the analysis by requiring the antisymmetry of \eqref{definitionpoissonbr} was found in \cite{Georgescu:2022iyx} for $T\bar{T}$.

%{\color{blue} which amounts to removing the field-dependent zero mode of $v$, by considering it to be a change in the intrinsic variation of $\bar{\mathcal{J}}$. This is the same conclusion that we obtained in the field theory analysis from \eqref{subsection:commutatorsfieldth}, after noticing the same problem regarding the Poisson brackets. Both in field theory and in the holographic analysis, the conserved charges depend only on the derivatives of $\phi$, so they do not see the zero mode. It is only at the level of the algebra that we notice that we need to remove the zero mode of $\phi$. In field theory this is related to an ambiguity in the definition of $v$, which is reflected in the bulk through the fact that we can remove the zero mode of $v$ induced by any transformation.Once again, we emphasize that the zero mode that we need to remove from $v$ for consistency is field-dependent, namely it corresponds to the zero mode of $\phi$ in field theory and to the zero mode of $\delta\J+\delta\bar{\J}$ in the bulk, while a field-independent zero mode in $v$ is allowed (and not interesting).}

\subsubsection*{Right pseudoconformal commutator $\{\bar{Q}_{\bar{f}},\bar{Q}_{\bar{g}}\}$}
We now compute the Poisson bracket between two right-moving pseudoconformal charges, parametrized by periodic functions $\bar{f}(v),\bar{g}(v)$:
\begin{align}
\{\bar{Q}_{\bar{f}},\bar{Q}_{\bar{g}}\}&=\delta_{\bar{g}}\bar{Q}_{\bar{f}}=-\int_0^R d\sigma\partial_{\sigma} v \bar{f}\bigg( \frac{\delta_{\bar{g}}\bar{\L}_{int}}{8\pi G\ell} +\frac{4\pi}{k}\bar{\J}\delta_{\bar{g}} \bar{\J}_{int}\bigg)
\end{align}
The variations that enter this computation are:
\begin{align}
\delta_{\bar{g}}\bar{\L}_{int}&=2\bar{\L}\bar{g}'+(\bar{g}-\mathcal{C}_{\bar{g}})\bar{\L}'-\frac{\ell^2}{2}\bar{g}'''\hspace{1cm}\delta_{\bar{g}}\bar{\J}_{int}=\bar{g}'\bar{\J}-\frac{w_{\bar{g}}}{R_v}+(\bar{g}-\mathcal{C}_{\bar{g}})\bar{\J}'
\end{align}
Plugging in these expressions, we obtain:
\begin{align}
&\{\bar{Q}_{\bar{f}},\bar{Q}_{\bar{g}}\}=-\int_0^R  d\sigma \partial_{\sigma} v (2\bar{f}\bar{g}')\bigg( \frac{\bar{\L}}{8\pi G\ell}+\frac{2\pi}{k}\bar{\J}^2\bigg)-\int_0^R d\sigma \partial_{\sigma} v \bar{f}(\bar{g}-\mathcal{C}_{\bar{g}})\bigg(\frac{\bar{\L}'}{8\pi G\ell}+\frac{4\pi}{k}\bar{\J}\bar{\J}'\bigg) +\nonumber\\
&+\frac{w_{\bar{g}}}{R_v}\frac{2}{\hat{k}}\int_0^R d\sigma\partial_{\sigma}v \bar{f}\bar{\J}+\frac{\ell}{16\pi G}\int_0^R d\sigma \partial_{\sigma} v \bar{f}\bar{g}'''=\nonumber\\
&=2 \frac{\bar{Q}_{\bar{f}\bar{g}'}}{R_v}-\frac{w_{\bar{g}}}{R_v}\frac{2}{\hat{k}}\bar{P}_{\bar{f}}+\int_0^R d\sigma \partial_{\sigma}\big(\bar{f}(\bar{g}-\mathcal{C}_{\bar{g}})\big)\bigg(\frac{\bar{\L}}{8\pi G \ell}+\frac{2\pi}{k}\bar{\J}^2\bigg)+\frac{\ell}{16\pi G}\int_0^R d\sigma \partial_{\sigma} v \bar{f}\bar{g}'''=\nonumber\\
&=2 \frac{\bar{Q}_{\bar{f}\bar{g}'}}{R_v}-\frac{w_{\bar{g}}}{R_v}\frac{2}{\hat{k}}\bar{P}_{\bar{f}} - \frac{\bar{Q}_{\bar{f}\bar{g}'}}{R_v}-\frac{\bar{Q}_{\bar{f}'\bar{g}}}{R_v}-\mathcal{C}_{\bar{g}}\int_0^R \partial_{\sigma}\bar{f}\bigg(\frac{\bar{\L}}{8\pi G \ell}+\frac{2\pi}{k}\bar{\J}^2\bigg)+\frac{\ell}{16\pi G}\int_0^R d\sigma \partial_{\sigma} v \bar{f}\bar{g}'''=\nonumber\\
&=\frac{\bar{Q}_{\bar{f}\bar{g}'-\bar{f}'\bar{g}}}{R_v}-\frac{w_{\bar{g}}}{R_v}\frac{2}{\hat{k}}\bar{P}_{\bar{f}} +\frac{\mathcal{C}_{\bar{g}}}{R_v}\bar{Q}_{\bar{f}'}+\frac{\ell}{16\pi G}\int_0^R d\sigma \partial_{\sigma} v \bar{f}\bar{g}'''
\end{align}
Requiring antisymmetry at $\bar{f}\leftrightarrow \bar{g}$ sets:
\begin{align}
\mathcal{C}_{\bar{g}}&=-\frac{\lambda R_v}{R_Q}\bar{P}_{\bar{g}}
\end{align}
Let us expand the functions in Fourier modes $\bar{f}=e^{-2\pi i m \hat{v}}$ and $\bar{g}=e^{-2\pi i n \hat{v}}$. We obtain:
\begin{align}
\{\bar{Q}_m,\bar{Q}_n\}&=\frac{2\pi i(m-n)\bar{Q}_{m+n}}{R_v} + \frac{2\pi i m \lambda}{R_Q R_v}\bar{Q}_m \bar{P}_n-\frac{2\pi i n \lambda}{R_Q R_v}\bar{Q}_n \bar{P}_m + \frac{c}{12} \frac{in^3}{(R_v/2\pi)^2}\delta_{m+n,0}
\end{align}
The result matches \eqref{algebranonlinearjtbar} up to a sign, due to the fact that in the holographic analysis the pseudoconformal charges have a different sign with respect to the field theory analysis and up to the central term which was not visible in the classical field theory analysis.

\subsubsection*{Right pseudoconformal - left conformal commutator $\{\bar{Q}_{\bar{f}},Q_f\}$}
We also compute the Poisson bracket between a right-moving pseudoconformal charge parametrized by $\bar{f}(v)$ and a left-moving conformal charge  parametrized by $f(U)$
\begin{align}
\{\bar{Q}_{\bar{f}},Q_f\}&=\delta_{\bar{f}}Q_f=\int_0^R d\sigma f \bigg(\frac{\delta_{\bar{f}}\L}{8\pi G\ell} +\frac{4\pi}{k}\J\delta_{\bar{f}}\J \bigg)=\frac{2}{\hat{k}}\int_0^R d\sigma f \J \frac{w_{\bar{f}}}{R}=\frac{2}{\hat{k}}\frac{w_{\bar{f}}}{R}P_f
\end{align}
where we used the fact that turning on $\bar{f}$ requires additional large affine transformations, such as $\eta=w_{\bar{f}}\hat{U}$ that enters this computation as
\begin{align}
\delta_{\bar{f}}\L=0\hspace{1cm}\delta_{\bar{f}}\J=\frac{w_{\bar{f}}}{R}
\end{align}
On the other hand, this non-ambiguous result should match:
\begin{align}
-\delta_f \bar{Q}_{\bar{f}}=-\mathcal{C}_f\int_0^R d\sigma \partial_{\sigma}v \bar{f} \bigg(\frac{\bar{\L}'}{8\pi G\ell} +\frac{2\pi}{k}(\bar{\J}^2)'\bigg)=-\frac{\mathcal{C}_f}{R_v}\bar{Q}_{\bar{f}'}
\end{align}
which sets
\begin{align}
\mathcal{C}_{f}&=\frac{\lambda R_v}{R R_Q}P_{f}
\end{align}
Finally, considering the Fourier modes, we obtain:
\begin{align}
\{ Q_m, \bar Q_n\}= \frac{2\pi i n \lambda}{R R_Q}\bar{Q}_n P_m
\end{align}

\subsubsection*{Right pseudoconformal-right affine commutator $\{\bar{Q}_{\bar{f}},\bar{P}_{\bar{\eta}}\}$}
Next, we compute the Poisson bracket between a right-moving pseudoconformal charge parametrized by $\bar{f}(v)$ and a right-moving affine charge parametrized by $\bar{\eta}(v)$
\begin{align}
\{\bar{Q}_{\bar{f}},\bar{P}_{\bar{\eta}}\}&=\delta_{\bar{\eta}}\bar{Q}_{\bar{f}}=-\frac{\hat{k}}{2}\int_0^R d\sigma \partial_{\sigma}v \bar{f} \bigg( \frac{\delta_{\bar{\eta}}\bar{\L}}{8\pi G\ell} +\frac{4\pi}{k}\bar{\J}\delta_{\bar{\eta}}\bar{\J} \bigg)=\nonumber\\
&=-\int_0^R d\sigma\partial_{\sigma}v \bar{f}\bar{\J}\bigg(\bar{\eta}'_p-\frac{w_{\eta}}{R_v}\bigg)+\frac{\hat{k}}{2}\mathcal{C}_{\bar{\eta}}\int_0^R d\sigma \partial_{\sigma}v \bar{f}\bigg(\frac{\bar{\L}'}{8\pi G\ell}+\frac{2\pi}{k}(\bar{\J}^2)'\bigg)=\nonumber\\
&=\frac{\bar{P}_{\bar{f}\bar{\eta}'_p}}{R_v}-\frac{w_{\eta}}{R_v}\bar{P}_{\bar{f}}-\frac{\hat{k}}{2}\mathcal{C}_{\bar{\eta}}\int_0^R d\sigma \partial_{\sigma}v \bar{f}'\bigg(\frac{\bar{\L}}{8\pi G\ell}+\frac{2\pi}{k}\bar{\J}^2\bigg)=\nonumber\\
&=\frac{\bar{P}_{\bar{f}\bar{\eta}'_p}}{R_v}-\frac{w_{\eta}}{R_v}\bar{P}_{\bar{f}}+\frac{\hat{k}}{2}\frac{\mathcal{C}_{\bar{\eta}}}{R_v}\bar{Q}_{\bar{f}'}=\frac{\bar{P}_{\bar{f}\bar{\eta}'_p}}{R_v}+\frac{\l}{R_Q}\bar{P}_{\bar{\eta}'}\bar{P}_{\bar{f}}+\frac{\hat{k}}{2}\frac{\mathcal{C}_{\bar{\eta}}}{R_v}\bar{Q}_{\bar{f}'}
\end{align}
where we used $\delta_{\bar{\eta}}\bar{\L}_{int}=-\mathcal{C}_{\bar{\eta}}\bar{\L}'$ and $\delta_{\bar{\eta}}\bar{\J}_{int}=\bar{\eta}'_p-\frac{w_{\eta}}{R_v}-\mathcal{C}_{\bar{\eta}}\bar{\J}'$. Expanding in Fourier modes:
\begin{align}
\{\bar{Q}_m,\bar{P}_n\}&=-\frac{2\pi i n }{R_v}\bar{P}_{m+n}-\frac{2\pi i n\lambda  }{R_Q R_v}\bar{P}_n\bar{P}_m-\frac{i m k \lambda}{R_Q}\bar{Q}_m\delta_{n,0}
\end{align}
where we replaced the value of $\mathcal{C}_{\bar{\eta}}$ for right affine transformations previously fixed \eqref{consistenalg}. 

\subsubsection*{Left affine - right pseudoconformal commutator $\{P_{\eta},\bar{Q}_{\bar{f}}\}$}
We compute the Poisson bracket between a left-moving affine charge parametrized by $\eta_p(U)$ and a right-moving pseudoconformal charge parametrized by $\bar{f}(v)$
\begin{align}
\{P_{\eta},\bar{Q}_{\bar{f}}\}&=\delta_{\bar{f}}P_{\eta}=\int_0^R d\sigma \eta_p \delta_{\bar{f}}\J
\end{align}
Using $\delta_{\bar{f}}\J=\frac{2}{\hat{k}}\frac{w_{\bar{f}}}{R}$, we obtain:
\begin{align}
\{P_{\eta},\bar{Q}_{\bar{f}}\}&=\frac{2}{\hat{k}}w_{\bar{f}}[\eta_p]_{zm}=-\frac{\l R_v}{R_Q}\bar{Q}_{\bar{f}'}[\eta_p]_{zm}
\end{align}
Alternatively, we obtain the same result from $\{P_{\eta},\bar{Q}_{\bar{f}}\}=-\delta_{\eta}\bar{Q}_{\bar{f}}=-\mathcal{C}_{\eta}\bar{Q}_{\bar{f}'}$, using \eqref{fixedvalueceta}.

The corresponding commutator for the Fourier modes is:
\begin{align}
\{P_m,\bar{Q}_{n}\}&=\frac{ 2\pi i n \l}{R_Q}\bar{Q}_{n}\delta_{m,0}
\end{align}

\subsubsection*{Left conformal - right affine commutator $\{Q_f,\bar{P}_{\bar{\eta}}\}$}
Finally, we compute the Poisson bracket between a left-moving conformal charge parametrized by $f(U)$ and a right-moving affine charge parametrized by $\bar{\eta}(v)$:
\begin{align}
\{Q_f,\bar{P}_{\bar{\eta}}\}&=\delta_{\bar{\eta}}Q_{f}=\int_0^R d\sigma f\J \delta_{\bar{\eta}}\J=\frac{w_{\eta}}{R}P_f=-\frac{\lambda  R_v}{R_Q R}\bar{P}_{\bar{\eta}'}P_f
\end{align}
Alternatively, we can compute:
\begin{align}
\{Q_f,\bar{P}_{\bar{\eta}}\}&=-\delta_f \bar{P}_{\bar{\eta}}=\int_0^R d\sigma \partial_{\sigma}v \bar{\eta} \delta_f\bar{\J}_{int}=-\mathcal{C}_f \int_0^R d\sigma \partial_{\sigma}v \bar{\eta}\bar{\J}'=-\frac{\mathcal{C}_f}{R_v}\bar{P}_{\bar{\eta}'}
\end{align}
The corresponding commutator for the Fourier modes is:
\begin{align}
\{Q_m,\bar{P}_{n}\}&=\frac{2\pi i n \lambda}{R_Q R} \bar{P}_{n}P_m
\end{align}

%\subsection*{Summary}
%To conclude, we managed to reproduce the $J\bar{T}$ symmetry algebra as the asymptotic symmetry algebra of its holographic dual, after carefully taking into account the effect of the transformations on the zero mode of the field-dependent coordinate. 

\section{Discussion}\label{section7:discussion}

One of the main motivations for studying $J\bar T$ - deformed CFTs in detail is their potential relevance as toy models for non-AdS holography and, in particular, for the so-called Kerr/``CFT'' correspondence. A particularly interesting feature of these theories is the coexistence of Virasoro symmetry and non-locality, a novel phanomenon that, beyond its holographic applications, also deserves a deeper field-theoretical understanding. 

In this article, we  performed a comprehensive study of the extended symmetries of classical $J\bar{T}$-deformed CFTs, from three different perspectives: Hamiltonian, Lagrangian and holographic. The corresponding generators span, in a particular basis associated to non-local symmetry transformations, two commuting copies of the Virasoro-Kac-Moody algebra. In the Hamiltonian formalism, this basis is singled out   by the flow operator associated to the adiabatic $J\bar T$ deformation. Nonetheless, when considering the action of the symmetry generators on the fields in the theory, a more natural basis is revealed, in which the induced symmetry transformations are more local and the symmetry algebra is non-linear. In the holographic analysis,  we showed how precisely these symmetries are realized as asymptotic symmetries of the gravitational dual  to a $J\bar{T}$-deformed CFT, %three dimensional gravity in asymptotically $AdS_3$ spacetime coupled to two $U(1)$ Chern-Simons gauge fields. With our proposed holographic dictionary, we obtained a precise matching between the bulk and the boundary for the spectrum, thermodynamics and symmetries. } \textcolor{red}{Rewrite.}
obtaining also an exact match of the full extended classical symmetry algebra.

There are many interesting directions for future research. On the field-theory side, it would be interesting to understand better the Lagrangian picture for the symmetries, including issues related to the invariance of the action and the off-shell significance of the compensating large affine terms. Another obvious question is how to implement the symmetries we discussed at the level of the quantum theory, for example in the path integral formulation of \cite{Anous:2019osb}. The peculiar transformation properties of the fields would also need to be properly understood  and  interpreted in the quantum theory - at full non-linear level, if possible -  and the physical significance and interplay of the two bases of symmetry generators would need to be clarified. While \cite{Guica:2021fkv} was able to show that the correlation functions of appropriately-defined operators in $J\bar T$ - deformed CFTs are entirely fixed in terms of the undeformed CFT correlators via a universal, non-local transform, much more work is required to understand the full structure of the Ward identities in $J\bar T$ - deformed CFTs, the general form of the correlation functions  of conserved currents (as \cite{Guica:2021fkv} only focused on deformed Virasoro primaries), as well as the physical properties of the various Lorentzian correlators that one could define and their implications for the causal structure of the theory. Moreover, given that the data that fixes the correlation functions in  $J\bar T$ - deformed CFTs  is in one-to-one correspondence to the  CFT data of the undeformed theory, one could  envisage setting up 
 a bootstrap programme for  the former, which could perhaps pave the way for  interesting generalisations of $J\bar T$ - deformed CFTs. 
 All these are very basic questions concerning the QFT structure of $J\bar T$ - deformed CFTs that would be worthwhile to  understand better, both from the purely QFT perspective of going beyond local quantum theories, and in view of refining the field theory predictions used to test the  holographic dictionary. 

On the holographic front, the main lesson of our analysis is the importance of having a consistent set of boundary conditions and - rather crucially - a correct parametrisation of the backgrounds\footnote{In the sense of knowing which quantities  can vary, and which ones are fixed.}, including the  identification, from the bulk point of view,  of the correct (field-independent) coordinates of the dual field theory. This identification is clearly very important for obtaining the correct deformed thermodynamics of the system, but not only. Once these basic data are provided -   which, in our case, were an input from the known dual field theory - 
 then the full structure of the asymptotic symmetry group follows almost algorithmically, at least when working  in the covariant phase space formalism\footnote{In the ``counterterm subtraction'' method presented in section \ref{section5:holography}, the construction of the conserved charges %{\color{ForestGreen} which involve only the periodic part of the associated functions} 
 - as opposed to the conserved currents 
 - from  first principles is less obvious, see discussion in footnote \ref{ctsubft}.  It would clearly be interesting to understand how to \emph{derive} the ASG also in this formalism.}:   the correlated variations of the metric and the gauge fields % that are implied by the boundary conditions conspire  
 (upon a large number of cancellations, see e.g. appendix \ref{detevalcc} for a sample) to produce simple,  integrable charges, while the field theory requirement that the $J_0, \bar J_0$ charges be fixed results into
 the presence of the large affine compensating transformations that yield the  non-linear terms in the algebra. Had we not been aware of this field-theoretical constraint, the ASG analysis would not have signaled anything unusual at the level of charge finiteness and integrability and  would have yielded, as in \cite{Bzowski:2018pcy,Guica:2020uhm},  two standard copies of the Virasoro-Kac-Moody algebra, which is not the correct result.   

In conclusion, certain minimal - but essential - information from the field theory was crucial for uncovering the correct asymptotic symmetry group. Unfortunately, in the standard situations in non-AdS holography where ASG methods are used, such additional information is usually not available, and oftentimes there are  several different choices of boundary conditions for which the conserved charges are finite and integrable. Since, as we also show in appendix \ref{appendixCSS}, the assumptions entering an ASG calculation can yield very different predictions for the nature and symmetries of the putative dual theory, having a carefully chosen set of criteria for determining the correct \footnote{Here we assume that 
in general, for theories with a holographic dual, 
  there is only one set of `correct'  boundary conditions  for a given  spacetime and bulk theory, unless there exists a field-theoretical mechanism - such as a double-trace deformation  - that can generate  additional correct possibilities.  The requirement of  finiteness and integrability of the conserved charges is only a necessary - but far from sufficient - condition,  as in principle  there are many other  types of constraints - perturbative or non-perturbative - on the phase space, as our  example involving the fixed value of  $J_0, \bar J_0$ shows.  } boundary conditions that, in turn,  determine the ASG,  is of utmost importance. %But, even more than the boundary conditions, knowing independently  which field-theory options are viable QFTs and which not, would also help guide the choice on the gravity side. 

Let us now discuss the  implications of our work for the Kerr/``CFT'' correspondence -  which was our initial motivation - including a provisional roadmap for how we envisage this problem to be solved. Since $J\bar T$ - deformed CFTs provide the only concrete QFT example so far of a theory that is both non-local and that possesses Virasoro symmetry, one would like to test the possibility that the holographic dual of warped AdS$_3$ shares their universal properties%\footnote{Note that we cannot expect an exact duality to $J\bar T$: the standard Smirnov-Zamolodchikov deformation constructed in \cite{} is double-trace, and thus can only correspond to mixed boundary conditions on AdS$_3$ that leave the local geometry unchanged, while the single-trace $J\bar{T}$ deformation, i.e. a symmetric product orbifold of $J\bar{T}$ deformed CFTs, is only defined for seed CFTs that are  symmetric product orbifolds themselves, whose bulk dual is highly stringy.  } 
. Note that the underlying assumption behind such a test is that there  exist generalisations of single-trace\footnote{In order for the asymptotics to change from AdS$_3$, the single-trace version of the $J\bar{T}$ deformation should be considered, which corresponds to the  symmetric product orbifold of $J\bar{T}$ deformed CFTs. However, the holographic dual of the latter is highly stringy, and thus yet another (intractable) deformation of it is necessary. } $J\bar T$ - deformed CFTs with the same universal properties as them, such as the modified Cardy formula and the symmetry algebra studied in \cite{Chakraborty:2023wel}. There is currently \emph{no} field-theoretical proposal for how to construct such irrelevant deformations of $2d$ CFTs, but their likely existence follows from decoupling arguments \cite{El-Showk:2011euy}. The only currently available method for studying their properties is  holography, with all its associated caveats regarding parametrisation and the choice of boundary conditions. 

One may nonetheless be optimistic, given recent progress in the qualitatively similar case of asymptotically linear dilaton  (ALD) holography. There, \cite{Georgescu:2024iam} uncovered a 21-parameter family of ALD-like decoupled backgrounds, which are holographically dual to (intractable) generalisations of the single-trace $T\bar T$ deformation. Nonetheless, upon dealing with extremely  subtle issues related to parametrisation, \cite{Georgescu:2024iam} was able to show that, in each case, the entropy-to-energy relation read off from holography was \emph{identical} to the corresponding universal relation in single-trace  $T\bar T$ - deformed CFTs. Moreover, \cite{Georgescu:2022iyx} found that the asymptotic symmetries of the  ALD background supported by purely NS flux are \emph{identical} to the extended symmetries of a single-trace $T\bar T$ - deformed CFT, and preliminary results \cite{silviaunpub} indicate this is also the case  for the more general backgrounds. Thus, even though there is currently no field-theoretical definition of these ``generalised single-trace  $T\bar T$ - deformed CFTs'', holography does support the idea that a class of such theories exists, which share certain universal properties with standard single-trace $T\bar T$.

In order to test for a possible link\footnote{The  match between the spectrum of long strings on a certain warped AdS$_3$ background and that of single-trace $J\bar T$ \cite{Apolo:2018qpq,Chakraborty:2018vja
} was one of the early reasons to propose a link between the two.  As emphasized in \cite{Chakraborty:2023wel}, since this match does not extend beyond the long string subsector, the relevance of the above result to the holographic problem at hand is unclear.
 %largely irrelevant for the holographic problem at hand. 
 } between warped AdS$_3$ and single-trace $J\bar T$, one should first ascertain whether the entropy-to-energy and charge relation follows the $J\bar T$ - deformed Cardy formula \eqref{entropyJTbarFT}, which differs from the charged Cardy formula for generic $U(1)$ charges.  We remind the reader that in the old literature on the Kerr/CFT correspondence, e.g. \cite{
Detournay:2012dz}, the evidence pointed only to the standard CFT Cardy formula, but it is unclear  whether the backgrounds under study were general enough for a difference between the two formulae to be noticeable.  More recently, \cite{Apolo:2019yfj} studied a  warped AdS$_3$ background  supported by pure NS-NS flux from the perspective of single-trace $J\bar T$ holography, finding a match to the $J\bar T$ - deformed    Cardy formula. Nonetheless, 
their analysis is only for a restricted set of charges, and we were unable to find a match to $J\bar T$ thermodynamics for generic ones \cite{unpublished}. The main  issue is how to  parametrize the backgrounds and, in particular,  how to identify the field-independent coordinates, which turns out to be surprisingly non-trivial task. %: \textcolor{blue}{ if, for example,  the field theory time coordinate is misidentified by an energy-dependent factor, the entire thermodynamics one  infers from the bulk can be incorrect.}

%This observation has a direct bearing on the Kerr/``CFT'' problem. It is probably a good idea to work with the string embedding of Kerr/CFT, which has a consistent quantum gravity backing, and we at least known which low-energy effective action to work with, and also allows us to avoid the no dynamics problem, by providing a decoupled warped AdS$_3$ geometry.

 Next, one needs to face the issue of which boundary conditions to impose, a recurrent problem in three-dimensional holography being that of an overabundance of possibly consistent choices. 
  A proposal for how to deal with this problem was put forth in \cite{Georgescu:2022iyx}: if the correct parametrization of the background for a generic set of (non-extremal) black hole solutions is known, then the allowed diffeomorphisms  can be found by requiring that the symplectic product between the change in the fields they generate and a perturbation within the space of allowed black hole solutions should be zero, so the flux though the boundary of the spacetime vanishes.  In \cite{Georgescu:2022iyx}, this simple physical prescription applied to the ALD black holes, whose parametrization was fixed by the decoupling limit, gave rise to precisely the single-trace version of the $T\bar T$ symmetry algebra, complete with non-linear terms. % -  at least to all orders to which we checked%\footnote{The corresponding boundary conditions on the bulk fields are not at all physically transparent, and it would have been virtually impossible to simply guess them.}
It would thus be interesting to apply this prescription to warped AdS$_3$ spacetimes and  re-evaluate their asymptotic symmetries, task for   which the results of the current work are highly relevant. Note, however, that this computation
  is only possible after the parametrisation issue is resolved.

%A series of works has shown that the spectrum of long strings excitations of this backgrounds are very similar to $J\bar T$, but this is irrelevant for the behaviour of the full theory. The best one could hope is, as in the $T\bar T$ case, that the relation between the entropy and the conserved charges is the same as in $J\bar T$ - deformed CFTs, which would suggest this ``QFT structure''. However, even this fact has not yet been established: while there are indeed claims in the literature that this is true \cite{},  The problem one faces is one of parametrization of the backgrounds, as these are obtained via just TsT, rather than a decoupling limit, and the naive TsT parametrisation can be  highly misleading in what regards identifying what are the fixed coordinates and parameters., so having a priori knowledge of how the field theory should behave is a very valuable input into the bulk analysis. 

The above roadmap is suggested by our experience with the link between  $T\bar T$ - deformed CFTs  and asymptotically linear dilaton backgrounds.  There may exist, however, simpler ways to approach the problem of the asymptotic symmetries of warped AdS  backgrounds, at least for those embedded in string theory. %, such as \cite{Apolo:2019yfj}. 
When an exact worldsheet description is available, it might be technically easier to use the method proposed in \cite{Du:2024tlu}.

As is well known and our work highlights,  uncovering the symmetries of the dual field theory  through an ASG computation is a subtle problem,  because
it involves two unknowns: 
  the dual field theory \emph{and} the holographic dictionary. In the case we studied,  where the dual field theory was known,  a minimum input from it  was essential for obtaining the correct ASG result. This suggests that any non-trivial progress in non-AdS holography would require parallel development of both the relevant non-gravitational  theories (whose consistency should be tested independently of any holographic applications) and of  the holographic dictionary to the spacetime of interest, whose construction would likely need some field theory input.  In the previous paragraphs, we sketched a possibly  self-standing  bulk procedure for inferring certain properties  of the dual field theory. It would be very interesting to devise a non-trivial test whether these are, indeed,   correct rules for non-AdS holography. The only way we can envisage such a test is by having more non-AdS/non-CFT dual pairs, where each side of the duality is independently defined.

%the asymptotic symmetries problem studied in this work reveals the necessity of additional information - ideally from the dual field theory - in determining the relevant boundary conditions and low-energy effective action for the bulk fields. Thus, we believe that any progress in non-AdS holography should rely 

%
%\bi
%\item comment on dropping the nonperiodic part of the symm param and how only cov ph sp formalism seems to work
%\ei

 \subsubsection*{Acknowledgements}

We are grateful to Glenn Barnich, Jan de Boer, Alejandra Castro, Jean-Fran\c{c}ois Fortin, Wolfgang Lerche and Ruben Monten for interesting conversations. SG's research  is supported in part by the Science 
Technology \& Facilities council under the
grant ST/X000753/1. 

\appendix

\section{Action of the symmetries on the free boson solution}\label{Appendixfreeboson}

%\subsection*{Making sense of the compensating transformation} 

%So far, we have found that while the field-dependent coordonate transformations are perfectly good symmetries of the $J\bar T$ - deformed free boson action, we do not like their action on $Q_0$, which is an additional field-dependent piece that is not clear to leave the action invariant. 

In order to  understand more concretely  the symmetries we uncovered, it is useful to work out their action  on a given solution to the equations of motion, which they should  map to a new solution. We discuss, for simplicity, the $J\bar T$ - deformed free boson, with $\hat{k}=1$.

The relevant equation of motion is  \eqref{jtbarscalareom}, and
the most general solution to it is given by
%\footnote{One can also consider the parametrization
%\be
%\phi = f(z) + g(\bar z - \l f(z)) + c
%\ee
%thta may appear more natural; however, the right-moving symmetries do not act simply on the Fourier coefficients of $g$.}

\be \label{solutioneomapp}
\phi =\varphi_0 +  f(\hat U) +g (\hat v)  \;,\;\; \;\;\;\;\;\hat U = \frac{U}{R} \;, \;\;\;\; \hat  v = \frac{ V - \l \phi +c}{R_v}
\ee
where $f,g$ are taken to have a Fourier expansion that only contains linear terms and non-zero Fourier modes

\be \label{expansionsphi}
f(\hat{U})= a_0 \hat{U} + \frac{i}{2\pi}\sum_{n\neq 0} \frac{a_n}{n} e^{-2\pi i n \hat{U}} \;, \;\;\;\;\;   
g(\hat{v})= b_0 \hat v -\frac{i}{2\pi} \sum_{n\neq 0} \frac{b_n}{n} e^{2\pi i n \hat v}
\ee 
and we have explicitly separated the zero mode, $\varphi_0$, of the solution from them.  %\textcolor{red}{Factors $2\pi$?}
%\textcolor{red}{\emph{Recheck! I think signs of $U, v$ should be different, also i}}
%
%do not contain a zero mode and are expanded in Fourier modes with coefficients $a_0, b_0, a_n/n, b_n/n$.
 The constant piece in $v$ is for now left arbitrary, as  it does not follow from the equations of motion. While it could be easily absorbed  into a redefinition of the $b_n$ and $\varphi_0$, we do find it useful to write it separately,  as above. 
Note \eqref{expansionsphi} is a classical expansion, in that operator ordering issues are to be ignored. 
% The reason we explicitly subtracted the zero mode of $\phi$ from $v$ is to make the latter invariant under constant shifts of $\phi$, which is equivalent with requiring that $Q_0$ commute with the field-dependent coordinate. 
Using 

\be
\p_U \phi  =  %{\color{ForestGreen}\frac{f'}{R}} -\frac{\l g'}{R_v} \p_U\phi = 
\frac{f'/R}{1+ \l g'/R_v}\;, \;\;\;\;\;\; \p_V \phi = %\frac{g'}{R_v} (1-\l \p_V \phi) = 
\frac{g'/R_v}{1+ \l g'/R_v}
\ee
%{\color{ForestGreen}(where the derivatives are wrt the hatted coordinates)}
where the prime denotes a derivative with respect to the corresponding argument of the functions $f,g$, one may easily check that 
%\emph{Check signs!}
%{\color{blue}
%\begin{align}
%\mathcal{K}=\mathcal{J}_+ + \frac{\lambda \mathcal{H}_R}{2}=\frac{\partial_U\phi}{1-\lambda\partial_V\phi}=\frac{f'}{R}\hspace{1cm}\bar{\mathcal{K}}=\mathcal{J}_- + \frac{\lambda \mathcal{H}_R}{2}=\frac{\partial_V\phi}{1-\lambda\partial_V\phi}=\frac{g'}{R_v}
%\end{align}}

\be \label{fourmodesapp}
P_n = \int_0^R \! \!d\s \frac{\partial_U\phi}{1-\lambda\partial_V\phi} \,  e^{2\pi i n \hat{U}}% =\int_0^R d\s \frac{f'}{R}  e^{i n \hat{U}}
= a_n \;, \;\;\;\;\;\;\;\; \bar P_n =  \int_0^R\!\! d\s   \frac{\partial_V\phi}{1-\lambda\partial_V\phi} (1-\l \phi')\, e^{-2\pi i n \hat{v}} = %\int_0^R d\s \frac{g'}{R_v}\partial_{\sigma} v e^{-i n \hat{v}} =
 b_n  
\ee 
%\textcolor{blue}{where in the last integral we used the fact that $1-\l \phi'=\p_\s v$.} 
In particular, using the notation map in the box in section \ref{defnbp}  with $E_R \r H_R$, we have

\be \label{linearlygrmodes}
a_0 = Q_K = J_0  + \frac{\l H_R}{2} \;, \;\;\;\;\; b_0 = - \bar Q_{\bar K} = -\bar J_0 - \frac{\l H_R}{2}
\ee
This dependence of the current zero modes on the quantized charges $J_0, \bar J_0 = \frac{1}{2} (Q_0\pm w)$ and the right-moving energy can be confirmed by direct computation of the winding

\be
w= \int_0^R d\s \p_\s \phi = a_0 +b_0
\ee
and of the shift charge \eqref{defchzm}
\be \label{undeformedchargeappendix}
Q_0 = \int_0^R d\s \frac{f'}{R} - \int_0^R d\s \frac{g'}{R_v} \left(1-\frac{\l f'}{R}\right) % = a_0 - \int d \s \p_\s v \frac{g'}{R_v} \left(1+\frac{\l g'}{R_v}\right) 
= a_0 - b_0 - \frac{\l}{R_v} \sum_n b_n b_{-n}
\ee
where we note that the quadratic term is  proportional to $H_R$ %\textcolor{red}%{Notation? $H_R$ vs $E_R$}%{\color{ForestGreen}(with $K_U$ given in  \eqref{defK})}

\be \label{expressionHRapp}
H_R = \int_0^R d\s \frac{(\p_V\phi)^2}{1-\l \p_V\phi} \left(1-\frac{\l \p_U \phi}{1-\l \p_V\phi}\right) = \int_0^R d\s  \p_\s v \frac{g'^2}{R_v^2}  = \frac{1}{R_v} \sum_n b_n b_{-n}
\ee
We may now import the results of the Hamiltionian analysis from section \ref{section3:Hamiltonian}, which states that the flowed charges
\be \label{flowedchargesapp}
\widetilde{P}_n = P_n - \frac{\l H_R}{2} \d_{n,0} = a_n  - \frac{\l H_R}{2} \d_{n,0} \;,\;\;\;\;\; \widetilde{\bar{P}}_n = \bar{P}_n - \frac{\l H_R}{2} \d_{n,0} = b_n  - \frac{\l H_R}{2} \d_{n,0} 
\ee
satisfy a Kac-Moody algebra
\be
\{\widetilde P_m, \widetilde P_n\} = \{\widetilde{\bar{P}}_m, \widetilde{\bar{P}}_n\} =-\frac{2\pi i m }{2}\delta_{m+n,0} \;, \;\;\;\;\; \{\widetilde P_m, \widetilde{\bar{P}}_n\} = 0
\ee 
For uniformity's sake, we may introduce the notation
%
%\be
% \widetilde a_0 =a_0- \frac{\l H_R}{2} \;, \;\;\;\;\; \widetilde b_0 =b_0+ \frac{\l H_R}{2}\;, \;\;\;\;\; \widetilde a_n = a_n\;, \;\;\widetilde b_n = b_n\;, n \neq 0
%\ee 
$\widetilde a_n = \widetilde P_n, \widetilde b_n = \widetilde{\bar{P}}_n$, whose Poisson brackets  are thus just the standard Kac-Moody ones; in particular, $\widetilde a_0 = J_0 $  and $\widetilde b_0 =-\bar J_0$ commute with  all the other generators. Using \eqref{variationundersymmtr},  these commutation relations simply tell us that 
\be
\d_{\eta_m} \widetilde a_n = \frac{2\pi i m}{2} \eta_m \d_{m+n}
\ee
where $\eta_m$  is the parameter of the transformation generated by $\widetilde P_m = P_m \; (m\neq 0) $, identified with  the corresponding Fourier  coefficient of the periodic function $\eta_p$. Since this is just a  shift of the constants parametrizing the solution, this transformation obviously takes solutions into solutions. Similarly, for the right-movers
\begin{align}
\d_{\bar{\eta}_m} \widetilde b_n = \frac{2\pi i m}{2} \bar{\eta}_m \d_{m+n}
\end{align}
Note that $\widetilde a_0=J_0, \widetilde b_0 = -\bar J_0$ are inert under these transformations. Nonetheless, given its expression \eqref{expressionHRapp}, $H_R$ must transform as 

\be
\d H_R = \frac{1}{ R_v} (2 b_0 \d b_0 + \bar P_{\bar \eta'})% = \frac{ \bar P_{\bar \eta'}}{ R_v (1+ \l b_0/R_v)}= \frac{\l \bar P_{\bar \eta'}}{2 R_Q}
\ee
%where we used the fact that $b_0$ must also transform as $\d b_0 = - \l \d H_R/2$. 
Using \eqref{flowedchargesapp} and the fact that $\widetilde a_0, \widetilde b_0$ are inert, we find   $a_0, b_0$ must transform as

\be \label{variationlinearapp}
\d a_0 = - \d b_0 = \frac{\l \d H_R}{2} = \frac{\l}{2 R_v} (2 b_0 \d b_0 + \bar P_{\bar \eta'}) = \frac{\l \bar P_{\bar \eta'}}{2 R_v (1+ \l b_0/R_v)}= \frac{\l \bar P_{\bar \eta'}}{2 R_Q}
\ee
where we used \eqref{linearlygrmodes} to write the penultimate relation. One immediately identifies this change in the coefficients of the linear terms with the coefficient of the ``compensating transformation" in \eqref{rightafftransf}. Note, moreover, that $v$ is to be held fixed under this transformation on the Fourier modes,  resulting in 
%{\color{ForestGreen}(the sign is plus if we use the conventions above, with the variation in 6.1)}
%
\be
\phi \r \phi'(U',V') = \phi (U,V) +\frac{1}{2} \left[\eta (\hat U) + \bar \eta(\hat v)\right] \;,\;\;\;\; U'=U\;, \;\;\;\; V'= V +\frac{\l}{2} \left[\eta(\hat U) + \bar \eta(\hat v)\right]
\ee
which precisely agrees with \eqref{actlmsymm}  and \eqref{rightafftransf} when $\hat{k}=1$. We thus see explicitly that the necessity of the large affine compensating transformation follows from the expression \eqref{linearlygrmodes} for the coefficients of the linearly growing modes, and the fact that the shift and topological charges are required to not vary under the field-dependent transformations. Note this argument need not invoke the full charge algebra result of the Hamiltonian analysis, but simply follows from the requirement of charge quantization, which is $\{Q_0,\bar P_n\} =0$, in addition to the obvious $\{w, \bar P_n\} =0$ . In other words, if we insist that the linearly growing terms in\eqref{rightafftransf}  not be present, then we immediately find that $\d a_0=\d b_0 =0$, which in turn implies $\d Q_0 = - \l \d H_R \neq 0$. This corresponds precisely the choice of \cite{Guica:2020uhm}, which violated charge quantization.

Note that also the momentum, $P=H_L-H_R$,  is now consistent with quantization, since under a purely right-moving affine transformation it changes as
%{\color{ForestGreen}
\be
\d P = \frac{2 a_0}{R} \d a_0 - \d H_R = - \d H_R \left(1- \frac{\l a_0}{R}\right) = - \d H_R \frac{R_Q}{R} = - \frac{1}{R} \bar P_{\bar \eta'}
\ee
Had we not taken into account the change in $a_0$ under the right-moving  transformation, the coefficient in front of $\bar P_{\bar \eta'}$ would have been the field-dependent radius, which would not have been compatible with momentum quantization on a cylinder of size $R$.

Finally, let us look at the pseudoconformal transformations. If we ignore the compensating affine  terms,  we have $\delta_f\phi=f \partial_U\phi, \delta_{\bar{f}}\phi=-\bar{f}\partial_V\phi $, with   $f,\bar{f}$  periodic. From 
%(we can delete this later) (mention that the Fourier expansion looks different for $f,\bar{f}$ than for $\eta,\bar{\eta}$ if we want to implement the shifts as above? by this I mean that $\phi$ does not have the standard Fourier expansion $\sum_n a_n e^{-in\hat{U}}$ in order for the currents to have it etc, and if the modes shift under affine as above it means that we take $\eta,\bar{\eta}$ also not to have a standard Fourier expansion. This is not the case for $f,\bar{f}$ since they are just diffeos.)
%\begin{align}
%\sum_n a_n e^{-i n \hat{U}}\bigg( f_0+i\sum_{m\neq 0}\frac{f_m}{m}e^{-im\hat{U}} \bigg)&=f_0 a_0 +i \sum_{m\neq 0}\frac{a_{-m}f_m}{m}+\sum_{p\neq 0} \frac{1}{p}\bigg( p f_0 a_p+ i\sum_{m\neq 0}\frac{a_{p-m}f_m p}{m} \bigg)e^{-i p\hat{U}}
%\end{align}
\begin{align}
\sum_n a_n e^{-2\pi i n \hat{U}}\sum_m f_m e^{-2\pi  i m \hat{U}}&=\sum_n \bigg(\sum_m a_m f_{n-m}\bigg)e^{-2\pi in\hat{U}} 
\end{align}
it is easy to read off the shifts of the modes. It is important to note that $a_0$ will not shift, since there is no term proportional to $\hat{U}$ above.  The analogous right-moving result is
%\begin{align}
%\sum_n b_n e^{i n \hat{v}}\bigg( \bar{f}_0-i\sum_{m\neq 0}\frac{\bar{f}_m}{m}e^{im\hat{v}} \bigg)&=\bar{f}_0 b_0 -i \sum_{m\neq 0}\frac{b_{-m}\bar{f}_m}{m}+\sum_{p\neq 0} \frac{1}{p}\bigg( p \bar{f}_0 b_p-i \sum_{m\neq 0}\frac{b_{p-m}\bar{f}_m p}{m} \bigg)e^{i p\hat{v}}
%\end{align}
\begin{align}
\sum_n b_n e^{2\pi i n \hat{v}}\sum_m \bar{f}_m e^{2\pi i m \hat{v}}&=\sum_n \bigg(\sum_m b_m \bar{f}_{n-m}\bigg)e^{2\pi in\hat{v}} 
\end{align}
which implies that $b_0$ will not shift, whereas  the  $b_{n\neq 0}$ generically will, resulting in a change in $H_R$ that will this time be proportional to $\bar{Q}_{\bar{f}'}$. The only way that $b_0$ can be fixed while $H_R$ varies is the the shift charge also changes, which is again inconsistent with charge quantization. Our   solution to this problem is to allow a change $\d b_0 = \frac{\l}{2} \d H_R$, which is precisely the compensating affine transformation \eqref{rightpseudocftr}. 

% which means again that $Q_0$ is not invariant. We want $Q_0$ to stay invariant, since it's quantized, so we need $a_0,b_0$ to change such that they absorb this change in the last term in \eqref{undeformedchargeappendix}. Same story as before, \eqref{variationlinearapp} still holds until the third equality where we need to consider now the variation under pseudoconformal and not affine. This just amounts for replacing $\bar{P}_{\bar{\eta}'}$ with $\bar{Q}_{\bar{f}'}$ (easy to see from the explicit expression of $\bar{Q}_{\bar{f}'}$, the variation of $\sum_n b_n b_{-n}$ gives exactly the same result). From here the conclusion is the same, exactly the same linear transformations, just the overall coefficient differs. 

 % We may hope, however, to be able to add a new term to the action, whose variation under the total $\bar f$ and compensating large gauge transformation equals the above, and which also doe snot affect the (solutions to) the equations of motion.   
 
%Notice that 
%
%\be
%\d \int dt Q_K = \d \int dt d\s \frac{\p_U \phi}{1-\l \p_V \phi}= \int dt d\s \left(\p_U (\eta Q) + \frac{\p_U \phi \l \p_V (\eta Q)}{1-\l \p_V \phi}+ \l \p_V (\eta Q) \frac{\p_U \phi}{1-\l \p_V \phi}
%\right)
%\ee 
% and thus, up to total derivatives $\d S= \frac{1}{\l} \d \int dt \,  Q_K$. We still need to check whether $Q_K$ varies under the old transformation (I think it doesn't).  However, now the total action $S- \l^{-1} \int Q_K$ appears trivial.  Another problem is that the variation is cancelled for any $\eta$, not just $\eta$ large and of a specific form. 

As we discuss in section \ref{subsection:commutatorsfieldth}, the charge algebra can also be used to fix the field dependence of the constant ambiguity that appears in  the definition of the field-dependent coordinate, $v$, in \eqref{solutioneomapp}.  Let us quickly run this argument at the level of the mode expansion \eqref{expansionsphi}. As argued above, charge quantization requires that $\d_{\bar P_n} Q_0 = \{ Q_0, \bar P_n\} =0$. 
%
%One may use the last equation above to show that the zero mode of $\phi$ needs to be subtracted from $v$, which fixes the field-dependence of the constant mode in . 
%
Since the charge  algebra should be  antisymmetric, this leads us to expect that $\d_{Q_0} \bar P_{\bar \eta} =0$. This implies, in particular, that $b_n$ in the expansion \eqref{expansionsphi} is invariant under the action of $Q_0$, which  generates a constant shift in the scalar. The only way that \eqref{solutioneomapp} will transform correctly under the constant shift is if $v$ is invariant under it, which is true if $c $ equals  $\l \times$ the zero mode of $\phi$. %\textcolor{violet}{On the other hand, if we did not subtract it, then $\bar P_{\bar \eta}$ would change by $\propto \l \bar P_{\bar \eta'}$ which would correspond, via antisymmetry, to a $Q_0$ that is not invariant under field-dependent affine transformations. }

To sum up, we saw explicitly in this simple example that in order to keep $Q_0$ fixed, we need to supplement the purely right-moving affine transformations parametrized by periodic functions $\bar{\eta}$ with terms linear in the coordinates, whose coefficients are charge dependent. Doing so is the only way of keeping $\widetilde{a}_0 = J_0$ and $\widetilde{b}_0 = - \bar J_0$ fixed. 
%\emph{How about pseudoconformal?} 

\section{Thermodynamic potentials in $J\bar T$ - deformed CFTs}
\label{Appendix:thermoJTbar}

In this appendix, we summarize the thermodynamic potentials for $J\bar T$ - deformed CFTs for the case of \emph{general} $U(1)$ charges, as only the chiral case appears to be treated in the previous literature.

%{\color{ForestGreen}(rechecked all, let's keep this parametrization, it looks better without introducing $h_{L,R}$)

The entropy in a $J\bar T$ - deformed CFT  can be obtained from the charged Cardy formula in the undeformed CFT (with central charge $c$) using adiabaticity and the relation \eqref{defspec} between the deformed and undeformed energies, which implies that $\hat{\mathcal{E}}_L \equiv R E_L - \frac{2\pi}{k} Q_L^2$ and $\hat{\mathcal{E}}_R \equiv R_v E_R - \frac{2\pi}{k} Q_R^2 $ are invariant under the $J\bar T$ flow\footnote{The relation between the boundary and the bulk notation is $\frac{12\pi}{c}\hat{\mathcal{E}}_L=\left(\frac{R}{\ell}\right)^2 \! \L$ and $\frac{12\pi}{c}\hat{\mathcal{E}}_R=\left(\frac{R_v}{\ell}\right)^2\bar{\L}$. }. The deformed entropy reads 

\be \label{entropyJTbarFT}
S  = \sqrt{\frac{\pi c}{3}}\bigg( \sqrt{R E_L - \frac{2\pi}{k}Q_L^2}+\sqrt{R_v E_R -\frac{2\pi}{k} Q_R^2} \bigg)
\ee
where $Q_{L,R}$ are related to the integer-quantized charges $J_{0}, \bar J_0$, via the standard expression \eqref{flowofcharges}. Since the latter charges are to be  fixed, it is natural to write   the first law of thermodynamics as
\begin{align}
\delta S&= \frac{1}{T_L}\delta E_L +\frac{1}{T_R}\delta E_R + \mu_L \delta J_0 + \mu_R \delta \bar{J}_0
\end{align}
where the temperatures and chemical potentials are given by

\begin{align}
\frac{1}{T_{L/R}}&=\frac{\partial S}{\partial E_{L/R}}\bigg|_{J_0,\bar{J}_0,E_{R/L}=fixed}%\hspace{0.5cm}\frac{1}{T_R}=\frac{\partial S}{\partial E_R}\bigg|_{J_0,\bar{J}_0,E_L=fixed}, 
\hspace{0.5cm}\mu_L=\frac{\partial S}{\partial J_0}\bigg|_{E_{L,R},\bar{J}_0=fixed}, \hspace{0.5cm}\mu_R=\frac{\partial S}{\partial \bar{J}_0}\bigg|_{E_{L,R},J_0=fixed}
\end{align}
The undeformed thermodynamic potentials are related to the  $J\bar T$ flow invariants $\hat{\mathcal{E}}_{L/R}$ introduced above and the charges as 
\begin{align}
\frac{1}{T_{L/R}^{[0]}}=\sqrt{\frac{\pi c}{12}}\frac{R}{\sqrt{\hat{\mathcal{E}}_{L/R}}}\;, %\hspace{0.7cm}\frac{1}{T_R^{[0]}}=\sqrt{\frac{\pi c}{12}}\frac{R}{\sqrt{\hat{\mathcal{E}}_R}}
\hspace{0.7cm}\mu_L^{[0]}=-\sqrt{\frac{\pi c}{12}}\frac{4\pi}{k}\frac{J_0}{\sqrt{\hat{\mathcal{E}}_L}}\;, \hspace{0.7cm}\mu_R^{[0]}=-\sqrt{\frac{\pi c}{12}}\frac{4\pi}{k}\frac{\bar{J}_0}{ \sqrt{\hat{\mathcal{E}}_R}} \label{undeftpot}
\end{align}
Using the expressions for the $U(1)$ charges
\begin{align}\label{flowofcharges}
Q_L&=J_0+\frac{\lambda k}{4\pi} E_R\hspace{1cm}Q_R=\bar{J}_0+\frac{\lambda k}{4\pi} E_R
\end{align}
and the $J\bar T$ flow invariants $\hat{\mathcal{E}}_{L,R}$, the deformed thermodynamic potentials  evaluate to 
%\textcolor{ForestGreen}{(rescale everything by $\sqrt{\frac{\pi c}{12}}$? leave only second expr for $\mu_R$?)}
\be\label{fieldtheorytempapp}
\frac{1}{T_L}=\sqrt{\frac{\pi c}{12}}\frac{R}{\sqrt{\hat{\mathcal{E}}_L}} = \frac{1}{T_L^{[0]}}\;, \hspace{0.71cm}\frac{1}{T_R}=\sqrt{\frac{\pi c}{12}}\bigg(\frac{R-\lambda Q_L}{\sqrt{\hat{\mathcal{E}}_R}}-\frac{\lambda Q_L}{\sqrt{\hat{\mathcal{E}}_L}}\bigg) = \frac{R-\l Q_L}{R T_R^{[0]}} - \frac{\l Q_L}{R T_L^{[0]}}
\ee
and 
\bea
\mu_L&=&\sqrt{\frac{\pi c}{12}}\bigg(-\frac{4\pi Q_L/k}{\sqrt{\hat{\mathcal{E}}_L}}-\frac{\lambda E_R}{\sqrt{\hat{\mathcal{E}}_R}}\bigg) =  \mu_L^{[0]}- \frac{\lambda E_R}{R} \bigg(\frac{1}{T_L^{[0]}}+\frac{1}{T_R^{[0]}}\bigg) \nonumber\\[4pt]
\mu_R&=&\sqrt{\frac{\pi c}{12}}\bigg(\frac{\lambda E_R - \frac{4\pi}{k}Q_R}{\sqrt{\hat{\mathcal{E}}_R}}\bigg)=-\sqrt{\frac{\pi c}{12}}\frac{4\pi}{k} \frac{\bar{J}_0}{\sqrt{\hat{\mathcal{E}}_R}} =\mu_R^{[0]}\label{fieldtheorypotentialapp}
\eea
where for the second set of equalities we used \eqref{undeftpot}. 
%If we denote by $[0]$ the quantities in the undeformed theory, which can be simply obtained by setting $\lambda=0$ above 
%
%we can see that the relations between the deformed and undeformed quantities can be written as
%\begin{align}\label{relationsdefundefapp}
%\frac{1}{T_L}=\frac{1}{T_L^{[0]}}\hspace{1cm}
%\frac{1}{T_R}&=\frac{R-\lambda Q_L}{R T_R^{[0]}}-\frac{\lambda Q_L}{R T_L^{[0]}}\hspace{1cm}\mu_L=\mu_L^{[0]}- \frac{\lambda E_R}{R} \bigg(\frac{1}{T_L^{[0]}}+\frac{1}{T_R^{[0]}}\bigg)\hspace{1cm}\mu_R=\mu_R^{[0]}
%\end{align}

\section{Computations in the covariant phase space formalism}\label{appendixcovph}

This appendix has two parts. In the first part,
we list the formulae we use to compute the conserved charges in the covariant phase space formalism, and in particular rederive the contributions originating from the
Chern-Simons terms in the action.  In the second part, we display certain technical details of the evaluation of these conserved charges, and in particular explain how the simple formulae \eqref{cscontribph} quoted  in the main text are derived.

% fields (the contribution from the metric for diffeos is standard \textcolor{red}{and can be found, for example, in \cite{Compere:2018aar}}). We work in the covariant phase space formalism, which provides an algorithmic way to construct conserved charges in theories with local symmetries (see \cite{Compere:2018aar} for a comprehensive review). 

The full action for the theory we consider is that of three-dimensional Einstein gravity with a negative cosmological constant coupled to a set of $U(1)$ Chern-Simons gauge fields
%
%Here we focus on a generic 3d CS theory defined by the action:
\begin{align}\label{actionapp}
&S=\int d^3 x \sqrt{-g} \left(R + \frac{2}{\ell^2} +  \frac{c_{ij}}{2}\epsilon^{\mu\nu\rho}A^i_{\mu}F_{\nu\rho}^j \right)
\end{align}
where we have let the Chern-Simons couplings be arbitrary. The particular coupling matrix we have for the two Chern-Simons gauge fields appearing in the main text is 
\be \label{particularcijnondiag}
c_{ij} =\frac{k}{8\pi}\begin{pmatrix}
-1 & 1\\
-1 & 1
\end{pmatrix}
\ee
 The contribution of the Einstein-Hilbert term to the conserved charges is well-known \cite{Iyer:1994ys}, and we quote it in \eqref{metriccontribph}; in the following, we derive the Chern-Simons contribution to the conserved charges for the generic action above. %\textcolor{red}{True the result is different from what I had with Geoffrey?} {\color{ForestGreen}(yes)}

%where $F^j_{\nu\rho}=\nabla_{\nu}A_{\rho}^j-\nabla_{\rho}A_{\nu}^j$ and $\epsilon^{\mu\nu\rho}=\frac{1}{\sqrt{-g}}\hat{\epsilon}^{\mu\nu\rho}$. 

\subsection{Chern-Simons contribution to the conserved charges}

The procedure for  obtaining these contributions is explained at length in \cite{Compere:2009dp,Compere:2018aar}. A general variation of the Chern-Simons Lagrangian in \eqref{actionapp} takes the form  
\be
\d L_{CS} = E^\mu_i \d A^i_\mu + \nabla_\mu \Theta^\mu
\ee
where $E^\mu_i$ are the equations of motion for the gauge fields
%
%From the variation of the action we extract the equations of motion, which force all field strengths to vanish:
\begin{align}
E^{\mu}_i=\frac{c_{ij}+c_{ji}}{2}\epsilon^{\mu\nu\rho}F_{\nu\rho}^j
\end{align}
and $\Theta^\mu $ is  the presymplectic potential, which is explicitly  given by
\begin{align}
\Theta^{\mu}&=-c_{ij}\epsilon^{\mu\nu\rho}A^i_{\nu}\delta A^j_{\rho}=c_{ij}\epsilon^{\mu\nu\rho}A^i_{\rho}\delta A^j_{\nu}
\end{align}
We would like to compute the conserved charges associated to (large) diffeomorphisms and gauge transformations, and thus we will particularize the variation to $\delta A^i=\mathcal{L}_{\xi}A^i$ and $\delta A^i=d \Lambda^i$, respectively.

\subsubsection*{Conserved charges associated with large diffeomorphisms}

Under the variation $\delta_{\xi} A^i=\mathcal{L}_{\xi}A^i$, the action is invariant up to a boundary term given by:
\begin{align}
M^{\mu}_{\xi}&=\xi^{\mu}\frac{c_{ij}}{2}\epsilon^{\beta\nu\rho}A^i_{\beta}F^{j}_{\nu\rho}
\end{align}
Evaluating $\Theta^{\mu}$ on the corresponding variation, we obtain:
\begin{align}
\Theta^{\mu}_{\xi}-M^{\mu}_{\xi}&=c_{ij}\epsilon^{\mu\nu\rho}A^i_{\rho}\mathcal{L}_{\xi}A^j_{\nu}-\xi^{\mu}\frac{c_{ij}}{2}\epsilon^{\beta\nu\rho}A^i_{\beta}F^{j}_{\nu\rho}
\end{align}
The on-shell vanishing Noether current is:
\begin{align}
S^{\mu}_{\xi}&=\frac{c_{ij}}{2}\epsilon^{\mu\nu\rho}\bigg(F_{\nu\rho}^j A_{\lambda}^i+F_{\nu\rho}^i A_{\lambda}^j\bigg)\xi^{\lambda}
\end{align}
From here, we can see that on-shell $
\Theta^{\mu}_{\xi}-M^{\mu}_{\xi}+S^{\mu}=\nabla_{\nu}\mathcal{Q}^{\mu\nu}$ with:
\begin{align}
\mathcal{Q}^{\mu\nu}_{\xi}&=c_{ij}\epsilon^{\mu\nu\rho}A^i_{\rho}\xi^{\lambda} A^j_{\lambda}
\end{align}
The charge variations associated to variations of the $U(1)$ Chern-Simons fields under diffeomorphisms are encoded in the 1-form:
\begin{align}
k^{CS}_{\xi}&=-\frac{1}{2}\epsilon_{\mu\nu\alpha}K^{\mu\nu}_{\xi,CS}dx^{\alpha}
\end{align}
where
\begin{align}
K^{\mu\nu}_{\xi,CS}&=-\delta \mathcal{Q}^{\mu\nu}_{\xi}+(i_{\xi}\Theta)^{\mu\nu}
\end{align}
Using the expressions above, we have:
\begin{align}
K^{\mu\nu}_{\xi,CS}&=-c_{ij}\epsilon^{\mu\nu\rho}\xi^{\lambda}A^j_{\lambda}\delta A^i_{\rho}-c_{ij}\epsilon^{\mu\nu\rho}\xi^{\lambda}A^i_{\rho}\delta A^j_{\lambda}+\xi^{\nu}c_{ij}\epsilon^{\mu\alpha\beta}A^i_{\beta}\delta A^j_{\alpha}-\xi^{\mu}c_{ij}\epsilon^{\nu\alpha\beta}A^i_{\beta}\delta A^j_{\alpha}
\end{align}
from which it follows that:
\begin{align}\label{csdiffcontribution}
k^{CS}_{\xi}&=-c_{ij}\xi^{\lambda}(A^j_{\lambda}\delta A^i_{\alpha}+A^i_{\lambda}\delta A^j_{\alpha})dx^{\alpha}
\end{align}
The final expression for the charge variations is obtained by integrating this 1-form on a circle of circumference $R$, whose coordinate we denote by $\sigma$ \footnote{Note that this expression differs from 2.43 from \cite{Compere:2014bia} by the fact that our result is symmetric under $i\leftrightarrow j$.} : 
\begin{align}
\slash{\!\!\!\delta} Q_{\xi}^{CS}&=-\int_0^R d\sigma c_{ij}\xi^{\lambda}(A^j_{\lambda}\delta A^i_{\sigma}+A^i_{\lambda}\delta A^j_{\sigma})
\end{align}
For our case of interest, of two Chern-Simons gauge fields, denoted by $A$ and $B$, with  $c_{ij}$ given in \eqref{particularcijnondiag}, we obtain: 

\begin{align}\label{chargeCSdif}
\slash{\!\!\!\delta} Q_{\xi}^{CS}&=\frac{k}{4\pi}\int_0^R d\sigma \bigg(\xi^{\lambda}A_{\lambda}\delta A_{\sigma}-\xi^{\lambda}B_{\lambda}\delta B_{\sigma}\bigg)
\end{align}
We note that, due to the fact that the sum of the off-diagonal terms in $c_{ij}$ is zero, the mixed terms dropped.

\subsubsection*{Conserved charges associated to (large) gauge transformations}
Under the variation $\delta A^i=d\Lambda^i$, the action is invariant up to a boundary term given by:
\begin{align}
M^{\mu}_{\Lambda}&=\frac{c_{ij}}{2}\epsilon^{\mu\nu\rho}\Lambda^i F^j_{\nu\rho}
\end{align}
while the on-shell vanishing Noether current is:
\begin{align}
S^{\mu}_{\Lambda}&=\frac{c_{ij}}{2}\epsilon^{\mu\nu\rho}(\Lambda^i F^j_{\nu\rho}+\Lambda^j F^i_{\nu\rho})
\end{align} 
We obtain that $\Theta^{\mu}_{\Lambda}-M^{\mu}_{\Lambda}+S^{\mu}_{\Lambda}=\nabla_{\nu}\mathcal{Q}^{\mu\nu}$, with
\begin{align}
\mathcal{Q}^{\mu\nu}_{\Lambda}&=c_{ij}\epsilon^{\mu\nu\rho}A^i_{\rho}\Lambda^j
\end{align}
In the case of gauge transformations, $K^{\mu\nu}_{\Lambda}=-\delta \mathcal{Q}^{\mu\nu}_{\Lambda}$, so the charge variations are trivially integrable in phase space and we obtain:
\begin{align}
\int_{\gamma} k_{CS}&=-c_{ij} A^i_{\alpha}\Lambda^j dx^{\alpha}
\end{align}
where we denoted by $\gamma$ the integration path in phase space. Finally, we obtain the charges associated to gauge transformations of the CS fields:
\begin{align}
Q_{\Lambda}&=-\int_0^R d\sigma c_{ij} \Lambda^j A^i_{\sigma}
\end{align}
For the particular case of two Chern-Simons gauge fields $A$ and $B$ with $c_{ij}$ given in \eqref{particularcijnondiag}, we obtain
\begin{align}
Q_{\Lambda_A,\Lambda_B}&=\frac{k}{8\pi}\int_0^R d\sigma (\Lambda_A-\Lambda_B)(A_{\sigma}+B_{\sigma}) \label{pbch}
\end{align}
while for a diagonal matrix $c_{ij}=\frac{k}{8\pi}\begin{pmatrix}
-1 & 0\\
0  & 1
\end{pmatrix}$ we obtain:
\begin{align}
Q_{\Lambda_A,\Lambda_B}&=\frac{k}{8\pi}\int_0^R d\sigma (\Lambda_A A_{\sigma}-\Lambda_B B_{\sigma}) \label{gaugech}
\end{align}
In this case, under a combination of diffeomorphism and gauge transformations we obtain

\begin{align}
\slash{\!\!\!\delta} Q_{\xi,\Lambda_A,\Lambda_B}&=\frac{k}{4\pi}\int_0^R d\sigma  \bigg( (\xi\cdot A+\Lambda_{A})\delta A_{\sigma}-(\xi\cdot B+\Lambda_{B})\delta B_{\sigma} \bigg)
\end{align}

\subsection{Details of the evaluation of the conserved charges \label{detevalcc} }
In this subsection, we present the details of the computation of the conserved charges associated to right-moving affine and pseudoconformal transformations. In both cases, the conserved charges receive contributions from both the metric and the two $U(1)$ Chern-Simons fields. Overall, these contributions can be split into a part proportional to the winding parameters entering the transformations and the ``periodic" ones, associated to the functions  $\bar{\eta}_p, \bar{f}$. We find it useful to evaluate these two types of contributions separately.

\subsubsection*{Right-moving affine charges}
In the case of the right-moving affine transformations, the periodic contribution to the charge variation is obtained by setting the winding to zero in \eqref{transformationsrmaff} and evaluating \eqref{metriccontribph},\eqref{cscontribph}, with the appropriate rescaling of $\bar{\eta}_p$, as explained in the main text
\begin{align}\label{initialexpressionrm}
\slash{\!\!\!\delta} \bar{P}_{\bar \eta}\big|_{\bar{\eta}_p}&=\int_0^R d\sigma\bigg(\frac{(\l\bar{\mathcal{J}}\delta \mathcal{J}-\partial_{\sigma}v \delta\bar{\J})}{1+\l\bar{\mathcal{J}}}\bar{\eta}_p-\frac{\l(\delta\mathcal{J}+\partial_{\sigma}v \delta\bar{\mathcal{J}})}{1+\l\bar{\mathcal{J}}}\int^{\hat{v}}\!\!\bar{\mathcal{J}}\bar{\eta}'_p\bigg)
\end{align}
where the primitive of $\bar \J \bar \eta_p'$ originates from the gauge parameter \eqref{transformationsrmaff}. It is useful to write explicitly $\delta\bar{\mathcal{J}}=\delta\bar{\mathcal{J}}_{int}+\bar{\mathcal{J}}'\delta \hat{v}$ and use the definition of $\delta v$ in order to express $(\delta \mathcal{J}+\partial_{\sigma}v\delta\bar{\mathcal{J}}_{int})$ in terms of $\partial_{\sigma}\delta v$. We will further manipulate \eqref{initialexpressionrm} by separating the periodic and the contributions that have potential winding 
\begin{align}\label{periodicrmaffine}
\slash{\!\!\!\delta} \bar{P}_{\bar \eta}\big|_{\bar{\eta}_p}&= -\int_0^R d\sigma \bar{\eta}_p \partial_{\sigma}v \delta\bar{\mathcal{J}}_{int}+ \int_0^R d\sigma\bigg[\frac{\l}{1+\l\bar{\mathcal{J}}}\big(\delta\mathcal{J}+\partial_{\sigma}v \delta\bar{\mathcal{J}}_{int}\big)\bigg(\bar{\mathcal{J}}\bar{\eta}_p-\int^{\hat{v}}\!\! \bar{\mathcal{J}}\bar{\eta}'_p\bigg) -\nonumber \\
& \hspace{1.5 cm}-\frac{\partial_{\sigma}v \delta \hat{v} \bar{\mathcal{J}}'\bar{\eta}_p}{1+\l\bar{\mathcal{J}}}-\frac{\l \partial_{\sigma}v \bar{\mathcal{J}}'\delta \hat{v}}{1+\l\bar{\mathcal{J}}}\int^{\hat{v}}\!\!\bar{\mathcal{J}}\bar{\eta}'_p\bigg]= \nonumber\\
& \hspace{-0.7 cm} =\delta\left(-\int_0^R d\sigma \bar{\eta}_p \bar{\mathcal{J}}\partial_{\sigma}v\right)+ \int_0^R d\sigma\bigg[-\bar{\mathcal{J}}\bar{\eta}_p\partial_{\sigma}\delta v+\partial_{\sigma}\delta v \int^{\hat{v}}\bar{\mathcal{J}}\bar{\eta}'_p - \frac{\l\partial_{\sigma}\bar{\mathcal{J}}\delta v}{1+\l\bar{\mathcal{J}}}\bar{\mathcal{J}}\bar{\eta}_p - \frac{\partial_{\sigma}v \delta \hat{v}}{1+\l\bar{\mathcal{J}}}\bar{\mathcal{J}}'\bar{\eta}_p\bigg]\nonumber\\
&  \hspace{-0.7 cm} =\delta\left(-\int_0^R d\sigma \bar{\eta}_p \bar{\mathcal{J}}\partial_{\sigma}v\right)+ \int_0^R d\sigma\bigg[\partial_{\sigma}\bigg(\delta v \int^{\hat{v}} \bar{\mathcal{J}}\bar{\eta}'_p\bigg)-\bar{\mathcal{J}}\partial_{\sigma}\bar{\eta}_p \delta v - \bar{\mathcal{J}}\bar{\eta}_p \partial_{\sigma}\delta v-\bar{\eta}_p \partial_{\sigma}\bar{\mathcal{J}}\delta v\bigg]\nonumber
\end{align}
\begin{align}
& \hspace{-0.7 cm} = \delta\left(-\int_0^R d\sigma \bar{\eta}_p \bar{\mathcal{J}}\partial_{\sigma}v\right)+ \int_0^R d\sigma\bigg[[\bar{\mathcal{J}}\bar{\eta}'_p]_{zm}\partial_{\sigma}\big(\hat{v}\delta v \big)-\partial_{\sigma}(\bar{\eta}_p\bar{\mathcal{J}}\delta v)\bigg]\nonumber\\
& \hspace{-0.7 cm} =\delta\left(-\int_0^R d\sigma \bar{\eta}_p \bar{\mathcal{J}}\partial_{\sigma}v\right)+ [\bar{\mathcal{J}}\bar{\eta}'_p]_{zm}\int d\sigma \partial_{\sigma}\big(\hat{v}\delta v \big)
\end{align}
where we emphasize that now all the derivatives are with respect to $\hat{v}$ and we dropped sigma derivatives of periodic functions.

Next, we evaluate the winding contribution, which is obtained by keeping only the terms proportional to the winding in \eqref{transformationsrmaff}
\begin{align}
&\slash{\!\!\!\delta} \bar{P}_{\bar \eta}\big|_{w_{\eta}}=w_{\eta}\int_0^R d\sigma\bigg[ 2\hat{\mathcal{T}}\delta \mathcal{J}+\frac{(\hat{v}+\lambda\int^{\hat{v}} \bar{\mathcal{J}})}{1+\lambda\bar{\mathcal{J}}}(\delta\mathcal{J}+\partial_{\sigma}v\delta\bar{\mathcal{J}})\bigg]=\nonumber\\
&=w_{\eta}\int_0^R d\sigma\bigg[2\hat{\mathcal{T}}\delta\mathcal{J}-\frac{(\hat{v}+\lambda\int^{\hat{v}}\bar{\mathcal{J}})}{\lambda}\bigg(\partial_{\sigma}\delta v + \frac{\lambda \delta v\partial_{\sigma}\bar{\mathcal{J}}}{1+\lambda\bar{\mathcal{J}}}\bigg)+\frac{(\hat{v}+\l\int^{\hat{v}}\bar{\mathcal{J}})}{1+\l\bar{\mathcal{J}}}\partial_{\sigma}\bar{\mathcal{J}}\delta v\bigg]=\nonumber\\
&=w_{\eta}\int_0^R d\sigma\bigg[2\hat{\mathcal{T}}\delta\mathcal{J}-\frac{\hat{v}}{\l}\partial_{\sigma}\delta v -\partial_{\sigma}\delta v \int^{\hat{v}}\bar{\mathcal{J}}\bigg]=\nonumber\\
&=w_{\eta}\int_0^R d\sigma\bigg(2\hat{\mathcal{T}}\delta\mathcal{J}-\partial_{\sigma}\left(\frac{1}{\l}\hat{v}\delta v\right)+\frac{1}{\l}\partial_{\sigma}\hat{v}\delta v -\partial_{\sigma}\left(\delta v \int^{\hat{v}}\bar{\mathcal{J}}\right)+ \bar{\mathcal{J}}\partial_{\sigma}\hat{v}\delta v\bigg)=\nonumber\\
&=w_{\eta}\int_0^R d\sigma\bigg(2\hat{\mathcal{T}}\delta\mathcal{J}+\frac{1}{\l}\partial_{\sigma}\hat{v} \delta v(1+\l\bar{\mathcal{J}})-\frac{1}{\l}[1+\l\bar{\mathcal{J}}]_{zm}\partial_{\sigma}\big(\hat{v}\delta v\big)\bigg)=\nonumber\\
&=w_{\eta}\int_0^R d\sigma\bigg(2\hat{\mathcal{T}}\delta\mathcal{J}+\frac{1}{\l}\partial_{\sigma}\hat{v} \delta v(1+\l\bar{\mathcal{J}})\bigg)-[\bar{\mathcal{J}}\bar{\eta}'_p]_{zm} \int_0^R d\sigma \partial_{\sigma}(\hat{v}\delta v)
\end{align}
The last term cancels the result of \eqref{periodicrmaffine}, such that overall we obtain:
\begin{align}\label{resultfinalrightaffine}
&\slash{\!\!\!\delta} \bar{P}_{\bar \eta}\big|_{\bar{\eta}_p}+\slash{\!\!\!\delta} \bar{P}_{\bar \eta}\big|_{w_{\eta}}=\delta\left(-\int_0^R d\sigma \bar{\eta}_p \bar{\mathcal{J}}\partial_{\sigma}v\right)+ w_{\eta}\int_0^R d\sigma\bigg[\big[2\hat{\mathcal{T}}\delta\mathcal{J}+\frac{\delta v}{\l}\partial_{\sigma}\hat{v} (1+\l\bar{\mathcal{J}})\big]\bigg]=\nonumber\\
&=\delta\left(-\int_0^R d\sigma \bar{\eta}_p \bar{\mathcal{J}}\partial_{\sigma}v\right)+ w_{\eta}\int_0^R d\sigma\bigg[\big[2\hat{\mathcal{T}}\delta\mathcal{J}+\frac{\delta \hat{v}}{\l}(1-\l \mathcal{J})\big]\bigg]=\nonumber\\
&=\delta\left(-\int_0^R d\sigma \bar{\eta}_p \bar{\mathcal{J}}\partial_{\sigma}v\right)+w_{\eta} \int_0^R d\sigma\delta\bigg[2\hat{\mathcal{T}}\mathcal{J}+\frac{\hat{v}}{\l}\bigg]
\end{align}

%{\color{blue}\emph{To rewrite:} [We remind that in field theory we obtained the charges:
%\begin{align}
%\bar{P}_{\bar{\eta}_p}&=-\int \bar{\eta}_p \frac{\partial_V\phi(1-\l\phi')}{1-\l\partial_V\phi}=-\int \bar{\eta}_p \bar{\mathcal{J}}\partial_{\sigma}v
%\end{align}
%The variations are given by (sign conventions?):
%\begin{align}\label{variationRMaff}
%\delta \bar{P}_{\bar{\eta}_p}&=-\int \bar{\eta}'_p\bar{\mathcal{J}}\partial_{\sigma}v \delta v-\int \bar{\eta}_p \bar{\mathcal{J}}'\partial_{\sigma} v \delta v-\int \bar{\eta}_p\delta\bar{\mathcal{J}}_{int}\partial_{\sigma}v-\int \bar{\eta}_p \bar{\mathcal{J}}\partial_{\sigma}\delta v=\nonumber\\
%&=-\int \partial_{\sigma}(\bar{\eta}_p\bar{\mathcal{J}}\delta v) - \int \bar{\eta}_p\partial_{\sigma}v\delta\bar{\mathcal{J}}_{int}=- \int \bar{\eta}_p\partial_{\sigma}v\delta\bar{\mathcal{J}}_{int}
%\end{align}
%since the function inside the sigma derivative is periodic.] Next, we isolate this term and write:}

\subsubsection*{Right-moving pseudoconformal charges}
In the case of the right-moving pseudoconformal transformations, let us introduce the following notations
\begin{align}
\mathcal{P}_{\bar{f}}&= \kappa \int^{\hat{v}}\!\! \bar{\mathcal{L}}\bar{f}'\hspace{1cm}\mathcal{P}_{\bar{J}}=\int^{\hat{v}}\!\! \bar{\mathcal{J}}
\end{align}
The contribution to the charge variation  obtained from the terms given by $\bar{f}$ in \eqref{rmpseudocftr} is 
\begin{align}
&\slash{\!\!\!\delta} \bar{Q}_{\bar{f}}\big|_{\bar{f}}=\int_0^R d\sigma\bigg[\frac{1}{8\pi G\ell}\bar{f}\delta\big(-\partial_{\sigma}v\bar{\mathcal{L}}\big)-\frac{2\pi}{k}\delta\big(\bar{f}\bar{\mathcal{J}}^2\partial_{\sigma}v\big)+\partial_{\sigma}\bigg(\frac{\l \ell}{32\pi G}\frac{\bar{f}'}{1+\l\bar{\mathcal{J}}}\delta\bar{\mathcal{J}}\bigg)-\nonumber\\
&-\frac{2\pi}{k}\frac{\l}{1+\l\bar{\mathcal{J}}}\mathcal{P}_{\bar{f}}(\delta\mathcal{J}+\partial_{\sigma}v\delta\bar{\mathcal{J}})-\frac{\ell}{16\pi G}\l R_v\bar{f}'\bigg(\frac{\l R_v\bar{\mathcal{J}}'}{(1+\l\bar{\mathcal{J}})^2}\delta\mathcal{J}+\frac{\l\mathcal{J}\partial_V\delta\bar{\mathcal{J}}+\partial_U\delta\bar{\mathcal{J}}}{1+\l\bar{\mathcal{J}}}\bigg)+\nonumber\\
&+\frac{4\pi}{k}[\bar{\mathcal{J}}(\mathcal{\bar{J}}\bar{f})']_{zm}\partial_{\sigma}\big(\hat{v}\delta v \big)\bigg]
\end{align}
The last term on the first line integrates to 0, so we drop it. The last term on the second line is 0, using the fact that $\l\J\partial_V+\partial_U$ annihilates all functions of $v$ only. The remaining terms are
\begin{align}
\slash{\!\!\!\delta} \bar{Q}_{\bar{f}}\big|_{\bar{f}}&=\int_0^R d\sigma\bigg[\delta\big(-\bar{f}(\frac{1}{8\pi G \ell}\bar{\mathcal{L}}+\frac{2\pi}{k}\bar{\mathcal{J}}^2)\partial_{\sigma}v\big)+\frac{1}{8\pi G\ell}\partial_{\sigma}v\bar{\mathcal{L}} \bar{f}' \delta \hat{v}+ \frac{2\pi}{k}\mathcal{P}_{\bar{f}}\partial_{\sigma}\delta v+\nonumber\\
&+\frac{4\pi}{k}[\bar{\mathcal{J}}(\mathcal{\bar{J}}\bar{f})']_{zm}\partial_{\sigma}\big(\hat{v}\delta v \big)\bigg]
\end{align}
We further group together the second and the third terms in a sigma derivative, from which we drop the periodic part
\begin{align}
&\slash{\!\!\!\delta} \bar{Q}_{\bar{f}}\big|_{\bar{f}}=\int_0^R d\sigma\bigg[\delta\bigg(-\bar{f}(\frac{1}{8\pi G \ell}\bar{\mathcal{L}}+\frac{2\pi}{k}\bar{\mathcal{J}}^2)\partial_{\sigma}v\bigg)+\frac{1}{8\pi G\ell}\partial_{\sigma}\big(\delta v \int^{\hat{v}} \bar{\mathcal{L}}\bar{f}'\big)+ \frac{4\pi}{k} [\bar{\mathcal{J}}(\mathcal{\bar{J}}\bar{f})']_{zm}\partial_{\sigma}\big(\hat{v}\delta v \big)\bigg]=\nonumber\\
&=\int_0^R d\sigma\bigg[\delta\bigg(-\bar{f}(\frac{1}{8\pi G \ell}\bar{\mathcal{L}}+\frac{2\pi}{k}\bar{\mathcal{J}}^2)\partial_{\sigma}v\bigg)+\bigg[(\frac{1}{8\pi G \ell}\bar{\mathcal{L}}+\frac{2\pi}{k}\bar{\mathcal{J}}^2)\bar{f}'\bigg]_{zm}\partial_{\sigma}(\hat{v}\delta v)\bigg]
\end{align}
Next, we compute the contribution from the  terms in \eqref{rmpseudocftr} proportional to the winding, which in this case is given by
\begin{align}
w_{\bar{f}}&=\frac{R_v}{R_Q}\frac{\lambda}{2}[(\kappa\bar{\mathcal{L}}+\bar{\mathcal{J}}^2)\bar{f}']_{zm}
\end{align}
We obtain 
\begin{align}
&\slash{\!\!\!\delta} \bar{Q}_{\bar{f}}\big|_{w_{\bar{f}}}=\frac{4\pi}{k}w_{\bar{f}}\int_0^R d\sigma\bigg[\hat{U}\delta\mathcal{J}+\frac{\mathcal{P}_{\bar{J}}\l(\delta\mathcal{J}+\partial_{\sigma}v\bar{\mathcal{J}})}{1+\l\bar{\mathcal{J}}}-\hat{v}\bigg(-\frac{\partial_{\sigma}v\delta\bar{\mathcal{J}}}{1+\l\bar{\mathcal{J}}}+\frac{\l\bar{\mathcal{J}}\delta\mathcal{J}}{1+\l\bar{\mathcal{J}}}\bigg)\bigg]=\nonumber\\
&=\frac{4\pi}{k}w_{\bar{f}}\int_0^R d\sigma\bigg[2\hat{\mathcal{T}}\delta\mathcal{J}-\mathcal{P}_{\bar{J}}\partial_{\sigma}\delta v + \hat{v}\frac{\delta\mathcal{J}+\partial_{\sigma}v\delta\bar{\mathcal{J}}}{1+\l\bar{\mathcal{J}}}\bigg]=\frac{4\pi}{k}w_{\bar{f}}\int_0^R d\sigma\bigg[2\hat{\mathcal{T}}\delta\mathcal{J}-\mathcal{P}_{\bar{J}}\partial_{\sigma}\delta v -\frac{\hat{v}}{\l}\partial_{\sigma}\delta v\bigg]=\nonumber\\
&=\frac{4\pi}{k} w_{\bar{f}}\int_0^R d\sigma\bigg[ 2\hat{\mathcal{T}}\delta\mathcal{J} -\partial_{\sigma}(\mathcal{P}_{\bar{J}}\delta v)+\bar{\mathcal{J}}\partial_{\sigma}\hat{v}\delta v -\frac{\partial_{\sigma}(\hat{v}\delta v)}{\l} +\frac{\partial_{\sigma}\hat{v} \delta v}{\l}\bigg]=\nonumber\\
&=\frac{4\pi}{k}w_{\bar{f}}\int_0^R d\sigma\bigg[ 2\hat{\mathcal{T}}\delta\mathcal{J} - \frac{[1+\l\bar{\mathcal{J}}]_{zm}}{\l}\partial_{\sigma}(\hat{v}\delta v )+\frac{1+\l\bar{\mathcal{J}}}{\l}\partial_{\sigma }\hat{v}\delta v\bigg]=\nonumber\\
&=\frac{4\pi}{k}w_{\bar{f}}\int_0^R d\sigma\bigg[2\hat{\mathcal{T}}\delta\mathcal{J}+\frac{1}{\l}(1-\l\mathcal{J})\delta v\bigg] -\frac{1}{\hat{k}}\int_0^R d\sigma[\kappa\bar{(\mathcal{L}}+\bar{\mathcal{J}}^2)\bar{f}']_{zm}\partial_{\sigma}(\hat{v}\delta v)
\end{align}
After adding the two contributions, our final result is
\begin{align}
\slash{\!\!\!\delta} \bar{Q}_{\bar{f}}=\slash{\!\!\!\delta} \bar{Q}_{\bar{f}}\big|_{\bar{f}}+\slash{\!\!\!\delta} \bar{Q}_{\bar{f}}\big|_{w_{\bar{f}}}=\int_0^R d\sigma
\bigg[ -\delta\bigg(\partial_{\sigma}v \bar{f} \big(\frac{\bar{\mathcal{L}}}{8\pi G\ell}+\frac{2\pi}{k}\bar{\mathcal{J}}^2\bigg)\big)+\frac{2}{\hat{k}} w_{\bar{f}} \delta\bigg(2\hat{\mathcal{T}}\mathcal{J}+\frac{\hat{v}}{\l}\bigg)\bigg]
\end{align}

\section{Relation to the Comp\`{e}re-Song-Strominger analysis }
\label{appendixCSS}

In this appendix, we would like to comment on the relation between our analysis and the one of \cite{Compere:2013bya}, who considered pure gravity in $AdS_3$ with non-standard, chiral boundary conditions  and obtained an asymptotic symmetry algebra consisting of a \emph{single} $U(1)$ Virasoro-Kac-Moody factor, with a background-dependent $U(1)$ level that is negative for black holes. 

The setup of  \cite{Compere:2013bya} is oftentimes used as a non-trivial holographic realisation of  a ``warped CFT'' \cite{Hofman:2011zj, Detournay:2012pc}, a putative %\footnote{Simple examples of such theories have been constructed in \cite{}. \textcolor{red}{Are any interacting examples known?}} 
two-dimensional QFT with $SL(2,\mathbb{R})_L \times U(1)_R$ global symmetry that gets enhanced to a left-moving Virasoro-Kac-Moody one, where the zero mode of the $U(1)$ symmetry corresponds to right-moving translations.  The unitarity issues related to the negative $U(1)$ level  need  then to be addressed, see e.g. \cite{Apolo:2018eky, Aggarwal:2022xfd} for discussions. 

The relation between our present work and \cite{Compere:2013bya} is that the backgrounds allowed by their boundary conditions  correspond to a truncation of the phase space we consider.  We can thus interpret their analysis   from the perspective of ours, which  has, by now, a very  solid foundation in $J\bar T$ holography. While the above-mentioned  truncation is consistent at the level of the metric (in the sense that it is obtained by fixing some of the independent functions that parametrize our solution), it is not so at the level of the gauge fields, which are simply absent in \cite{Compere:2013bya}, but do not become zero on our restricted solution. In particular, we can link the appearance of the background-dependent $U(1)$ level in \cite{Compere:2013bya} to missing gauge field contributions to the conserved charges. The reason that, in \cite{Compere:2013bya}, these missing contributions do   not lead to inconsistent results is   the very restricted nature of the phase space variations considered.  Nonetheless, the price to pay for these restricted variations is an important loss of dynamics of the theory.  

Let us start by comparing the two sets of backgrounds. To obtain a match, we set $\bar{\mathcal{J}}=0$ in our analysis and  consider constant\footnote{Note that, according to our analysis in section \ref{section5:holography}, it is generally not consistent to  fully set $\bar{\mathcal{J}}=0$, since any fluctuation in $E_R \leftrightarrow \bar \L$ can change its zero mode if $\bar J_0$ is fixed. We will ignore such subtle issues here. 
% This, however, is not a problem here, since  $\bar{\L}$ is also held fixed. \textcolor{red}{We do actually vary it...}.
  Note also that we could, in principle, still obtain a match of the two metrics  for $\bar{\mathcal{J}}$ set to an arbitrary constant, by rescaling  $V$.  We do not consider this possibility.  }  $\bar{\mathcal{L}}$. 
Comparing $(2.10)$ in \cite{Compere:2013bya} with our \eqref{deformedmetricsec6}, we find that the two metrics agree if we identify 
\begin{align}
\Delta \leftrightarrow \bar{\mathcal{L}}\;, \hspace{1.5cm}\bar{L}(t^+)\leftrightarrow\mathcal{L}(U)\;, \hspace{1.5cm}\partial_+\bar{P}(t^+)\leftrightarrow -\lambda\mathcal{J}(U)
\end{align}
where we dropped an irrelevant factor of   $ \ell/4G$. %The factor of $\l$ could be absorbed into the definition of $\J$. 
%
% {\color{ForestGreen}(we keep $k$ for our level)} \textcolor{red}{What is $\l$ for them?} {\color{ForestGreen}($\lambda$ does not appear in their analysis, we could also absorb it in a redefinition of $\J$ and $\l \bar{\mathcal{J}}$, since all we do is a coordinate tr $V-\l\phi$ so we can absorb $\l$)}
Note, however, that the Chern-Simons gauge fields \eqref{paramgaugefields} do not vanish for these choices of parameters, whereas in the  analysis of \cite{Compere:2013bya} they are simply absent. %Of course, in our analysis it would be impossible to discard the Chern-Simons terms, as their variations are intimately linked to those of the metric, via the boundary conditions that define the theory.  %the gravitational contributions to the charges {\color{ForestGreen}are not integrable}  and only the total charge is well-defined; {\color{ForestGreen}(the CS fields are part of the definition of the theory via boundary conditions)} however, since \cite{Compere:2013bya} consider very restricted variations of only $\L$ and $\J$, this inconsistency can be ignored. 

The allowed transformations on the restricted phase space of \cite{Compere:2013bya} consists of  left affine and  left conformal transformations, as well as a constant right-moving  translation. %\textcolor{red}{Are our restricted diffeos the same as theirs?} {\color{ForestGreen}(if we just set $\bar{\J}=0$, we can do left affine, left conformal and some combination of right affine and right pseudoconformal such that $\delta\bar{\J}=0$. They additionally fix $\bar{\L}$, which kills right pseudoconformal except for constant ones and automatically kills right affine except for constant ones).}
 As already remarked in \cite{Bzowski:2018pcy}, since the Chern-Simons fields are absent in  \cite{Compere:2013bya}, the only piece of the left affine transformation that is visible in their analysis is the accompanying shift \eqref{accshift} of the right-moving coordinate.  
To understand how their calculation works from our perspective, we  separately write  below the contributions from the metric and from the gauge fields to our final expressions for the conserved charges $Q_f,P_{\eta}$. For the left-moving affine transformations we obtain, for $\bar{\J}=0$
\begin{align}
[\slash{\!\!\!\delta} P_{\eta}]_{metric} &=-\int_0^R d\sigma\frac{\lambda \eta(U)}{2}\bigg( 2\lambda \bar{\mathcal{L}}\delta\mathcal{J}-(1-\lambda\mathcal{J})\delta\bar{\mathcal{L}} \bigg)\\
[\slash{\!\!\!\delta} P_{\eta}]_{CSdiff}&=\int_0^R d\sigma\frac{\lambda^2\eta(U)}{2}\bar{\mathcal{L}}\delta\mathcal{J}\\
[\slash{\!\!\!\delta} P_{\eta}]_{CSgauge}&=\int_0^R d\sigma\frac{\eta(U)}{2}\bigg(-\lambda(1-\lambda\mathcal{J})\delta\bar{\mathcal{L}} +(2+\lambda^2\bar{\mathcal{L}})\delta\mathcal{J}\bigg)
\end{align}
where we split the contributions from the Chern-Simons gauge fields into the part associated to the diffeomorphisms and the part associated to the gauge transformations. %Note the sum of the three is the simple expression \eqref{resultleftaffineph}, which leads to the same $U(1)$ level as in the undeformed CFT. {\color{ForestGreen}(here the same comment applies as in the main text, we probably need to rescale, but it will be for sure the same $k$)} 
If  we only consider the metric contribution, we note that the charge variations are not integrable and we are forced to fix $\bar{\mathcal{L}}$ (or $\mathcal{J}$, but we do not consider this possibility), %\textcolor{red}{So, no issue with finiteness?} {\color{ForestGreen}(no, they are separately finite)}
which corresponds to keeping $\Delta$ fixed in \cite{Compere:2013bya}. This seems overly restrictive, though, and the authors contrive a way to make the charges integrable for $\Delta$ a variable constant. The resulting asymptotic symmetry algebra is a $U(1)$ Kac-Moody algebra with a $\Delta$ ($=\bar \L$) - dependent level.

If, on the other hand, we add in  the contribution from the gauge fields, the terms proportional to $\delta\bar{\mathcal{L}}$ and $\bar{\mathcal{L}} \d \J$  automatically cancel, so that there is no issue regarding integrability and we can freely vary $\bar{\mathcal{L}}$ in the phase space. 
%Moreover, the terms proportional to $\bar{\mathcal{L}} \d \J$ cancel and 
The final variation is given by the simple expression \eqref{resultleftaffineph}, which leads to a Kac-Moody algebra with the same $U(1)$ level as in the undeformed CFT.

 %not energy-dependent, namely we obtain a standard energy-independent $U(1)$ level. (I'd like to discuss the interpretation. Does this mean that there is no meaning regarding the energy-dependent level in their set-up because it can be traded for gauge fields? \textcolor{red}{yes, that's the point of this appendix})

%After setting $\delta\bar{\mathcal{L}}=0$, we have $[P_{\eta}]_{metric}=\frac{4\pi}{k}\lambda \bar{\mathcal{L}}\int \eta(U)(-\lambda\mathcal{J})$, up to some integration constant in phase space, which matches 2.16 {\color{ForestGreen}(rescalings)}, up to an overall factor of $\lambda$ {\color{ForestGreen}(this comes from the fact that they absorbed $\l$ in $\J$ in our language, so we should also absorb it in $\eta$ which shifts $\J$)}. Since the variation gives the algebra, it is clear that the level that we obtain in this case depends on $\bar{\mathcal{L}}$, namely it is energy-dependent. 

%In this case, the only non-trivial right-moving charge is $E_R$ and we obtain also a Virasoro-Kac-Moody algebra (left-moving in our conventions), spanned by the flowed versions of the charges $P_{\eta},Q_{f}$.

For  the left-moving conformal symmetries, the separate metric and Chern-Simons contributions to the associated conserved charges are, for $\bar{\J}=0$  
\begin{align}
[\slash{\!\!\!\delta} Q_f]_{metric} &=\frac{1}{8\pi G \ell}\int_0^R d\sigma f\bigg(\delta \mathcal{L}-2\l^2 \bar{\mathcal{L}}\J\delta\mathcal{J}+\lambda\J(1-\l\J)\delta\bar{\L}\bigg)\\
[\slash{\!\!\!\delta} Q_f]_{CSdiff}&=\frac{2\pi}{k}\int_0^R d\sigma f\J\bigg(-\l \kappa (1-\l\J)\delta\bar{\mathcal{L}} + 2(1+\kappa\l^2 \bar{\L})\delta\J \bigg)\\
[\slash{\!\!\!\delta} Q_f]_{CSgauge}&=\mathcal{O}(z^2)
\end{align}
where   $\kappa=\frac{k}{16\pi^2 \ell}$. Again,  the metric contribution is not integrable by itself, unless we keep $\bar{\mathcal{L}}$ fixed. Upon doing so, the expression agrees with equation $(2.15)$ in \cite{Compere:2013bya}, using the identifications that we specified. As before, the gauge fields contribute non-trivially,  and $\bar{\mathcal{L}}$ does not appear in our final result \eqref{resultleftconformal}. %The expression for the global $U(1)$ is trivially the same, assuming we ignore the fact that $\bar \L$ doesn't vary in order to perform the computation. \textcolor{red}{How should we say this?}

Finally, for a global right translation  generated by $\xi_R = - \p_V$, the non-zero contributions to the right-moving energy for $\bar \J=0$ are
\begin{align}
[\slash{\!\!\!\delta} E_R]_{metric}&=-\frac{1}{8\pi G \ell}\int_0^R d\sigma \bigg(2\l \bar{\L}\delta\J -(1-\l\J)\delta\bar{\L}\bigg)\\
[\slash{\!\!\!\delta}  E_R]_{CSdiff}&=\frac{1}{8\pi G \ell}\int_0^R d\sigma \l\bar{\L}\delta \J 
\end{align}
The sum is trivially integrable for $\bar \L, \J$ unrestricted and gives  $\frac{1}{8\pi G \ell}\bar{\L}(1-\l\J)$, which agrees with $\bar \L \p_\s v$ for $\bar \J=0$.

%. If we consider only the metric contribution we need to fix $\bar{\L}$ for integrability as before and we need a negative level in order to have positive energy. \textcolor{red}{Where does the level enter in this expression? Does the sign agree with CSS?}

The interpretation of this analysis is as follows. As the authors themselves emphasize, an essential working \emph{assumption} of \cite{Compere:2013bya} was that the semiclassical analysis of the conserved charges in pure $3d$ gravity with the given chiral boundary conditions does lift to a computation in a full theory of quantum gravity, which has a consistent holographic dual. Then, the semiclassical conserved charge analysis allows one to \emph{infer} certain properties of this putative holographic dual. First, integrability of the conserved charges computed with just the Einstein action requires that $\bar \L$ be fixed (or, at most a varying constant), which is a very severe restriction on the dynamics of the theory:  in a CFT, this would correspond to requiring that all the non-zero Fourier modes of the right-moving stress tensor  vanish. In particular, 
 states of the form $\O (U, V) | 0\rangle $, which would yield  a non-trivial profile for the expectation value of the momentum-space right-moving stress tensor, would not be allowed for any operator $\O$ that has a non-trivial right-moving scaling. How to include dynamical matter fields in this phase space is thus not entirely clear.  For these reasons, the construction of the phase space put forth in \cite{Compere:2013bya} appears somewhat unnatural.

% If the resulting restricted theory is just one of conserved currents, it is not even clear whether the Bekenstein-Hawking entropy of its black holes  has a statistical interpretation \cite{}. 
 
 A second prediction of the semiclassical analysis of  \cite{Compere:2013bya} is the negative $U(1)$ level for the black hole states of most interest, which violates unitarity. While this could be viewed as a feature with potential applications to the physics of ergoregions, to our knowledge a fully consistent interpretation of this prediction has yet to emerge. 

The way that our analysis (whose departure point  are the \emph{independently}-defined $J\bar T$ - deformed CFTs, from which the dual bulk action and boundary conditions are  then  \emph{derived}) solves both problems  is  that  the low-energy effective action for  the  gravitational theory  contains other contributions to the conserved charges besides the metric, namely from the  Chern-Simons gauge fields. Their
%
 %\emph{independent} QFT definition. \textcolor{red}{Understandable?} Using this field-theoretical definition, we are able to the dual gravitational low-energy action and boundary conditions for the dual bulk fields.  In particular, the action must contain 
%
%Our analysis (see also \cite{Bzowski:2018pcy}) suggests a very simple solution to both problems: that the low-energy description of a gravitational theory allowing for such  non-standard boundary conditions must contain other contributions to the conserved charges besides the metric, for example from a
% Chern-Simons gauge fields whose
  inclusion%, from the point of view of the analysis of \cite{Compere:2013bya}
  : i) permits  relaxing the boundary conditions, so that arbitrary matter field excitations (with reasonable near-boundary behaviour)  are allowed, thus solving the dynamics problem, and ii)  gives additional contributions to the charges, so that the negative background-dependent $U(1)$ level never shows up.  This points towards the conclusion that  the framework  of pure $3d$ Einstein gravity  is simply not sufficient
  to properly model a consistent phase space 
 for  the  non-standard boundary conditions  put forth in \cite{Compere:2013bya}  
  % 
%  It is not unreasonable to conclude that, in presence of such non-standard boundary conditions,  the inclusion of other fields in the low-energy action besides the metric is necessary in order to have a 
  %
  that \emph{a priori} allows for unrestricted dynamics of the theory. %, and which also happens to resolve some of the unitarity issues. 
 The dual field-theoretical interpretation of this statement would be that, in a theory with no left-moving $U(1)$ currents, there does not exist a well-defined double-trace irrelevant operator constructed from just the stress tensor that could holographically drive the deformation of the boundary conditions for the dual metric from Dirichlet to CSS. 
 
 % It thus appears  that the result of the ASG analysis is highly sensitive  the choice of low-energy effective action 
 
% , unlike in AdS/CFT with standard boundary conditions, 
 
 %reasonable to propose that, whenever a gravitational theory allows for non-standard, CSS-type boundary conditions in the metric sector,  must contain other contributions to the conserved charges besides the metric

 Note that our proposal immediately  implies that   the asymptotic symmetry algebra is no longer that of a warped CFT, with its characteristic enhancement of right translations to a left-moving $U(1)$ Kac-Moody algebra,  but more likely that of a $J\bar T$ - deformed CFT, where the $U(1)_L$ Kac-Moody symmetry is the $J\bar T$ flow  of the  affine symmetry already present in the undeformed CFT. Note this reinterpretation of how the right-moving energy fits inside the symmetry algebra completely changes how the asymptotic density of states is to be derived from a modified notion of modular invariance, from \cite{Detournay:2012pc} to \cite{Aharony:2018ics}. It also implies that the dual theory is non-local once right-moving excitations are allowed, even though this fact may not be fully obvious from the behaviour of the holographic currents. All these   features, with their rather different implications and interpretation,  follow from a re-evaluation of what the appropriate boundary conditions for the system should be, and from considering a low-energy effective action that is actually capable of accomodating such  boundary conditions. %This point of view suggests that a careful definition of what it means to have a consistent phase space is necessary 
%before performing%/ drawing conclusions from 
%any asymptotic symmetry group computation. 
 
Of course, our argument above does not demonstrate that the boundary  conditions put forth in \cite{Compere:2013bya}  are inconsistent. Rather, we simply point out that certain problems and features associated with them - namely, the dynamical coupling to matter problem and the negative Kac-Moody level - disappear upon a very simple change of framework.  The new  dual interpretation is within a class of field theories that are significantly better defined and understood than warped CFTs, for which the only known concrete examples are free  \cite{Hofman:2014loa}. The ultimate test whether a given set of boundary conditions in a low-energy effective field theory - assumed, as usual, to admit a consistent UV completion - are correct or not is whether they match the predictions from an independently-defined and consistent type of dual field theory, which would likely impose many more constraints on the gravitational phase space than are currently known.

\end{document}